\documentclass{aastex}
\usepackage{spr-astr-addons}
\usepackage{url}
\usepackage{graphicx}
\usepackage{longtable}
\usepackage{comment}
\usepackage{natbib}
\usepackage{rotating}
\usepackage{threeparttable}
\usepackage{wasysym}
\usepackage{txfonts}
\usepackage{multirow}
\usepackage{appendix}
\usepackage{enumitem}

\RequirePackage{color}

\newcommand{\emaila}{harold.penaherazo@unito.it}

\newcommand{\mywidth}{9.0cm}

\usepackage{aas_macros}
\usepackage{lscape}

\begin{document}

\title{Optical Spectroscopic Observations of Gamma-Ray Blazar Candidates. VII. \\ Follow-up Campaign in the Southern Hemisphere}
\shorttitle{Southern}
\shortauthors{SOMEONE et al.}

\author{
Pe\~na-Herazo, H. A.\altaffilmark{1,2,3}, Marchesini, E. J.\altaffilmark{1,2,4,5,6}, \'Alvarez Crespo, N.\altaffilmark{1,2}, Ricci, F. \altaffilmark{7}, Massaro, F. \altaffilmark{1,2}, Chavushyan, V. \altaffilmark{3}, Landoni, M. \altaffilmark{8}, Strader, J. \altaffilmark{9}, Chomiuk, L. \altaffilmark{9}, Cheung, C. C. \altaffilmark{10}, Masetti, N. \altaffilmark{6,11}, Jim\'enez-Bail\'on, E. \altaffilmark{12}, D'Abrusco, R. \altaffilmark{14}, Paggi, A. \altaffilmark{13}, Milisavljevic, D \altaffilmark{13},  La Franca, F. \altaffilmark{7},  Smith, H. A. \altaffilmark{13}, Tosti, G. \altaffilmark{15}
} 
\email{\emaila}

\altaffiltext{1}{Dipartimento di Fisica, Universit\`a degli Studi di Torino, via Pietro Giuria 1, I-10125 Torino, Italy}
\altaffiltext{2}{Istituto Nazionale di Fisica Nucleare, Sezione di Torino, I-10125 Torino, Italy}
\altaffiltext{3}{Instituto Nacional de Astrof\'{i}sica, \'Optica y Electr\'onica, Apartado Postal 51-216, 72840 Puebla, M\'exico}
\altaffiltext{4}{Facultad de Ciencias Astron\'{o}micas y Geof\'{i}­sicas, Universidad Nacional de La Plata, Paseo del Bosque, B1900FWA, La Plata, Argentina}
\altaffiltext{5}{Instituto de Astrof\'{i}sica de La Plata, CONICET-UNLP, CCT La Plata, Paseo del Bosque S/N, B1900FWA, La Plata, Argentina}
\altaffiltext{6}{INAF - Istituto di Astrofisica Spaziale e Fisica Cosmica di Bologna, via Gobetti 101, 40129, Bologna, Italy}
\altaffiltext{7}{Dipartimento di Matematica e Fisica, Universit\`a Roma Tre, via della Vasca Navale 84, I-00146, Roma, Italy}
\altaffiltext{8}{INAF-Osservatorio Astronomico di Brera, Via Emilio Bianchi 46, I-23807 Merate, Italy}
\altaffiltext{9}{Center for Data Intensive and Time Domain Astronomy, Department of Physics and Astronomy, Michigan State University, East Lansing, MI 48824, USA}
\altaffiltext{10}{Space Science Division, Naval Research Laboratory, Washington, DC 20375-5352, USA}
\altaffiltext{11}{Departamento de Ciencias F\'{\i}sicas, Universidad Andr\'es Bello, Fern\'andez Concha 700, Las Condes, Santiago, Chile}
\altaffiltext{12}{Instituto de Astronom\'{\i}a, Universidad Nacional Aut\'onoma de M\'exico, Apdo. Postal 877, Ensenada, 22800 Baja California, M\'exico}
\altaffiltext{13}{Harvard - Smithsonian Center for Astrophysics, 60 Garden Street, Cambridge, MA 02138, USA}
\altaffiltext{14}{Department of Physical Sciences, University of Napoli Federico II, via Cinthia 9, 80126 Napoli, Italy}
\altaffiltext{15}{Dipartimento di Fisica, Universit\`a degli Studi di Perugia, I-06123 Perugia, Italy}

\begin{abstract}
Searching for low energy counterparts of $\gamma$-rays sources is one of the major challenges in modern $\gamma$-ray astronomy. In the third Fermi source catalog about 30 $\%$ of detected sources are unidentified/unassociated Gamma-ray Sources (UGSs). We recently started an optical spectroscopic follow up campaign to confirm the blazar-like nature of candidates counterparts of UGSs.
Here we report the spectra of \textbf{61} targets collected with the Southern Astrophysical Research Telescope (SOAR) between 2014 and the 2017. Our sample includes \textbf{33} potential counterparts of UGSs, selected on the basis of WISE colors, and \textbf{27} blazar candidates of uncertain type associated with gamma-ray sources of the last release of the Fermi catalog.
We confirm the BZB nature of \textbf{20} sources lying within the positional uncertainty region of the UGSs. All the observed BCUs show blazar-like spectra, classified as 2 BZQs and \textbf{25} BZBs, for which we obtained 6 redshift estimates. Within the BCUs observations we report the redshift estimate for the BZB associated with, 3FGL J1106.4-3643 that is the second most distant BL Lac known to date, at $z \geq 1.084$.
\end{abstract}


\section{Introduction}
\label{sec:intro}
Since the first release of the source catalog based on observations of the Energetic Gamma Ray Experiment Telescope (EGRET) (\citeauthor{fichtel94}~\citeyear{fichtel94}), the nature of unidentified/unassociated gamma-ray sources (UGS) was under debate. With greater effective area and better angular resolution than previous $\gamma$-ray satellites, the Large Area Telescope (LAT), onboard of the Fermi Gamma-ray Space Telescope (\citeauthor{atwood09}~\citeyear{atwood09}), increased the number of detected sources by an order of magnitude, revolutionizing our understanding of the gamma-ray sky in the energy range of tens of MeV to hundreds of GeV, and significantly improving the association of $\gamma$-ray sources with their low energy counterparts (see e.g. \citeauthor{review}~\citeyear{review} for a recent review).\\

However, in the first $Fermi$-LAT source catalog (1FGL) about $ 40 \, \%$ of sources were still unassociated (\citeauthor{abdo10a}~\citeyear{abdo10a}). Then the UGS population in the Fermi-LAT second source catalog (2FGL) decreased to about $ 30 \,  \%$ \citep{nolan12} and remained almost constant in the Fermi-LAT third source catalog (3FGL) \citep{acero15}. Even if the number of UGSs is still decreasing there is still a significant fraction of them that could hide new potential discoveries (\citeauthor{rwbls}~\citeyear{rwbls}).\\


Blazars show high flux and spectral variability from optical to $\gamma$-rays (in time scales from week to minutes) coupled with superluminal motions and high bolometric luminosities (up to $10^{46}-10^{47} $ erg s$^{-1}$) and with an emission dominated by non-thermal process over the entire electromagnetic spectrum \citep{urry95}. The spectral energy distribution of blazars mainly shows two bumps, the first one peaking at low frequencies between the infrared and the ultraviolet range and the second one dominating in the $\gamma$-rays \citep{abdo10b,fossati98,giommi95}. The well entertained unification scenario of active galaxies explains the observed characteristics of blazars as AGN with relativistic jets pointing at a small angle with respect to our line of sight \citep{blandford78,urry95}. \\

Blazars are distinguished on the basis of their optical spectra in two classes: BL Lacs, objects having emission lines with rest frame equivalent width $EW\leqslant 5 \, \AA$ \citep{stickel91}, and Flat Spectrum Radio Quasars (FSRQs), showing optical features as in normal quasar spectra. Here we follow the Roma-BZCAT nomenclature, where BL Lac objects are label as BZB and FSRQs are labelled as BZQ \citep{massaro09,massaro15b}. It is worth mentioning that emission and/or absorption lines in BL Lacs spectra could be hidden due to their flux variations in the optical continuum in blazars, making their redshift estimate challenging \citep{refined}.\\

It is worth highlighting that blazars are the dominant class of sources in the $\gamma$-ray extragalactic sky, being about $85 \, \%$ of the associated sources in 3FGL \citep{acero15}, therefore we expect a significant fraction of UGSs at high Galactic latitudes ($\left|b \right| > 30 ^{\circ}$) to be blazars. Thus, in recent years, several methods and follow up observations were developed and used to search for blazar-like sources that, lying within the positional uncertainty region of UGSs, could be their low-energy counterparts \citep{archival,blarch,refined}. \\


Several methods were developed to find blazar-like potential counterparts of UGSs. For example, at radio frequencies, searching for compact radio sources (\citeauthor{petrov13} \citeyear{petrov13}; \citeauthor{schinzel15} \citeyear{schinzel15}) or objects with flat radio spectra \citep{crates} or even at low radio frequencies (i.e., below 1 GHz \citep{ugs3lowfrq,nori14,giroletti16}), or using follow up observations at high frequencies in X-rays (\citeauthor{paggi13}\citeyear{paggi13}; \citeauthor{takeuchi13}\citeyear{takeuchi13}; \citeauthor{acero13}\citeyear{acero13}). In addition, at IR frequencies, using WISE colors \citeauthor{massaro12c} \citeyearpar{massaro12c} developed a method to identify $\gamma$-ray blazar candidates that could find potential counterparts of UGSs thanks to an infrared $\gamma$-ray connection \citep{gir,paper1}. Several catalogs of UGSs potential counterparts were built on the basis of this IR analysis \citep{massaro12a}.\\

However none of these methods offer a conclusive way to clarify the nature of the selected potential counterparts, unless there is an optical spectroscopic confirmation. Thus, in 2013 we started an optical follow up campaign to obtain the spectra of selected blazar like counterparts for UGSs. In addition to the UGSs, we also observed Blazar Candidates of Uncertain Type (BCUs), as defined and associated in the Fermi catalogs i.e. Fermi sources with an assigned counterpart showing a multifrequency behavior similar to blazars but lacking a spectroscopic classification \citep{acero15,ackermann15a}. The BCU definition in the 3FGL and in the third LAT catalog of AGNs (3LAT) corresponds to the old class of Active Galaxies of Uncertain type (AGUs) listed in the previous 1FGL and 2FGL Fermi catalogs \citep{optbcu}. To date we identified 223 blazars of which 173 are BZBs, additionally we have measured 49 redshifts and found 2 BZBs with $z>1$, see \citeauthor{quest16} \citeyearpar{quest16} for a recent review. \\

In this paper we present the last results of the spectroscopic observations of our campaign, focused in the southern hemisphere and acquired with the Southern Astrophysical Research Telescope (SOAR) between 2014-2017. The paper is organised as follows. In section~\ref{sec:sample} we present the sample description, while in section~\ref{sec:obs} we describe the observations and the data reduction procedure. Results of our spectroscopic observations are presented in section~\ref{sec:results}. Finally section~\ref{sec:summary} is devoted to the summary and conclusions.\\

Throughout this work we used cgs units unless stated otherwise and we regard as flat spectrum those sources with spectral index $\alpha<0.5$, where $\alpha$ is defined as the flux density $S_{\nu} \propto \nu^{-\alpha}$.

\section{Sample Description}
\label{sec:sample}

The main aim of our campaign is to clarify the nature of blazar-like sources lying within the positional uncertainty region of UGSs and BCUs listed in the latest release of the Fermi catalog via optical spectroscopic observations. Sources observed during our follow up campaign were selected from the catalogs of WISE potential counterparts \citep{ugs1met,wibrals1} on the basis of their visibility during the available nights and with an airmass lower than 1.5 \citep{opt2,opt3}.\\

The sample presented in the current work consists of \textbf{61} sources grouped in three categories: UGSs, BCUs and known blazars, as described below.\\

\begin{enumerate}
\item \textbf{Thirty-three} of our sources are blazar-like potential counterparts of UGSs selected with the WISE colors, all of them are listed in the 3FGL catalog with the exception of 1FGL J1129.2-0528.
\item Nearly half of our sample (\textbf{27} out of 61) are BCUs, with counterparts in the X-rays and/or flat radio spectrum, all of them are included in the 3FGL catalog except the source 2FGL J1922.6-7454. 
\item We pointed an additional source, the BZCAT object \textbf{5BZB J0814-1012} (3FGL J0814.1-1012), which is associated with the radio source NVSS J081411-101208 in 3FGL. This source already observed and classified by \citeauthor{crespo16b} \citeyearpar{crespo16b} was pointed because being a BL Lac we tried to get a redshift estimate hoping to observe it during a low flux state. \\
\end{enumerate}

It is worth mentioning that 37 of our selected targets have Galactic latitude $\left|b \right| < 30 ^{\circ}$, and 7 of them have $\left|b \right| < 10 ^{\circ}$. All the sources are listed in Table~\ref{tab:main}, including its Fermi name, counterpart name and the observational log. For those we were able to estimate its redshift we reported it and for those we were not able we marked them with a quotation mark. Additionally we report its classification and finally show multifrequency notes for each objects to point up in their broad band detections.

\section{Observations and Data Reduction}
\label{sec:obs}

The strategy for our follow up campaign consists in observing small samples of potential counterparts each observing run to minimize the impact on telescope schedules. A detailed explanation of the observing strategy used during the campaign is presented in \citeauthor{quest16} \citeyearpar{quest16}.\\

All the spectra reported in this work were acquired at the Southern Astrophysical Research Telescope (SOAR) 4.1 m telescope, at Cerro Pach\'on, Chile. We performed both visitor and remote mode observations. We used the Goodman High Throughput spectrograph \citep{clemens04} to acquire the spectra in single slit mode, with a slit width of $1\farcs0$ and the 400 l mm$^{-1}$ grating, obtaining a dispersion of $\sim 2 \, \AA $ pixel$^{-1}$. The observations were taken in a time span between December 2014 up to January 2017 as shown in the Table~\ref{tab:main}.\\

\begin{figure*}{}
\begin{center}$
\begin{array}{cc}
\includegraphics[width=\mywidth,angle=0]{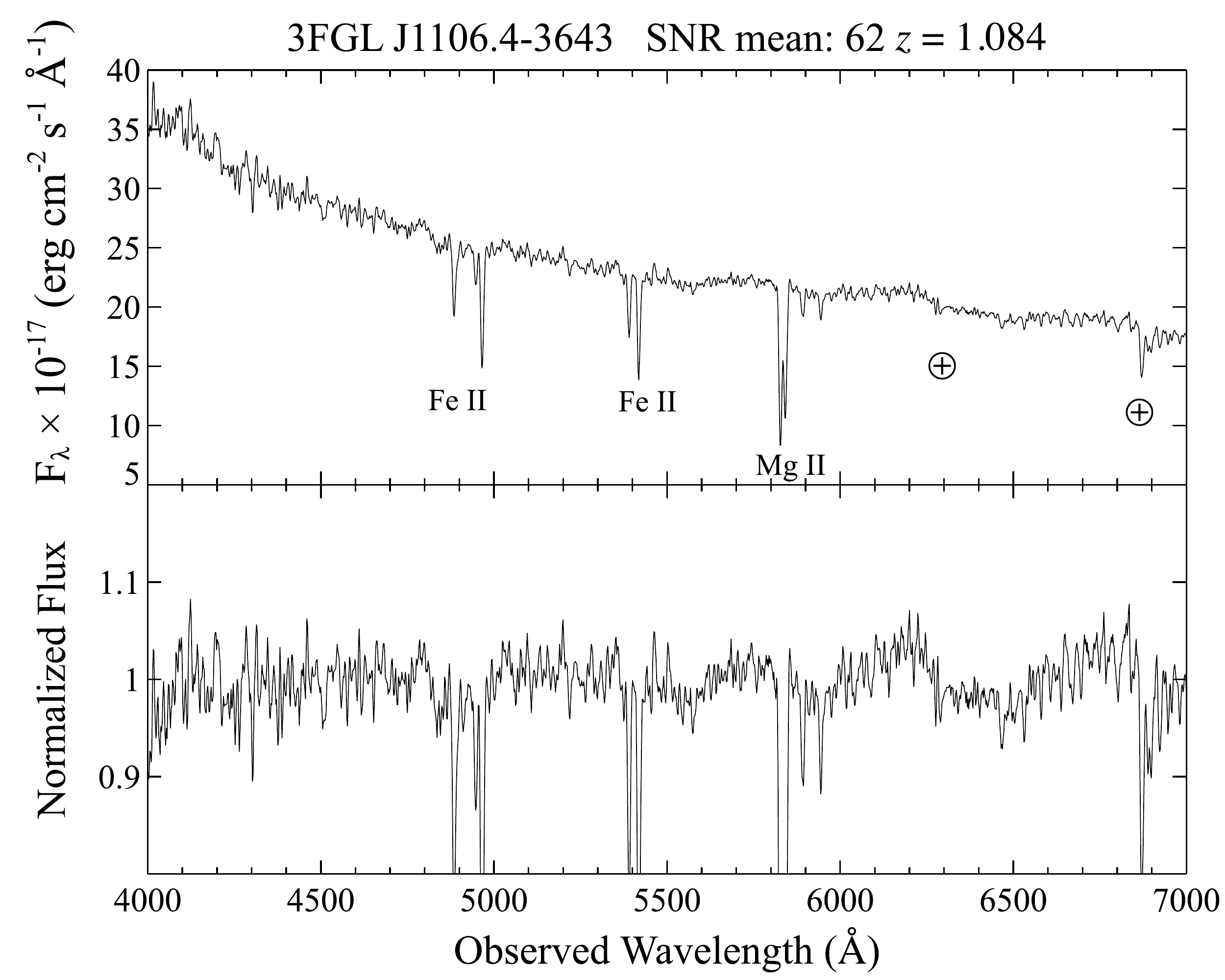} &
\includegraphics[trim=4cm 0cm 4cm 0cm, clip=true, width=7cm,angle=0]{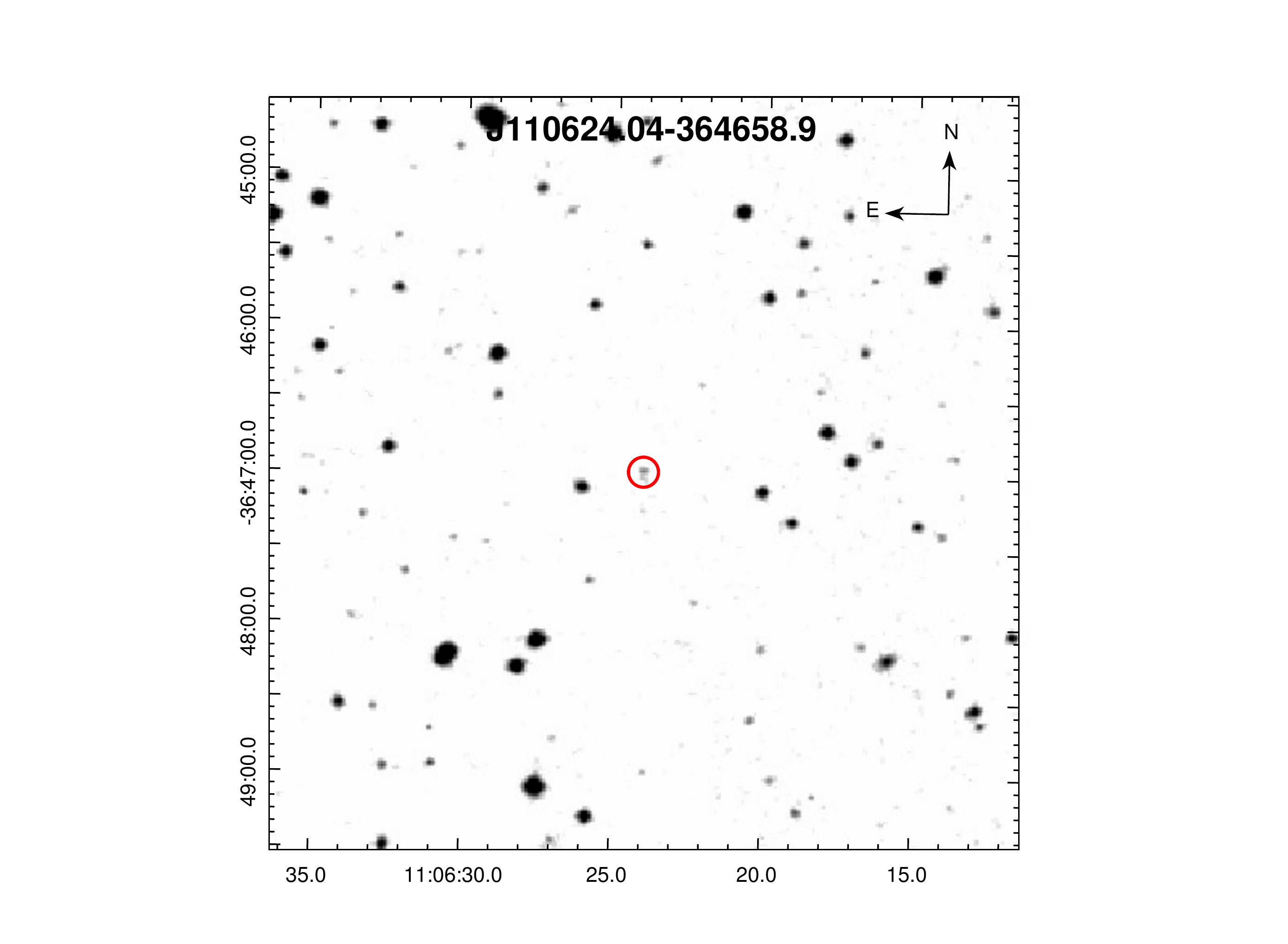} \\
\end{array}$
\end{center}
\caption{(Left panel) Optical spectrum of  WISE J110624.04-364658.9 associated with 3FGL J1106.4-3643, in the upper part it is shown the Signal-to-Noise Ratio of the spectrum. (Right panel) The finding chart ( $5'\times 5'$ ) retrieved from the Digital Sky Survey highlighting the location of the counterpart: WISE J110624.04-364658.9 (red circle).}
\label{fig:J1106}
\end{figure*}

\begin{figure*}{}
\begin{center}$
\begin{array}{cc}
\includegraphics[width=\mywidth,angle=0]{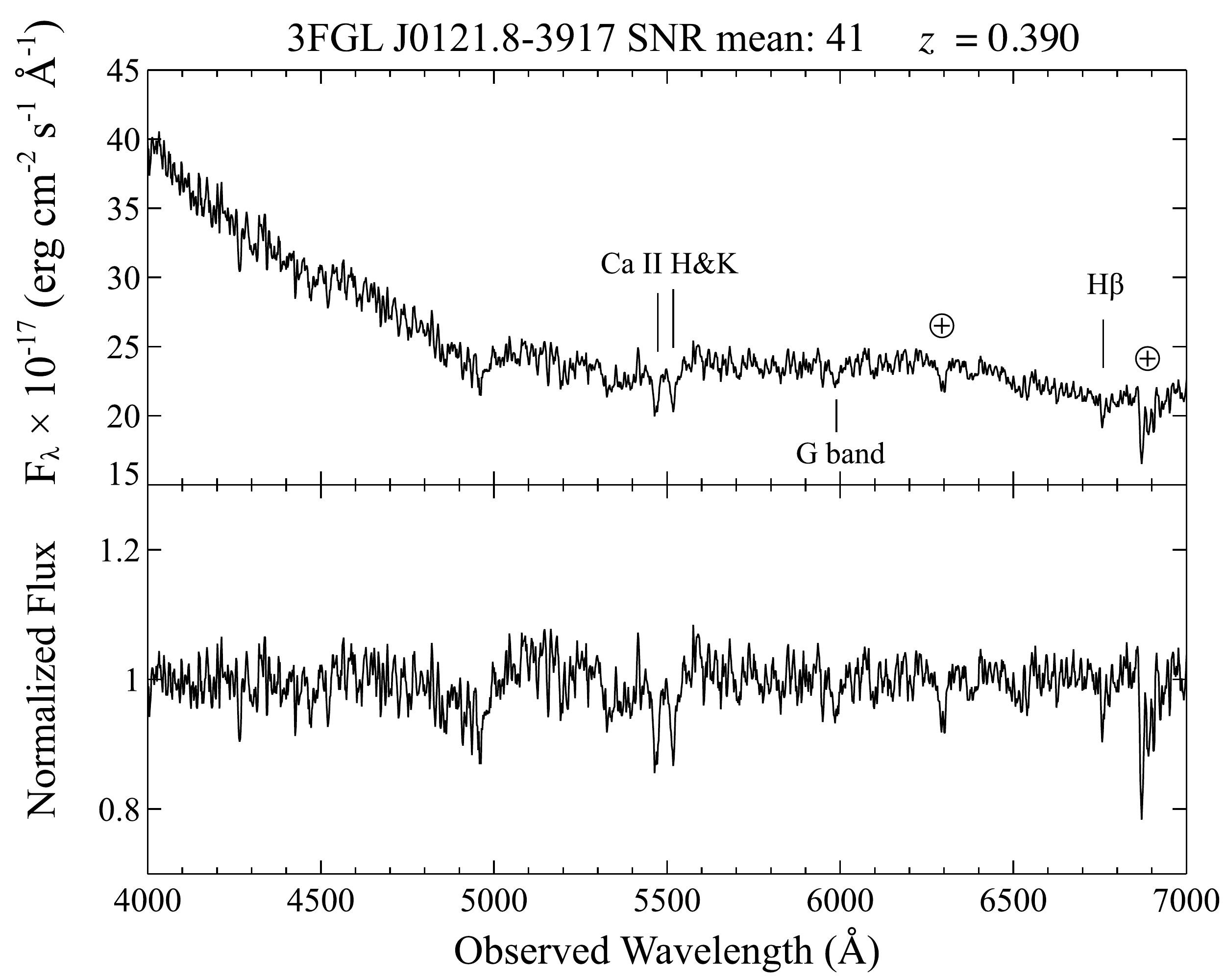} &
\includegraphics[trim=4cm 0cm 4cm 0cm, clip=true, width=7cm,angle=0]{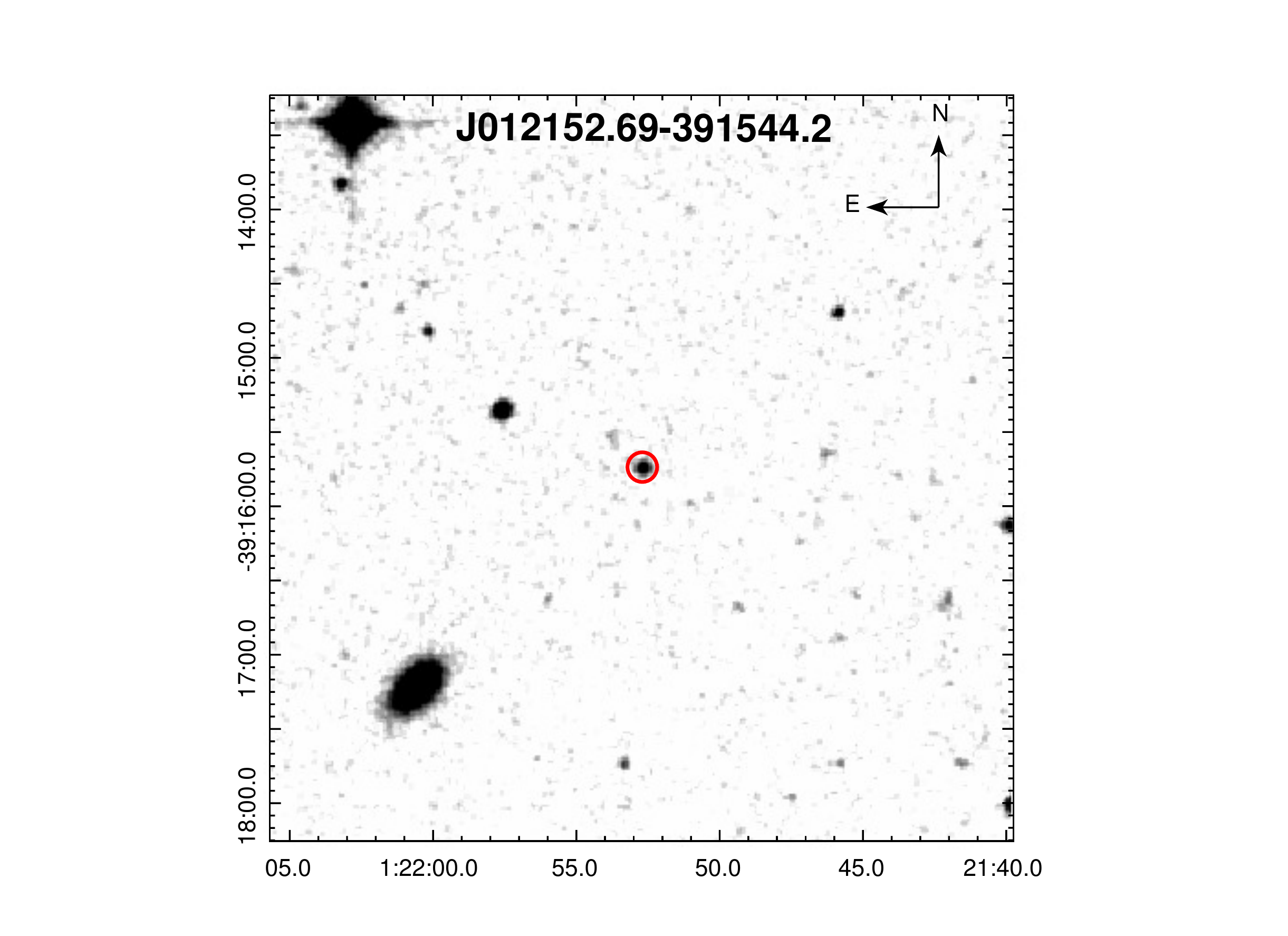} \\
\end{array}$
\end{center}
\caption{(Left panel) Optical spectrum of  WISE J012152.69-391544.2 associated with 3FGL J0121.8-3917. Signal-to-noise ratio is reported in the Figure. (Right panel) The finding chart ( $5'\times 5'$ ) retrieved from the Digital Sky Survey highlighting the location of the potential source: WISE J012152.69-391544.2 (red circle).}
\label{fig:J0121}
\end{figure*}

\begin{figure*}{}
\begin{center}$
\begin{array}{cc}
\includegraphics[width=\mywidth,angle=0]{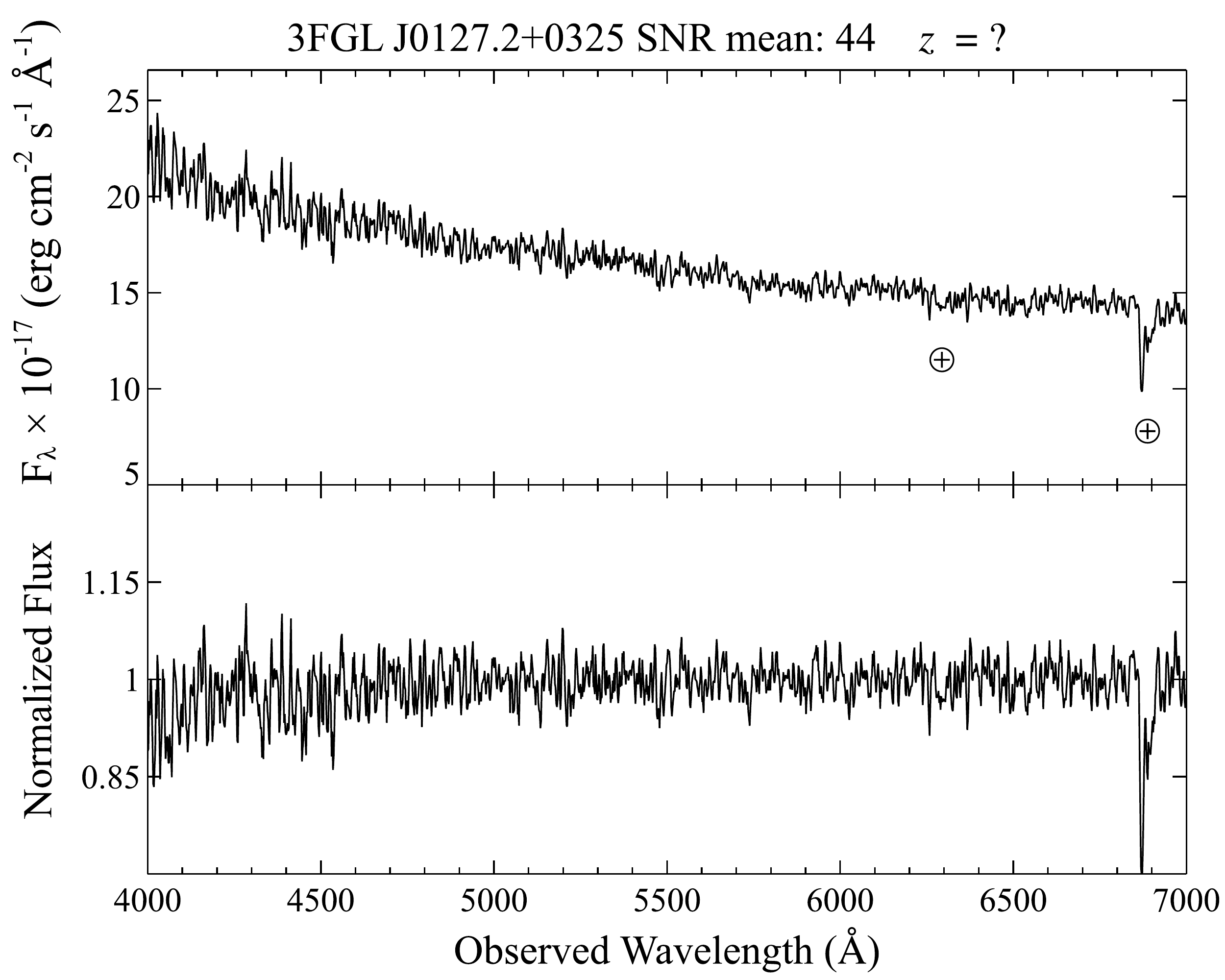} &
\includegraphics[trim=4cm 0cm 4cm 0cm, clip=true, width=7cm,angle=0]{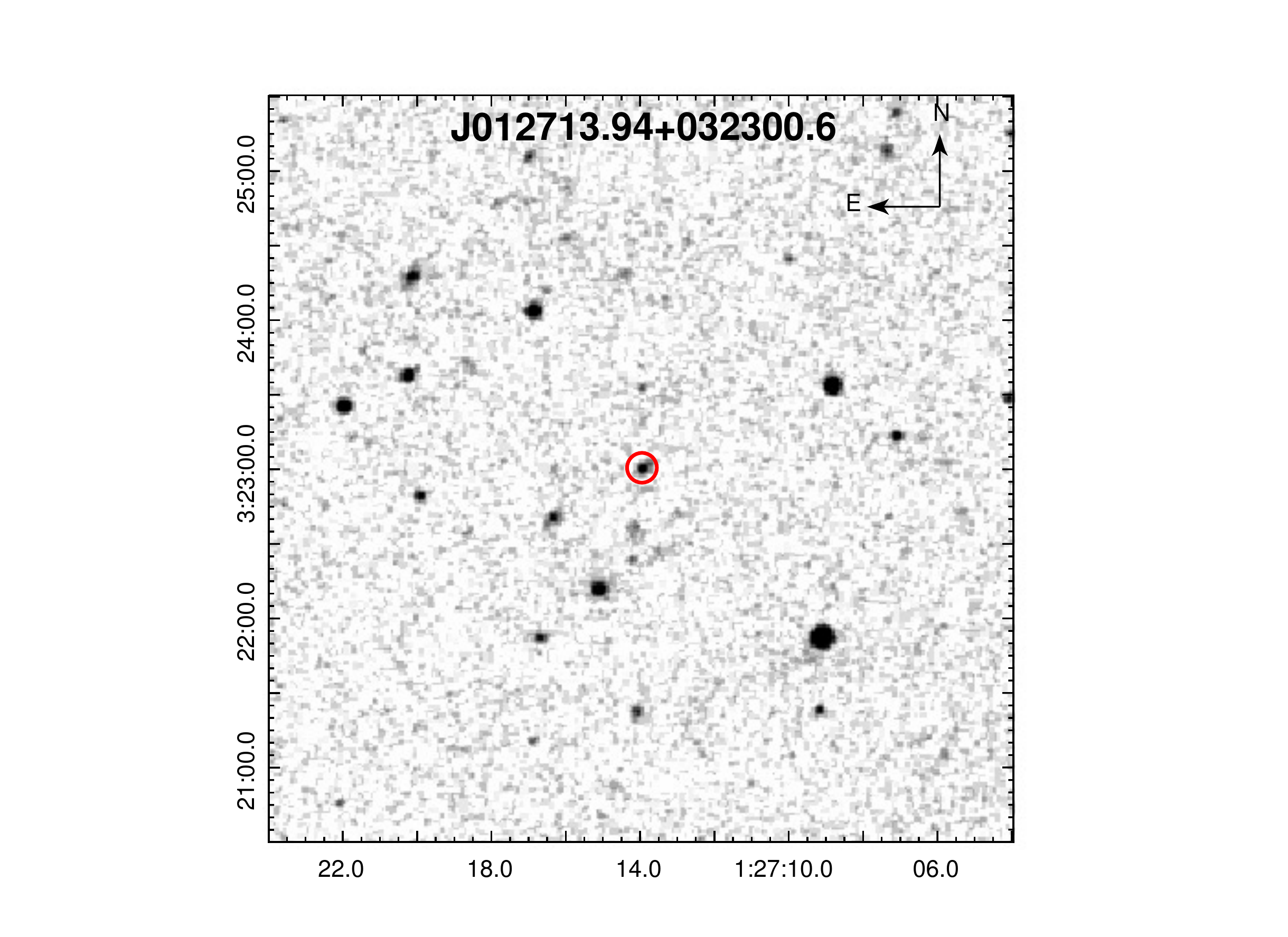} \\
\end{array}$
\end{center}
\caption{(Left panel) Optical spectrum of  WISE J012713.94+032300.6 associated with 3FGL J0127.2+0325, in the upper part it is shown the Signal-to-Noise Ratio of the spectrum. (Right panel) The finding chart ( $5'\times 5'$ ) retrieved from the Digital Sky Survey highlighting the location of the counterpart: WISE J012713.94+032300.6 (red circle).}
\label{fig:J0127}
\end{figure*}

\begin{figure*}{}
\begin{center}$
\begin{array}{cc}
\includegraphics[width=\mywidth,angle=0]{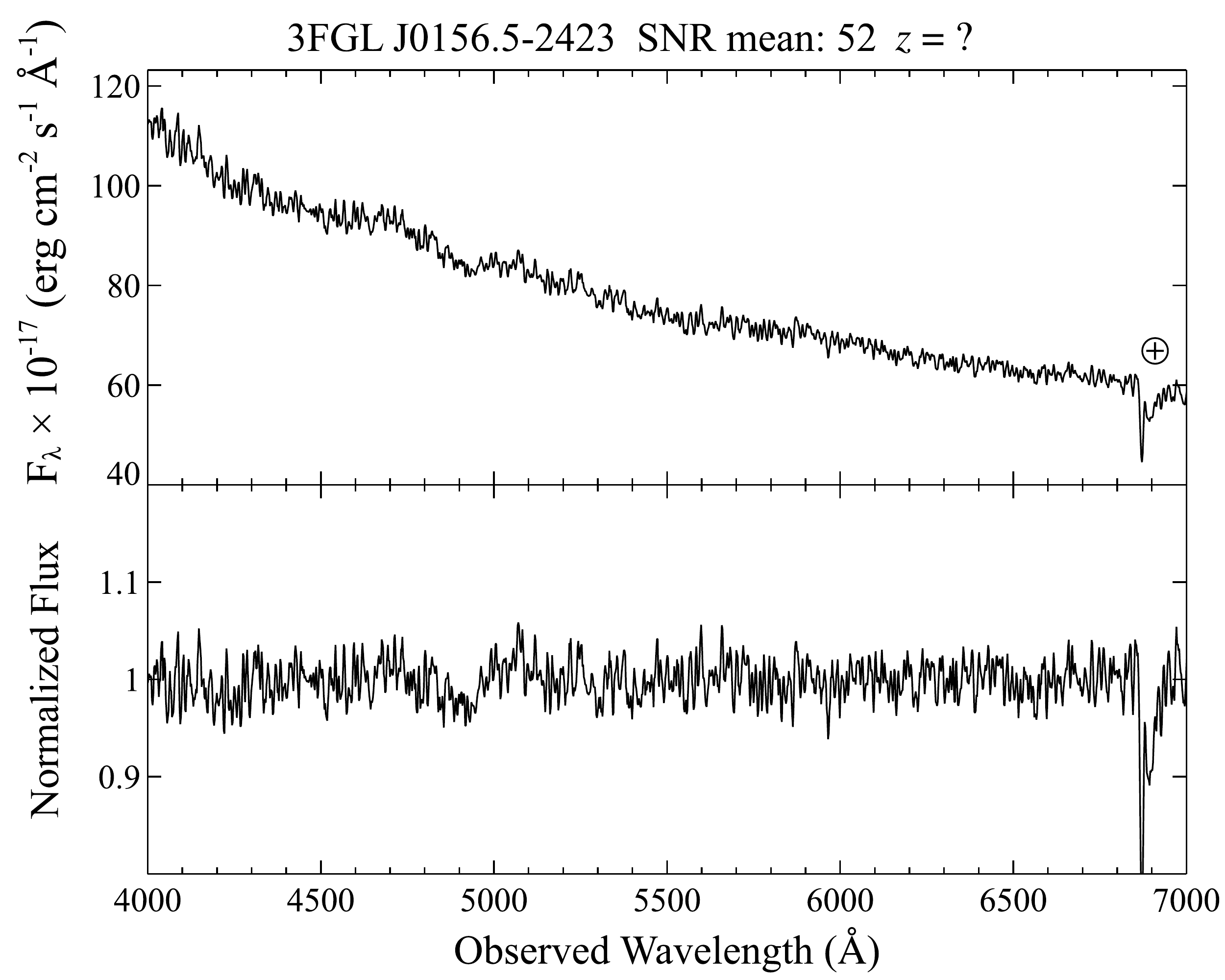} &
\includegraphics[trim=4cm 0cm 4cm 0cm, clip=true, width=7cm,angle=0]{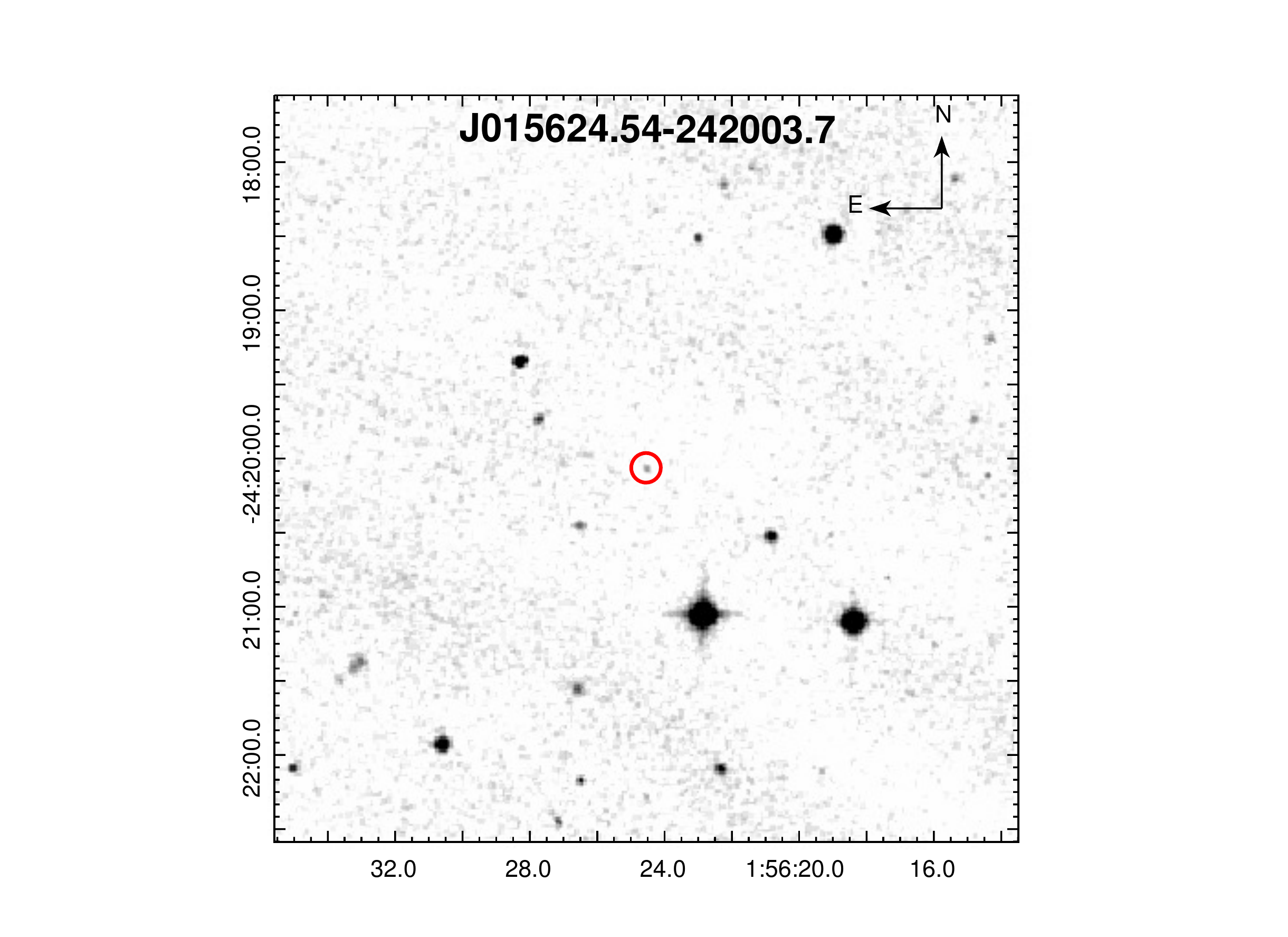} \\
\end{array}$
\end{center}
\caption{(Left panel) Optical spectrum of  WISE J015624.54-242003.7 associated with 3FGL J0156.5-2423, in the upper part it is shown the Signal-to-Noise Ratio of the spectrum. (Right panel) The finding chart ( $5'\times 5'$ ) retrieved from the Digital Sky Survey highlighting the location of the counterpart: WISE J015624.54-242003.7 (red circle).}
\label{fig:J0156}
\end{figure*}

\begin{figure*}{}
\begin{center}$
\begin{array}{cc}
\includegraphics[width=\mywidth,angle=0]{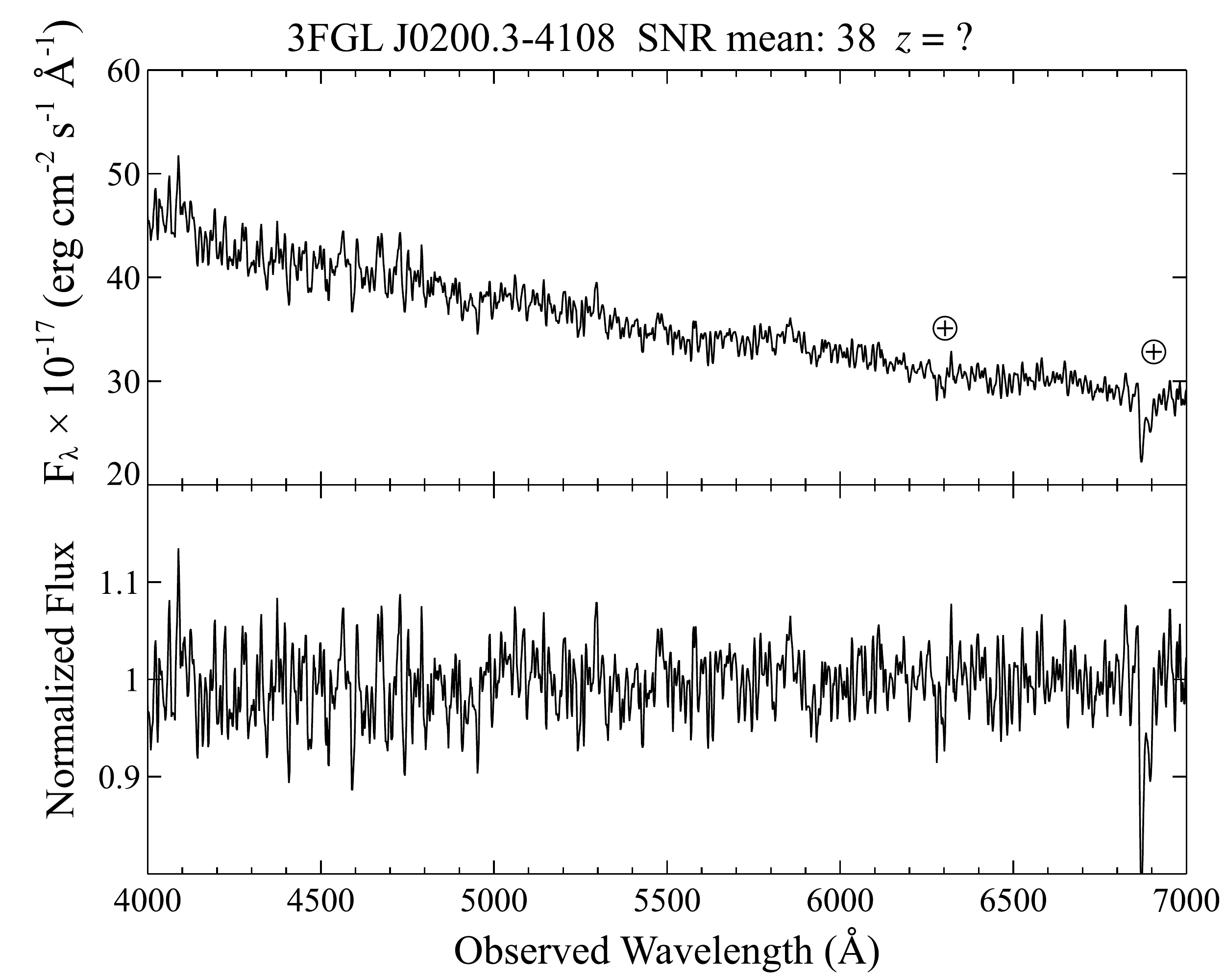} &
\includegraphics[trim=4cm 0cm 4cm 0cm, clip=true, width=7cm,angle=0]{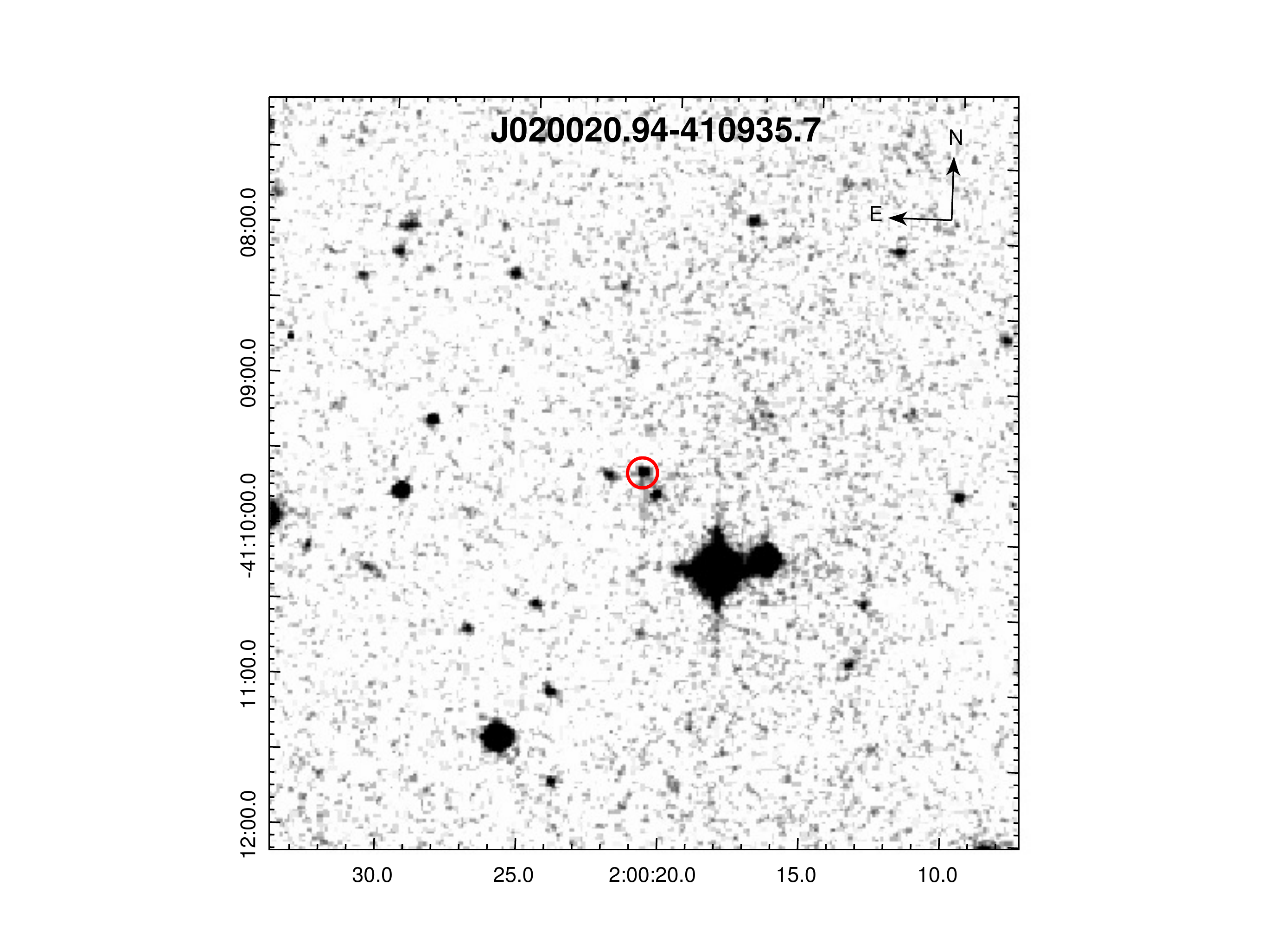} \\
\end{array}$
\end{center}
\caption{(Left panel) Optical spectrum of  WISE J020020.94-410935.7 associated with 3FGL J0200.3-4108. Signal-to-noise ratio is reported in the Figure. (Right panel) The finding chart ( $5'\times 5'$ ) retrieved from the Digital Sky Survey highlighting the location of the potential source: WISE J020020.94-410935.7 (red circle).}
\label{fig:J0200}
\end{figure*}

\begin{figure*}{}
\begin{center}$
\begin{array}{cc}
\includegraphics[width=\mywidth,angle=0]{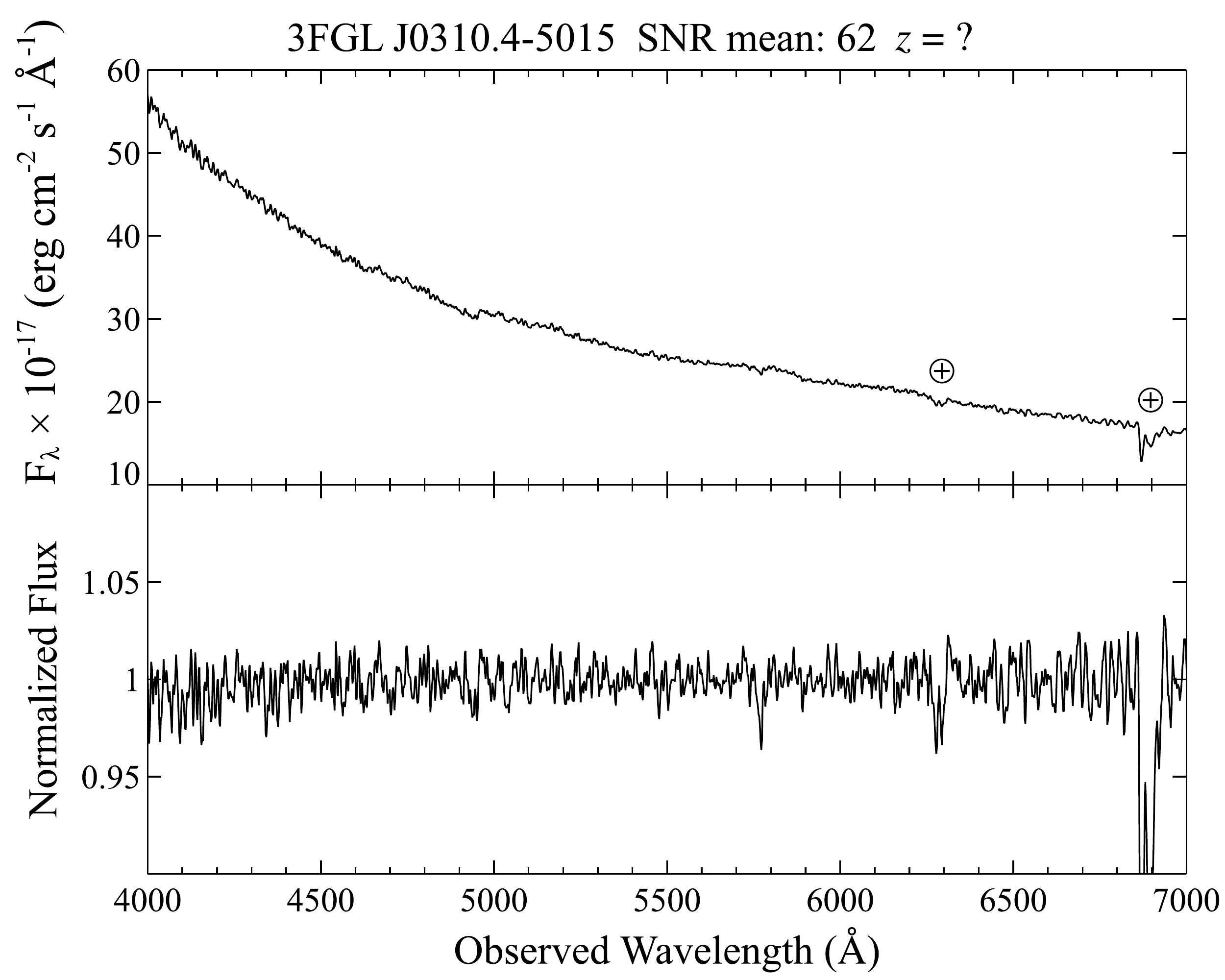} &
\includegraphics[trim=4cm 0cm 4cm 0cm, clip=true, width=7cm,angle=0]{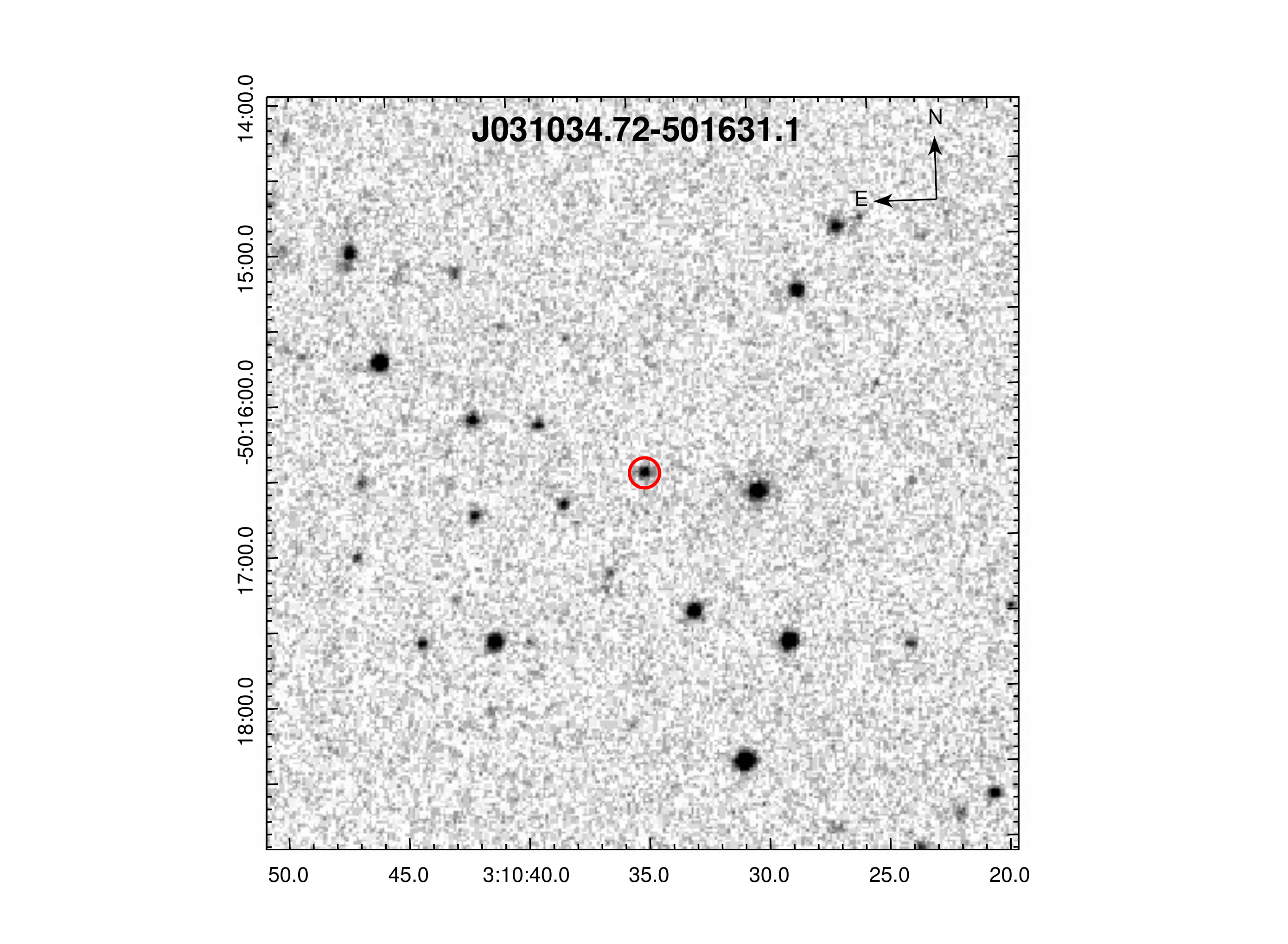} \\
\end{array}$
\end{center}
\caption{(Left panel) Optical spectrum of WISE J031034.72-501631.1, associated with 3FGL J0310.4-5015. Signal-to-noise ratio is reported in the Figure. (Right panel) The finding chart ( $5'\times 5'$ ) retrieved from the Digital Sky Survey highlighting the location of the potential source: WISE J031034.72-501631.1 (red circle).}
\label{fig:J0310}
\end{figure*}

\begin{figure*}{}
\begin{center}$
\begin{array}{cc}
\includegraphics[width=\mywidth,angle=0]{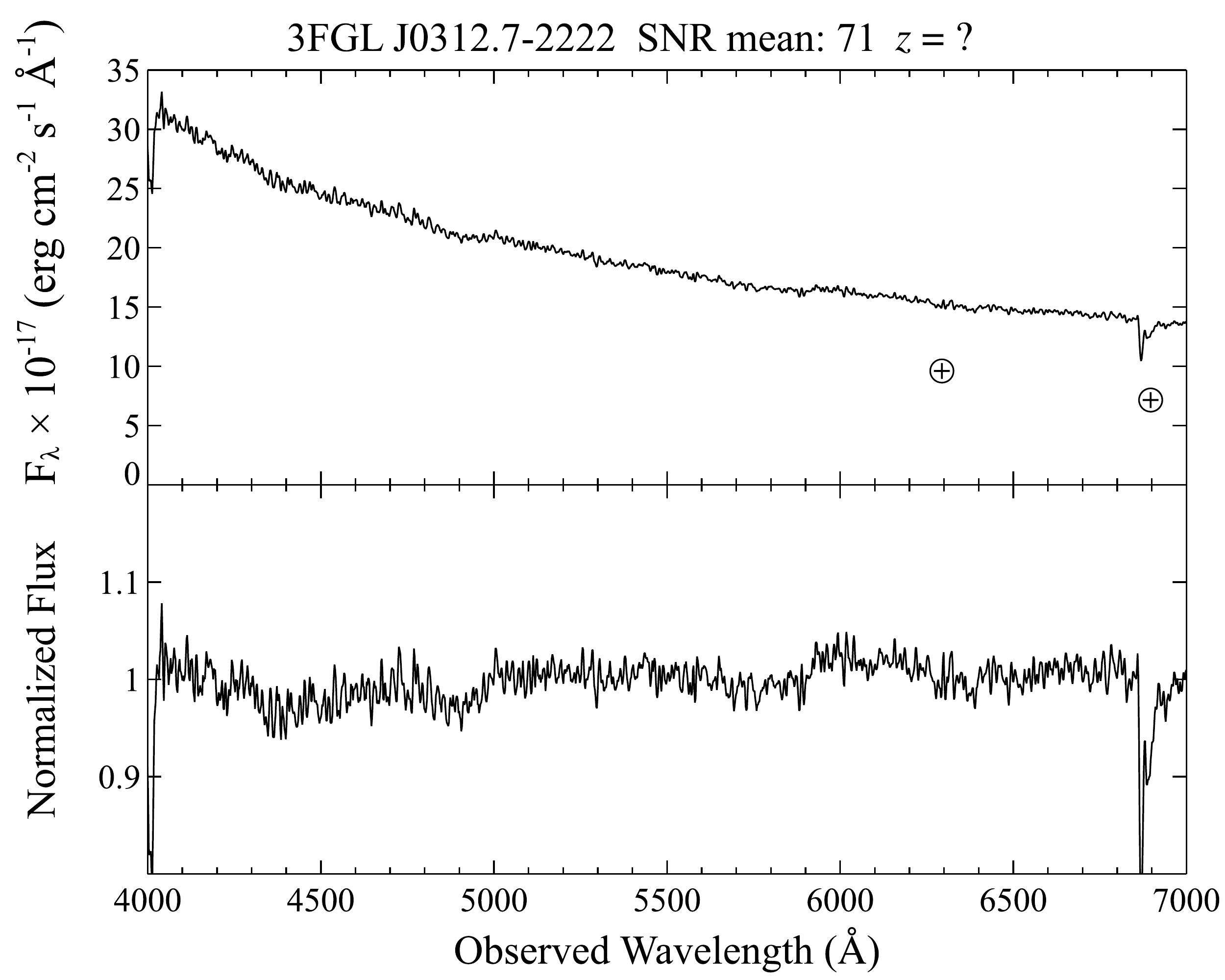} &
\includegraphics[trim=4cm 0cm 4cm 0cm, clip=true, width=7cm,angle=0]{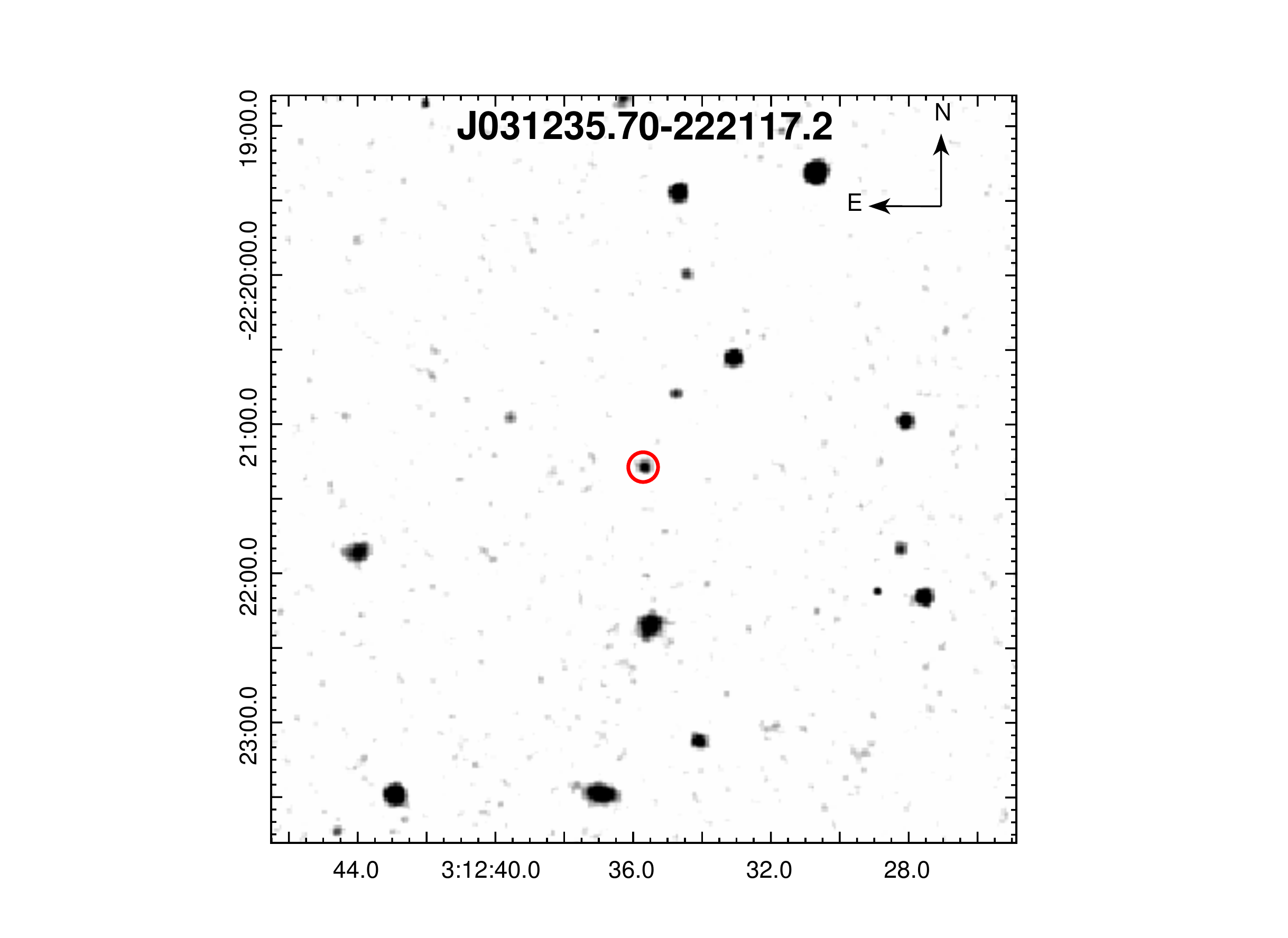} \\
\end{array}$
\end{center}
\caption{(Left panel) Optical spectrum of  WISE J031235.70-222117.2 associated with 3FGL J0312.7-2222. Signal-to-noise ratio is reported in the Figure. (Right panel) The finding chart ( $5'\times 5'$ ) retrieved from the Digital Sky Survey highlighting the location of the potential source: WISE J031235.70-222117.2 (red circle).}
\label{fig:J0312}
\end{figure*}

\begin{figure*}{}
\begin{center}$
\begin{array}{cc}
\includegraphics[width=\mywidth,angle=0]{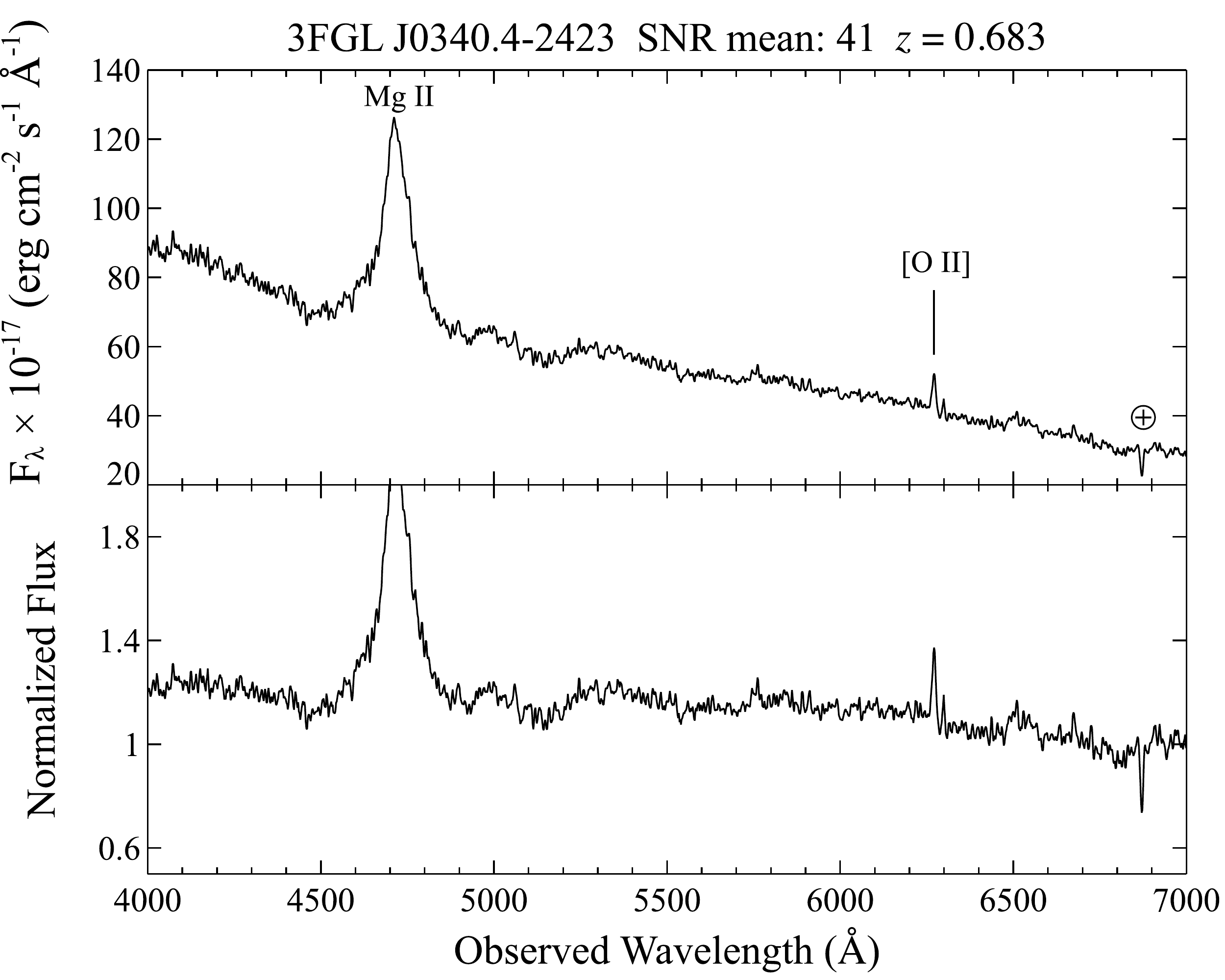} &
\includegraphics[trim=4cm 0cm 4cm 0cm, clip=true, width=7cm,angle=0]{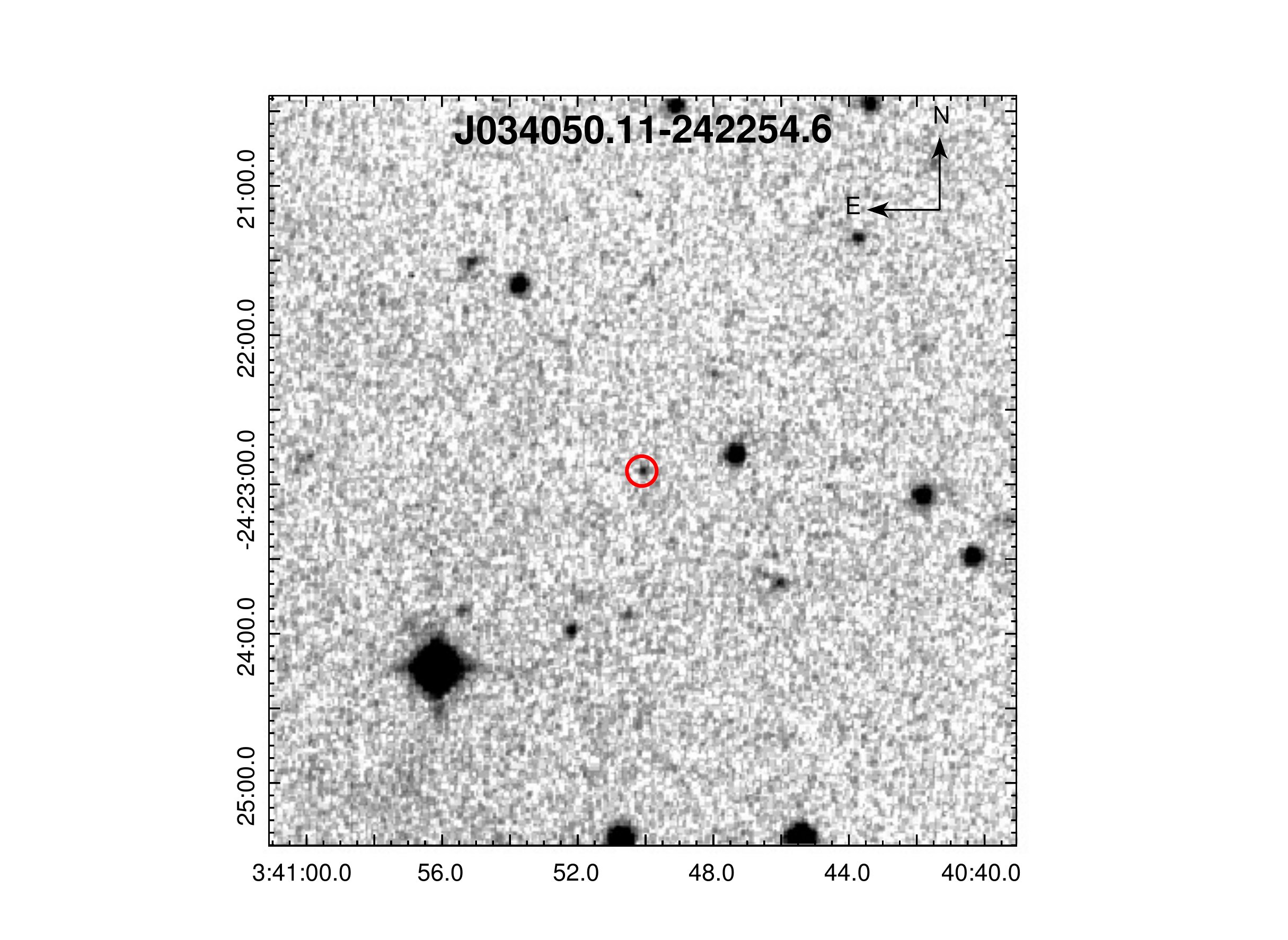} \\
\end{array}$
\end{center}
\caption{(Left panel) Optical spectrum of  WISE J034050.11-242254.6 associated with 3FGL J0340.4-2423. Signal-to-noise ratio is reported in the Figure. (Right panel) The finding chart ( $5'\times 5'$ ) retrieved from the Digital Sky Survey highlighting the location of the potential source: WISE J034050.11-242254.6 (red circle).}
\label{fig:j03407}
\end{figure*}

\begin{figure*}{}
\begin{center}$
\begin{array}{cc}
\includegraphics[width=\mywidth,angle=0]{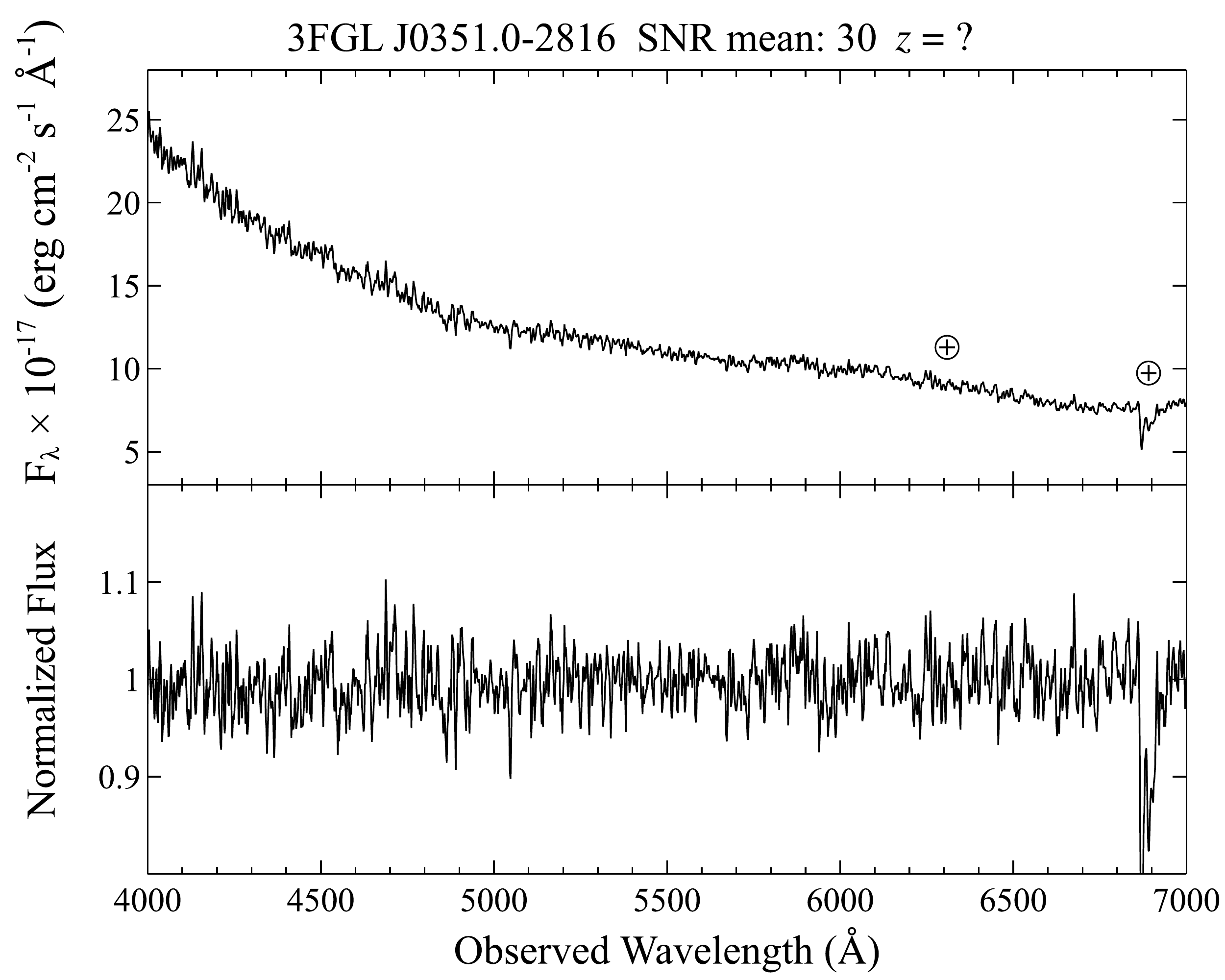} &
\includegraphics[trim=4cm 0cm 4cm 0cm, clip=true, width=7cm,angle=0]{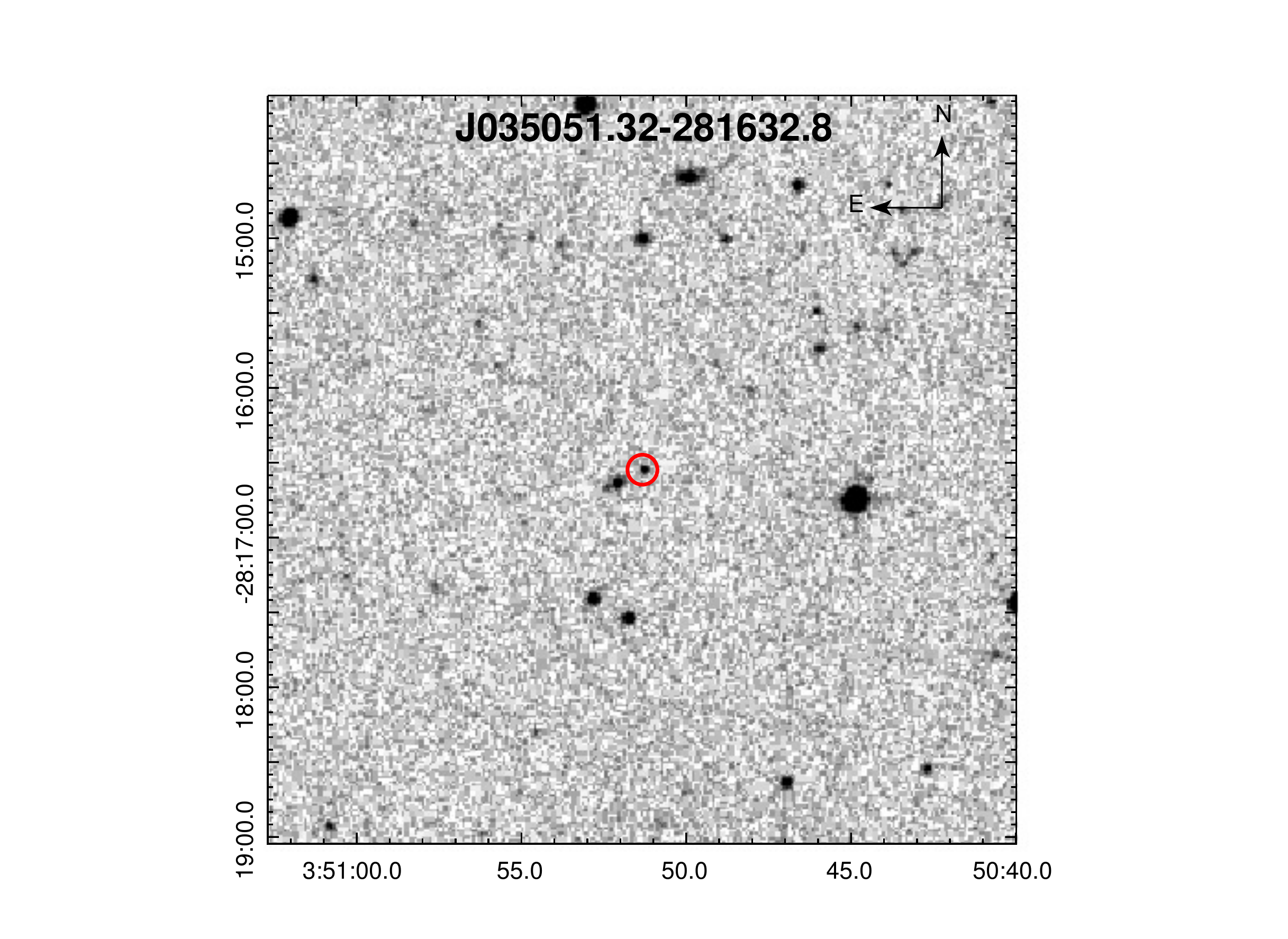} \\
\end{array}$
\end{center}
\caption{(Left panel) Optical spectrum of  WISE J035051.32-281632.8 associated with 3FGL J0351.0-2816. Signal-to-noise ratio is reported in the Figure. (Right panel) The finding chart ( $5'\times 5'$ ) retrieved from the Digital Sky Survey highlighting the location of the potential source: WISE J035051.32-281632.8 (red circle).}
\label{fig:J0351}
\end{figure*}

\begin{figure*}{}
\begin{center}$
\begin{array}{cc}
\includegraphics[width=\mywidth,angle=0]{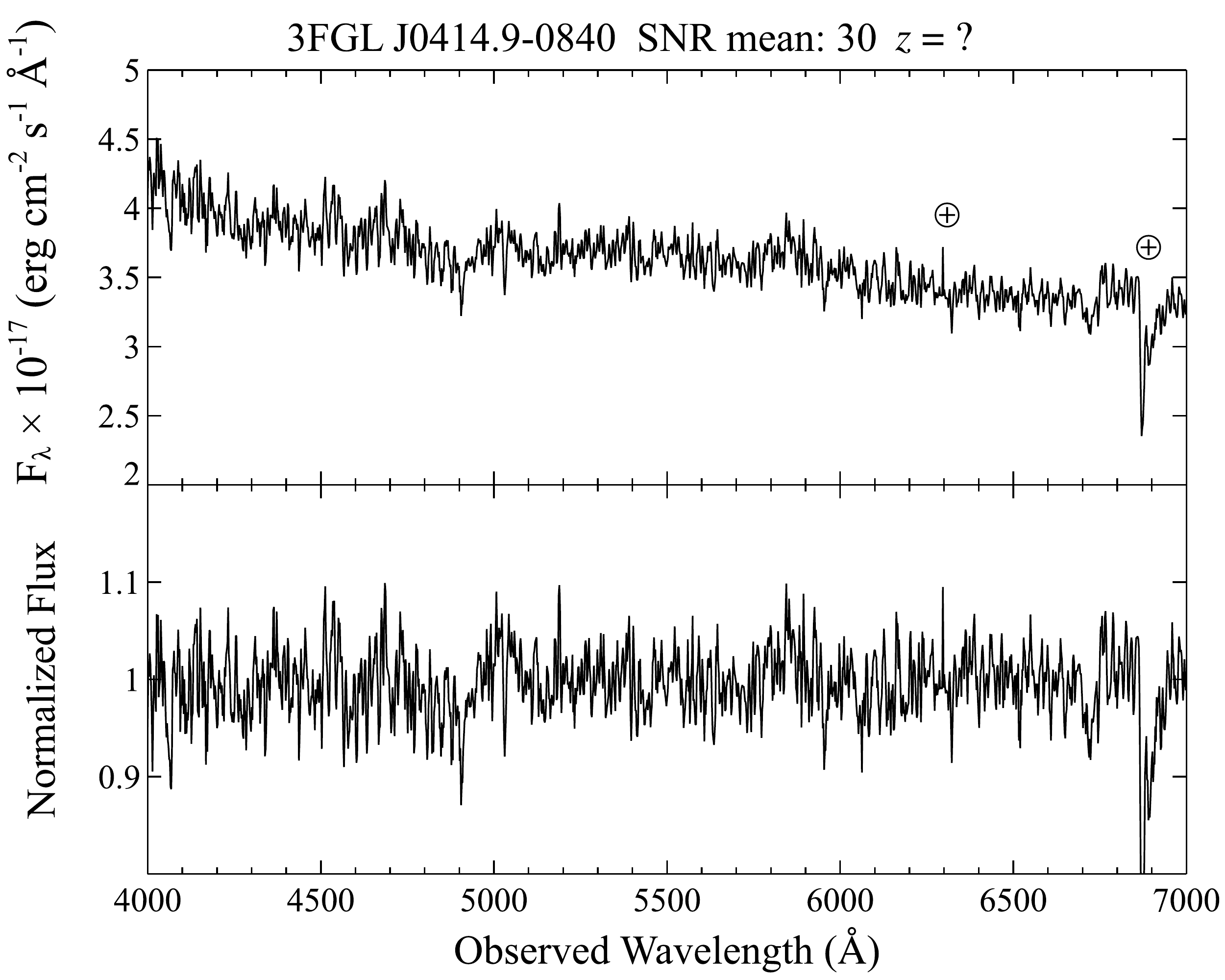} &
\includegraphics[trim=4cm 0cm 4cm 0cm, clip=true, width=7cm,angle=0]{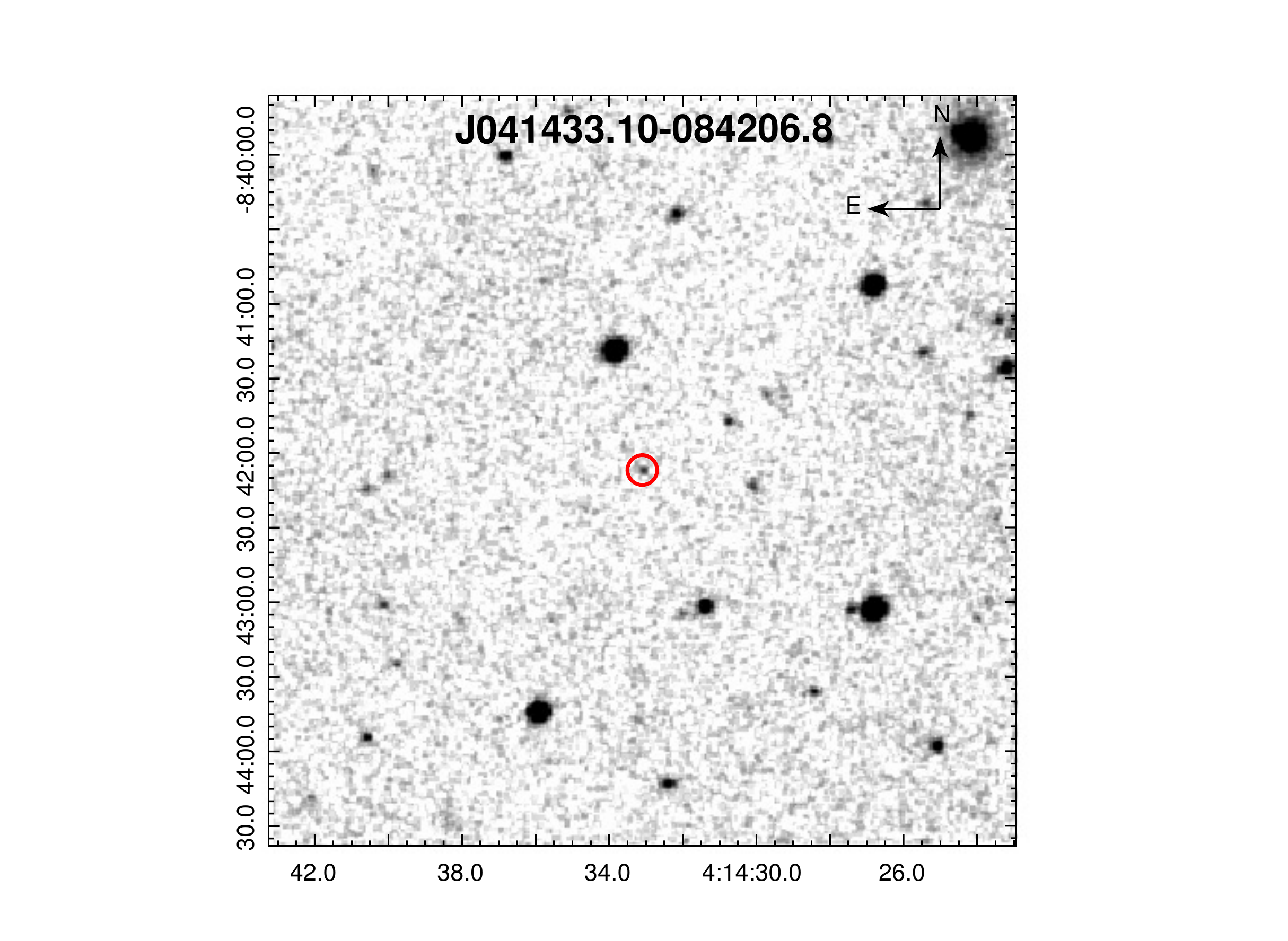} \\
\end{array}$
\end{center}
\caption{(Left panel) Optical spectrum of  WISE J041433.10-084206.8, associated with 3FGL J0414.9-0840. Signal-to-noise ratio is reported in the Figure. (Right panel) The finding chart ( $5'\times 5'$ ) retrieved from the Digital Sky Survey highlighting the location of the potential source: WISE J041433.10-084206.8 (red circle).}

\label{fig:J0414}
\end{figure*}

\begin{figure*}{}
\begin{center}$
\begin{array}{cc}
\includegraphics[width=\mywidth,angle=0]{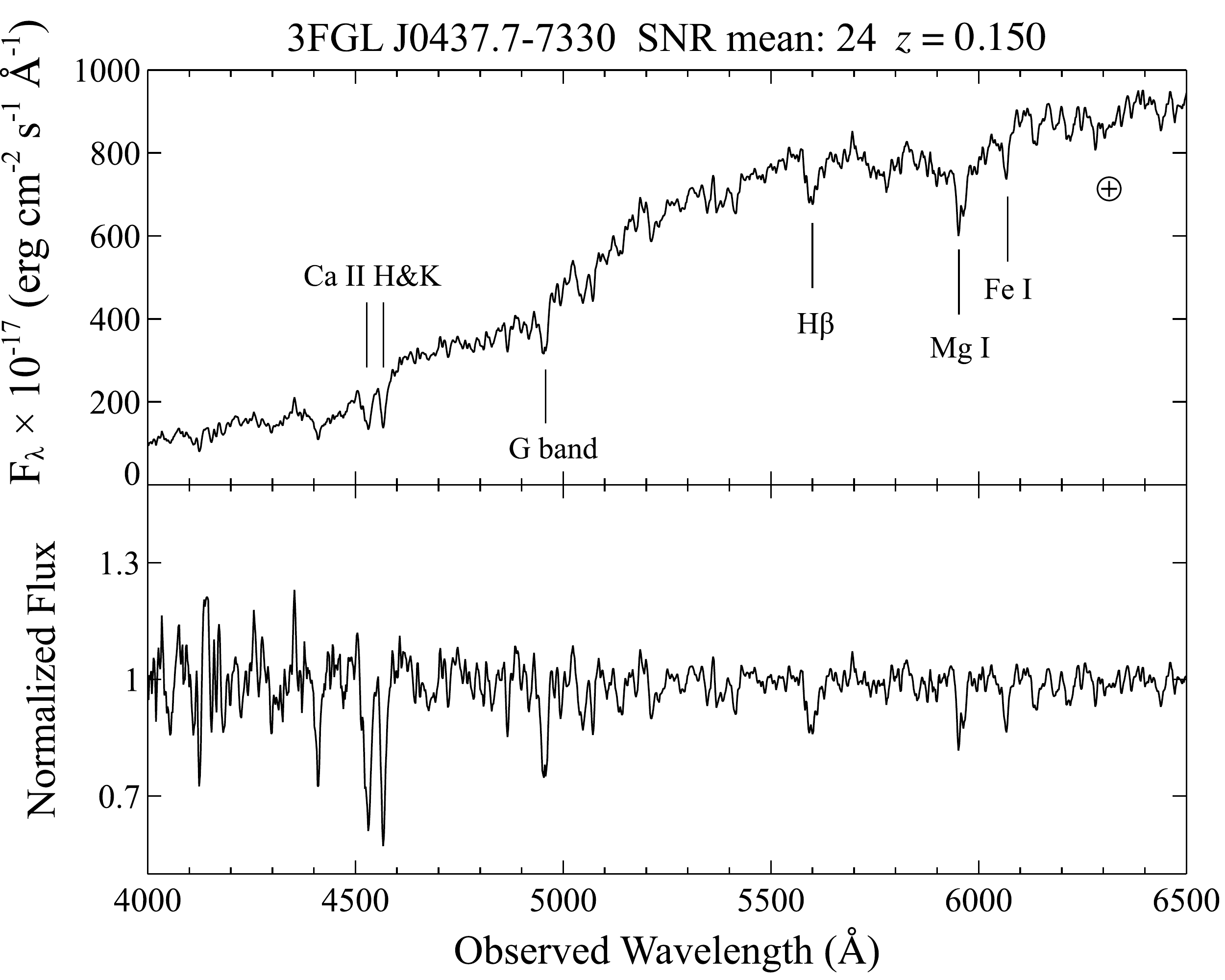} &
\includegraphics[trim=4cm 0cm 4cm 0cm, clip=true, width=7cm,angle=0]{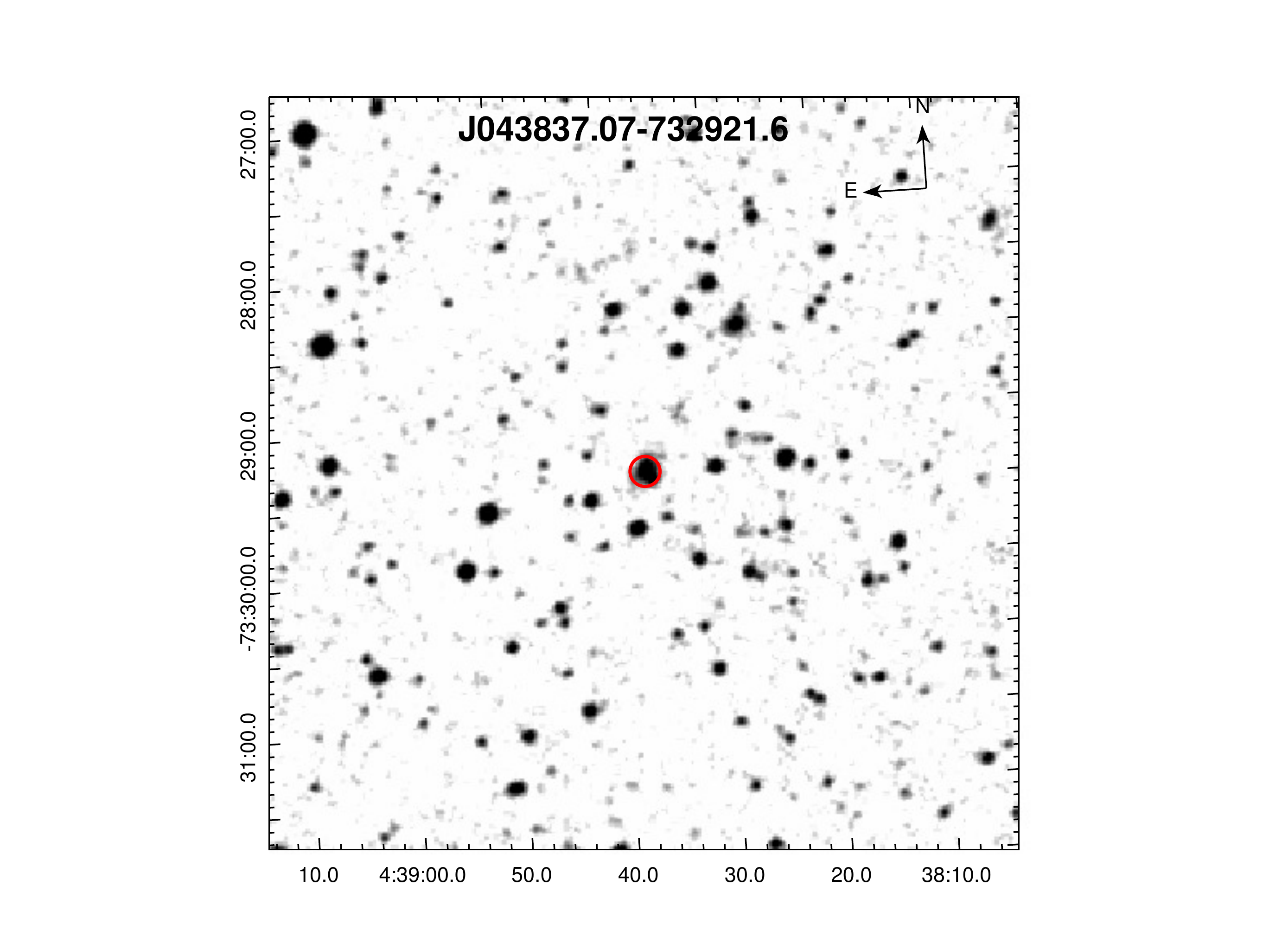} \\
\end{array}$
\end{center}
\caption{(Left panel) Optical spectrum of  WISE J043837.07-732921.6, associated with 3FGL J0437.7-7330. Signal-to-noise ratio is reported in the Figure. (Right panel) The finding chart ( $5'\times 5'$ ) retrieved from the Digital Sky Survey highlighting the location of the potential source: WISE J043837.07-732921.6 (red circle).}
\label{fig:J0438}
\end{figure*}

\begin{figure*}{}
\begin{center}$
\begin{array}{cc}
\includegraphics[width=\mywidth,angle=0]{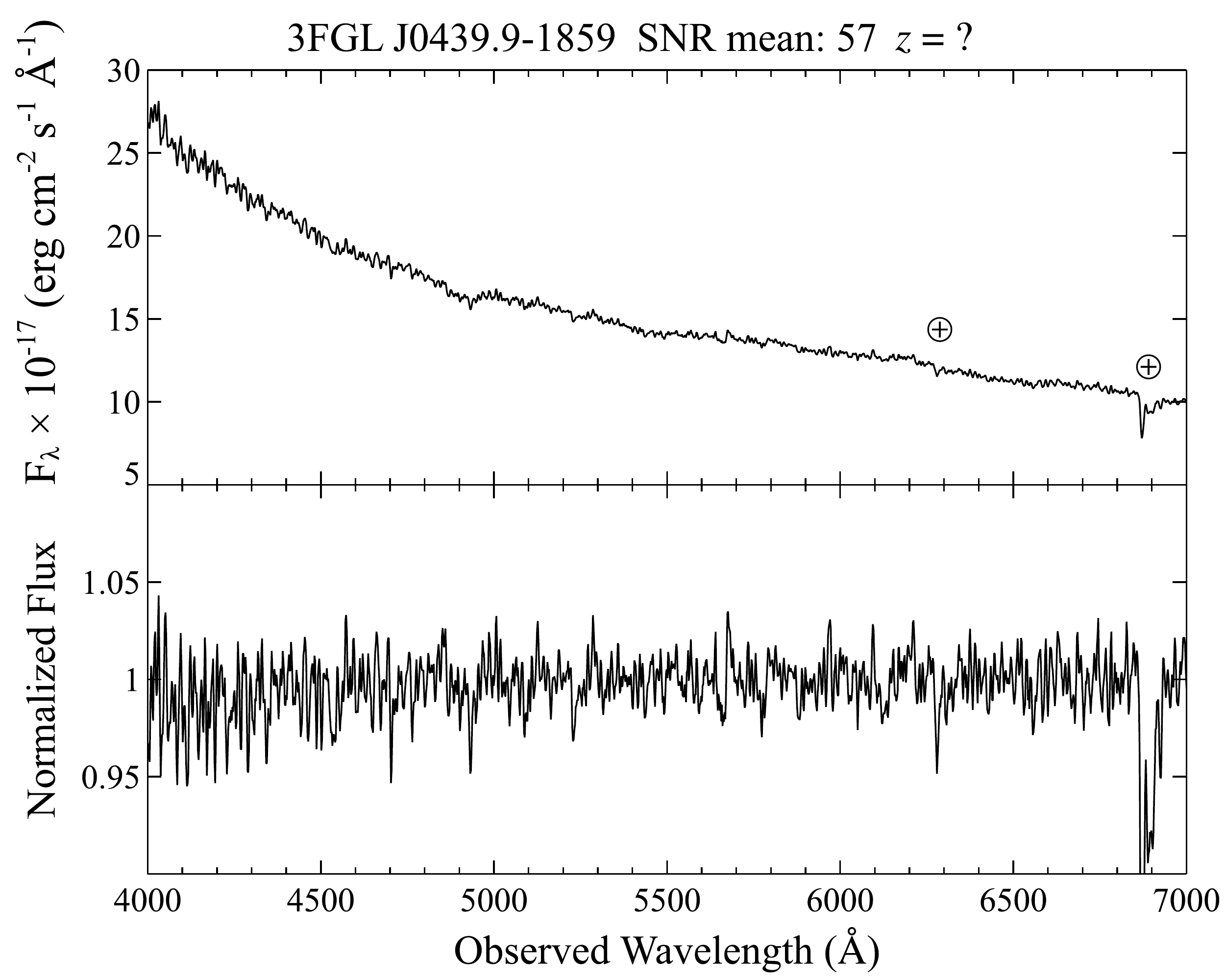} &
\includegraphics[trim=4cm 0cm 4cm 0cm, clip=true, width=7.5cm,angle=0]{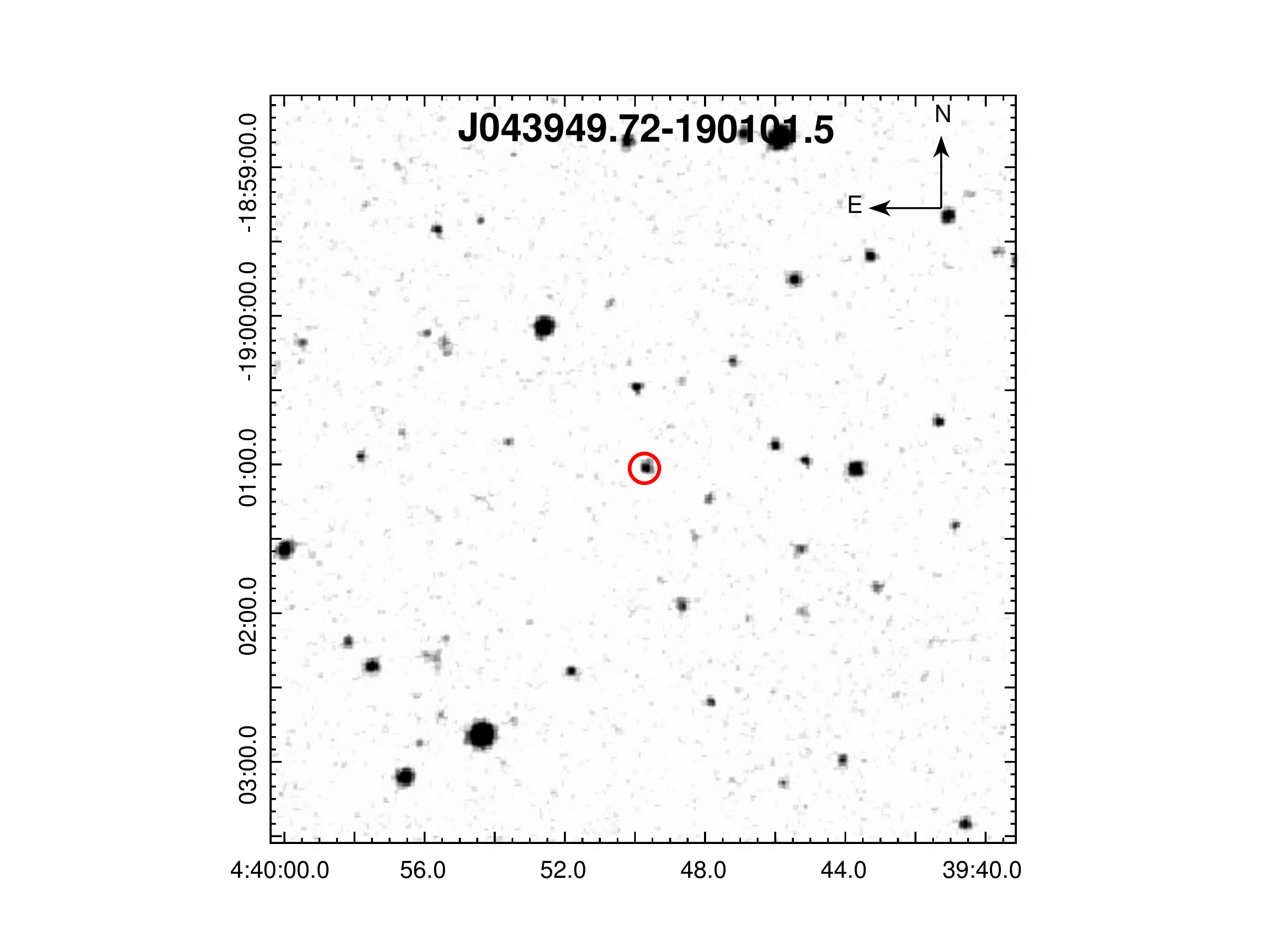} \\
\end{array}$
\end{center}
\caption{(Left panel) Optical spectrum of  WISE J043949.72-190101.5 associated with 3FGL J0439.9-1859, in the upper part it is shown the Signal-to-Noise Ratio of the spectrum. (Right panel) The finding chart ( $5'\times 5'$ ) retrieved from the Digital Sky Survey highlighting the location of the counterpart: WISE J043949.72-190101.5 (red circle).}
\label{fig:J0439}
\end{figure*}

\begin{figure*}{}
\begin{center}$
\begin{array}{cc}
\includegraphics[width=\mywidth,angle=0]{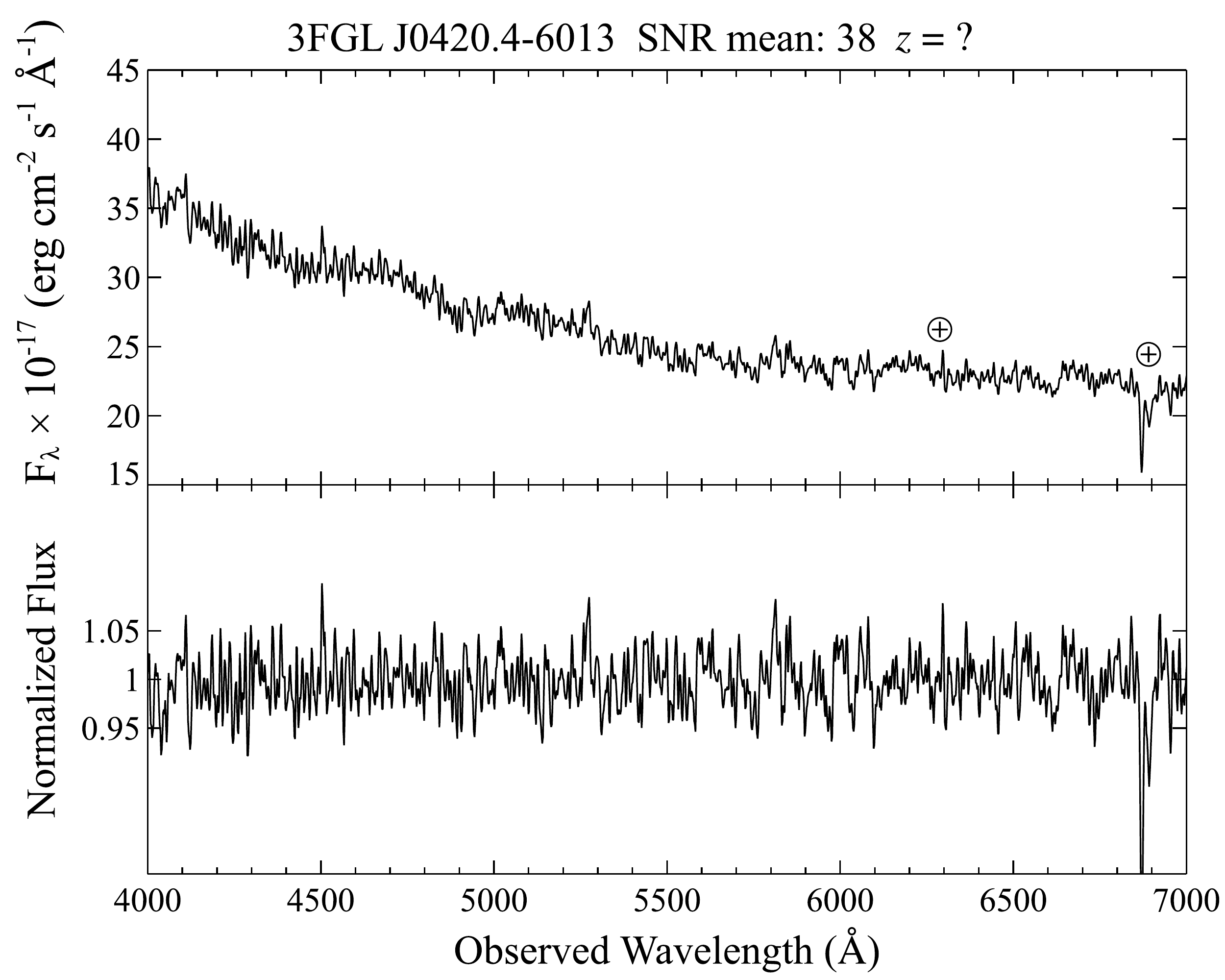} &
\includegraphics[trim=4cm 0cm 4cm 0cm, clip=true, width=7cm,angle=0]{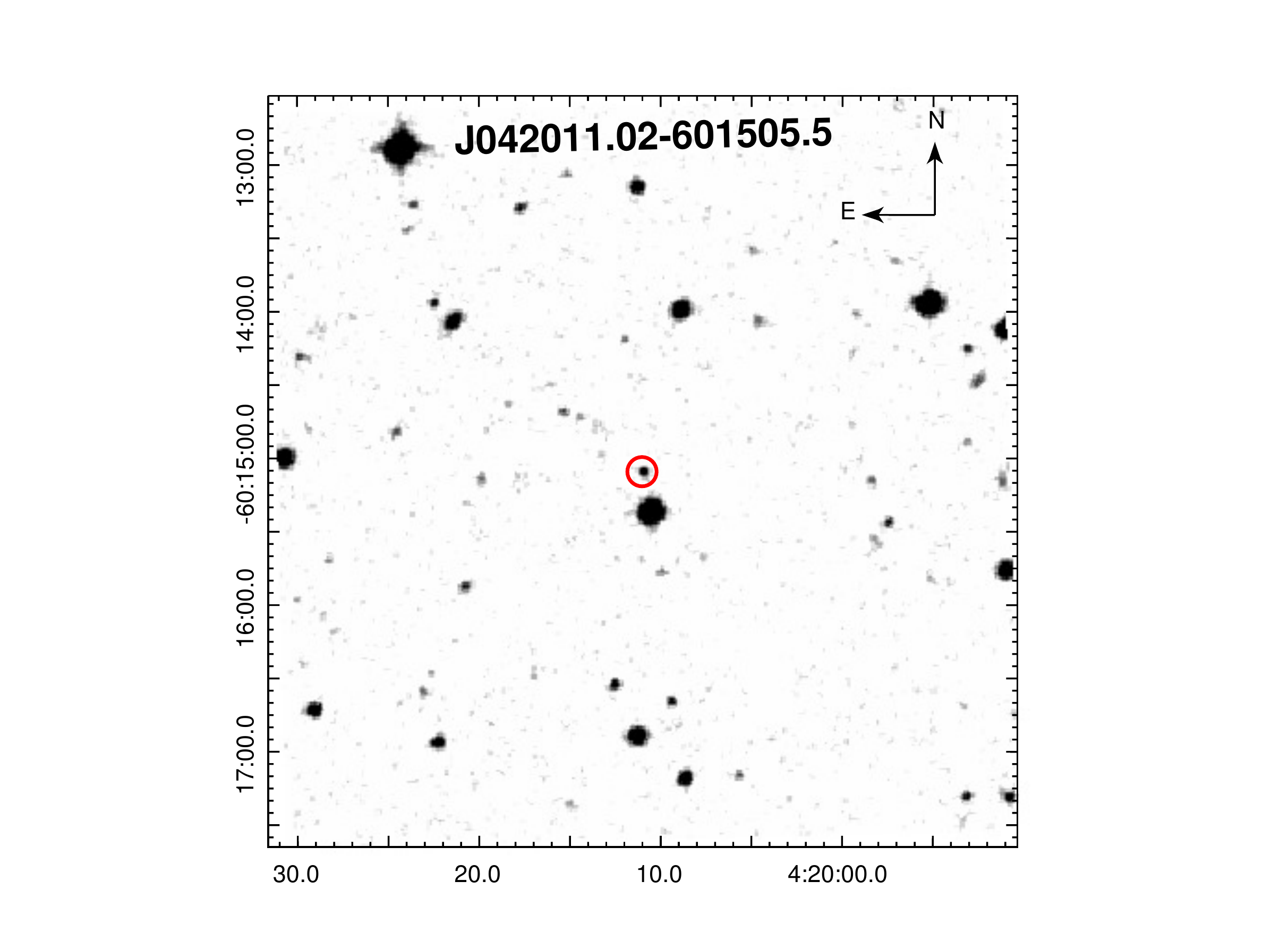} \\
\end{array}$
\end{center}
\caption{(Left panel) Optical spectrum of  WISE J042011.02-601505.5, associated with 3FGL J0420.4-6013. Signal-to-noise ratio is reported in the Figure. (Right panel) The finding chart ( $5'\times 5'$ ) retrieved from the Digital Sky Survey highlighting the location of the potential source: WISE J042011.02-601505.5 (red circle).}
\label{fig:J0420}
\end{figure*}

\begin{figure*}{}
\begin{center}$
\begin{array}{cc}
\includegraphics[width=\mywidth,angle=0]{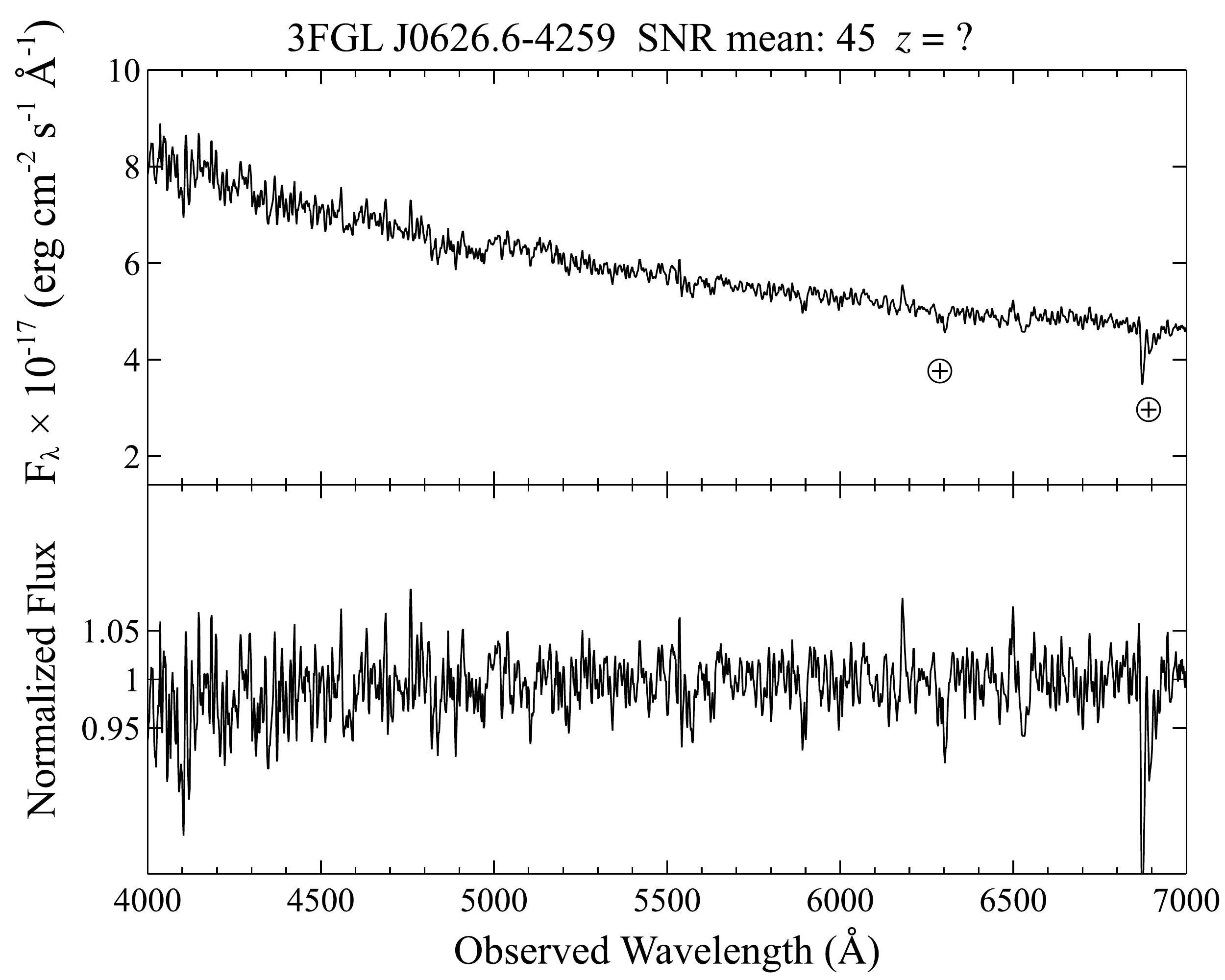} &
\includegraphics[trim=4cm 0cm 4cm 0cm, clip=true, width=7cm,angle=0]{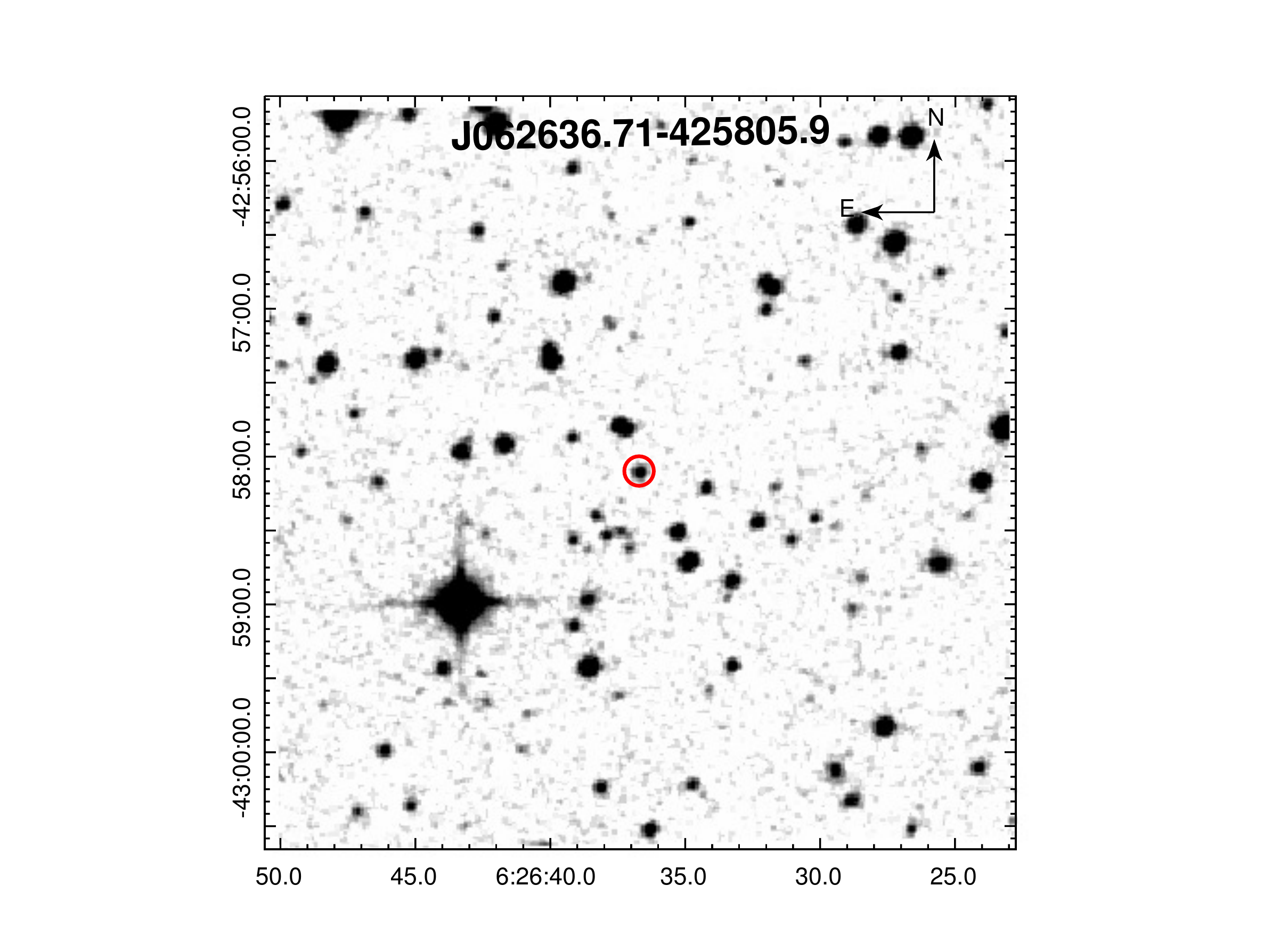} \\
\end{array}$
\end{center}
\caption{(Left panel) Optical spectrum of  WISE J062636.71-425805.9 associated with 3FGL J0626.6-4259, in the upper part it is shown the Signal-to-Noise Ratio of the spectrum. (Right panel) The finding chart ( $5'\times 5'$ ) retrieved from the Digital Sky Survey highlighting the location of the counterpart: WISE J062636.71-425805.9 (red circle).}
\label{fig:J0626}
\end{figure*}

\begin{figure*}{}
\begin{center}$
\begin{array}{cc}
\includegraphics[width=\mywidth,angle=0]{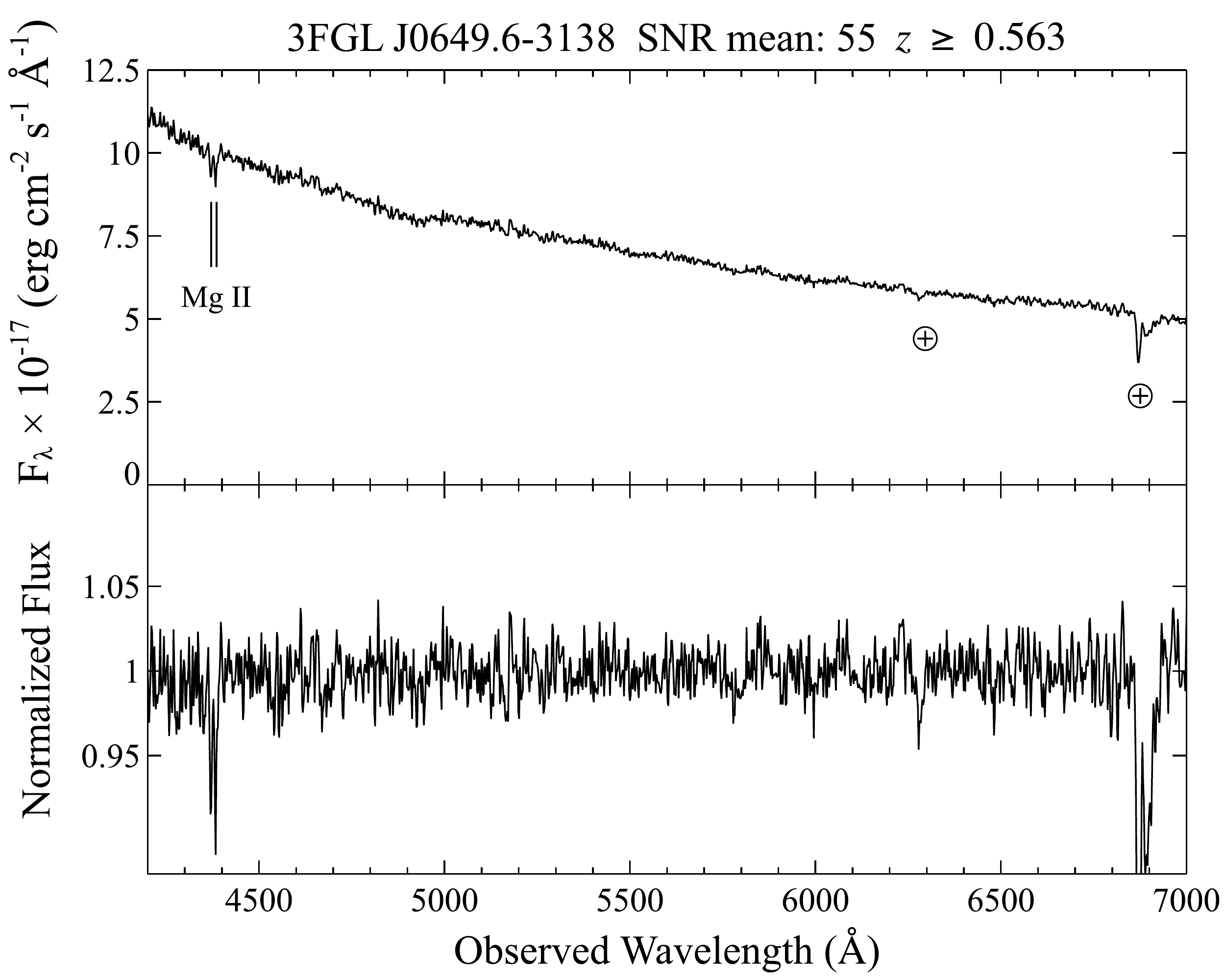} &
\includegraphics[trim=4cm 0cm 4cm 0cm, clip=true, width=7cm,angle=0]{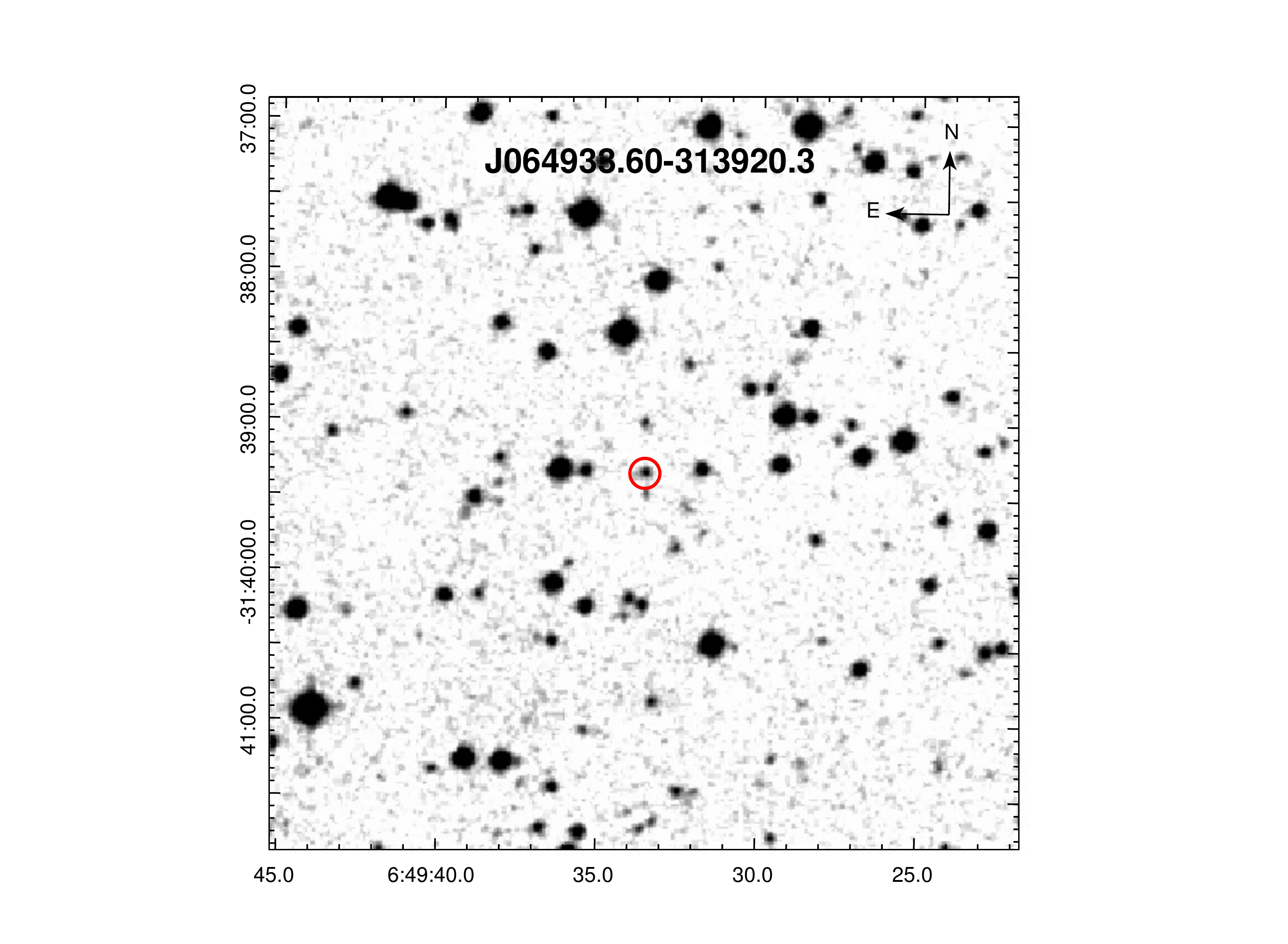} \\
\end{array}$
\end{center}
\caption{(Left panel) Optical spectrum of  WISE J064933.60-313920.3 associated with 3FGL J0649.6-3138, in the upper part it is shown the Signal-to-Noise Ratio of the spectrum. (Right panel) The finding chart ( $5'\times 5'$ ) retrieved from the Digital Sky Survey highlighting the location of the counterpart: WISE J064933.60-313920.3 (red circle).}
\label{fig:J0649}
\end{figure*}

\begin{figure*}{}
\begin{center}$
\begin{array}{cc}
\includegraphics[width=\mywidth,angle=0]{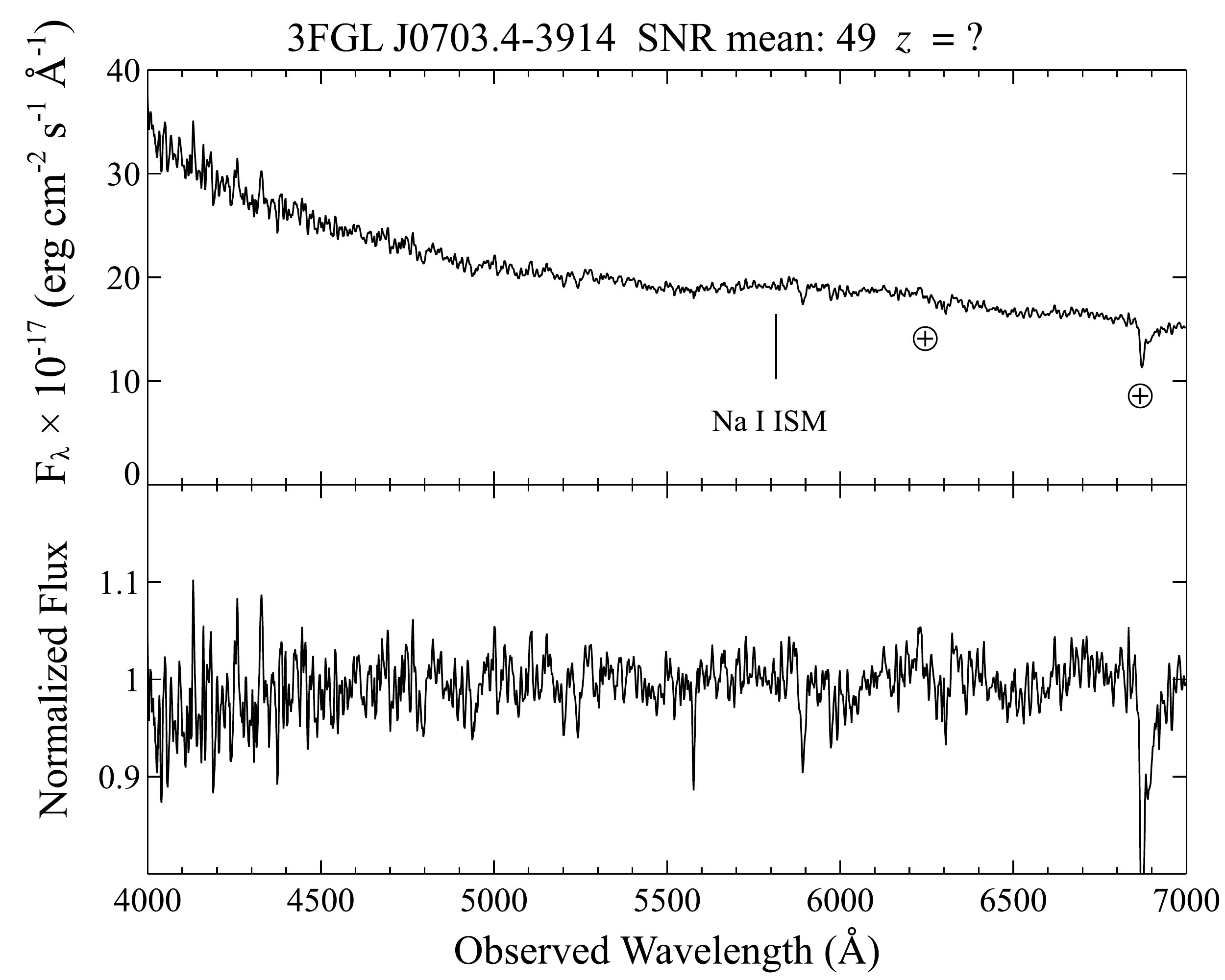} &
\includegraphics[trim=4cm 0cm 4cm 0cm, clip=true, width=7cm,angle=0]{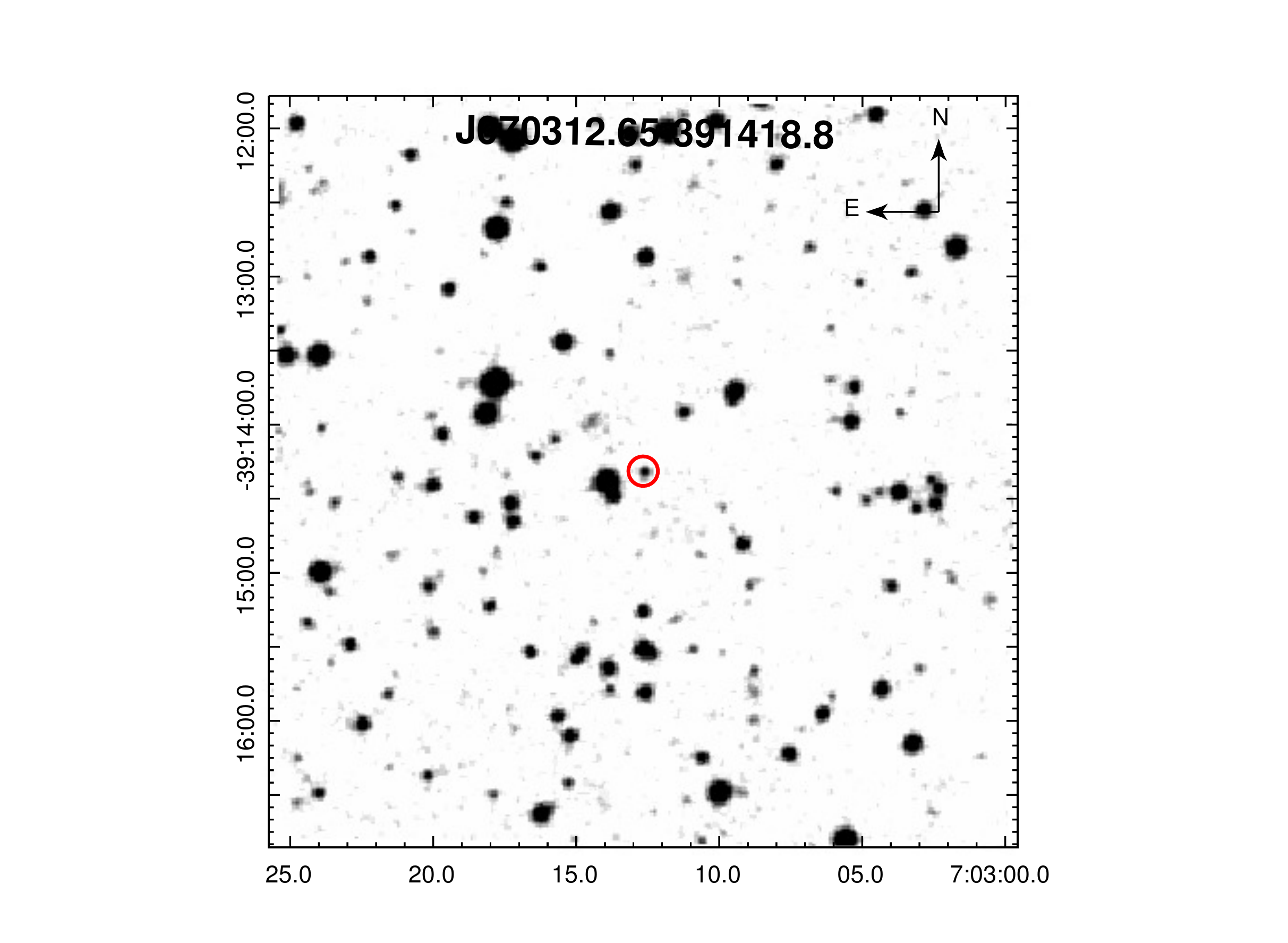} \\
\end{array}$
\end{center}
\caption{(Left panel) Optical spectrum of  WISE J070312.65-391418.8 associated with 3FGL J0703.4-3914, in the upper part it is shown the Signal-to-Noise Ratio of the spectrum. (Right panel) The finding chart ( $5'\times 5'$ ) retrieved from the Digital Sky Survey highlighting the location of the counterpart: WISE J070312.65-391418.8 (red circle).}
\label{fig:J0703}
\end{figure*}

\begin{figure*}{}
\begin{center}$
\begin{array}{cc}
\includegraphics[width=\mywidth,angle=0]{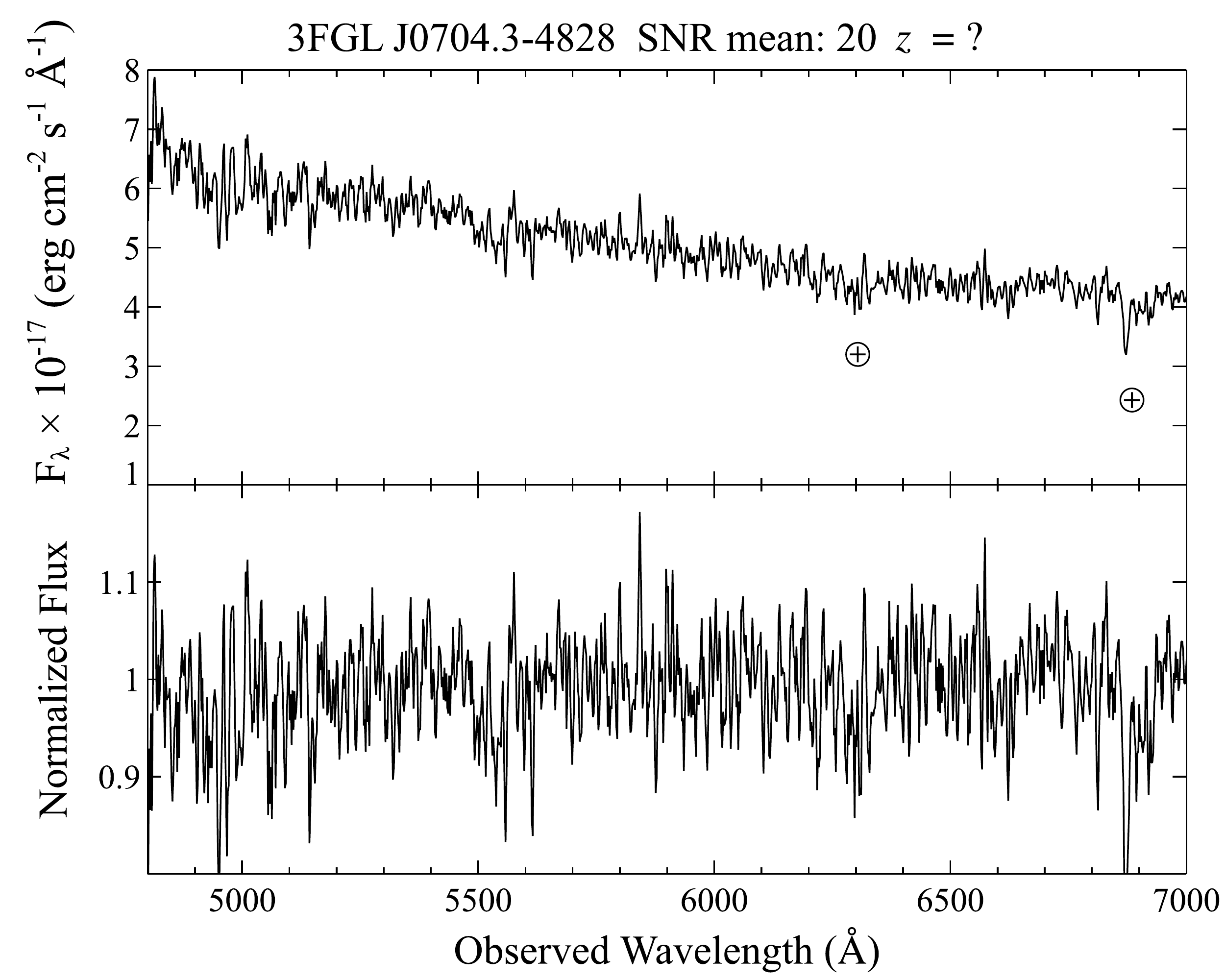} &
\includegraphics[trim=4cm 0cm 4cm 0cm, clip=true, width=7cm,angle=0]{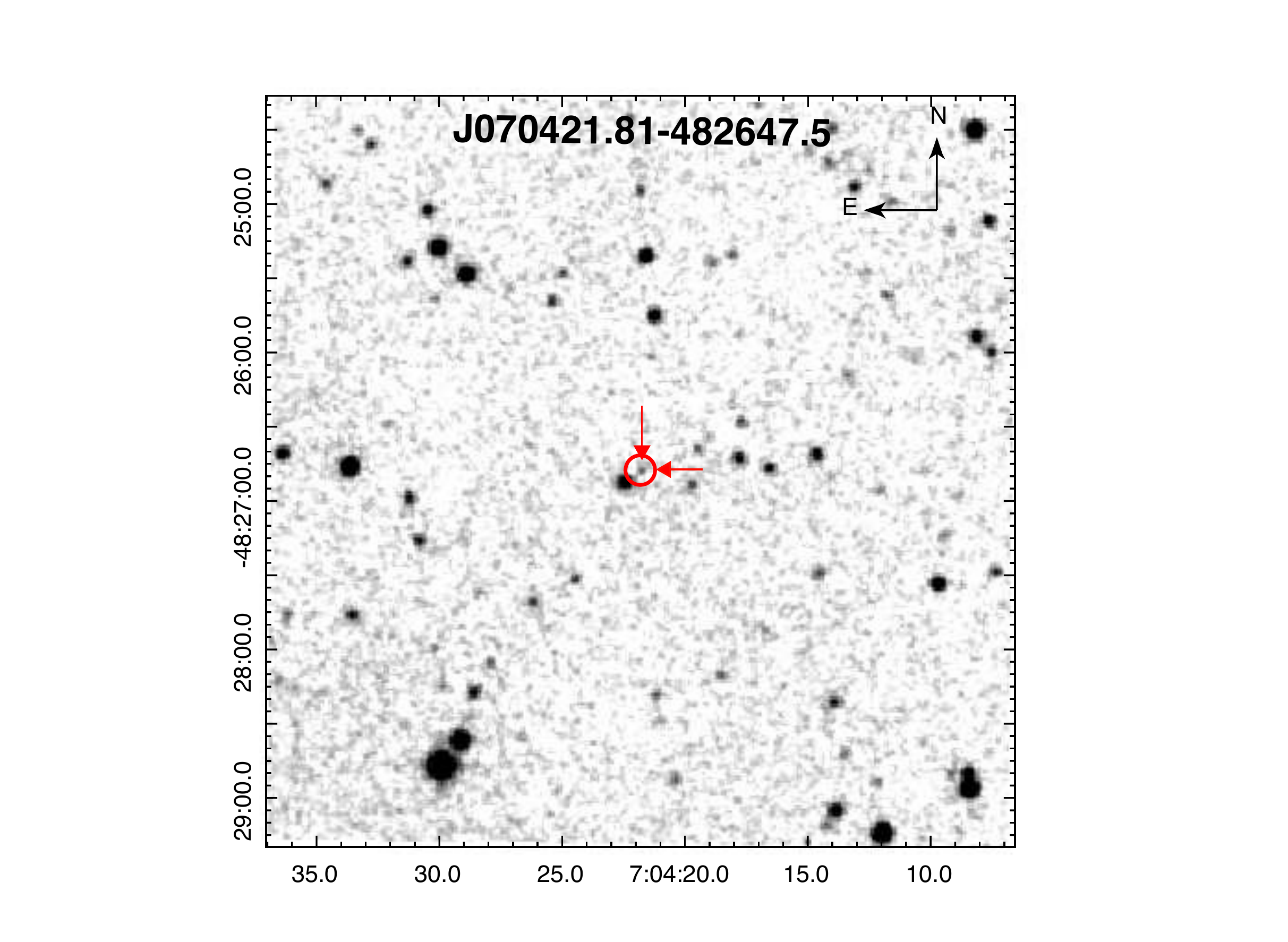} \\
\end{array}$
\end{center}
\caption{(Left panel) Optical spectrum of  WISE J070421.81-482647.5, associated with 3FGL J0704.3-4828. Signal-to-noise ratio is reported in the Figure. (Right panel) The finding chart ( $5'\times 5'$ ) retrieved from the Digital Sky Survey highlighting the location of the potential source: WISE J070421.81-482647.5 (red circle).}
\label{fig:J0704}
\end{figure*}

\begin{figure*}{}
\begin{center}$
\begin{array}{cc}
\includegraphics[width=\mywidth,angle=0]{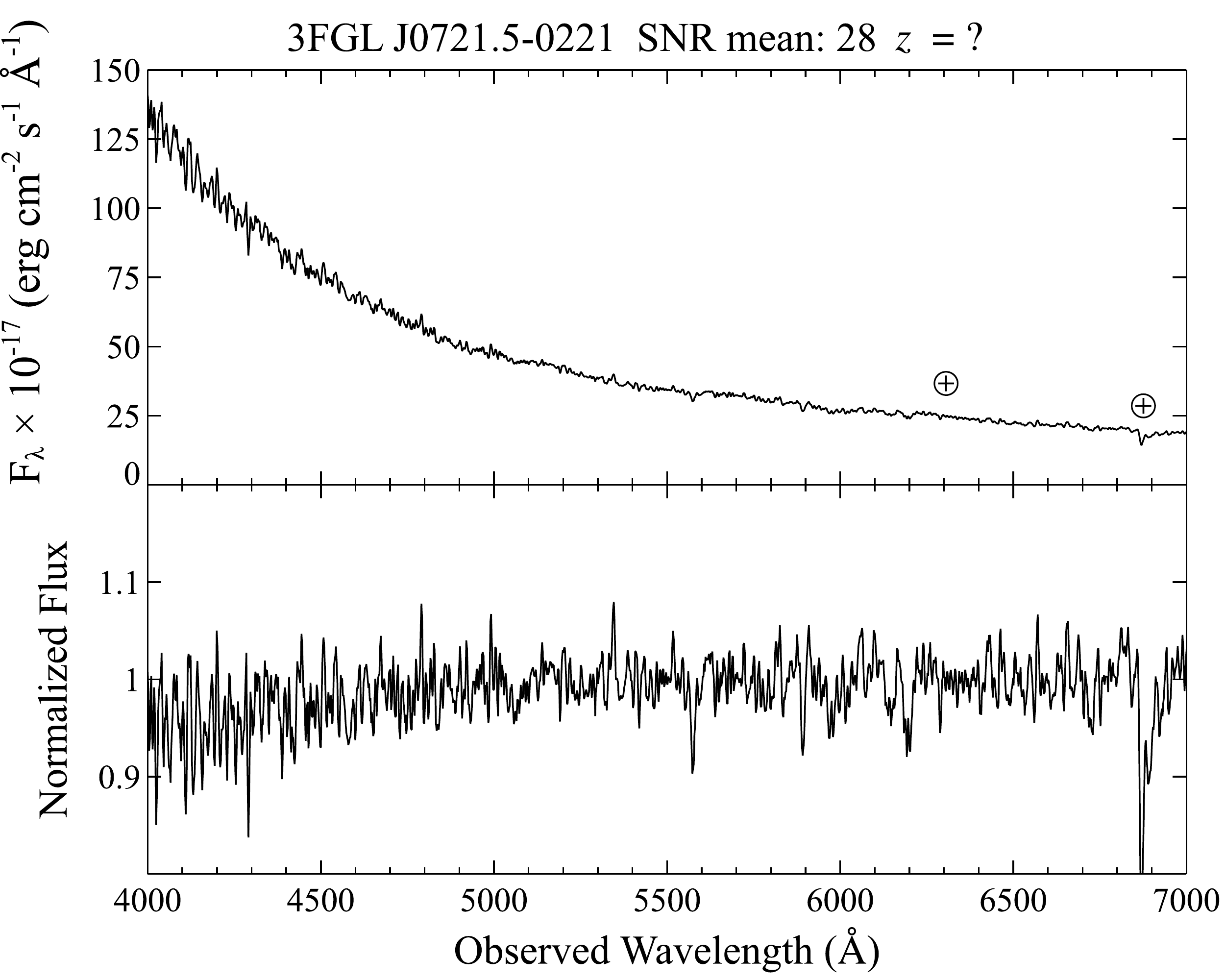} &
\includegraphics[trim=4cm 0cm 4cm 0cm, clip=true, width=7cm,angle=0]{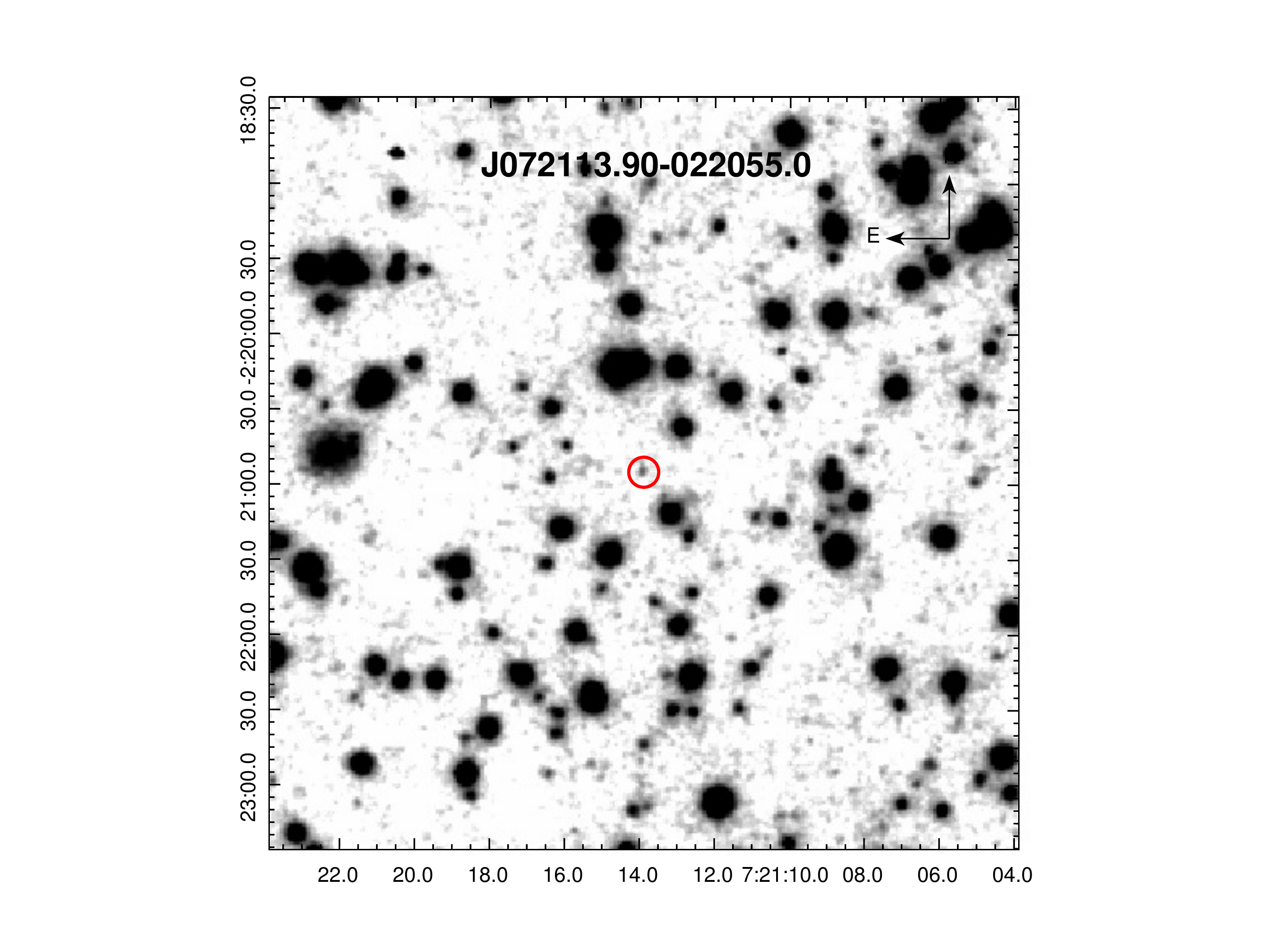} \\
\end{array}$
\end{center}
\caption{(Left panel) Optical spectrum of  WISE J072113.90-022055.0 associated with 3FGL J0721.5-0221. Signal-to-noise ratio is reported in the Figure. (Right panel) The finding chart ( $5'\times 5'$ ) retrieved from the Digital Sky Survey highlighting the location of the potential source: WISE J072113.90-022055.0 (red circle).}
\label{fig:J0721}
\end{figure*}

\begin{figure*}{}
\begin{center}$
\begin{array}{cc}
\includegraphics[width=\mywidth,angle=0]{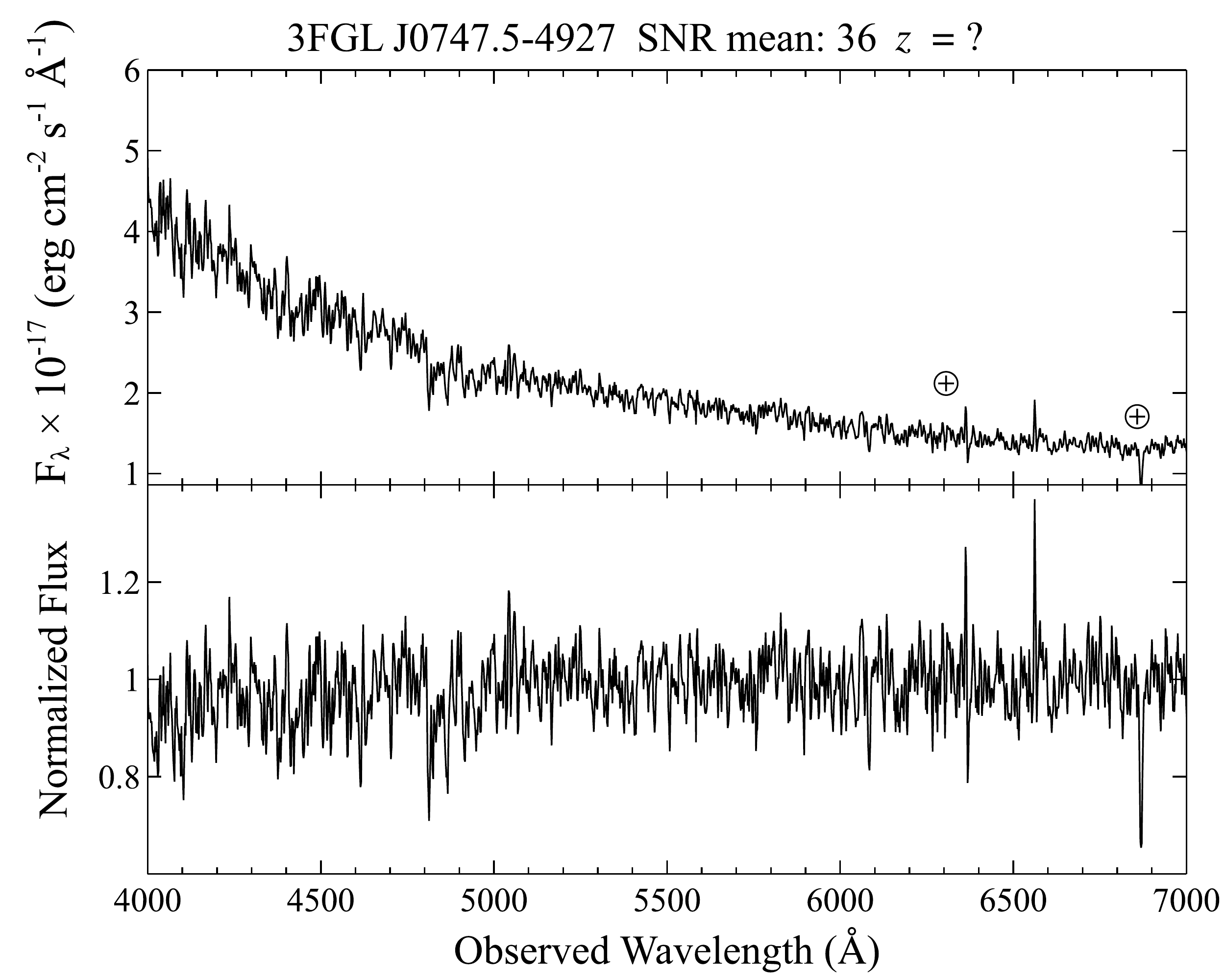} &
\includegraphics[trim=4cm 0cm 4cm 0cm, clip=true, width=7cm,angle=0]{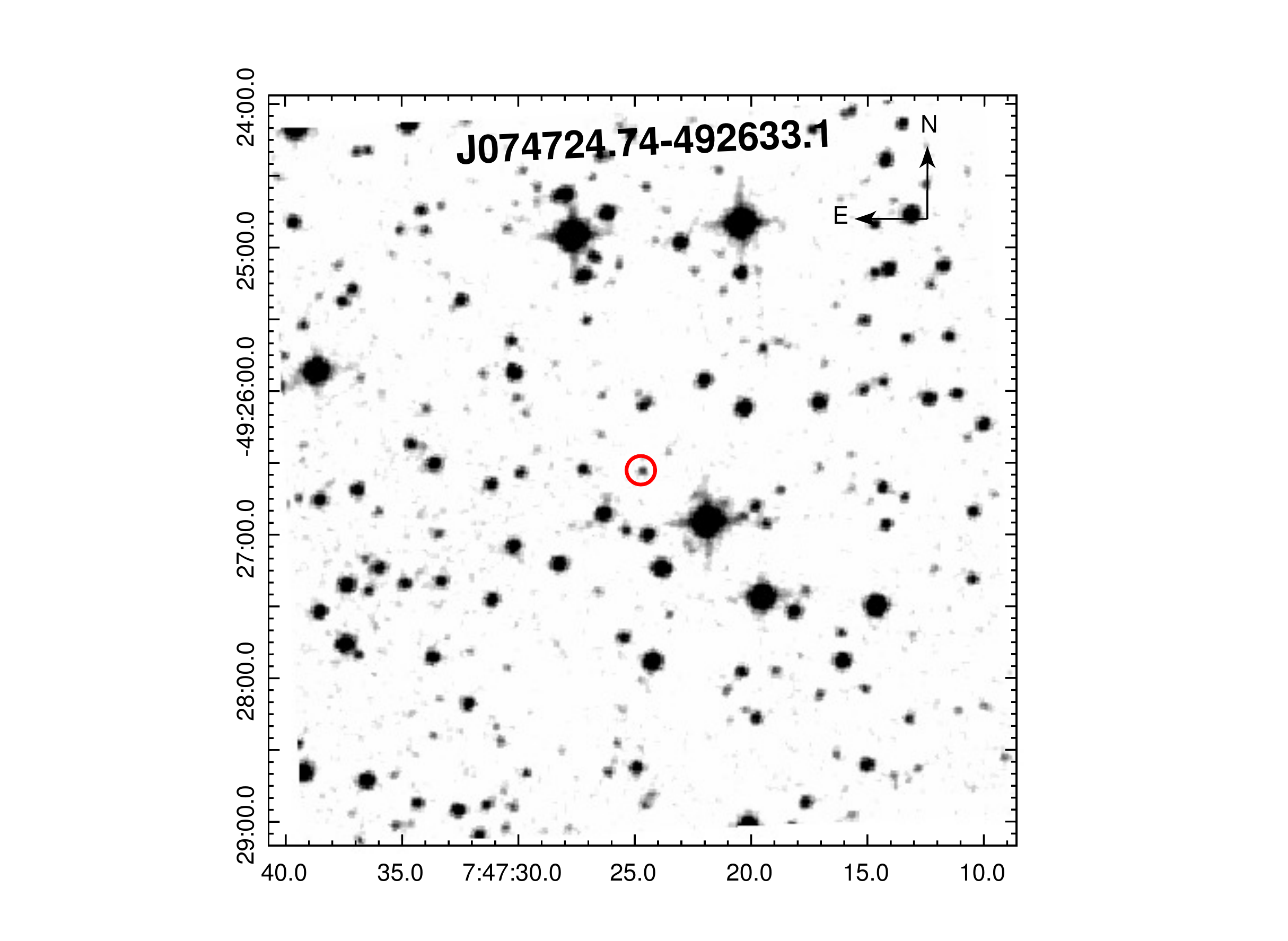} \\
\end{array}$
\end{center}
\caption{(Left panel) Optical spectrum of  WISE J074724.74-492633.1 associated with 3FGL J0747.5-4927. Signal-to-noise ratio is reported in the Figure. (Right panel) The finding chart ( $5'\times 5'$ ) retrieved from the Digital Sky Survey highlighting the location of the potential source: WISE J074724.74-492633.1 (red circle).}
\label{fig:J0747}
\end{figure*}

\begin{figure*}{}
\begin{center}$
\begin{array}{cc}
\includegraphics[width=\mywidth,angle=0]{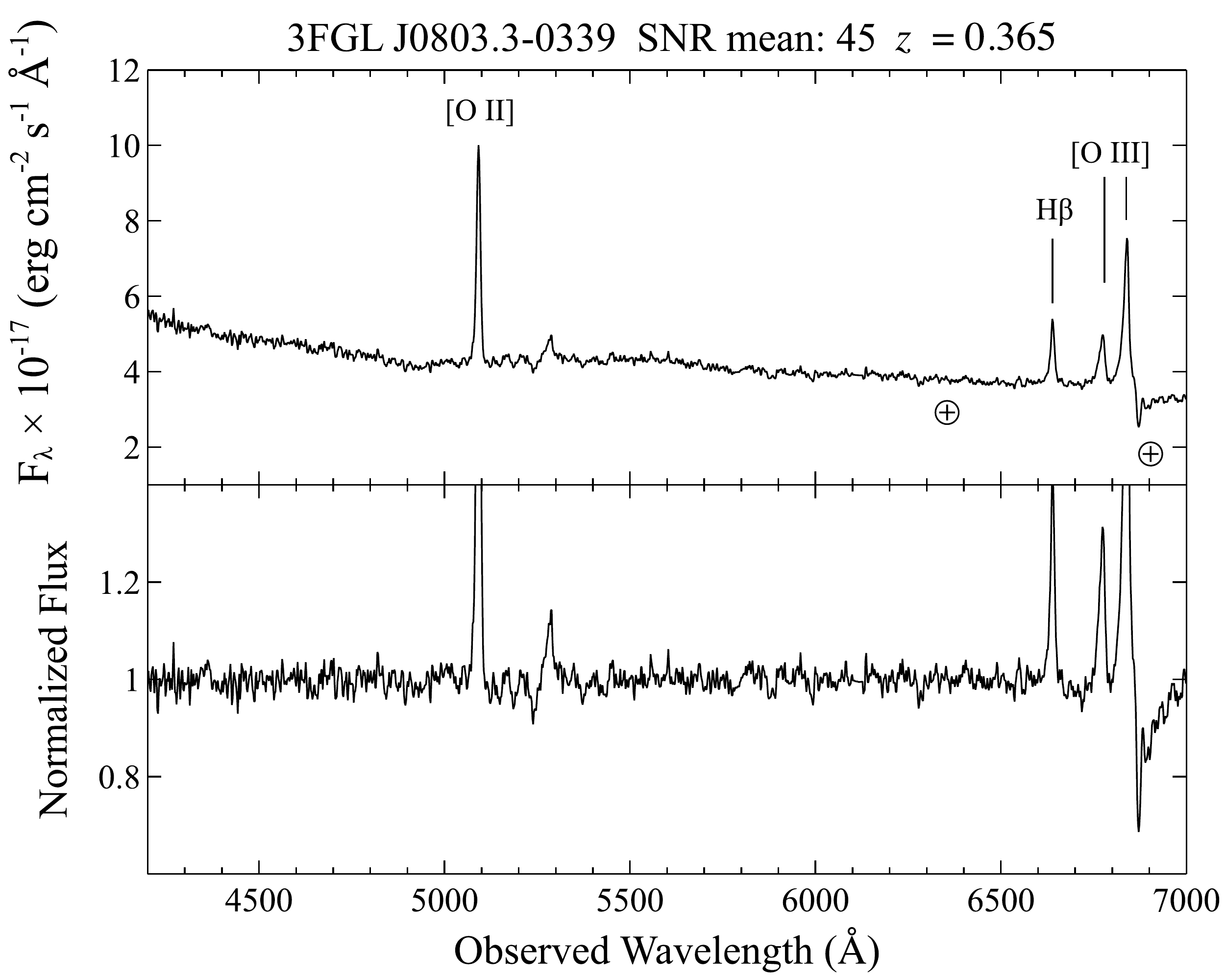} &
\includegraphics[trim=4cm 0cm 4cm 0cm, clip=true, width=7cm,angle=0]{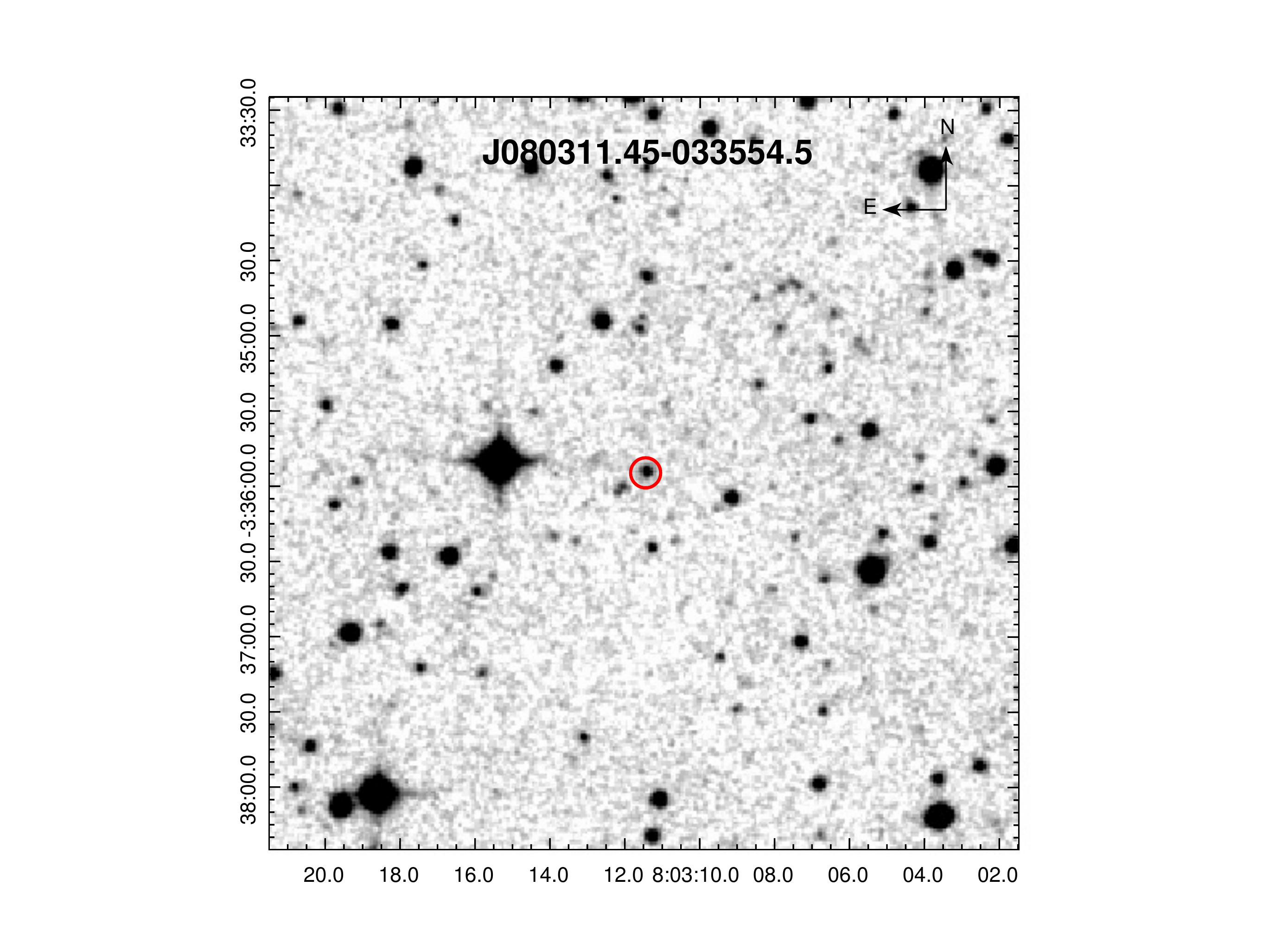} \\
\end{array}$
\end{center}
\caption{(Left panel) Optical spectrum of  WISE J080311.45-033554.5 associated with 3FGL J0803.3-0339. Signal-to-noise ratio is reported in the Figure. (Right panel) The finding chart ( $5'\times 5'$ ) retrieved from the Digital Sky Survey highlighting the location of the counterpart: WISE J080311.45-033554.5 (red circle).}
\label{fig:J0803}
\end{figure*}

\begin{figure*}{}
\begin{center}$
\begin{array}{cc}
\includegraphics[width=\mywidth,angle=0]{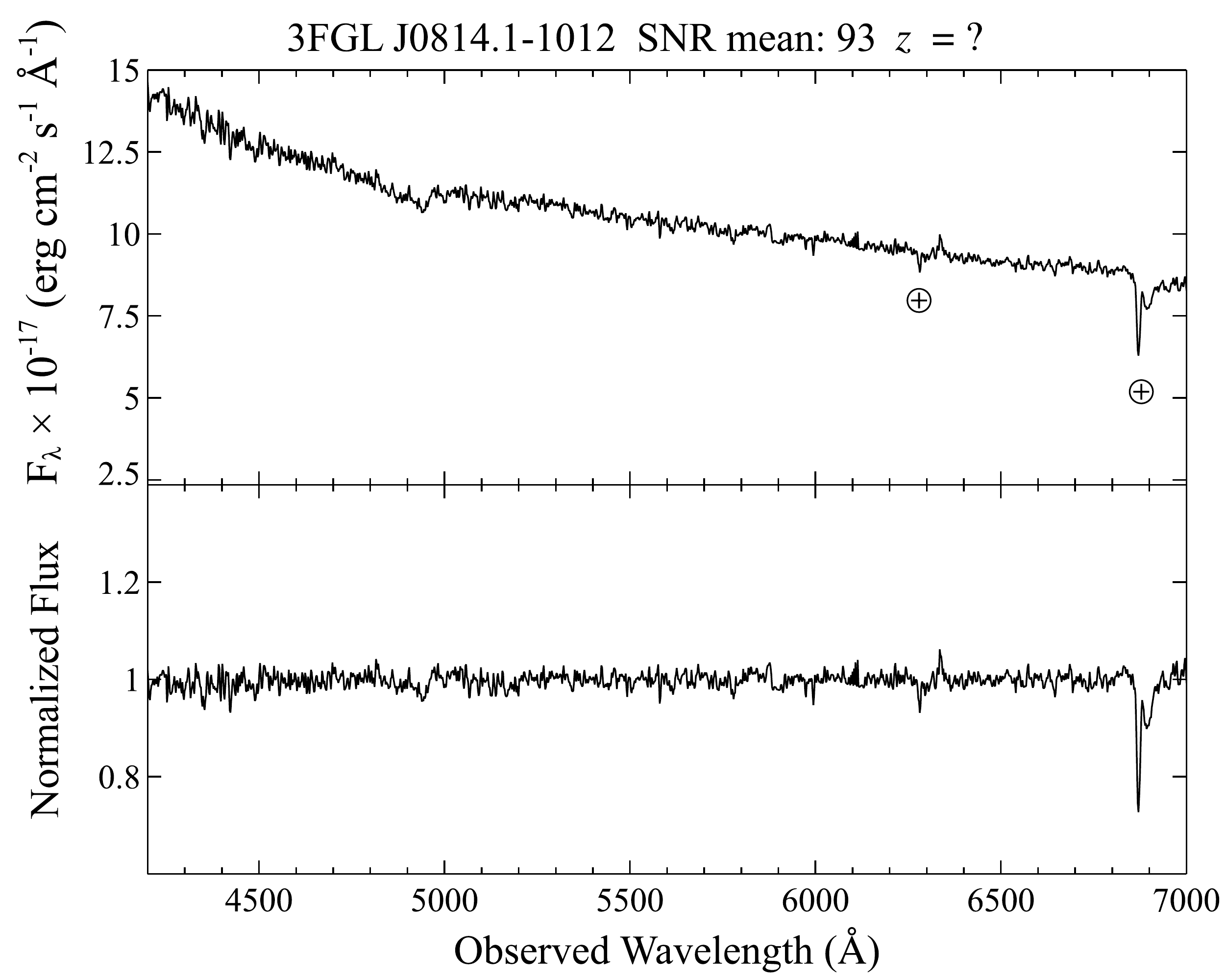} &
\includegraphics[trim=4cm 0cm 4cm 0cm, clip=true, width=7cm,angle=0]{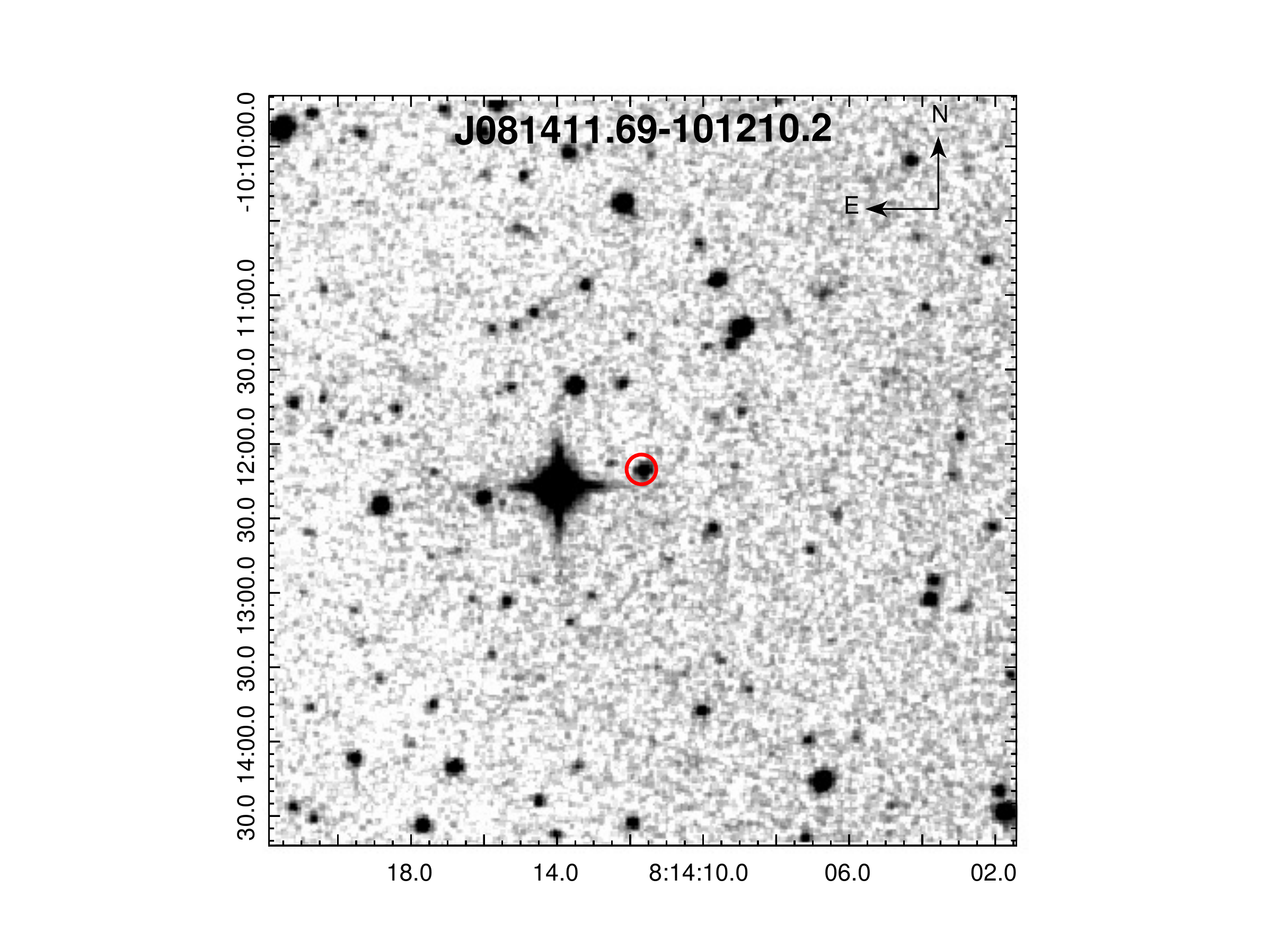} \\
\end{array}$
\end{center}
\caption{(Left panel) Optical spectrum of WISE J081411.69-101210.2 associated with 3FGL J0814.1-1012, in the upper part it is shown the Signal-to-Noise Ratio of the spectrum. (Right panel) The finding chart ( $5'\times 5'$ ) retrieved from the Digital Sky Survey highlighting the location of the counterpart: WISE J081411.69-101210.2 (red circle).}
\label{fig:J08141}
\end{figure*}

\begin{figure*}{}
\begin{center}$
\begin{array}{cc}
\includegraphics[width=\mywidth,angle=0]{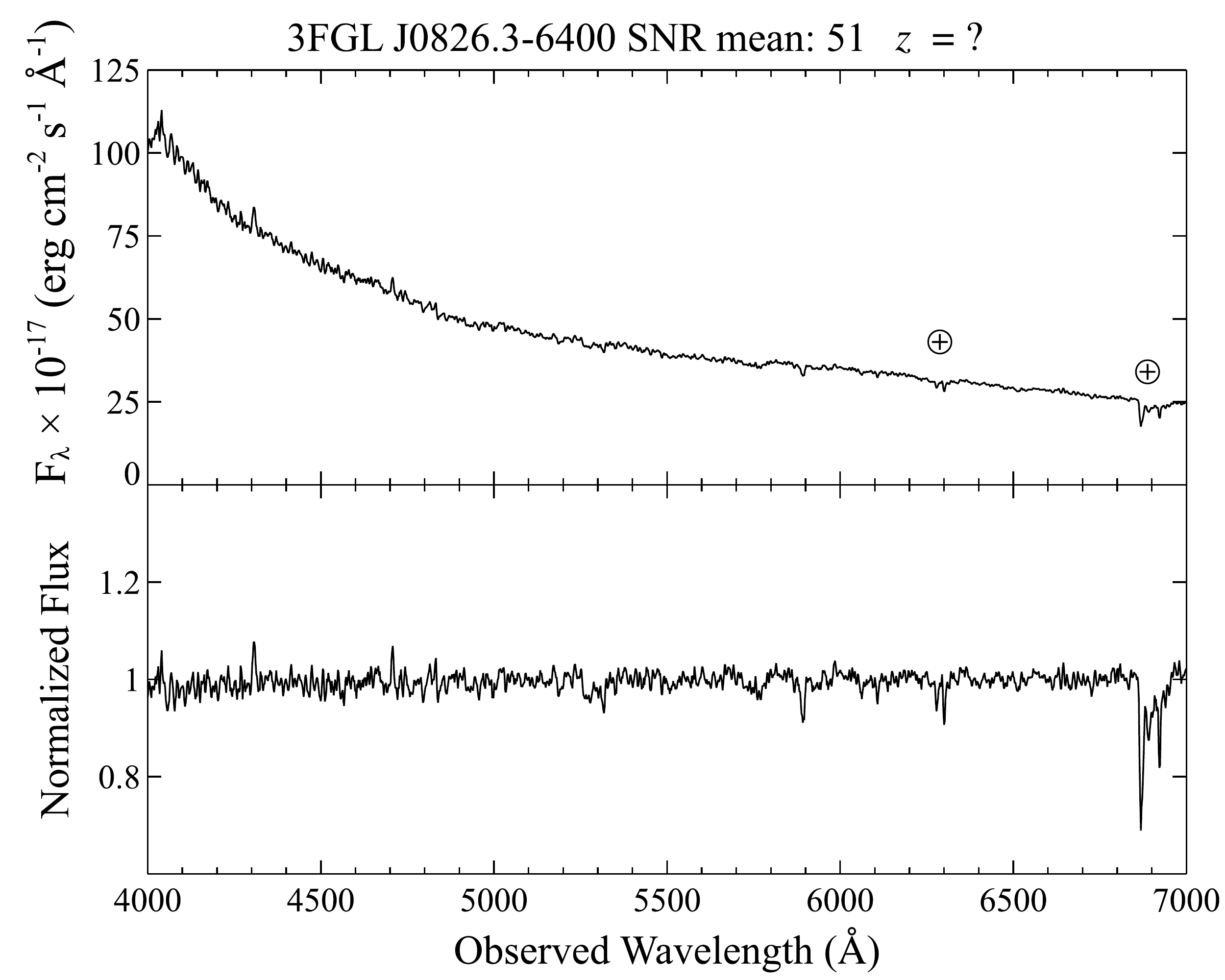} &
\includegraphics[trim=4cm 0cm 4cm 0cm, clip=true, width=7cm,angle=0]{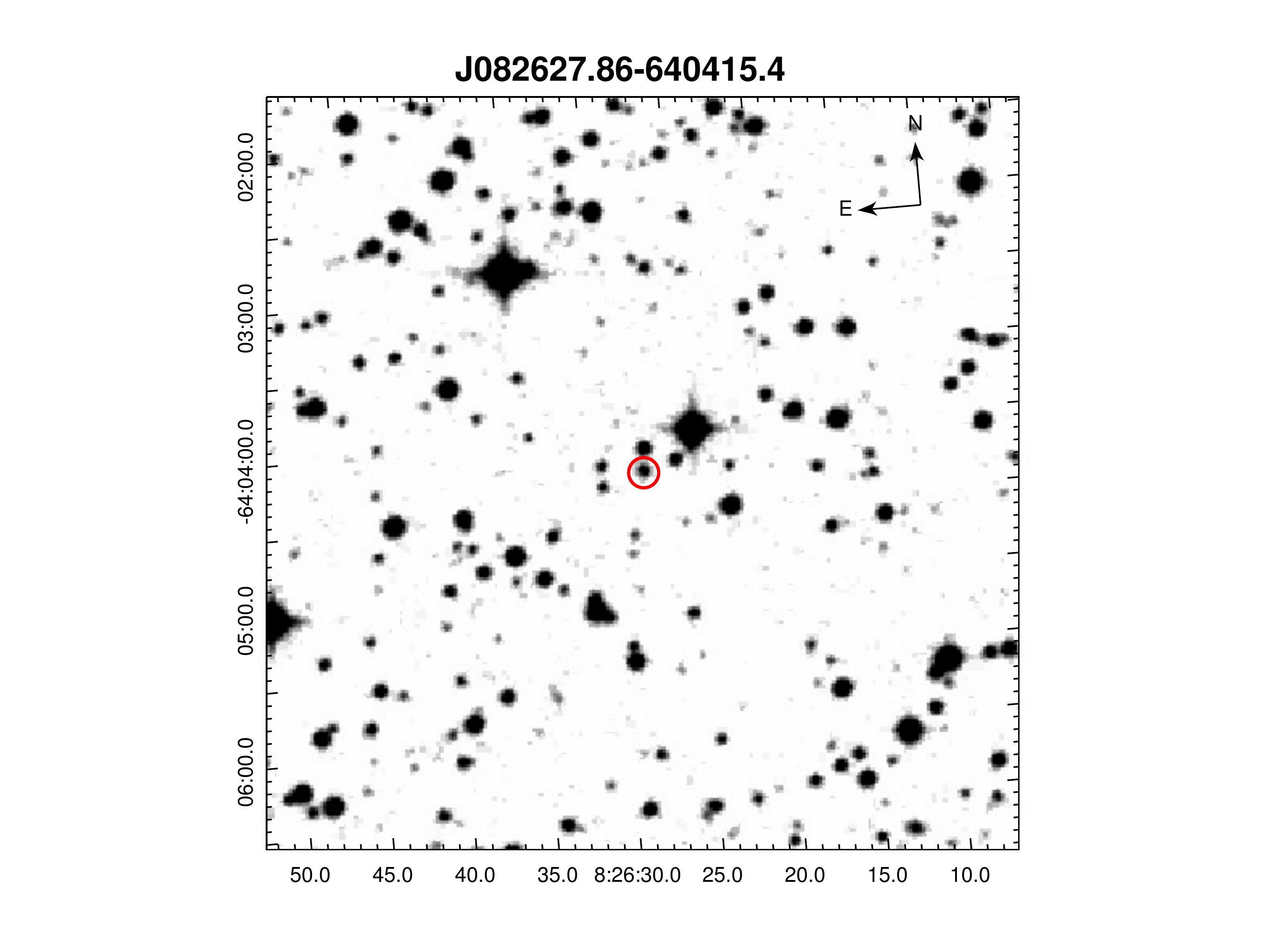} \\
\end{array}$
\end{center}
\caption{(Left panel) Optical spectrum of  WISE J082627.86-640415.4 associated with 3FGL J0826.3-6400. Signal-to-noise ratio is reported in the Figure. (Right panel) The finding chart ( $5'\times 5'$ ) retrieved from the Digital Sky Survey highlighting the location of the potential source: WISE J082627.86-640415.4 (red circle).}
\label{fig:J0826}
\end{figure*}

\begin{figure*}{}
\begin{center}$
\begin{array}{cc}
\includegraphics[width=\mywidth,angle=0]{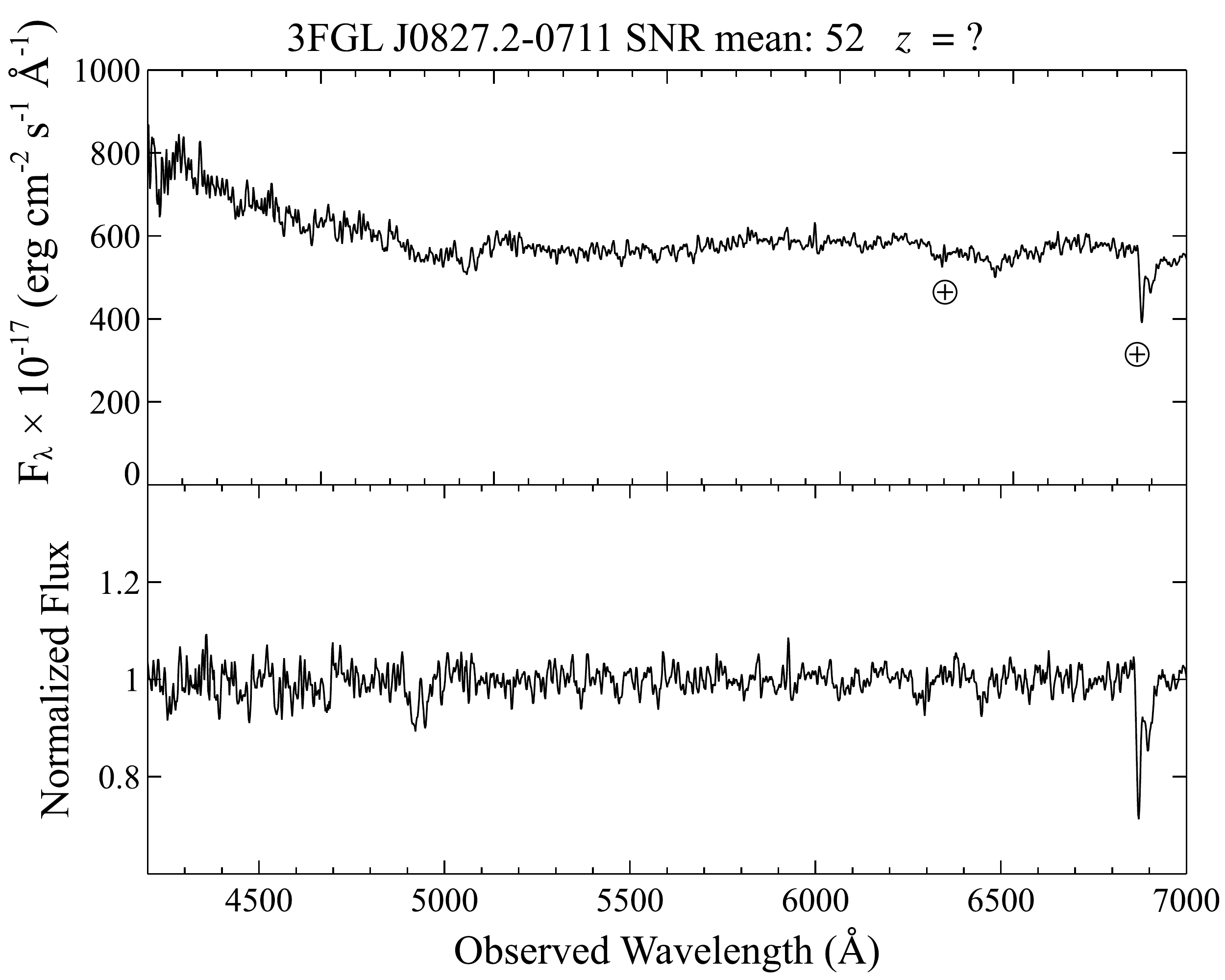} &
\includegraphics[trim=4cm 0cm 4cm 0cm, clip=true, width=7cm,angle=0]{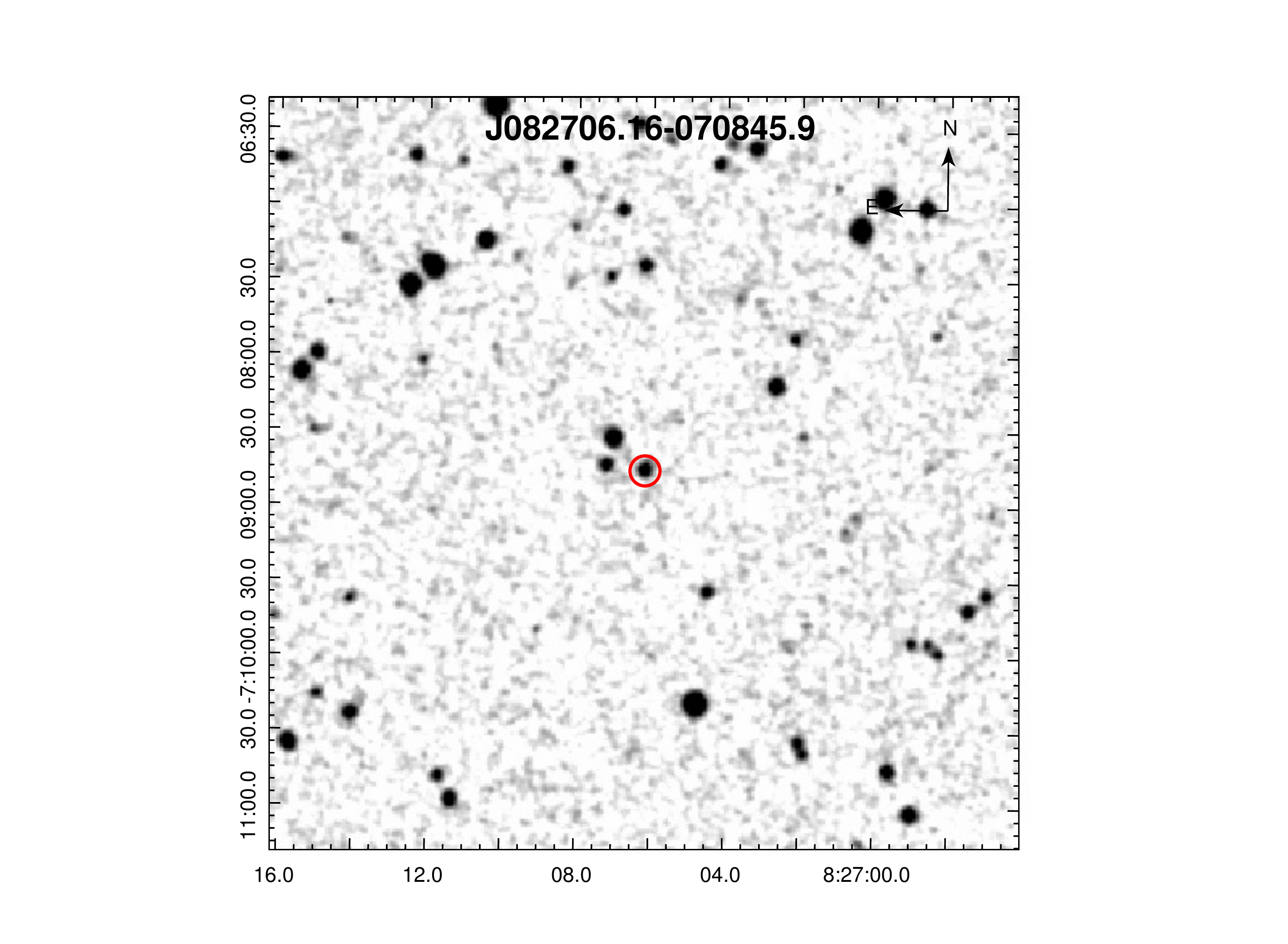} \\
\end{array}$
\end{center}
\caption{(Left panel) Optical spectrum of  WISE J082706.16-070845.9 associated with 3FGL J0827.2-0711, in the upper part it is shown the Signal-to-Noise Ratio of the spectrum. (Right panel) The finding chart ( $5'\times 5'$ ) retrieved from the Digital Sky Survey highlighting the location of the counterpart: WISE J082706.16-070845.9 (red circle).}
\label{fig:J0827}
\end{figure*}

\begin{figure*}{}
\begin{center}$
\begin{array}{cc}
\includegraphics[width=\mywidth,angle=0]{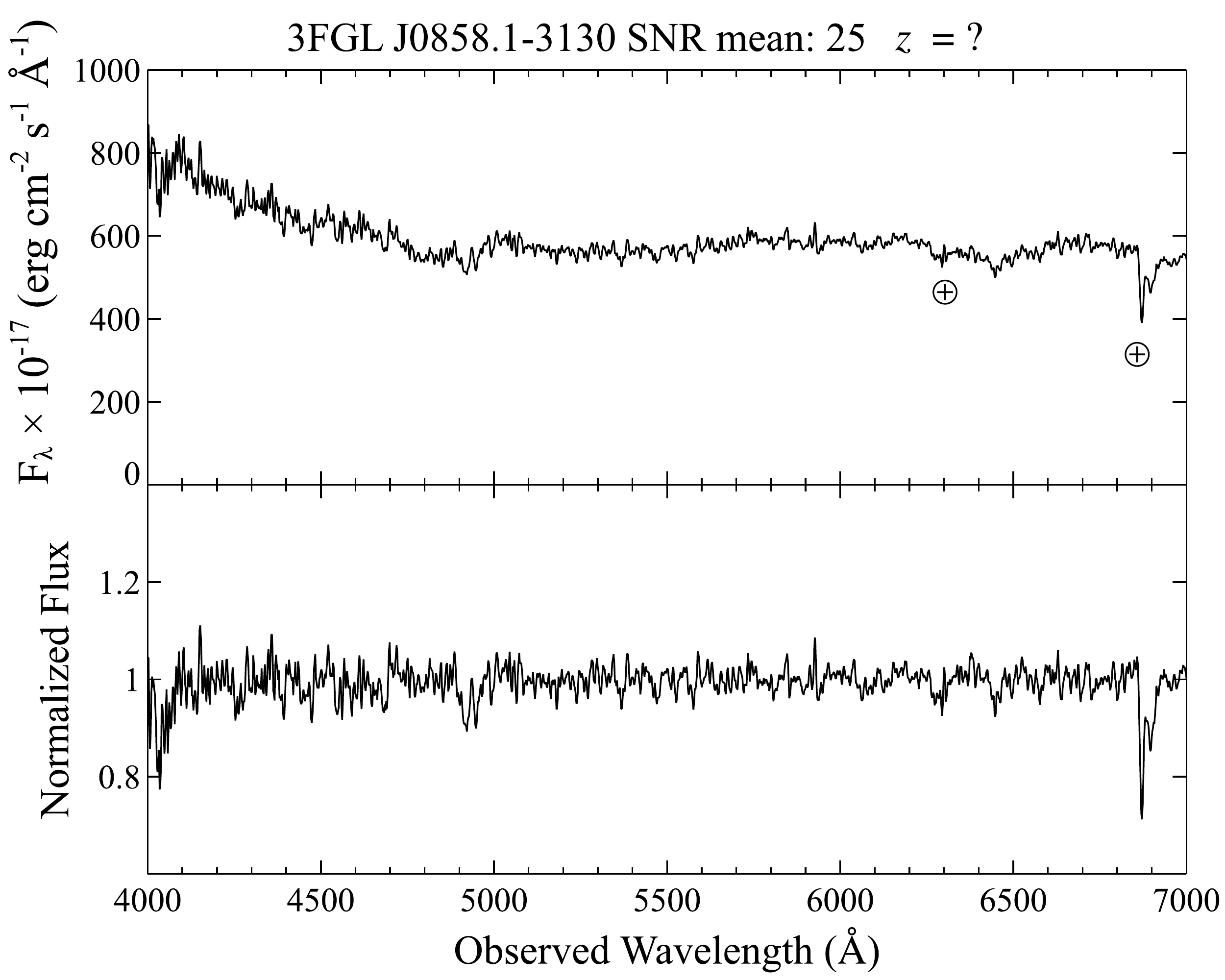} &
\includegraphics[trim=4cm 0cm 4cm 0cm, clip=true, width=7cm,angle=0]{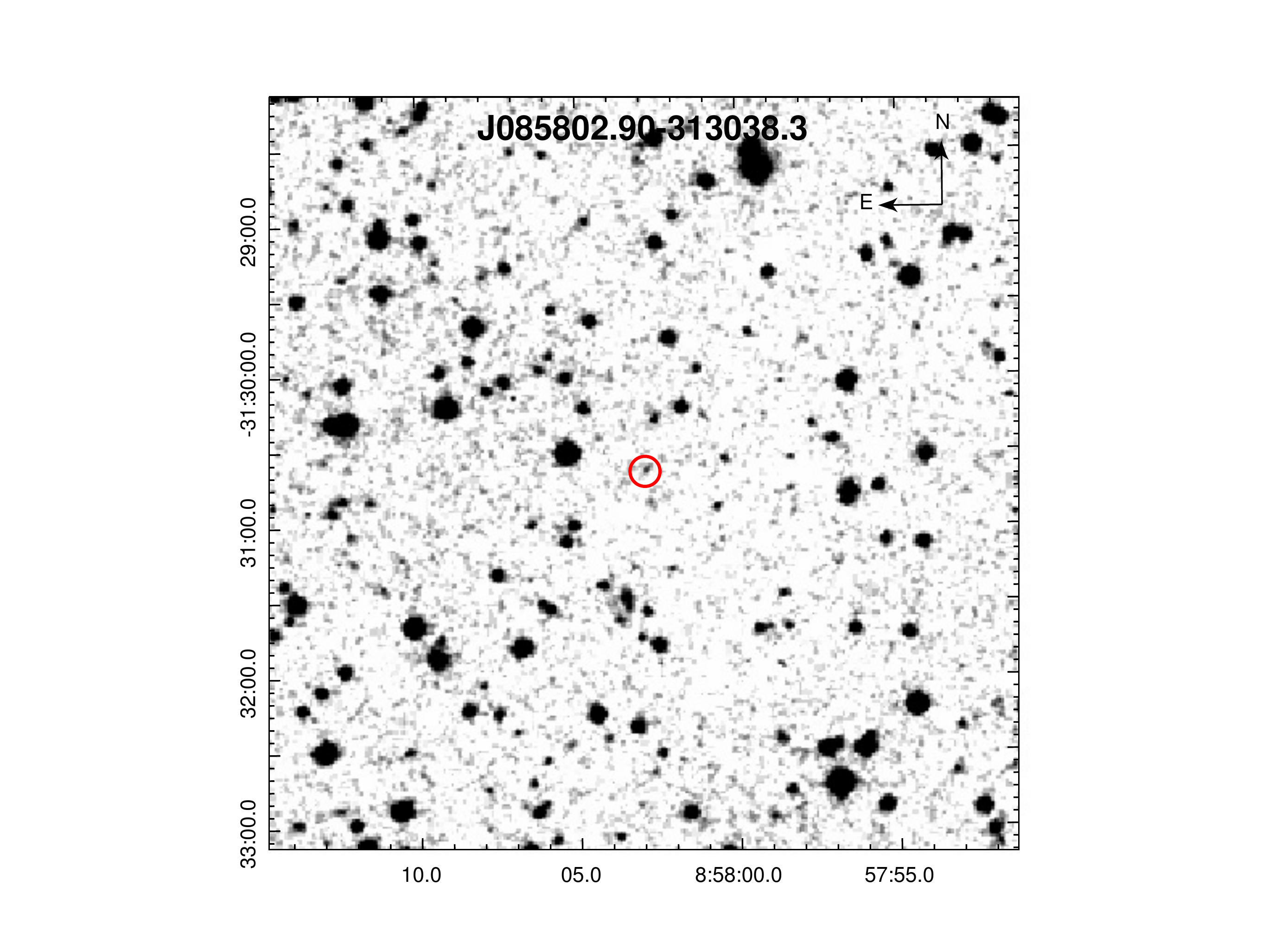} \\
\end{array}$
\end{center}
\caption{(Left panel) Optical spectrum of  WISE J085802.90-313038.3 associated with 3FGL J0858.1-3130, in the upper part it is shown the Signal-to-Noise Ratio of the spectrum. (Right panel) The finding chart ( $5'\times 5'$ ) retrieved from the Digital Sky Survey highlighting the location of the counterpart: WISE J085802.90-313038.3 (red circle).}
\label{fig:J0858}
\end{figure*}

\begin{figure*}{}
\begin{center}$
\begin{array}{cc}
\includegraphics[width=\mywidth,angle=0]{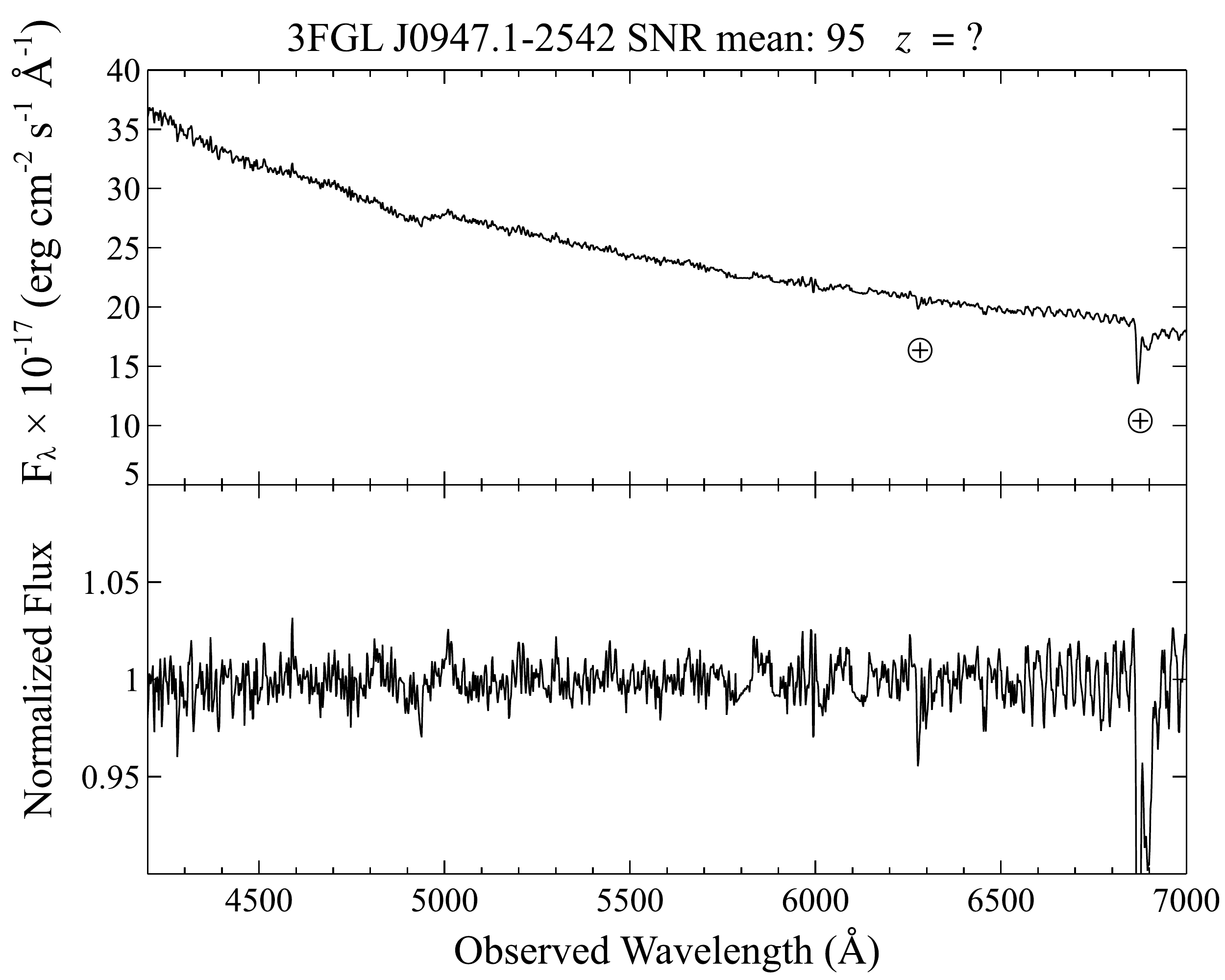} &
\includegraphics[trim=4cm 0cm 4cm 0cm, clip=true, width=7cm,angle=0]{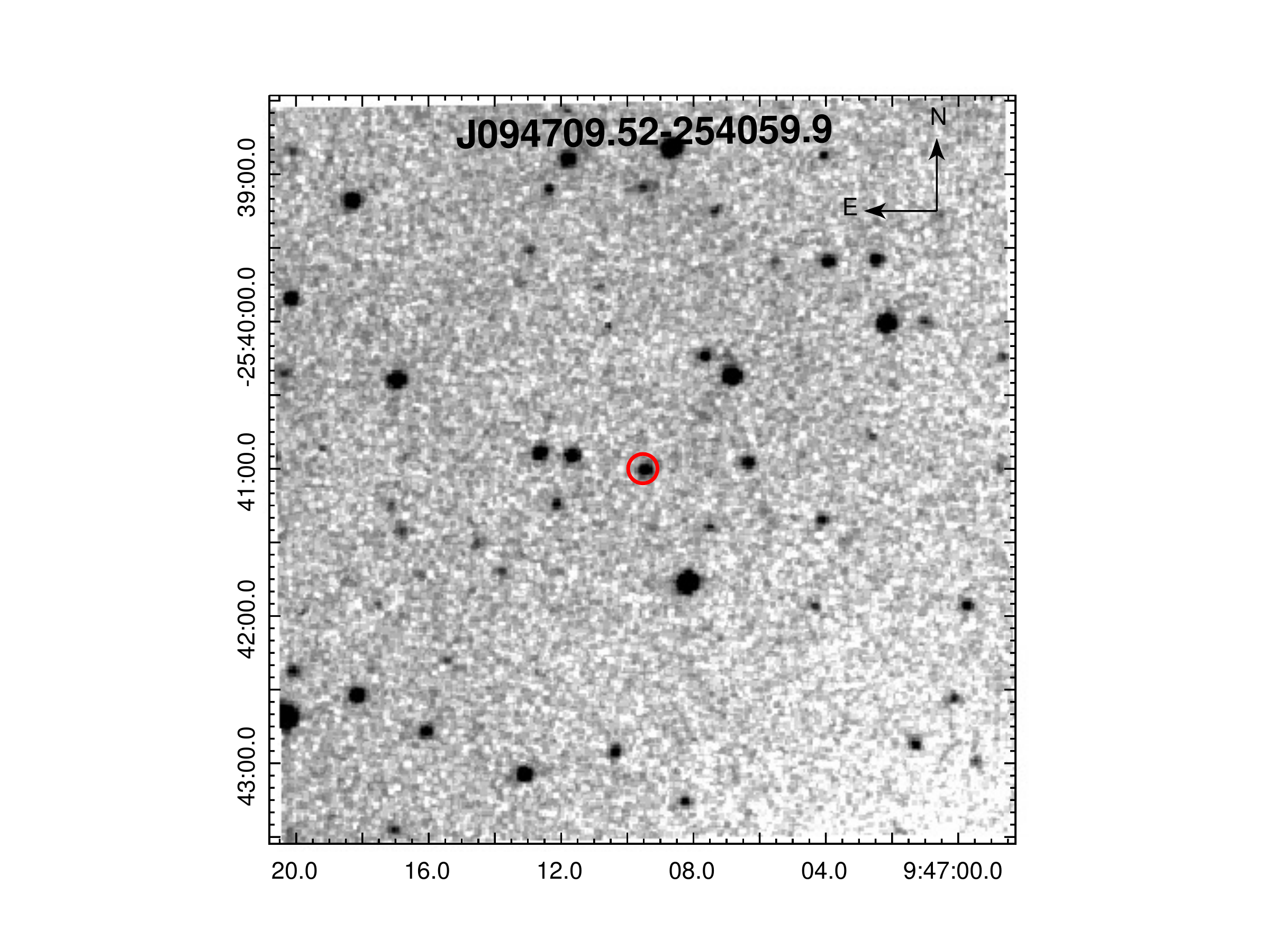} \\
\end{array}$
\end{center}
\caption{(Left panel) Optical spectrum of  WISE J094709.52-254059.9 associated with 3FGL J0947.1-2542, in the upper part it is shown the Signal-to-Noise Ratio of the spectrum. (Right panel) The finding chart ( $5'\times 5'$ ) retrieved from the Digital Sky Survey highlighting the location of the counterpart: WISE J094709.52-254059.9 (red circle).}
\label{fig:J0947}
\end{figure*}

\begin{figure*}{}
\begin{center}$
\begin{array}{cc}
\includegraphics[width=\mywidth,angle=0]{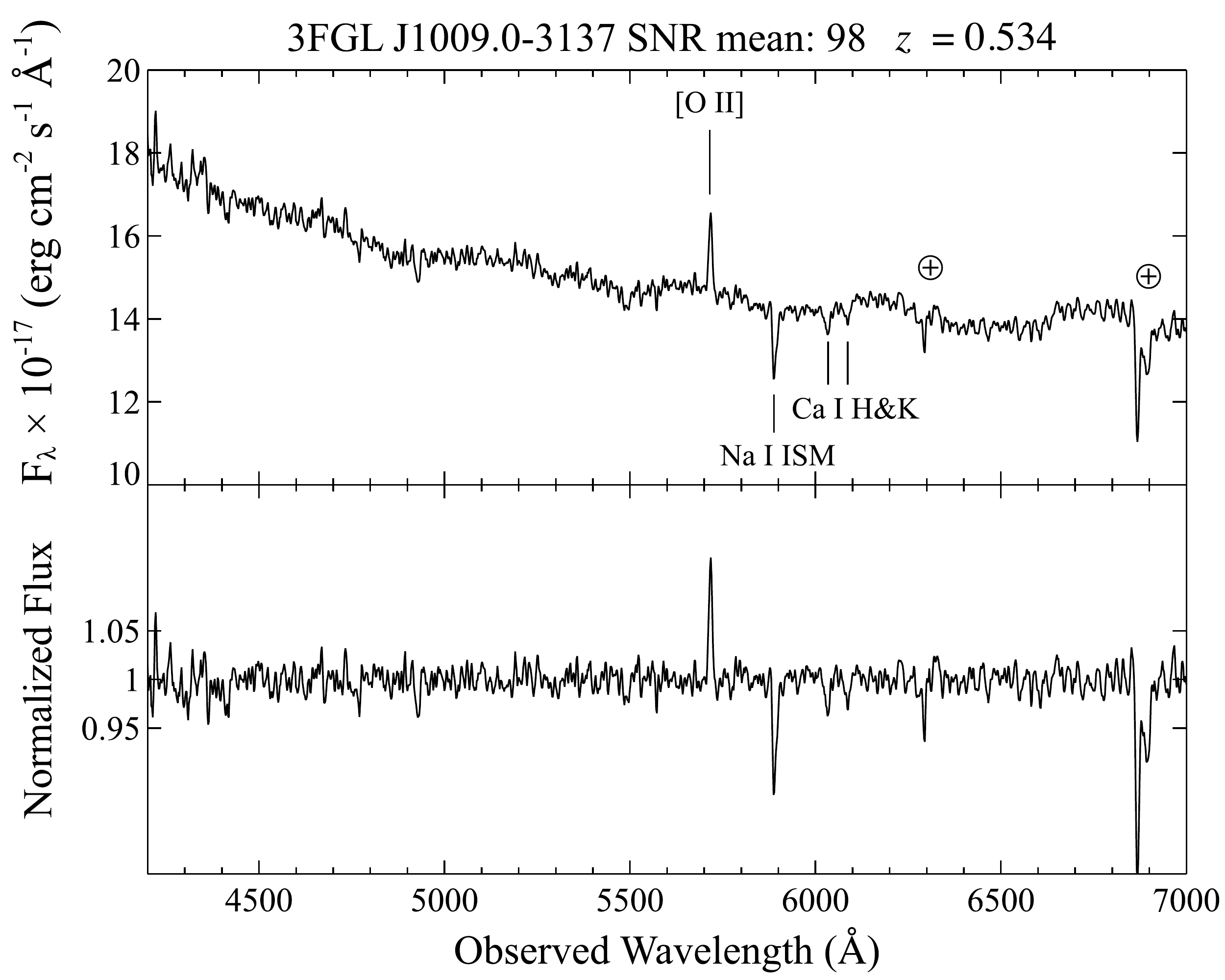} &
\includegraphics[trim=4cm 0cm 4cm 0cm, clip=true, width=7cm,angle=0]{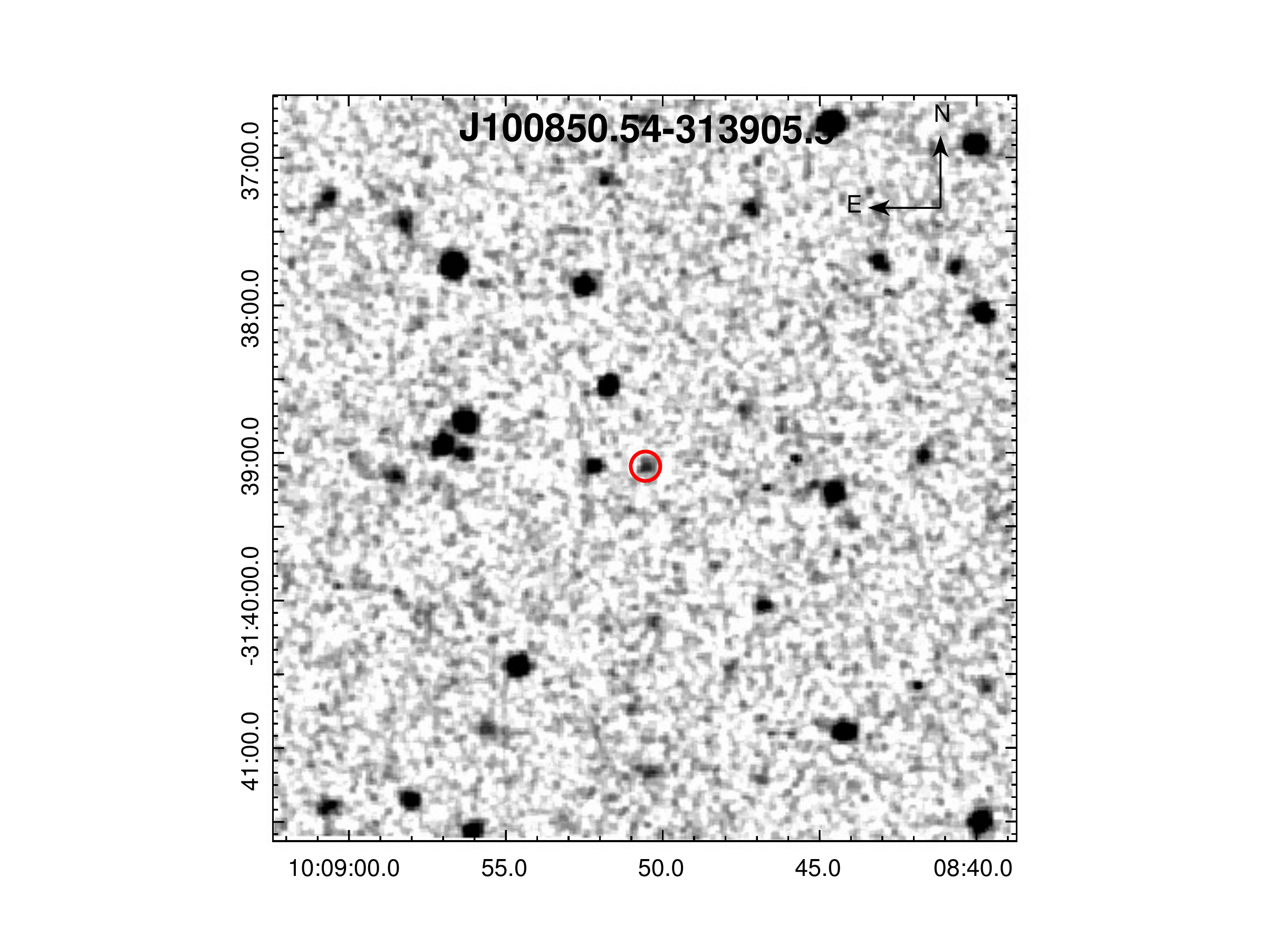} \\
\end{array}$
\end{center}
\caption{(Left panel) Optical spectrum of  WISE J100850.54-313905.5 associated with 3FGL J1009.0-3137, in the upper part it is shown the Signal-to-Noise Ratio of the spectrum. (Right panel) The finding chart ( $5'\times 5'$ ) retrieved from the Digital Sky Survey highlighting the location of the counterpart: WISE J100850.54-313905.5 (red circle).}
\label{fig:J1009}
\end{figure*}

\begin{figure*}{}
\begin{center}$
\begin{array}{cc}
\includegraphics[width=\mywidth,angle=0]{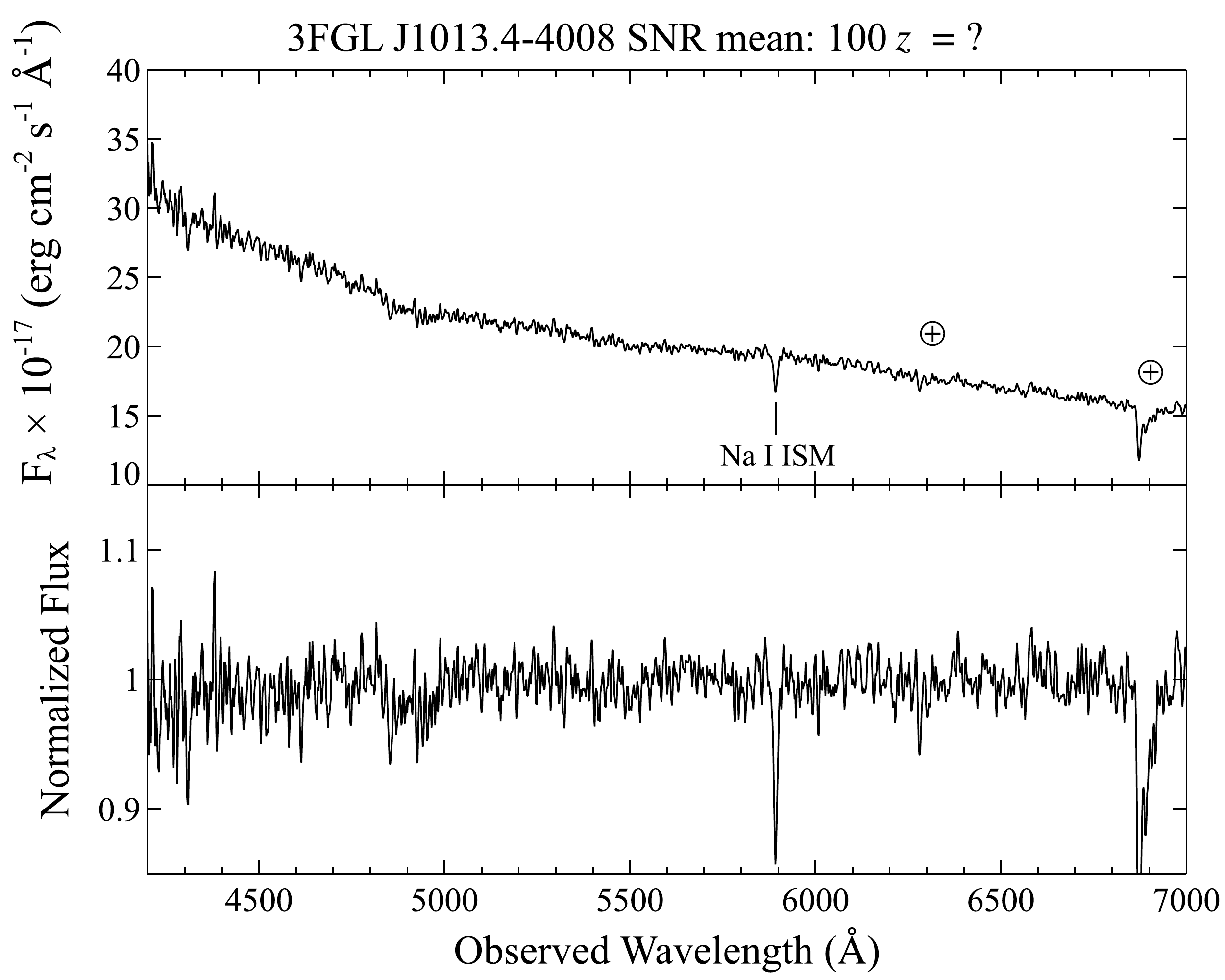} &
\includegraphics[trim=4cm 0cm 4cm 0cm, clip=true, width=7cm,angle=0]{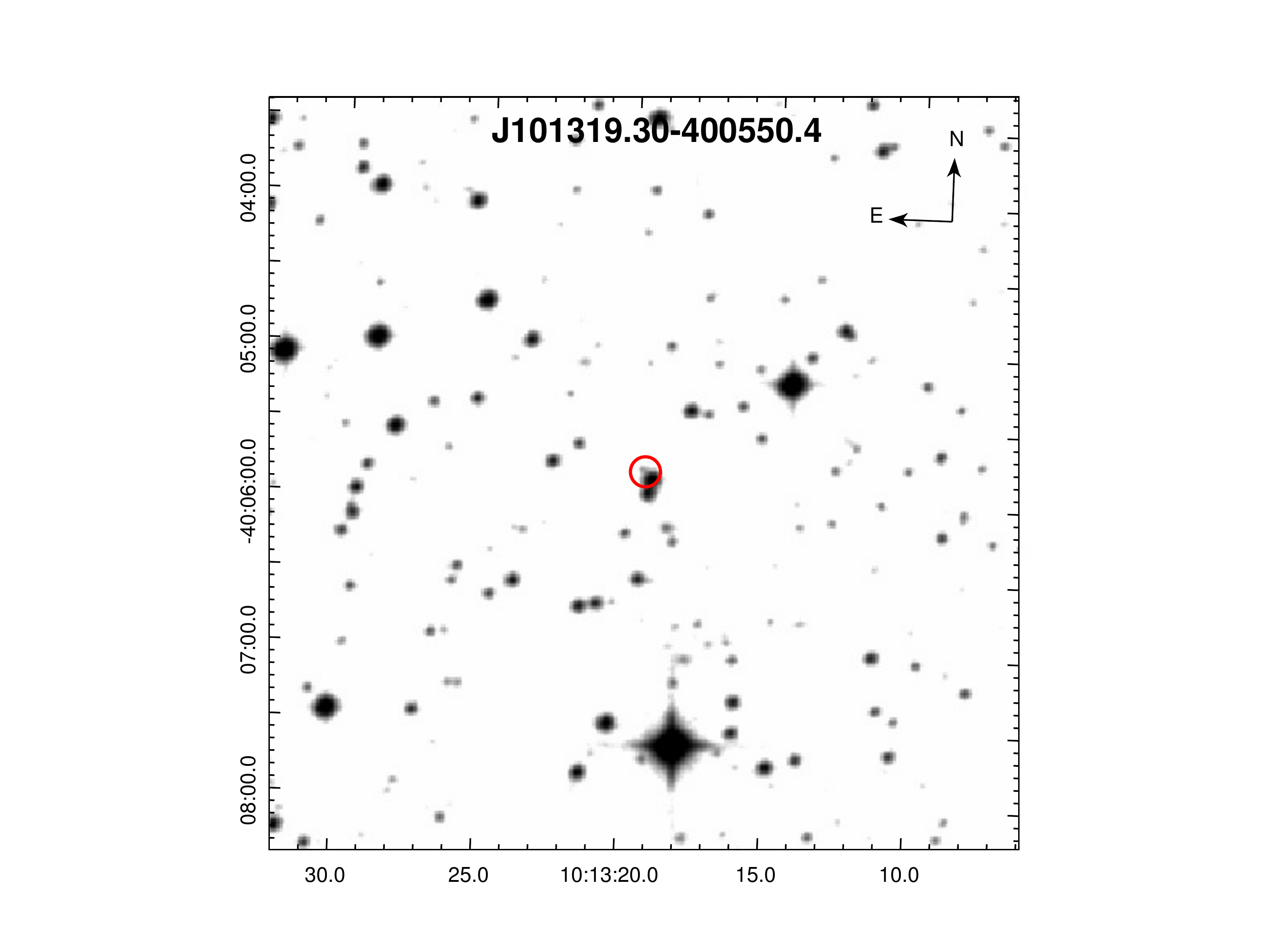} \\
\end{array}$
\end{center}
\caption{(Left panel) Optical spectrum of  WISE J101319.30-400550.4 associated with 3FGL J1013.4-4008. Signal-to-noise ratio is reported in the Figure. (Right panel) The finding chart ( $5'\times 5'$ ) retrieved from the Digital Sky Survey highlighting the location of the potential source: WISE J101319.30-400550.4 (red circle).}
\label{fig:J1013}
\end{figure*}

\begin{figure*}{}
\begin{center}$
\begin{array}{cc}
\includegraphics[width=\mywidth,angle=0]{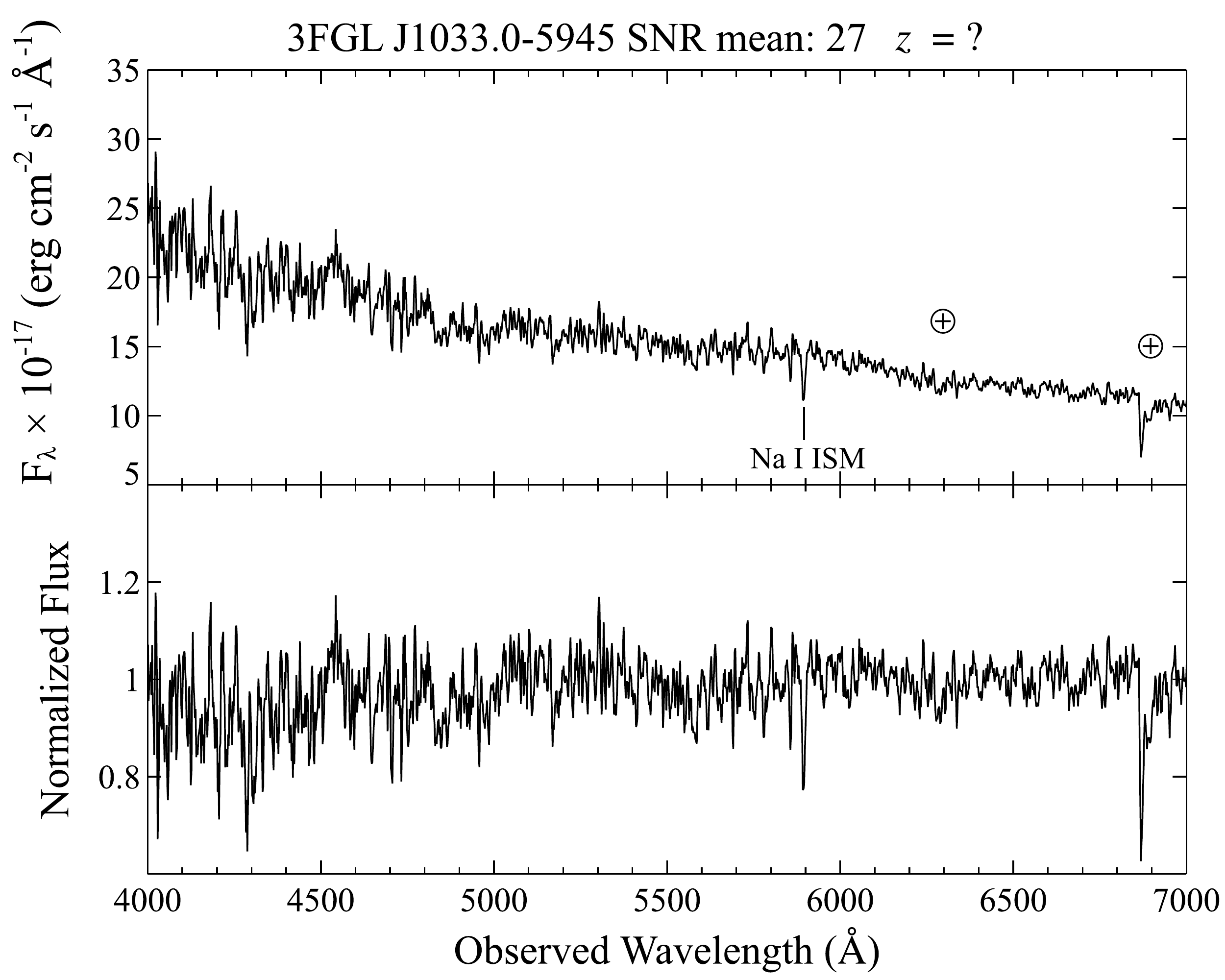} &
\includegraphics[trim=4cm 0cm 4cm 0cm, clip=true, width=7cm,angle=0]{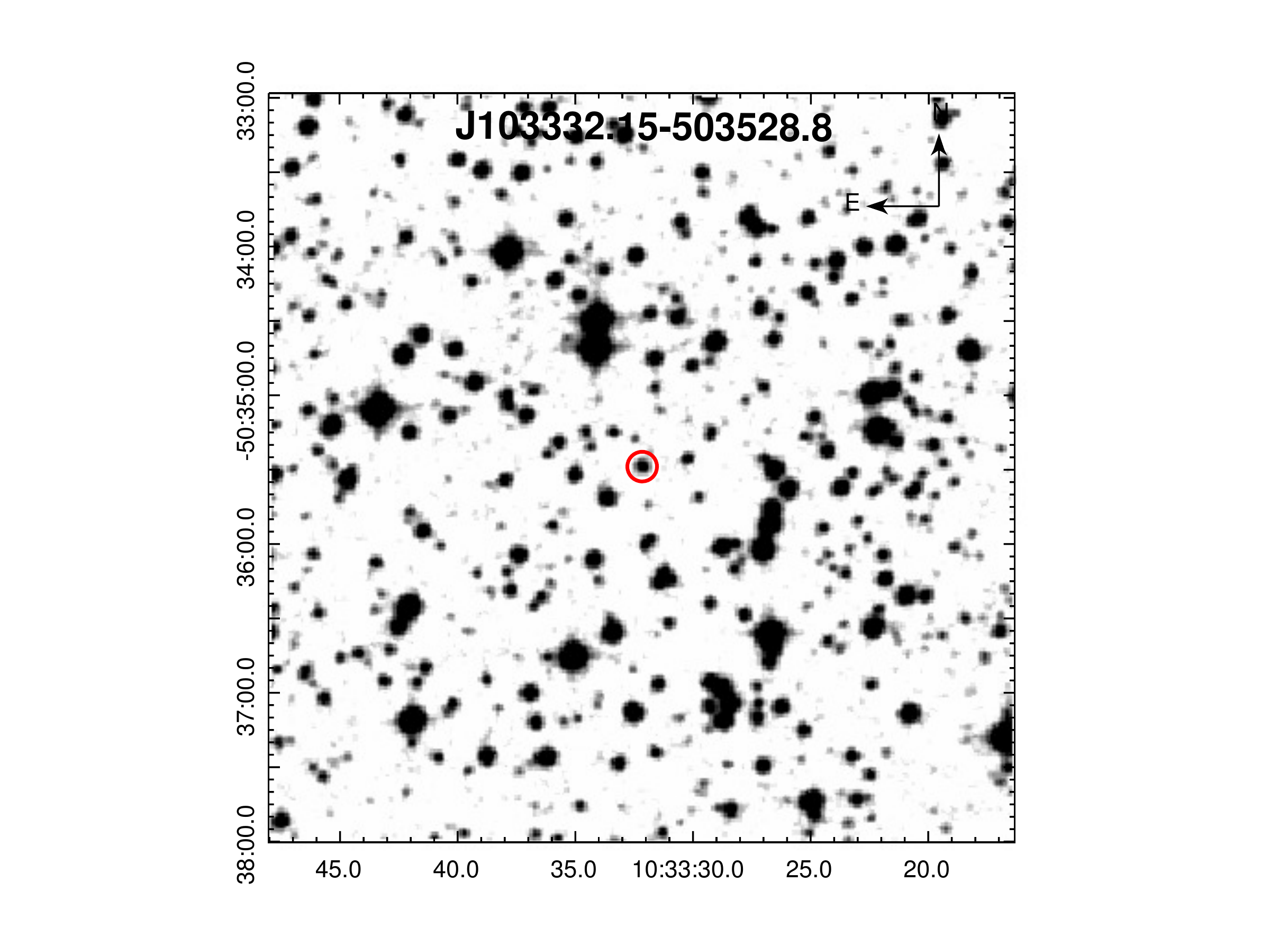} \\
\end{array}$
\end{center}
\caption{(Left panel) Optical spectrum of  WISE J103332.15-503528.8 associated with 3FGL J1033.0-5945, in the upper part it is shown the Signal-to-Noise Ratio of the spectrum. (Right panel) The finding chart ( $5'\times 5'$ ) retrieved from the Digital Sky Survey highlighting the location of the counterpart: WISE J103332.15-503528.8 (red circle).}
\label{fig:J1033}
\end{figure*}

\begin{figure*}{}
\begin{center}$
\begin{array}{cc}
\includegraphics[width=\mywidth,angle=0]{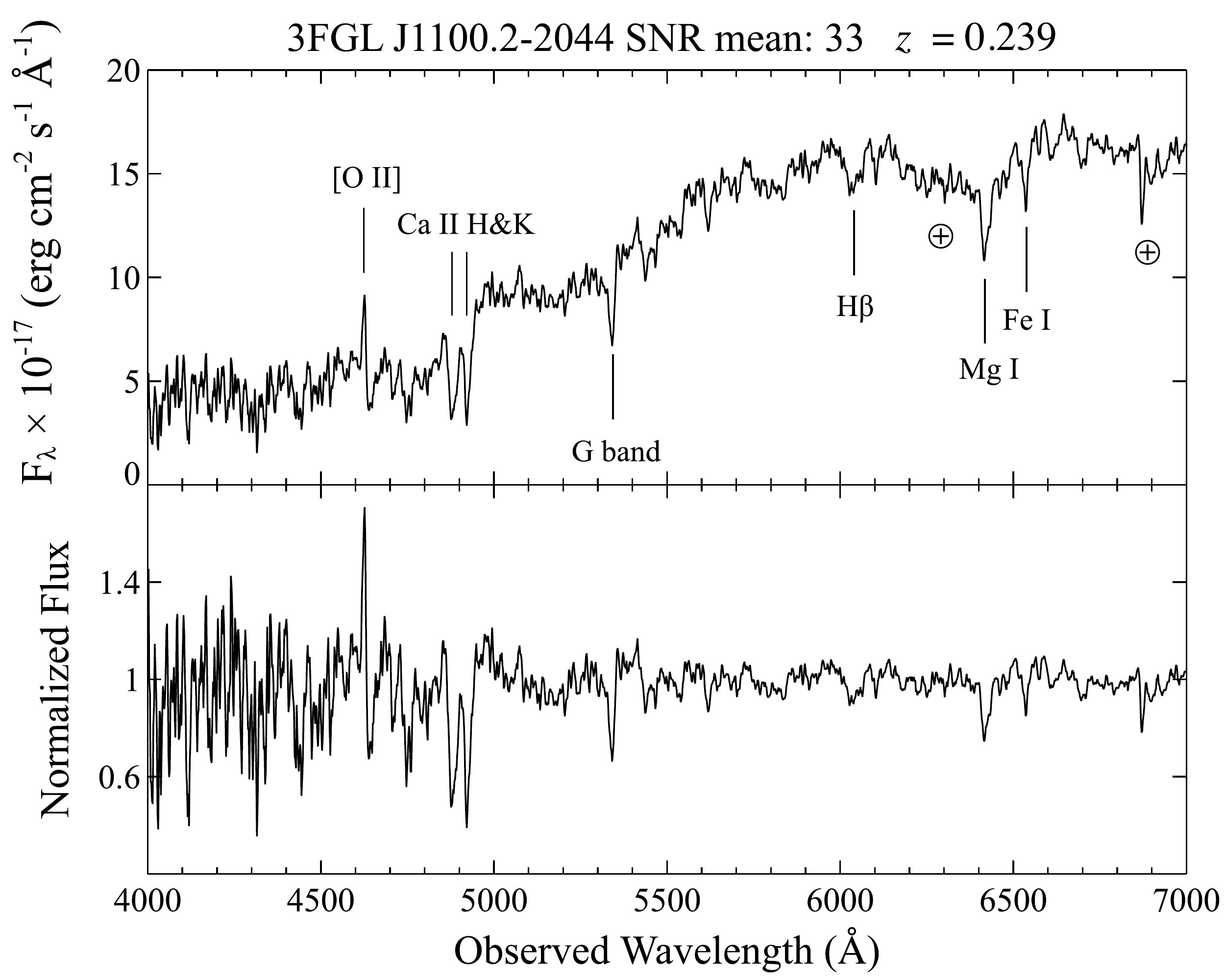} &
\includegraphics[trim=4cm 0cm 4cm 0cm, clip=true, width=7cm,angle=0]{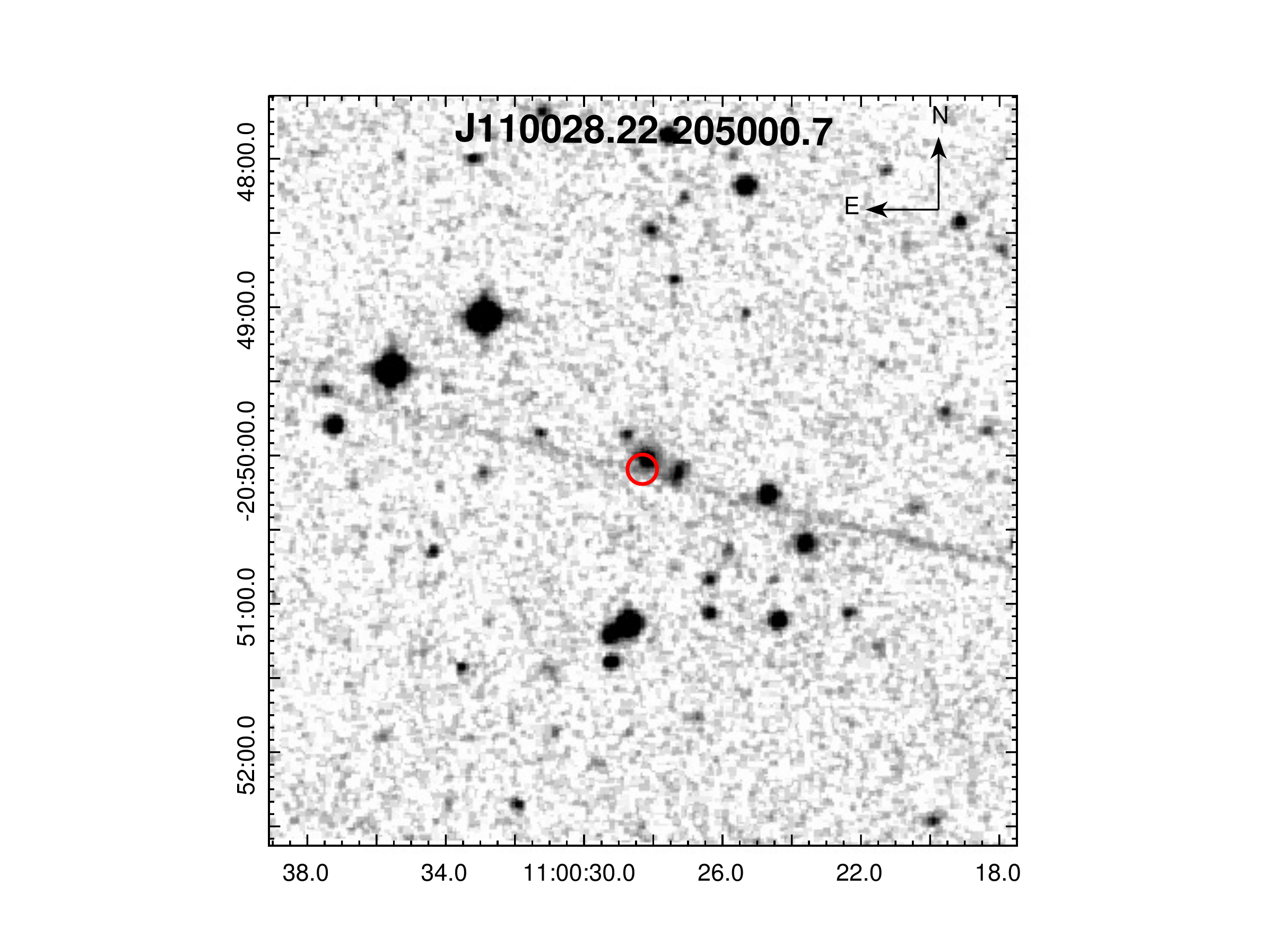} \\
\end{array}$
\end{center}
\caption{(Left panel) Optical spectrum of  WISE J110028.22-205000.7 associated with 3FGL J1100.2-2044. Signal-to-noise ratio is reported in the Figure. (Right panel) The finding chart ( $5'\times 5'$ ) retrieved from the Digital Sky Survey highlighting the location of the potential source: WISE J110028.22-205000.7 (red circle).}
\label{fig:J1100}
\end{figure*}

\begin{figure*}{}
\begin{center}$
\begin{array}{cc}
\includegraphics[width=\mywidth,angle=0]{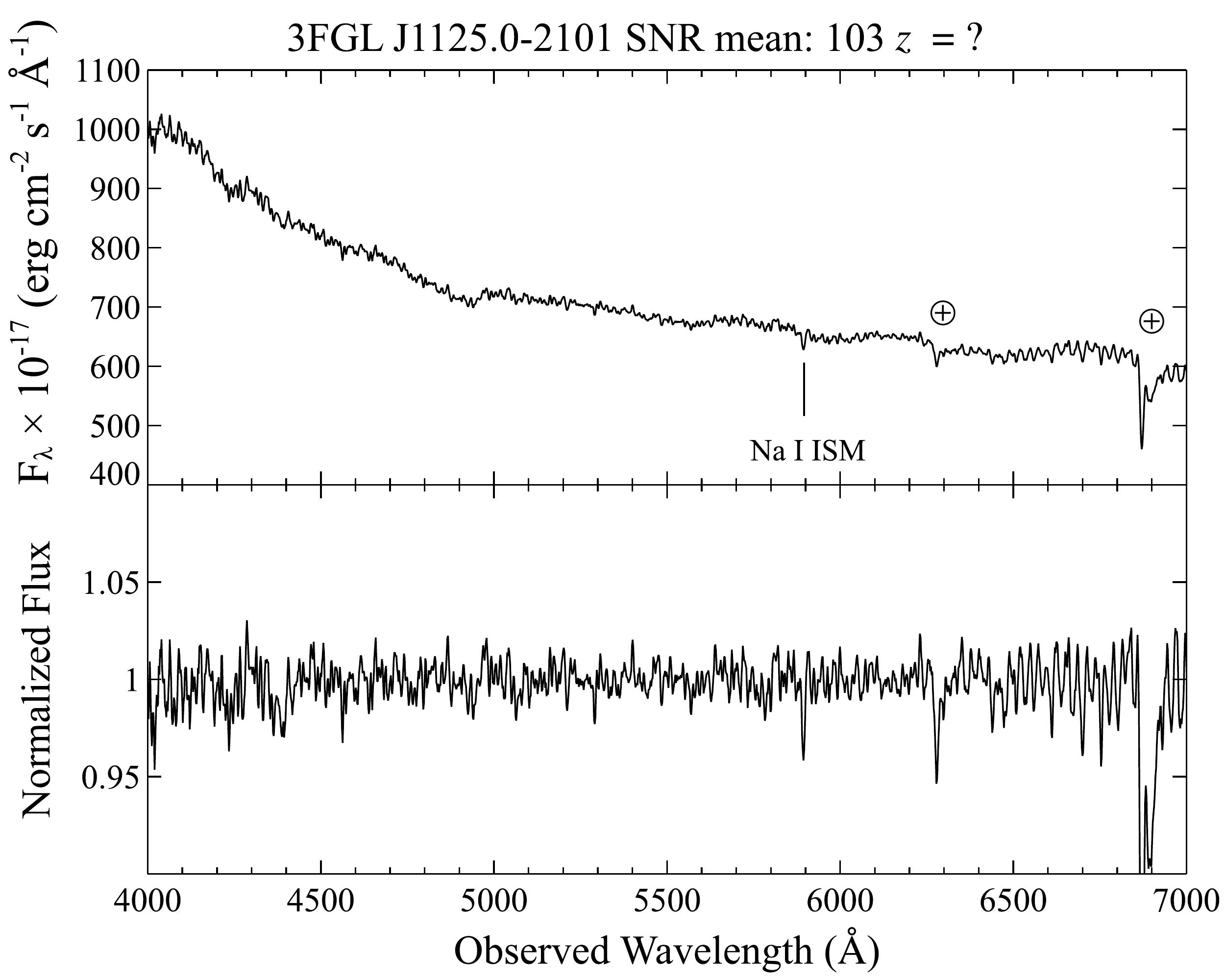} &
\includegraphics[trim=4cm 0cm 4cm 0cm, clip=true, width=7cm,angle=0]{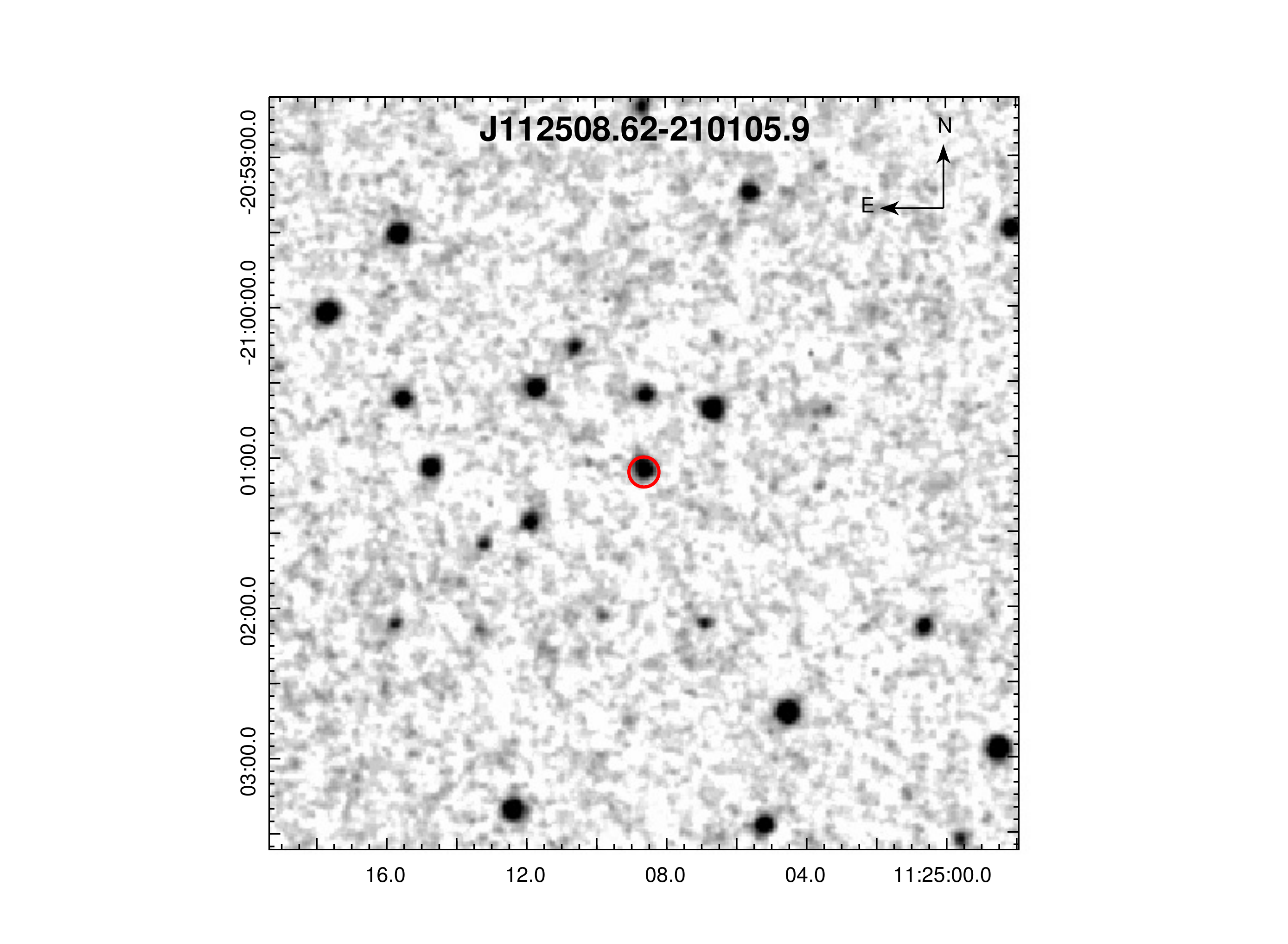} \\
\end{array}$
\end{center}
\caption{(Left panel) Optical spectrum of  WISE J112508.62-210105.9 associated with 3FGL J1125.0-2101, in the upper part it is shown the Signal-to-Noise Ratio of the spectrum. (Right panel) The finding chart ( $5'\times 5'$ ) retrieved from the Digital Sky Survey highlighting the location of the counterpart: WISE J112508.62-210105.9 (red circle).}
\label{fig:J1125}
\end{figure*}

\begin{figure*}{}
\begin{center}$
\begin{array}{cc}
\includegraphics[width=\mywidth,angle=0]{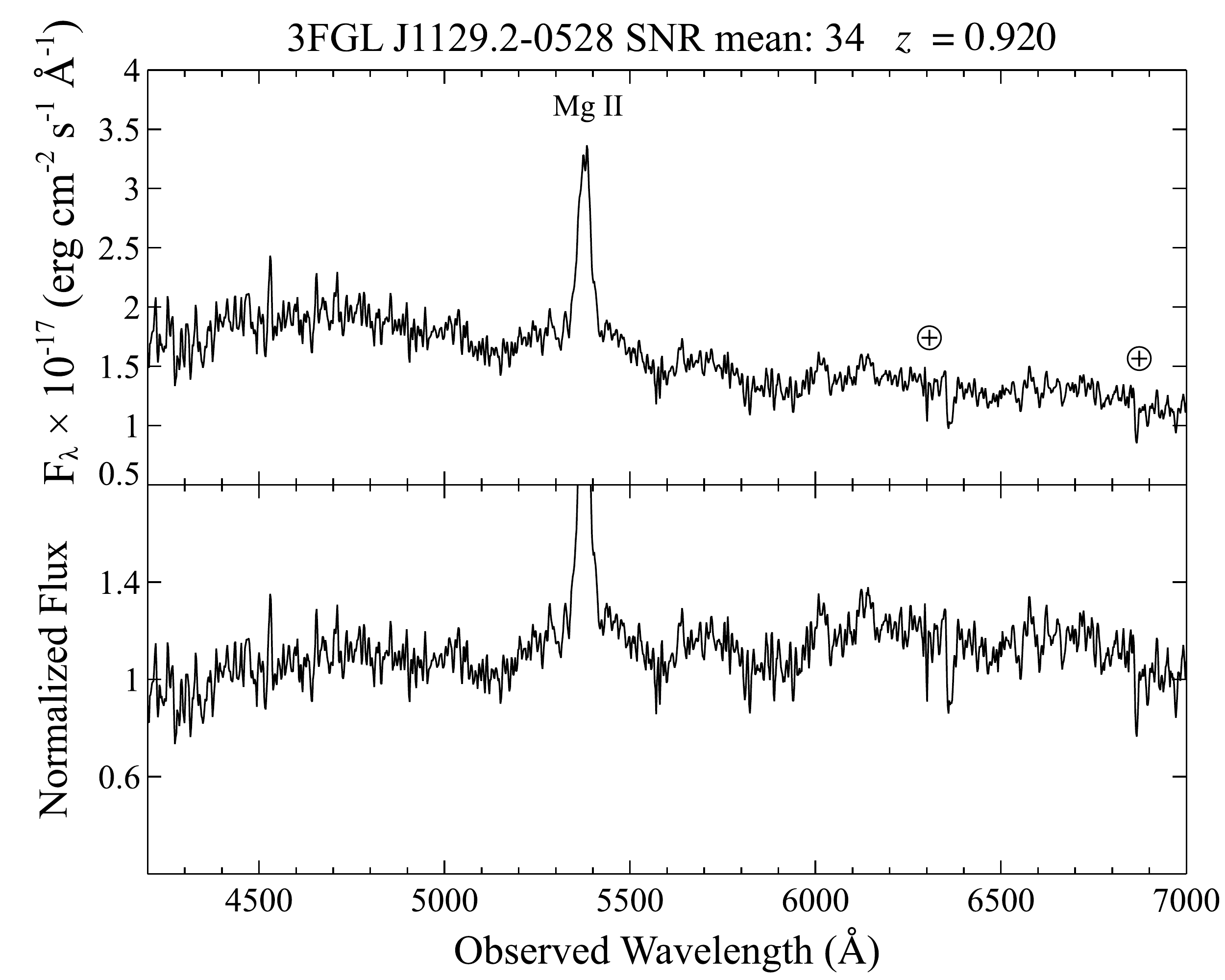} &
\includegraphics[trim=4cm 0cm 4cm 0cm, clip=true, width=7cm,angle=0]{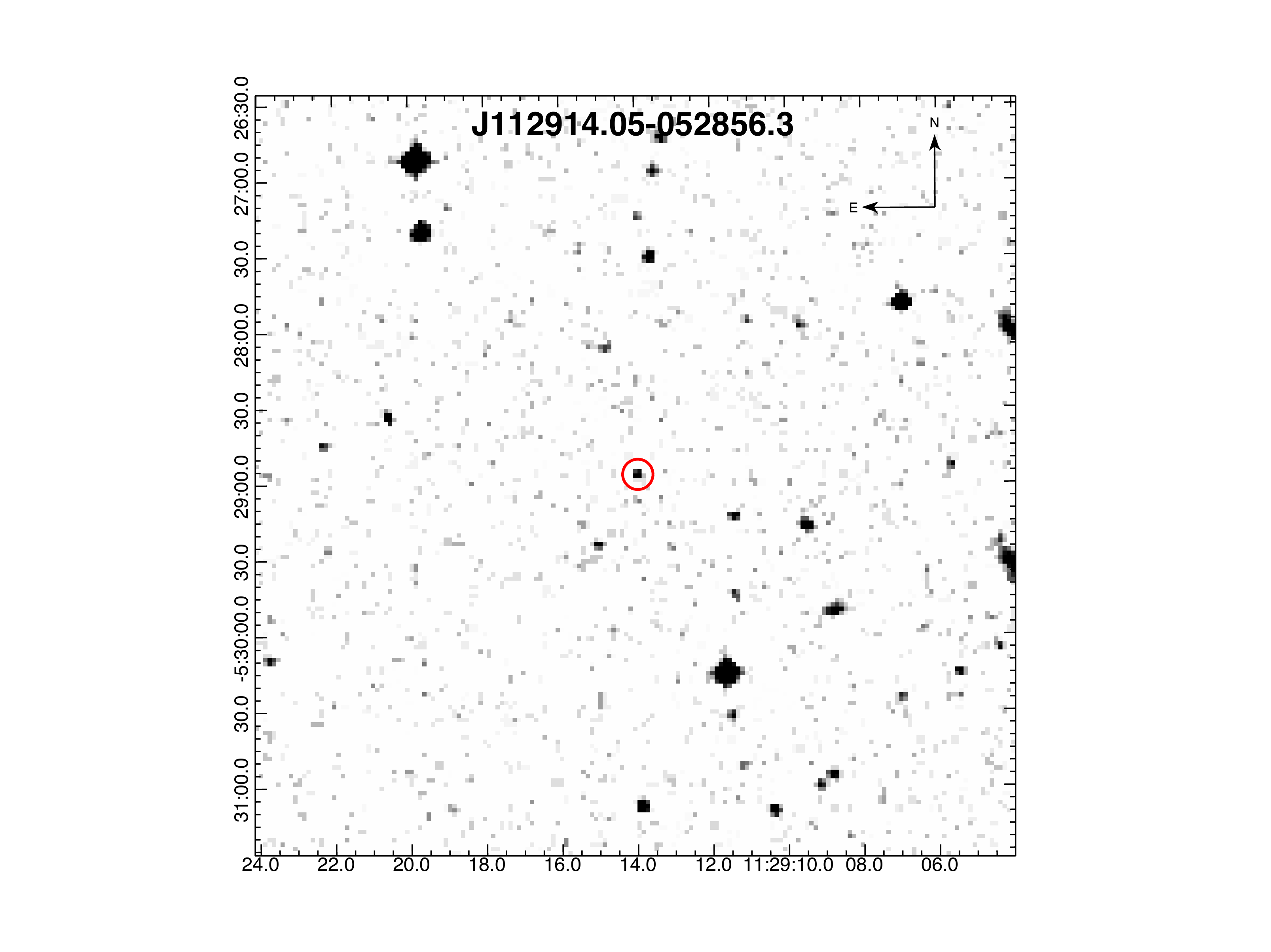} \\
\end{array}$
\end{center}
\caption{(Left panel) Optical spectrum of  WISE J112914.05-052856.3 associated with 3FGL J1129.2-0528. Signal-to-noise ratio is reported in the Figure. (Right panel) The finding chart ( $5'\times 5'$ ) retrieved from the Digital Sky Survey highlighting the location of the potential source: WISE J112914.05-052856.3 (red circle).}
\label{fig:J1129}
\end{figure*}

\begin{figure*}{}
\begin{center}$
\begin{array}{cc}
\includegraphics[width=\mywidth,angle=0]{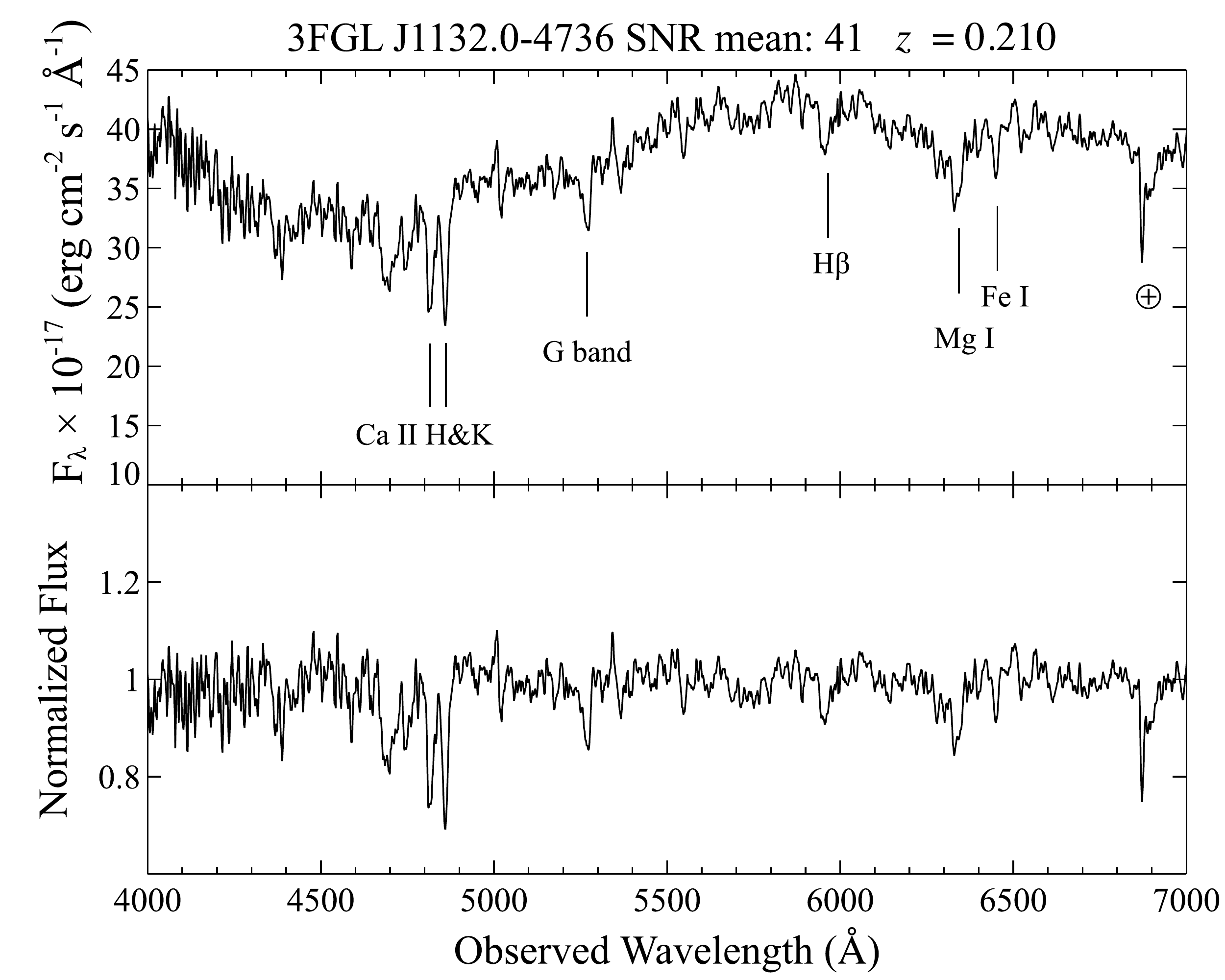} &
\includegraphics[trim=4cm 0cm 4cm 0cm, clip=true, width=7cm,angle=0]{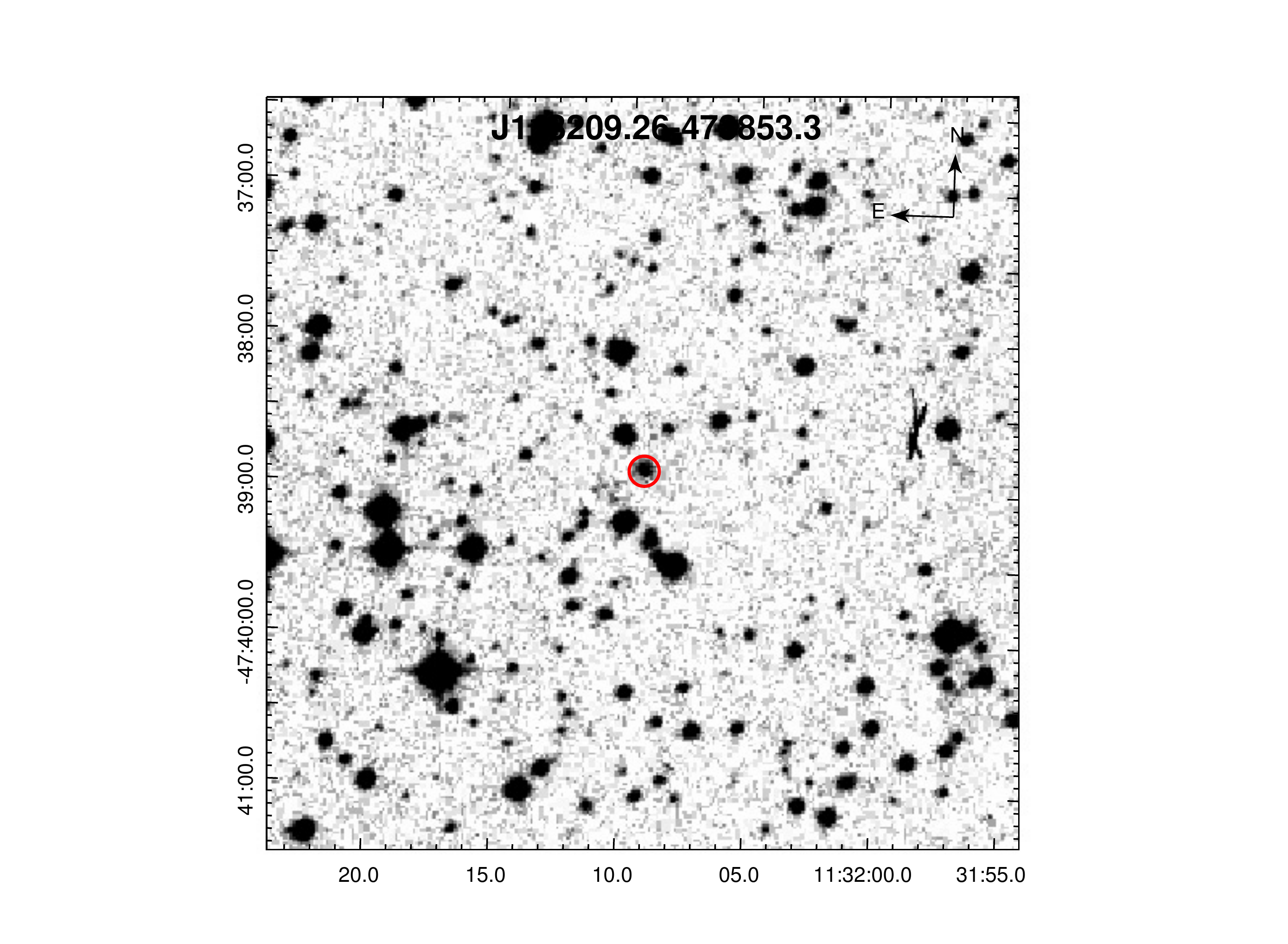} \\
\end{array}$
\end{center}
\caption{(Left panel) Optical spectrum of  WISE J113209.26-473853.3 associated with 3FGL J1132.0-4736. Signal-to-noise ratio is reported in the Figure. (Right panel) The finding chart ( $5'\times 5'$ ) retrieved from the Digital Sky Survey highlighting the location of the potential source: WISE J113209.26-473853.3 (red circle).}
\label{fig:J1132}
\end{figure*}

\begin{figure*}{}
\begin{center}$
\begin{array}{cc}
\includegraphics[width=\mywidth,angle=0]{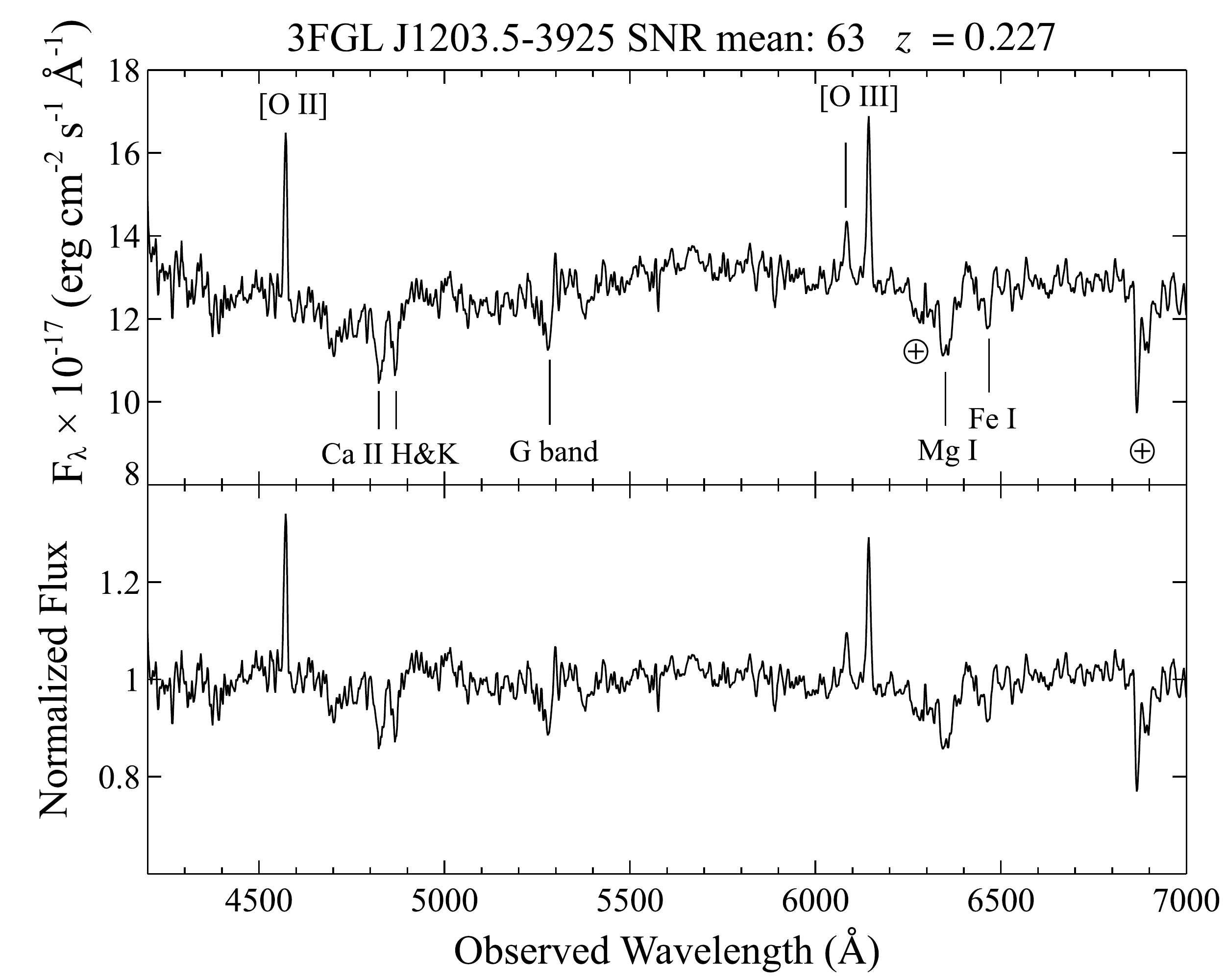} &
\includegraphics[trim=4cm 0cm 4cm 0cm, clip=true, width=7cm,angle=0]{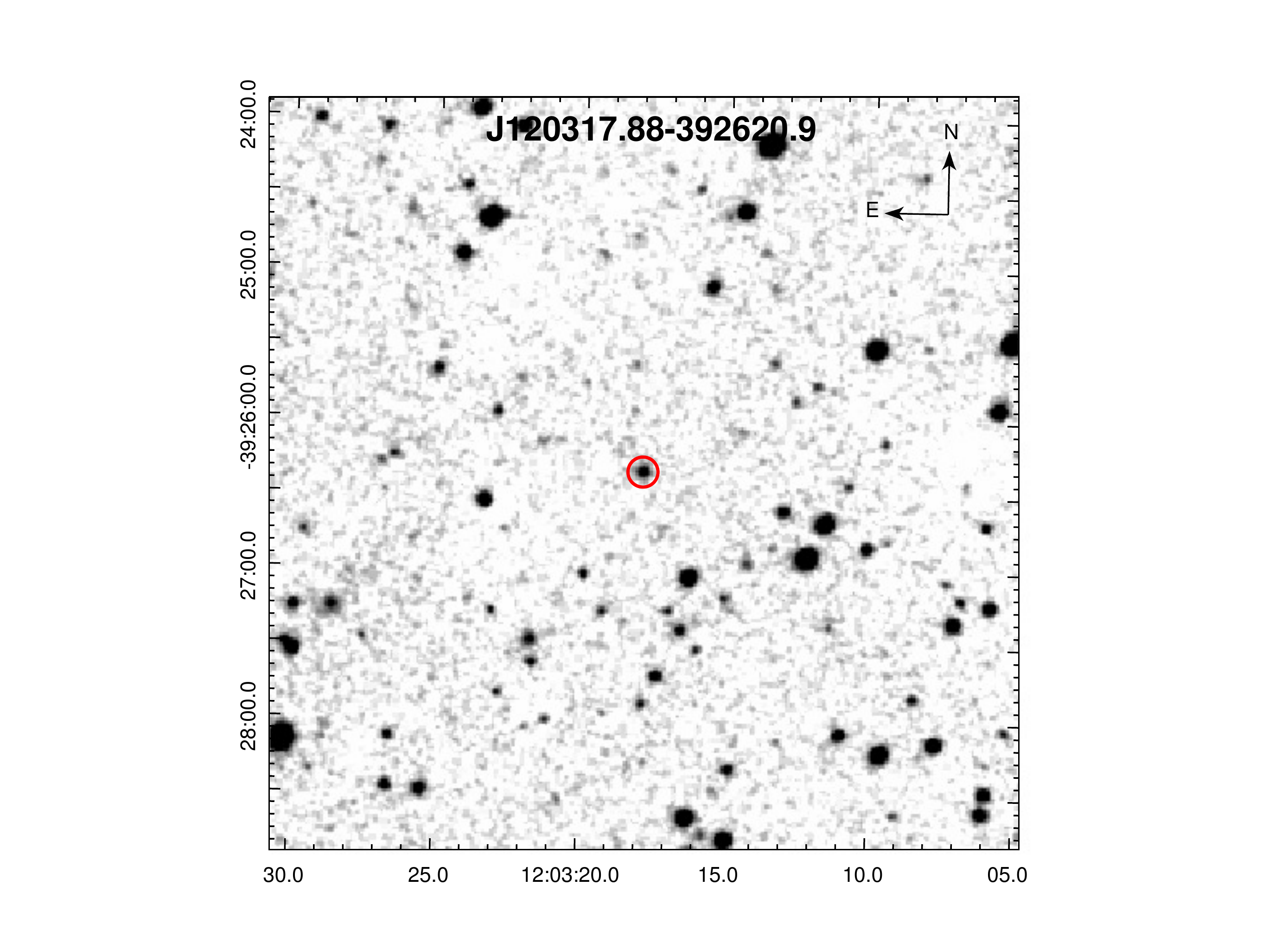} \\
\end{array}$
\end{center}
\caption{(Left panel) Optical spectrum of  WISE J120317.88-392620.9 associated with 3FGL J1203.5-3925, in the upper part it is shown the Signal-to-Noise Ratio of the spectrum. (Right panel) The finding chart ( $5'\times 5'$ ) retrieved from the Digital Sky Survey highlighting the location of the counterpart: WISE J120317.88-392620.9 (red circle).}
\label{fig:J1203}
\end{figure*}

\begin{figure*}{}
\begin{center}$
\begin{array}{cc}
\includegraphics[width=\mywidth,angle=0]{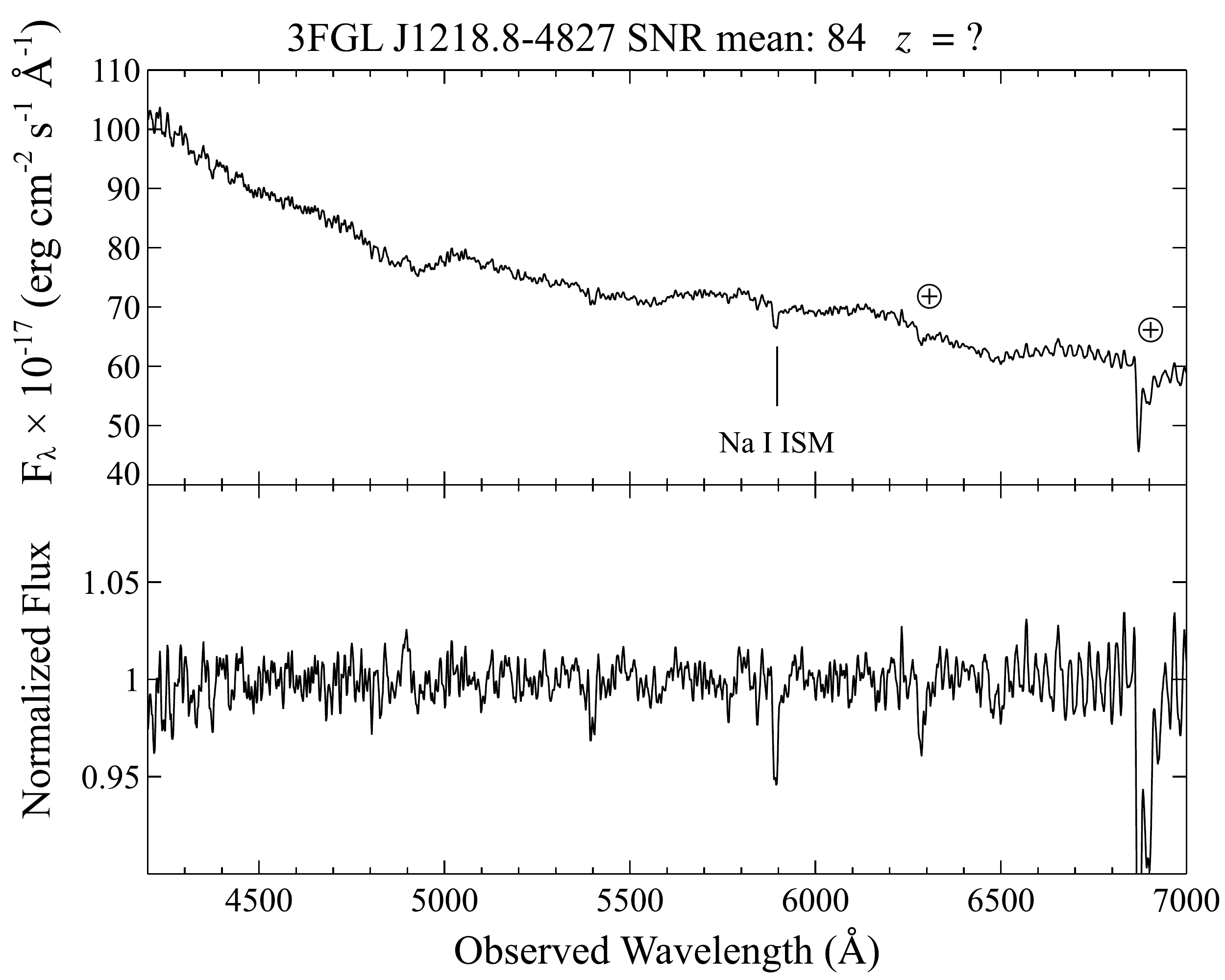} &
\includegraphics[trim=4cm 0cm 4cm 0cm, clip=true, width=7cm,angle=0]{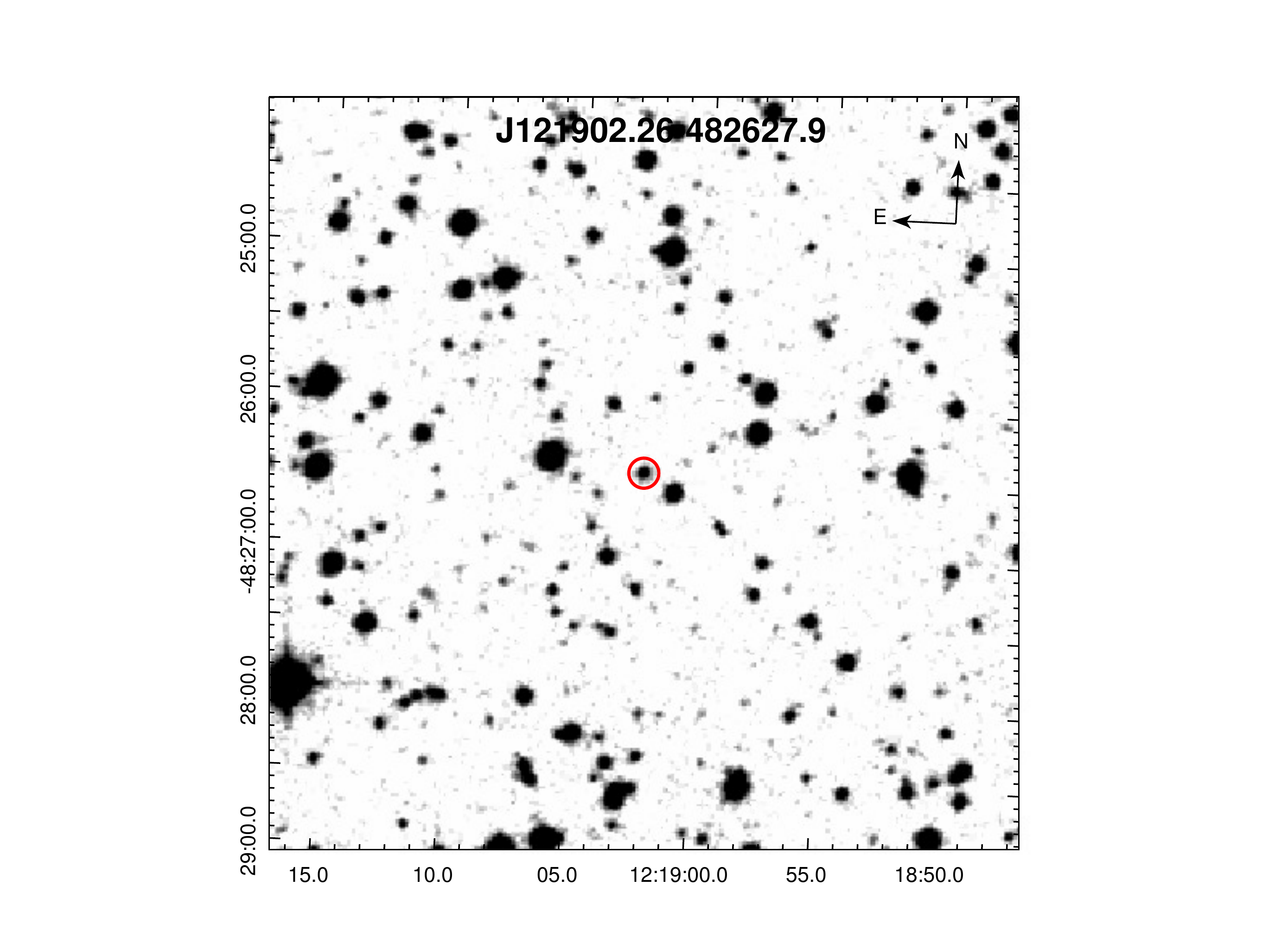} \\
\end{array}$
\end{center}
\caption{(Left panel) Optical spectrum of  WISE J121902.26-482627.9 associated with 3FGL J1218.8-4827, in the upper part it is shown the Signal-to-Noise Ratio of the spectrum. (Right panel) The finding chart ( $5'\times 5'$ ) retrieved from the Digital Sky Survey highlighting the location of the counterpart: WISE J121902.26-482627.9 (red circle).}
\label{fig:J1218}
\end{figure*}

\begin{figure*}{}
\begin{center}$
\begin{array}{cc}
\includegraphics[width=\mywidth,angle=0]{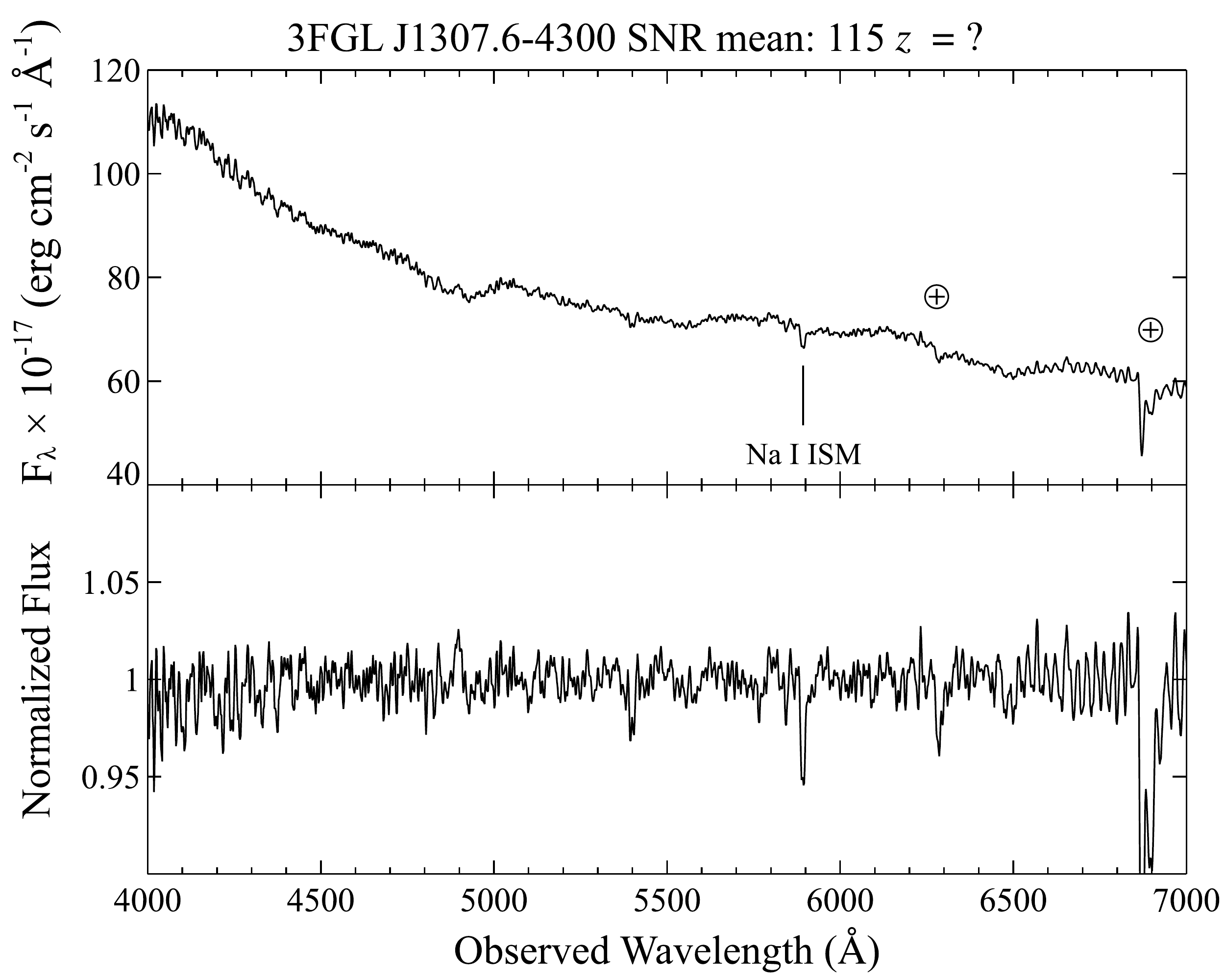} &
\includegraphics[trim=4cm 0cm 4cm 0cm, clip=true, width=7cm,angle=0]{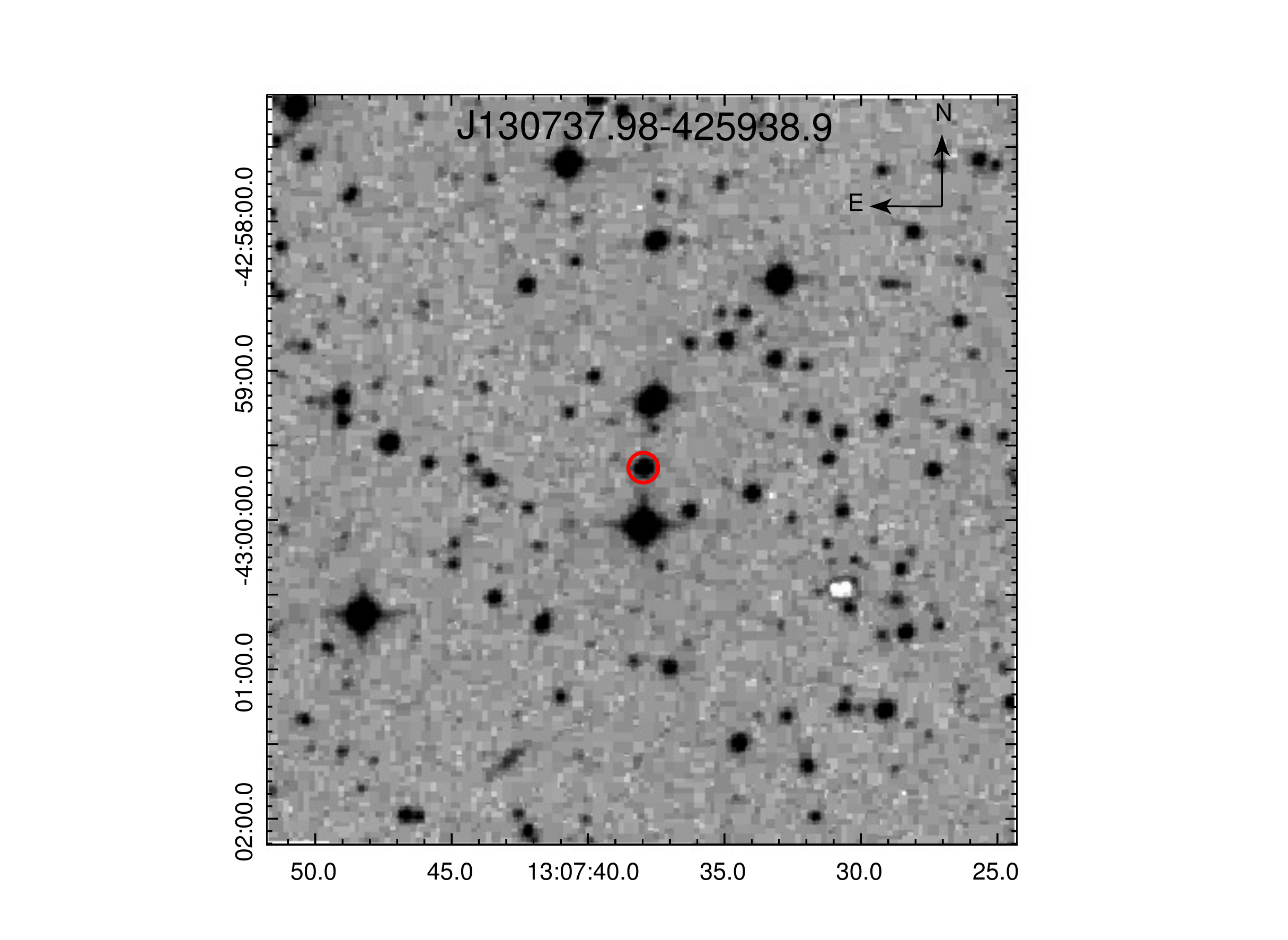} \\
\end{array}$
\end{center}
\caption{(Left panel) Optical spectrum of  WISE J130737.98-425938.9 associated with 3FGL J1307.6-4300, in the upper part it is shown the Signal-to-Noise Ratio of the spectrum. (Right panel) The finding chart ( $5'\times 5'$ ) retrieved from the Digital Sky Survey highlighting the location of the counterpart: WISE J130737.98-425938.9 (red circle).}
\label{fig:J1307}
\end{figure*}

\begin{figure*}{}
\begin{center}$
\begin{array}{cc}
\includegraphics[width=\mywidth,angle=0]{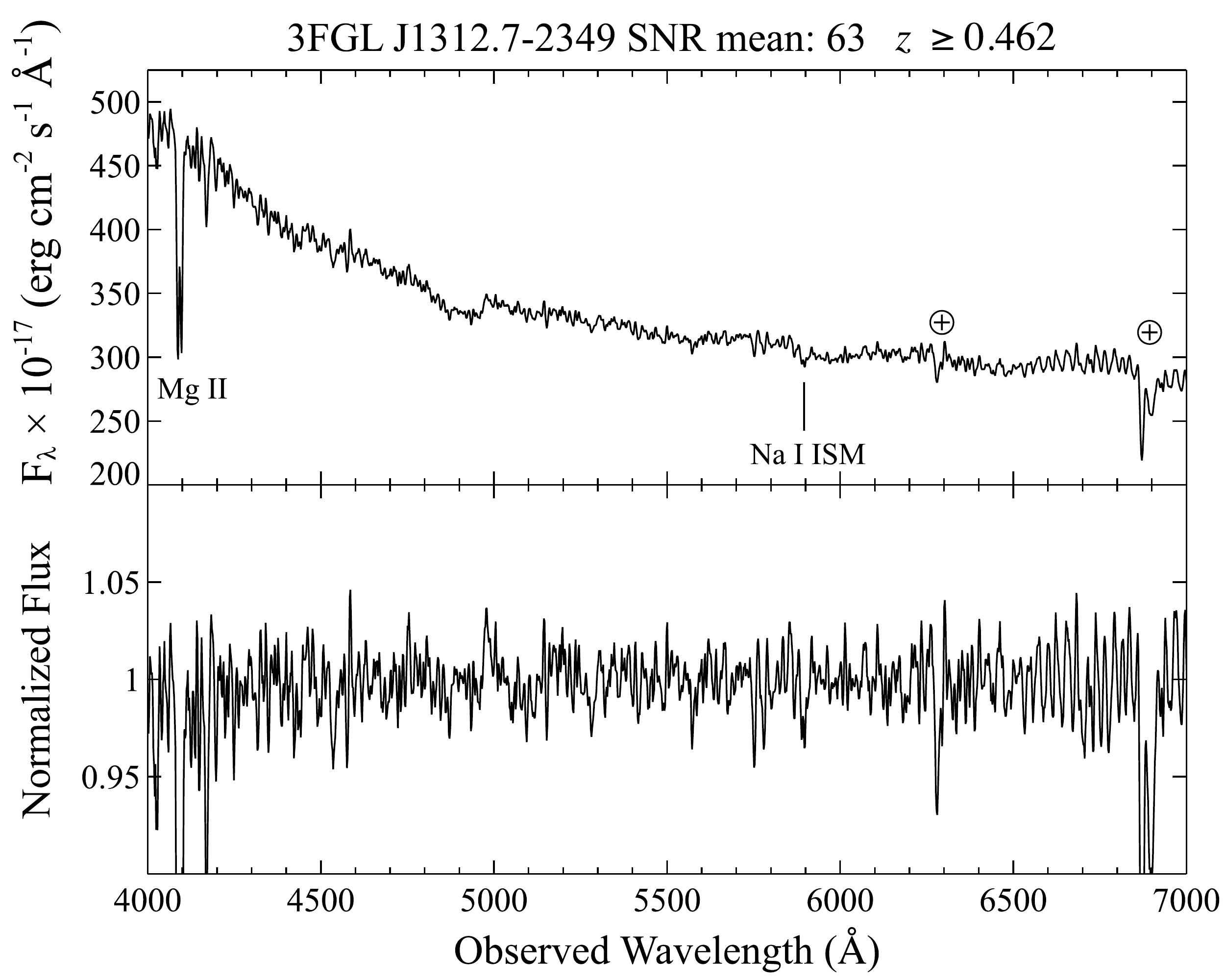} &
\includegraphics[trim=4cm 0cm 4cm 0cm, clip=true, width=7cm,angle=0]{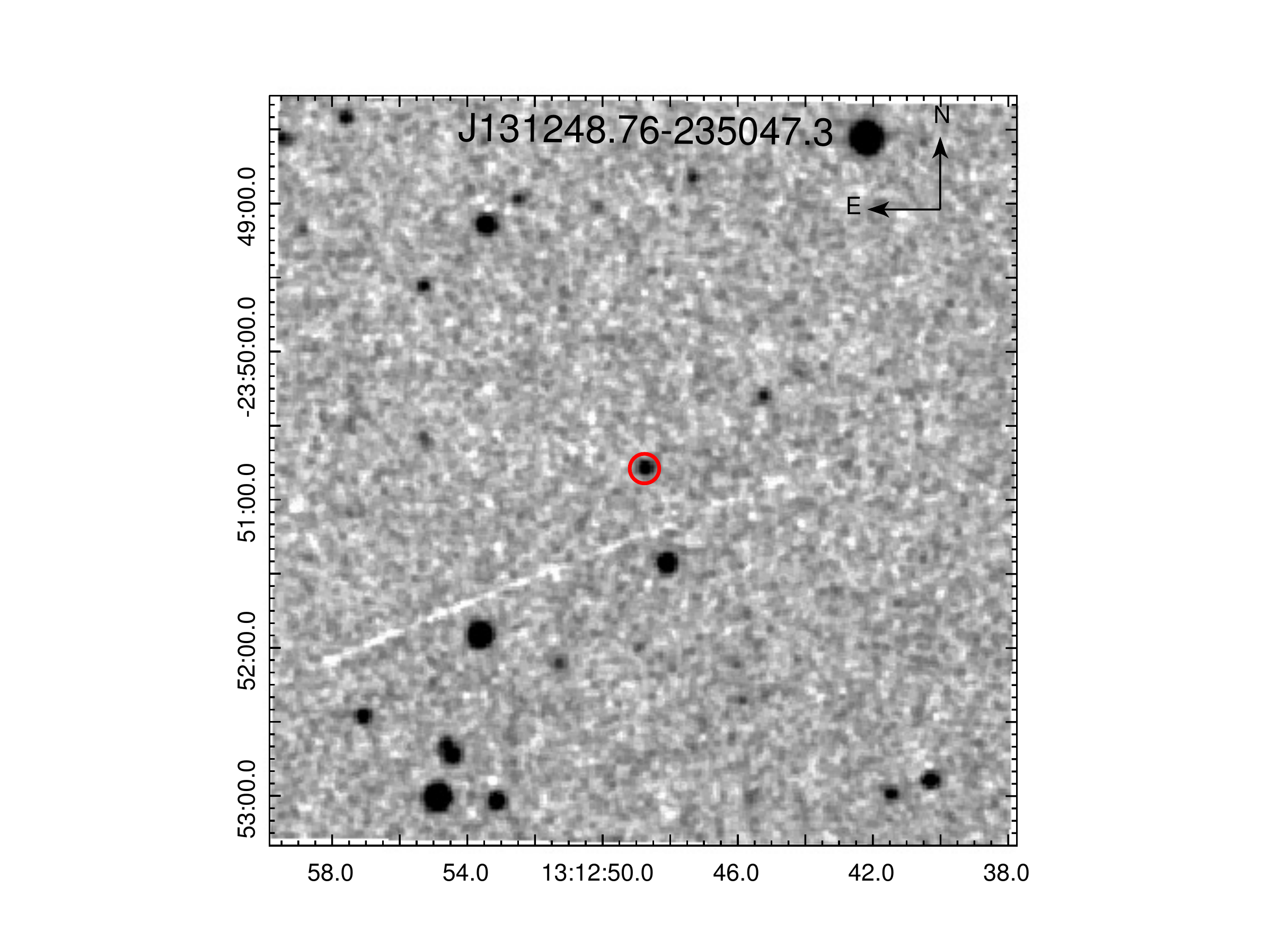} \\
\end{array}$
\end{center}
\caption{(Left panel) Optical spectrum of  WISE J131248.76-235047.3 associated with 3FGL J1312.7-2349, in the upper part it is shown the Signal-to-Noise Ratio of the spectrum. (Right panel) The finding chart ( $5'\times 5'$ ) retrieved from the Digital Sky Survey highlighting the location of the counterpart: WISE J131248.76-235047.3 (red circle).}
\label{fig:J1312}
\end{figure*}

\begin{figure*}{}
\begin{center}$
\begin{array}{cc}
\includegraphics[width=\mywidth,angle=0]{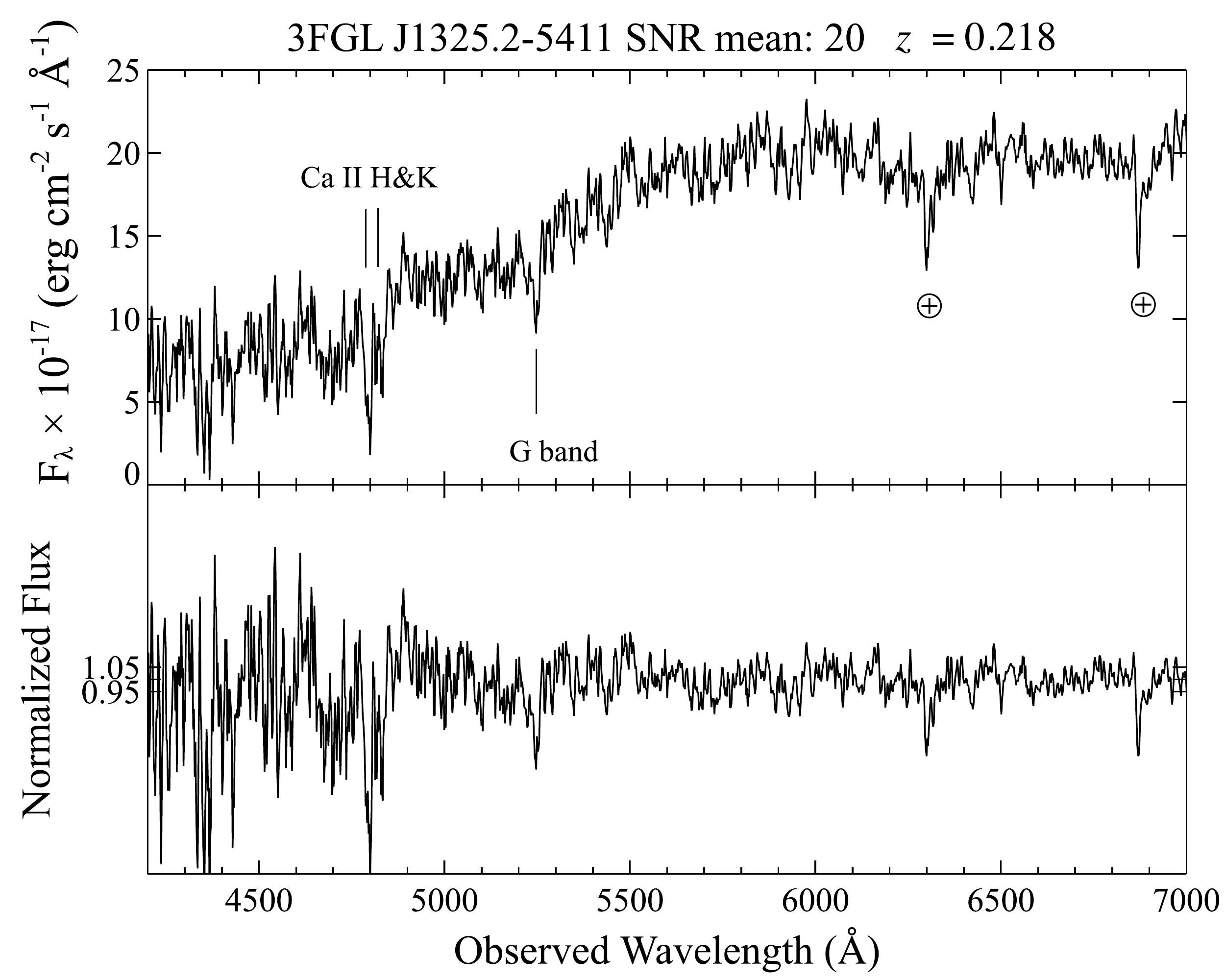} &
\includegraphics[trim=4cm 0cm 4cm 0cm, clip=true, width=7cm,angle=0]{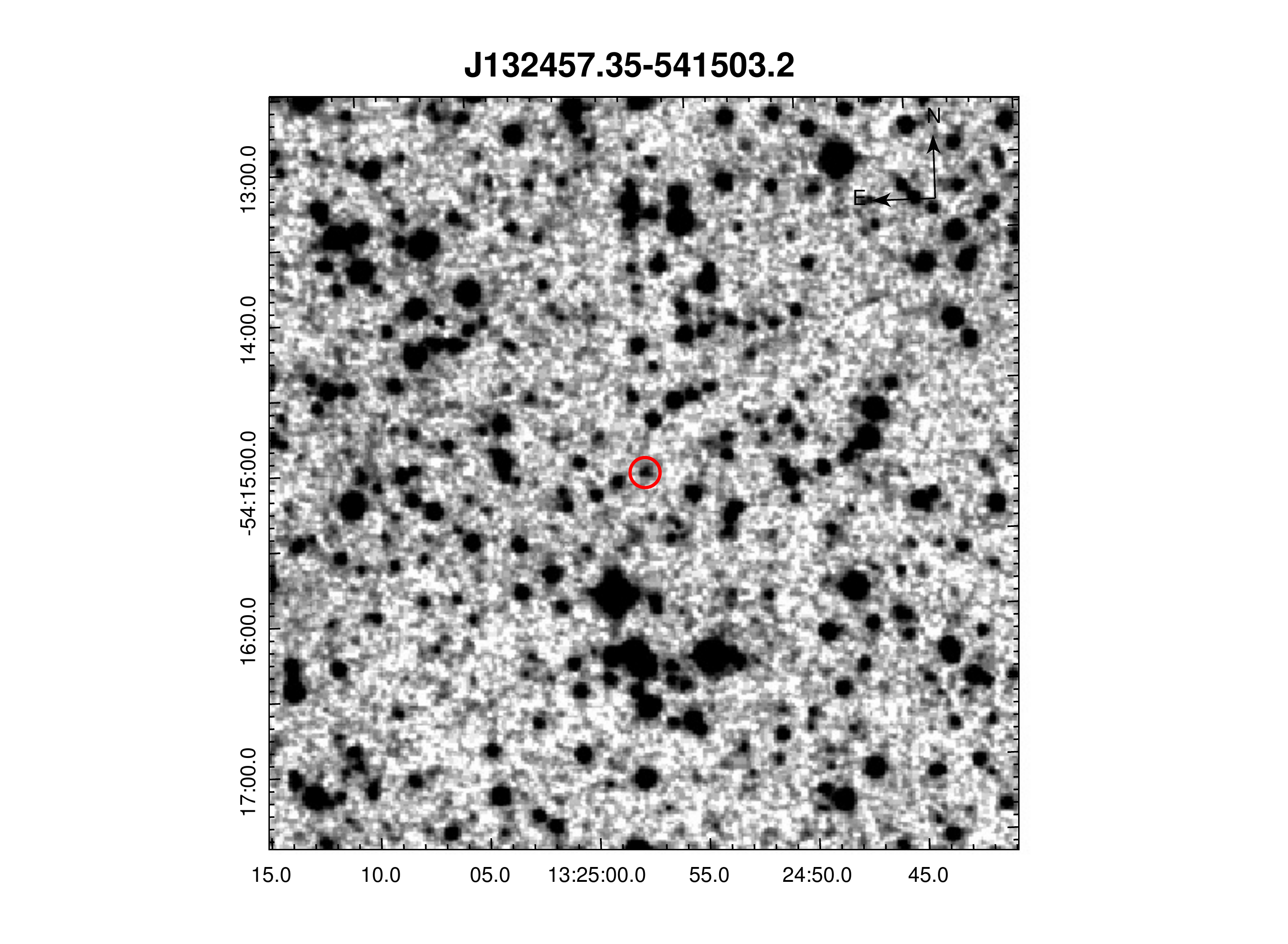} \\
\end{array}$
\end{center}
\caption{(Left panel) Optical spectrum of  WISE J132457.35-541503.2 associated with 3FGL J1325.2-5411. Signal-to-noise ratio is reported in the Figure. (Right panel) The finding chart ( $5'\times 5'$ ) retrieved from the Digital Sky Survey highlighting the location of the potential source: WISE J132457.35-541503.2 (red circle).}
\label{fig:J1325}
\end{figure*}

\begin{figure*}{}
\begin{center}$
\begin{array}{cc}
\includegraphics[width=\mywidth,angle=0]{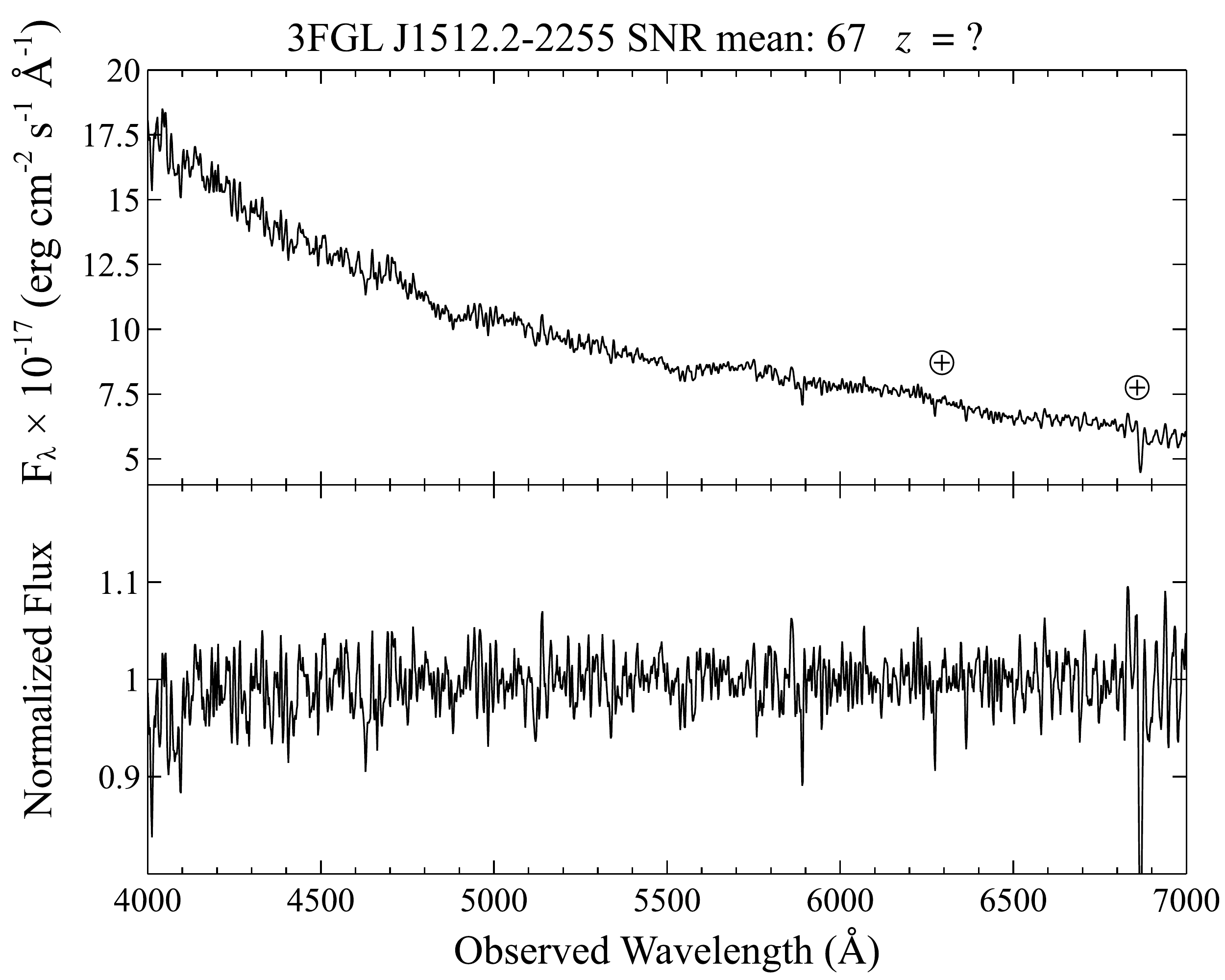} &
\includegraphics[trim=4cm 0cm 4cm 0cm, clip=true, width=7cm,angle=0]{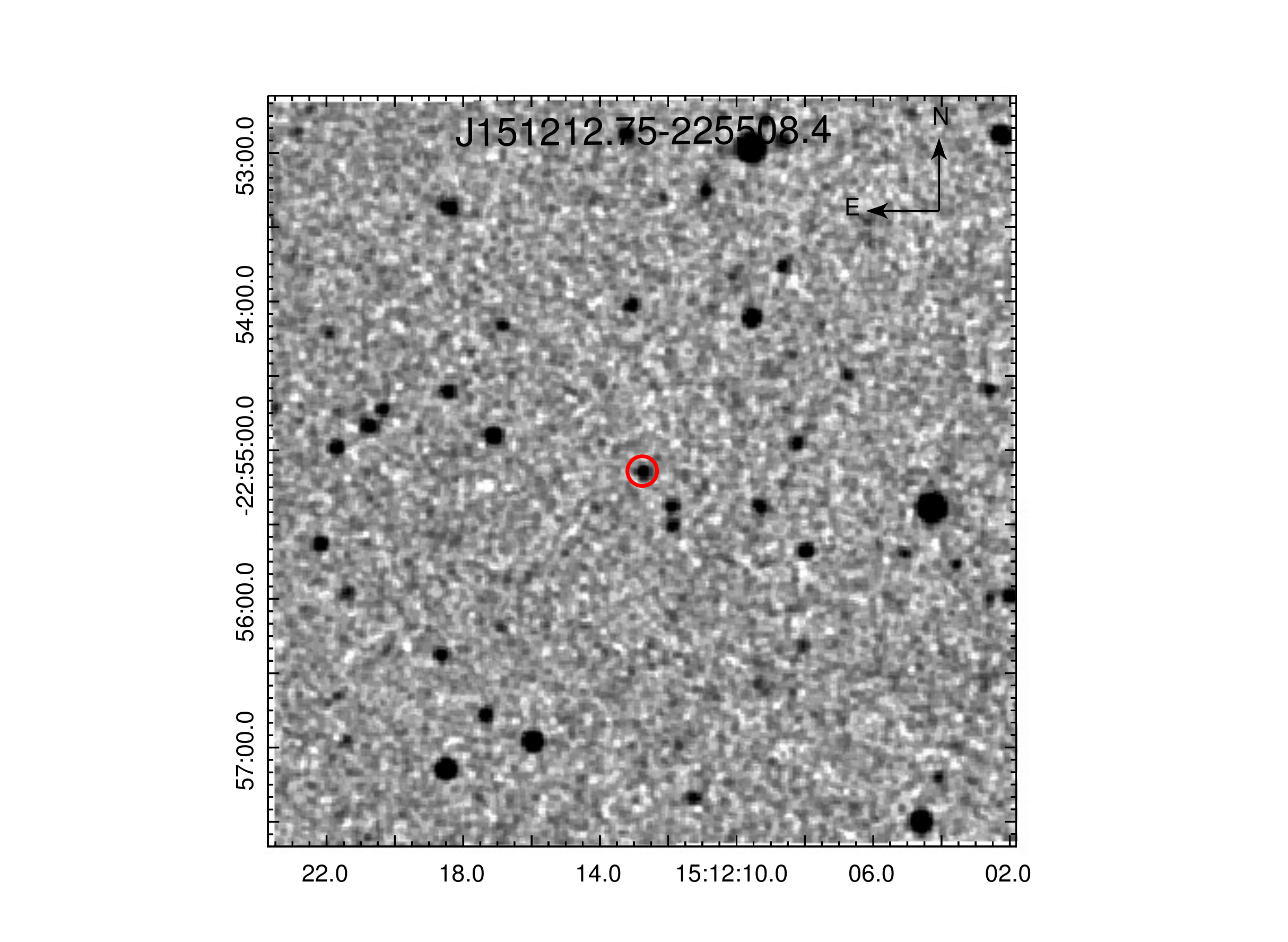} \\
\end{array}$
\end{center}
\caption{(Left panel) Optical spectrum of  WISE J151212.75-225508.4 associated with 3FGL J1512.2-2255, in the upper part it is shown the Signal-to-Noise Ratio of the spectrum. (Right panel) The finding chart ( $5'\times 5'$ ) retrieved from the Digital Sky Survey highlighting the location of the counterpart: WISE J151212.75-225508.4 (red circle).}
\label{fig:J1512}
\end{figure*}

\begin{figure*}{}
\begin{center}$
\begin{array}{cc}
\includegraphics[width=\mywidth,angle=0]{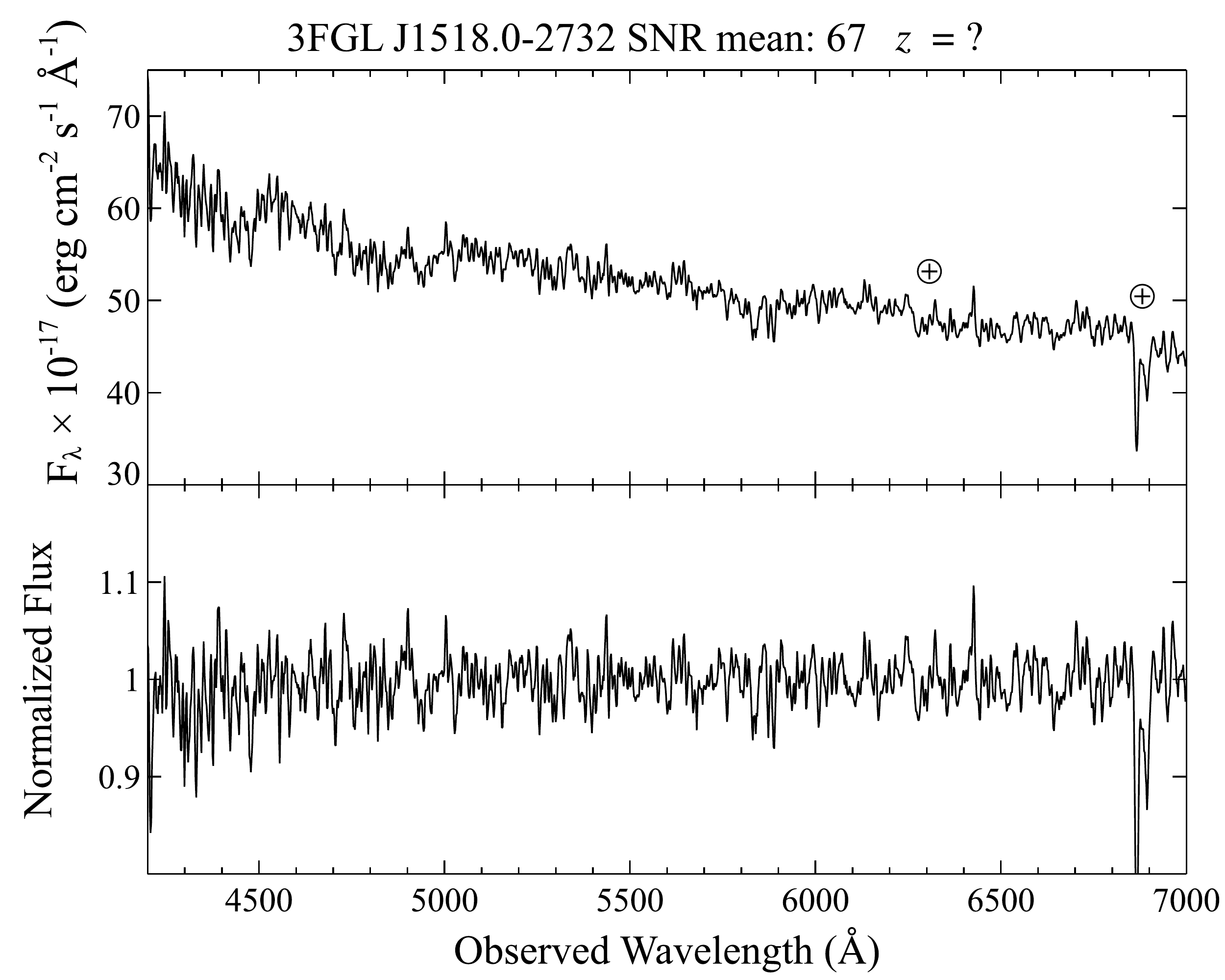} &
\includegraphics[trim=4cm 0cm 4cm 0cm, clip=true, width=7cm,angle=0]{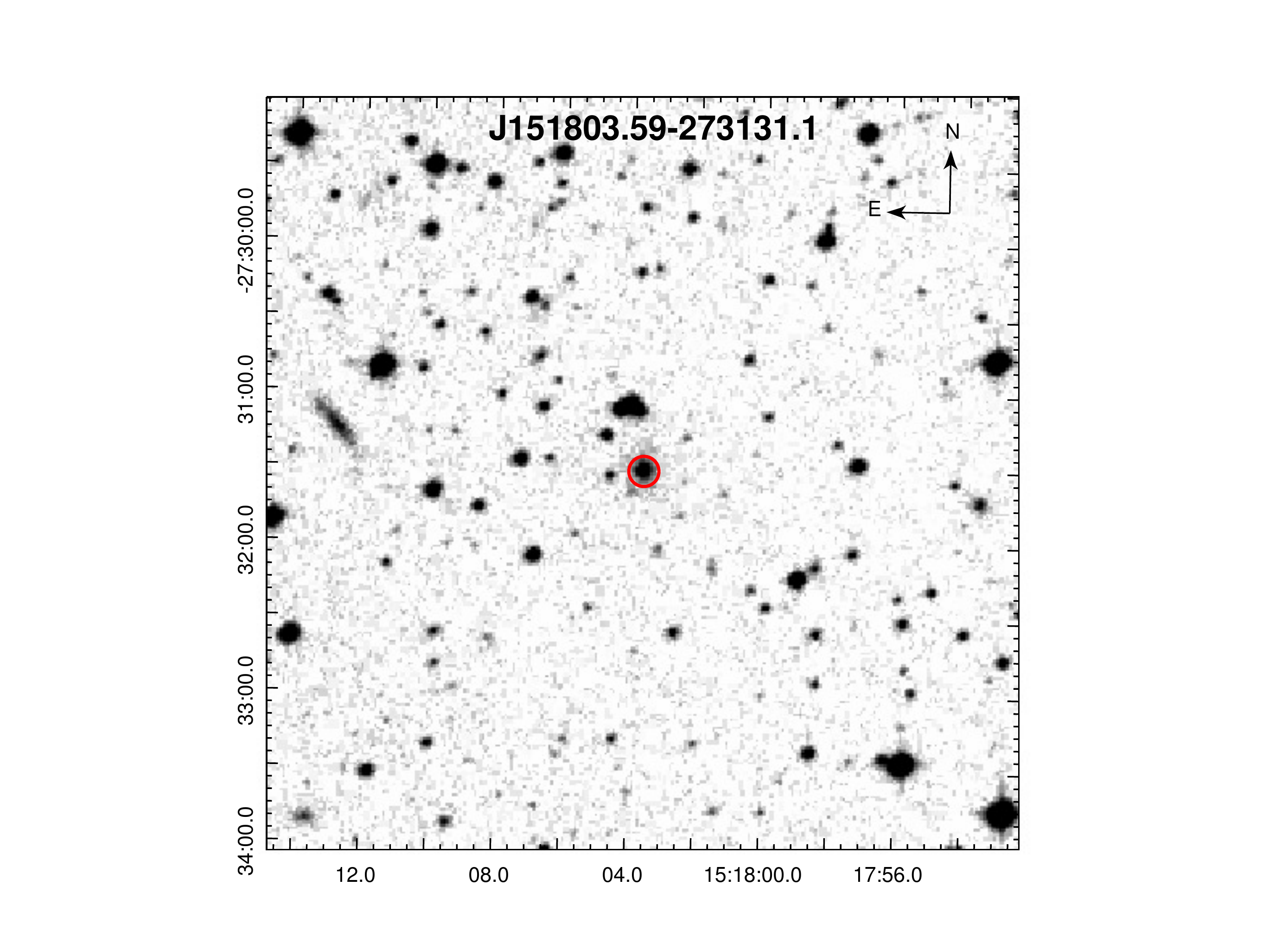} \\
\end{array}$
\end{center}
\caption{(Left panel) Optical spectrum of  WISE J151803.59-273131.1 associated with 3FGL J1518.0-2732, in the upper part it is shown the Signal-to-Noise Ratio of the spectrum. (Right panel) The finding chart ( $5'\times 5'$ ) retrieved from the Digital Sky Survey highlighting the location of the counterpart: WISE J151803.59-273131.1 (red circle).}
\label{fig:J1518}
\end{figure*}

\begin{figure*}{}
\begin{center}$
\begin{array}{cc}
\includegraphics[width=\mywidth,angle=0]{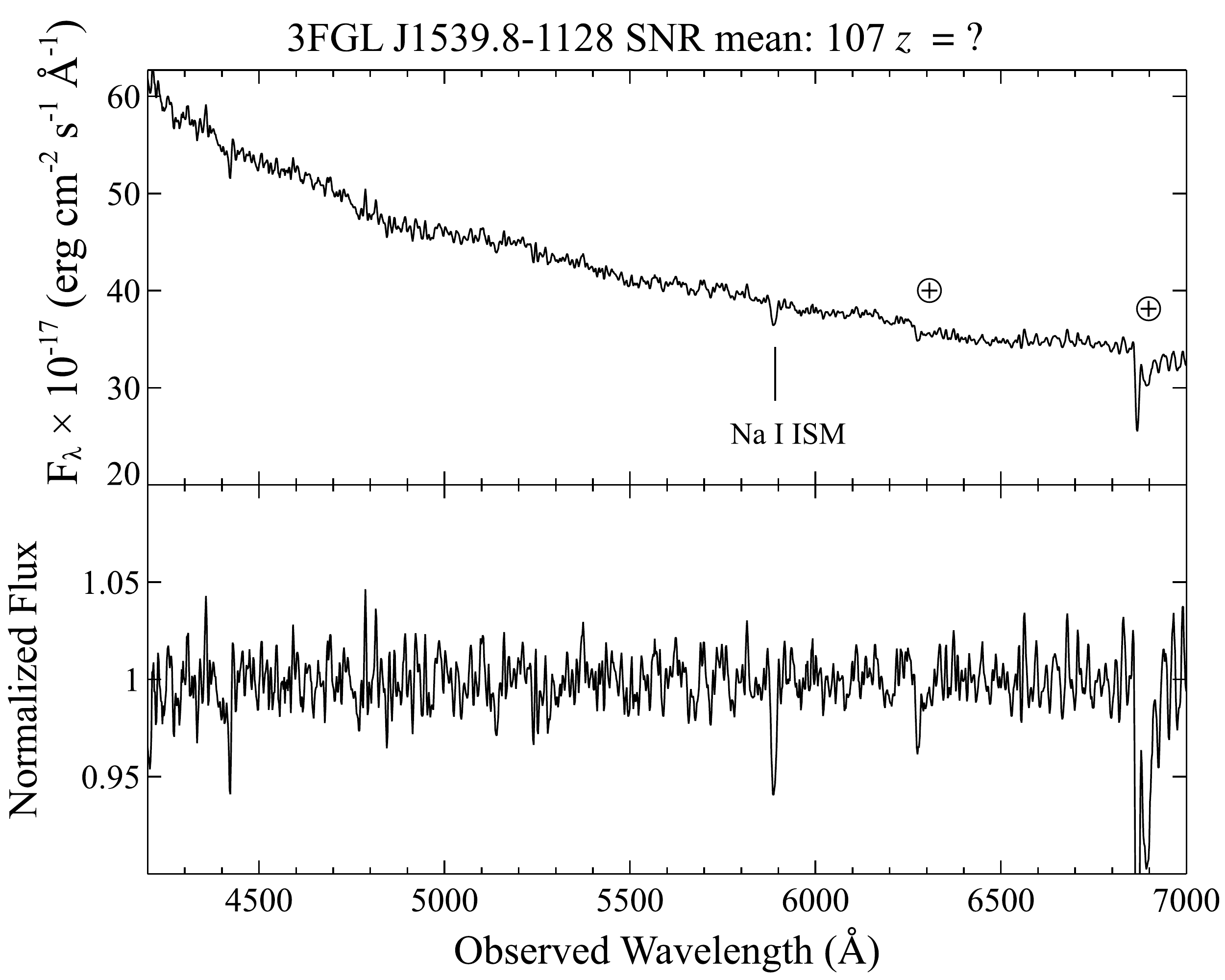} &
\includegraphics[trim=4cm 0cm 4cm 0cm, clip=true, width=7cm,angle=0]{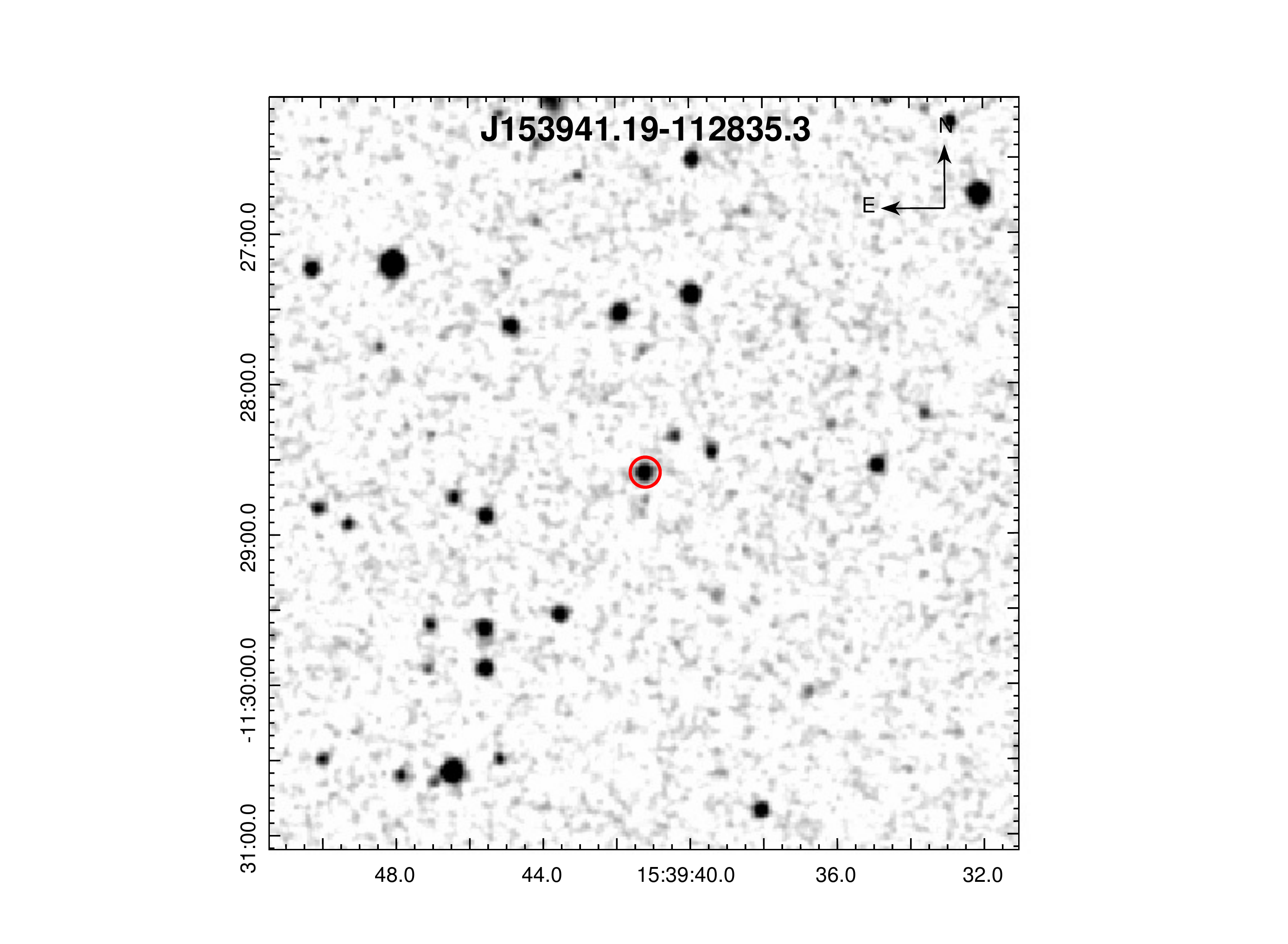} \\
\end{array}$
\end{center}
\caption{(Left panel) Optical spectrum of  WISE J153941.19-112835.3 associated with 3FGL J1539.8-1128, in the upper part it is shown the Signal-to-Noise Ratio of the spectrum. (Right panel) The finding chart ( $5'\times 5'$ ) retrieved from the Digital Sky Survey highlighting the location of the counterpart: WISE J153941.19-112835.3 (red circle).}
\label{fig:J1539}
\end{figure*}

\begin{figure*}{}
\begin{center}$
\begin{array}{cc}
\includegraphics[width=\mywidth,angle=0]{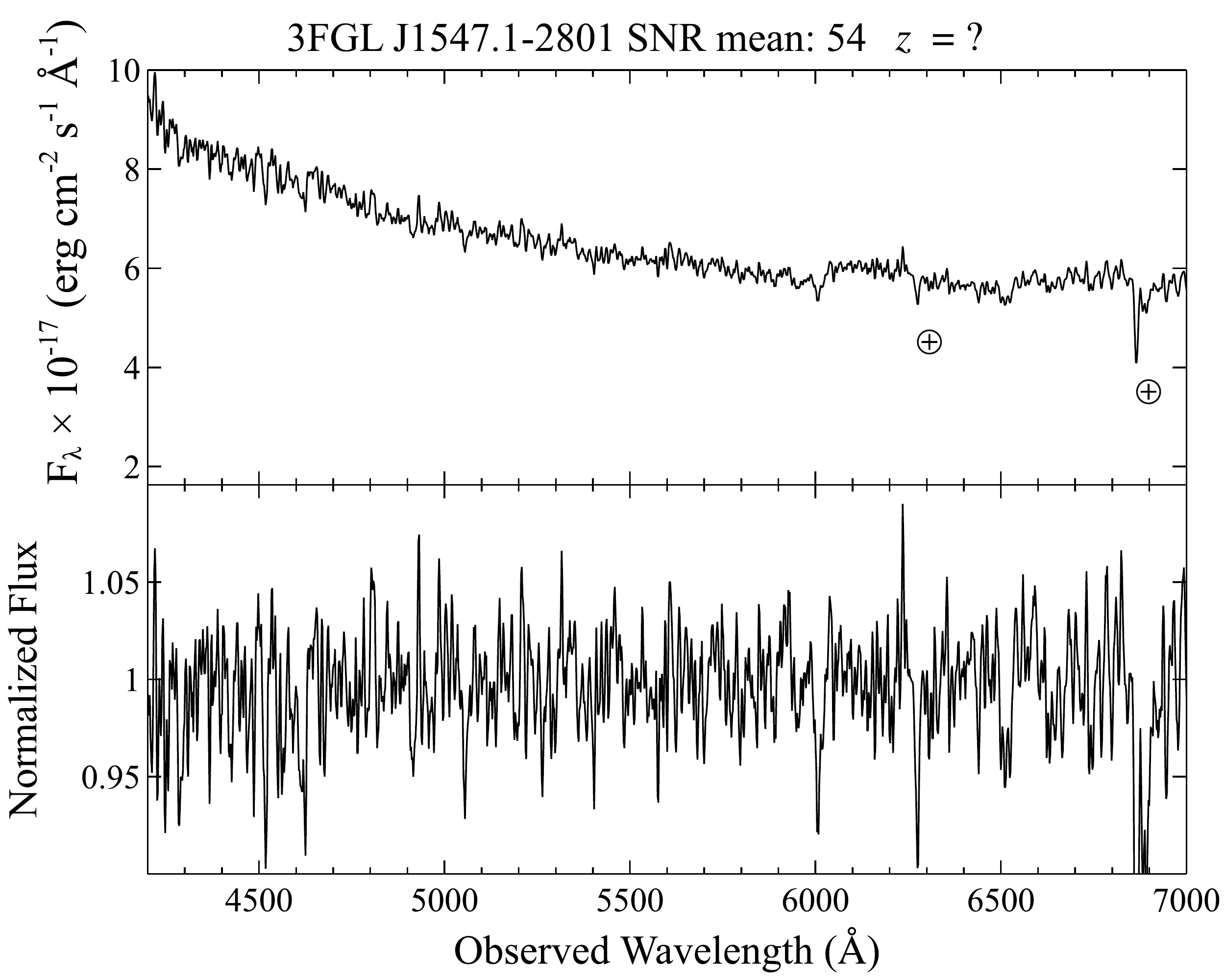} &
\includegraphics[trim=4cm 0cm 4cm 0cm, clip=true, width=7cm,angle=0]{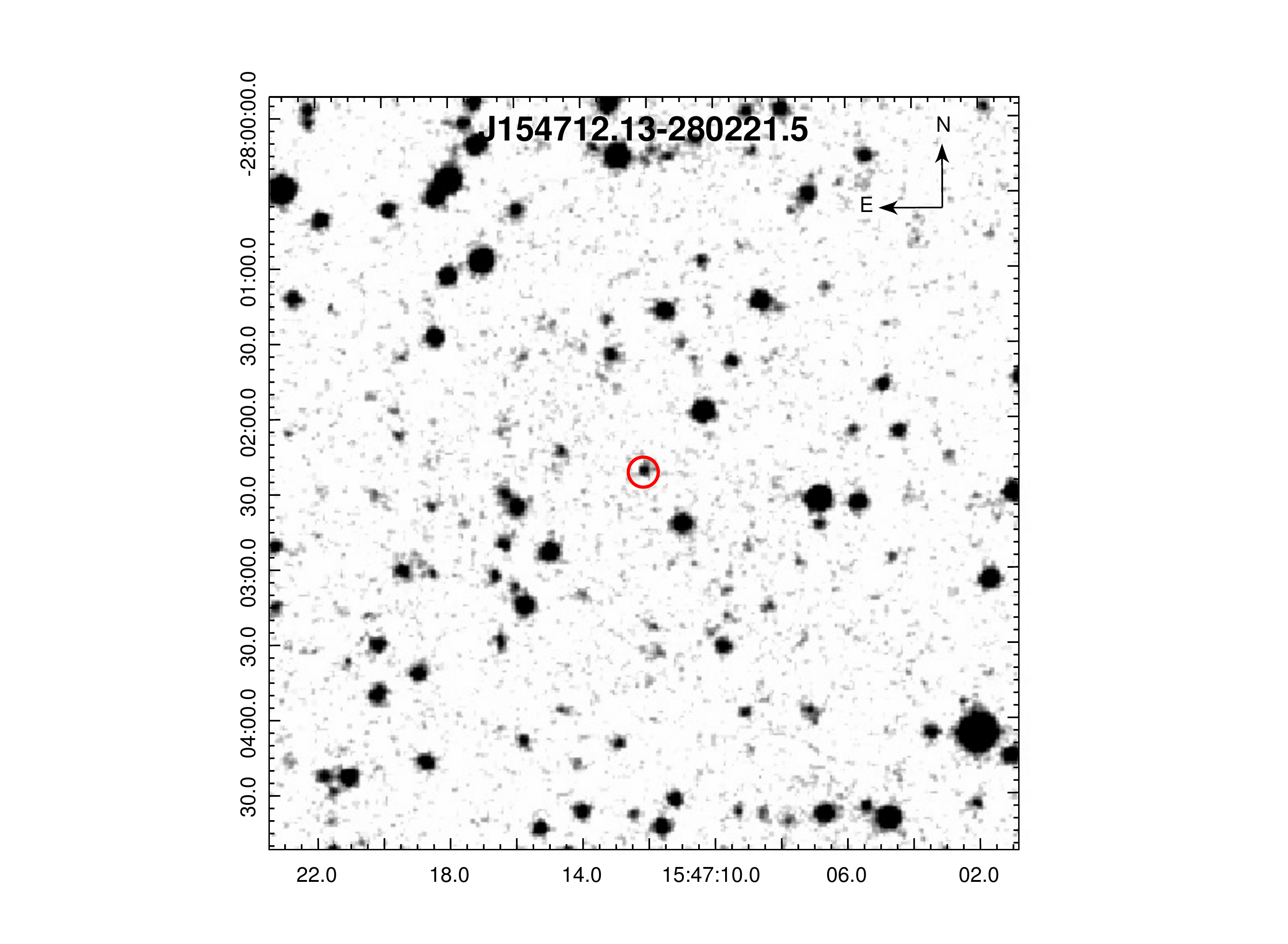} \\
\end{array}$
\end{center}
\caption{(Left panel) Optical spectrum of  WISE J154712.13-280221.5 associated with 3FGL J1547.1-2801, in the upper part it is shown the Signal-to-Noise Ratio of the spectrum. (Right panel) The finding chart ( $5'\times 5'$ ) retrieved from the Digital Sky Survey highlighting the location of the counterpart: WISE J154712.13-280221.5 (red circle).}
\label{fig:J1547}
\end{figure*}

\begin{figure*}{}
\begin{center}$
\begin{array}{cc}
\includegraphics[width=\mywidth,angle=0]{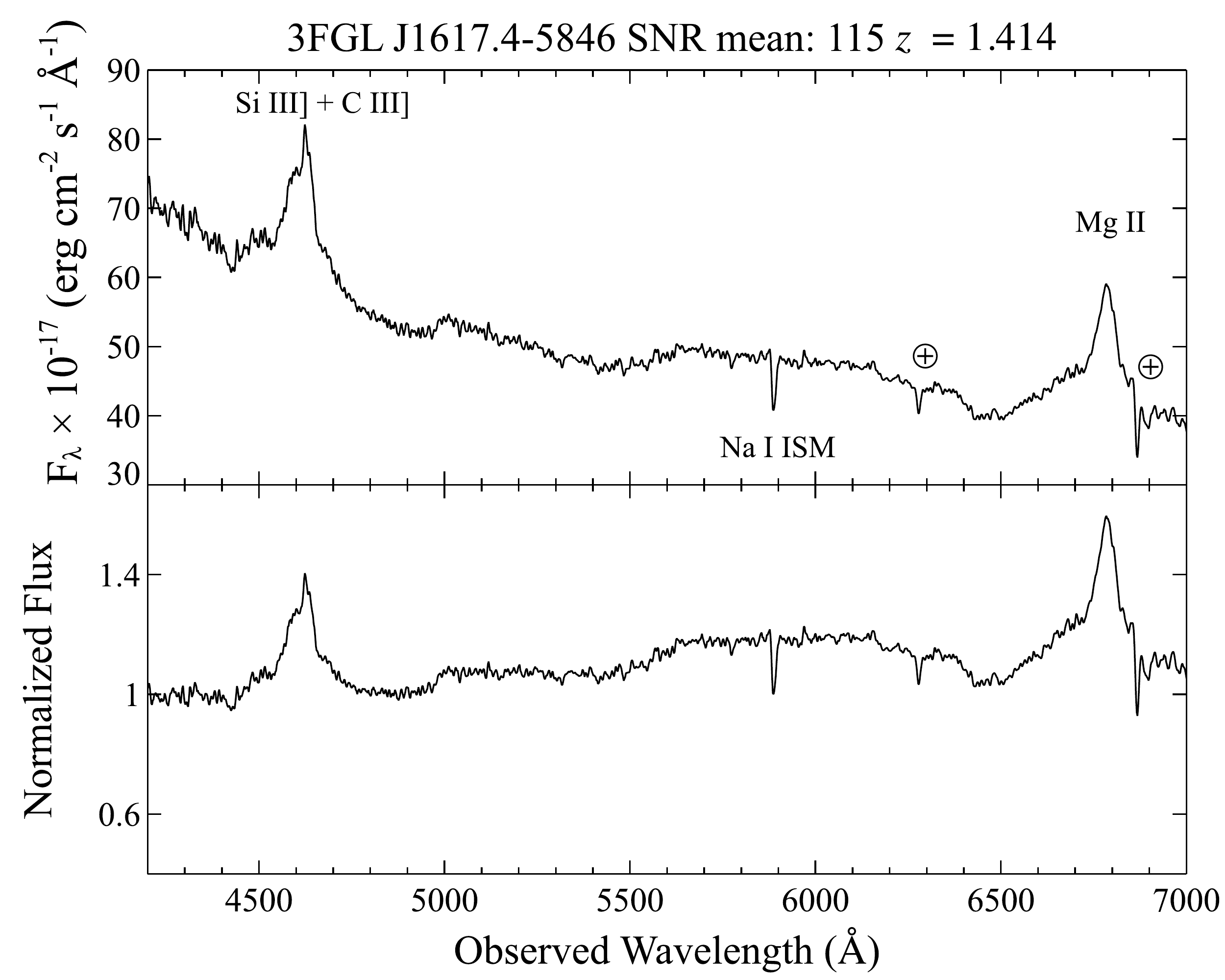} &
\includegraphics[trim=4cm 0cm 4cm 0cm, clip=true, width=7cm,angle=0]{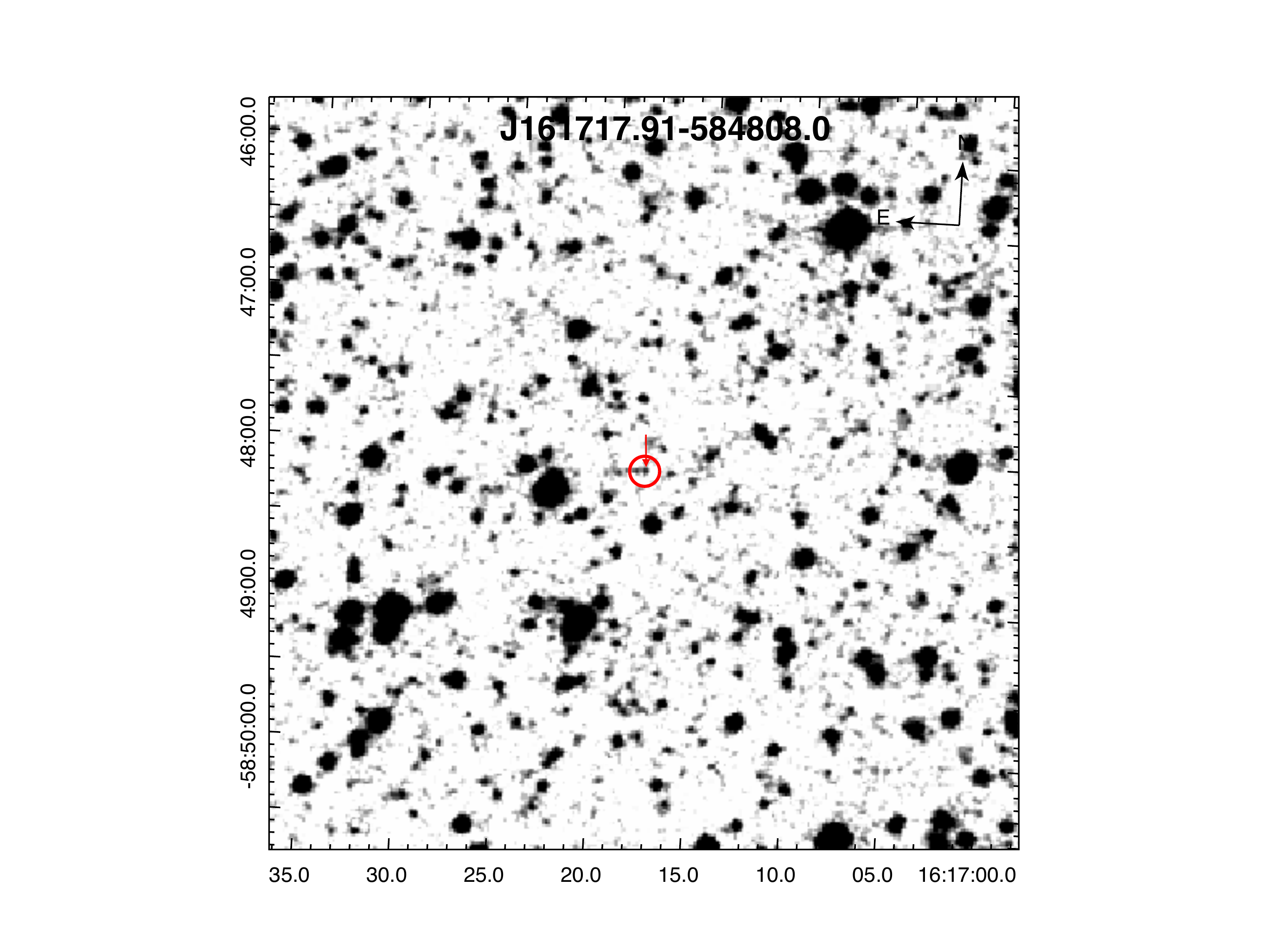} \\
\end{array}$
\end{center}
\caption{(Left panel) Optical spectrum of  WISE J161717.91-584808.0 associated with 3FGL J1617.4-5846. Signal-to-noise ratio is reported in the Figure. (Right panel) The finding chart ( $5'\times 5'$ ) retrieved from the Digital Sky Survey highlighting the location of the counterpart: WISE J161717.91-584808.0 (red circle).}
\label{fig:J1617}
\end{figure*}

\begin{figure*}{}
\begin{center}$
\begin{array}{cc}
\includegraphics[width=\mywidth,angle=0]{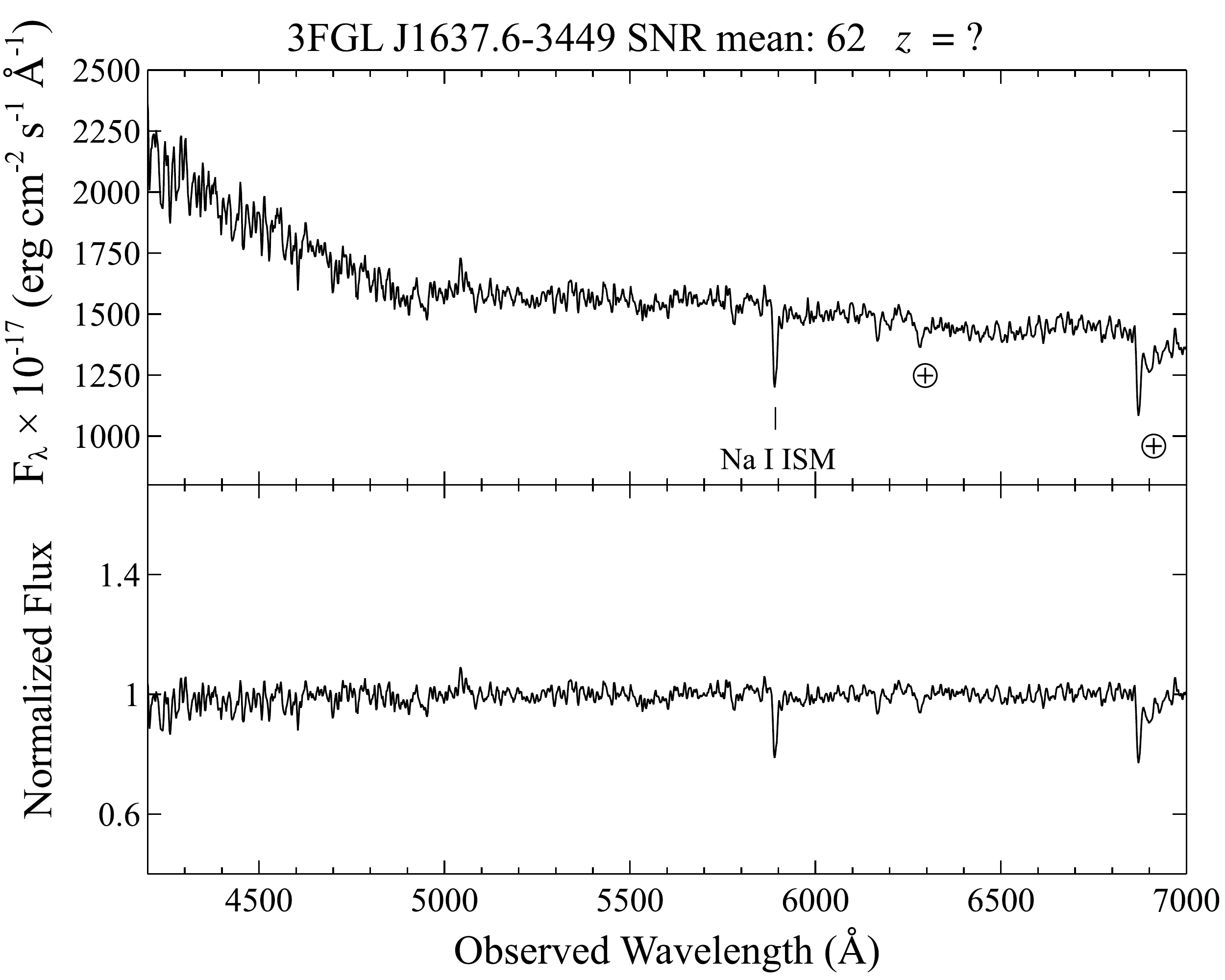} &
\includegraphics[trim=4cm 0cm 4cm 0cm, clip=true, width=7cm,angle=0]{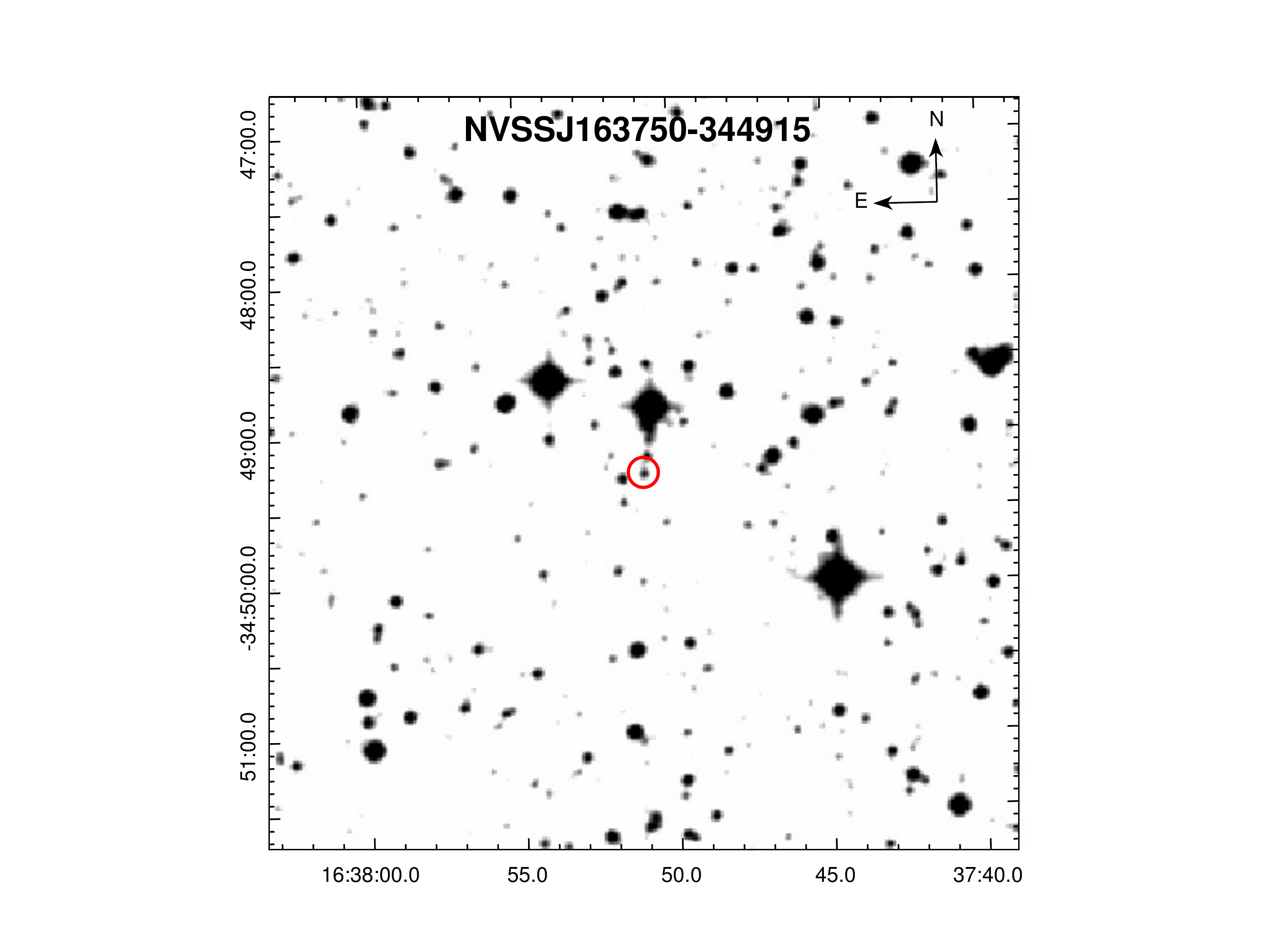} \\
\end{array}$
\end{center}
\caption{(Left panel) Optical spectrum of  NVSS J163750-344915 associated with 3FGL J1637.6-3449, in the upper part it is shown the Signal-to-Noise Ratio of the spectrum. (Right panel) The finding chart ( $5'\times 5'$ ) retrieved from the Digital Sky Survey highlighting the location of the counterpart: NVSS J163750-344915 (red circle).}
\label{fig:J1637}
\end{figure*}

\begin{figure*}{}
\begin{center}$
\begin{array}{cc}
\includegraphics[width=\mywidth,angle=0]{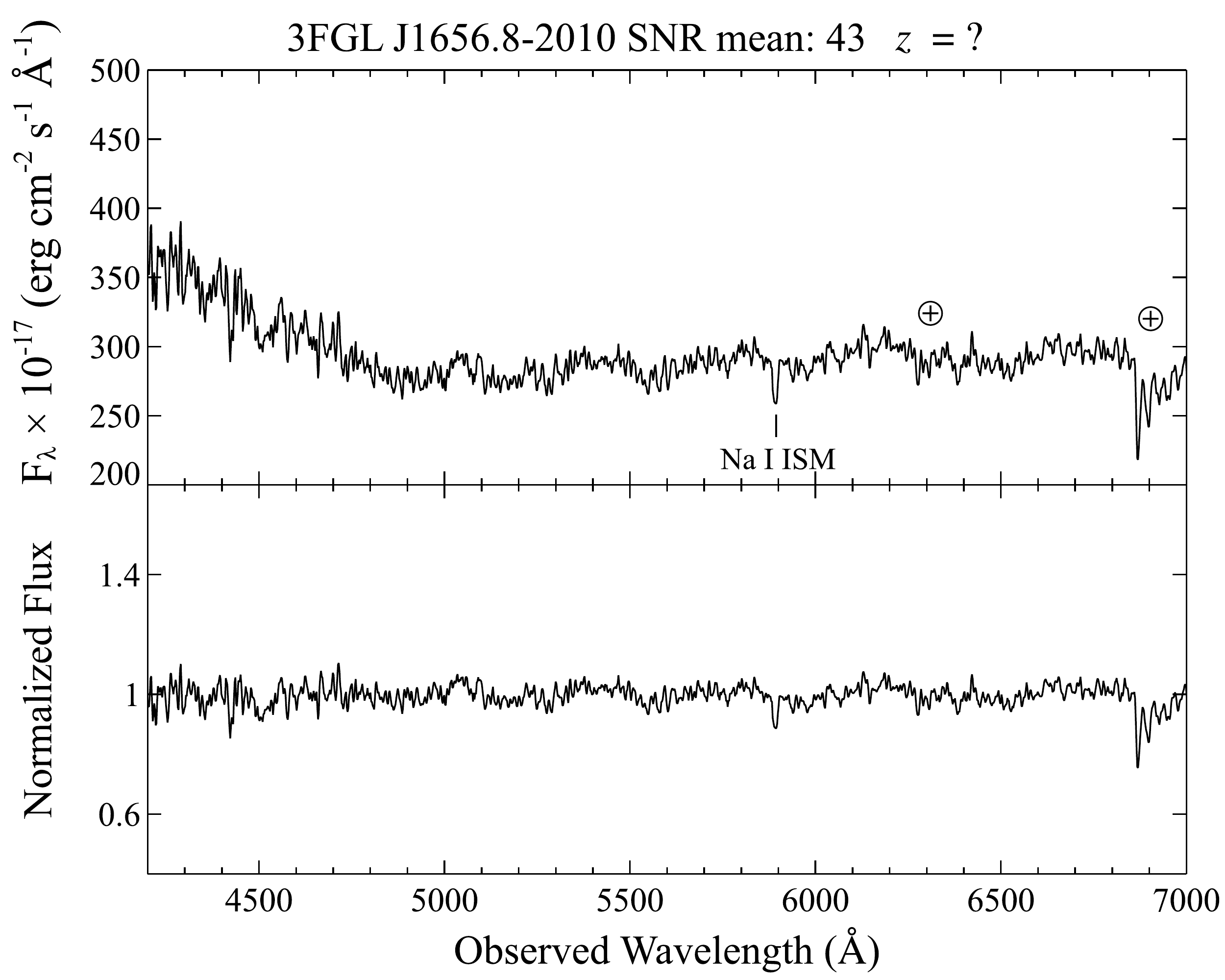} &
\includegraphics[trim=4cm 0cm 4cm 0cm, clip=true, width=7cm,angle=0]{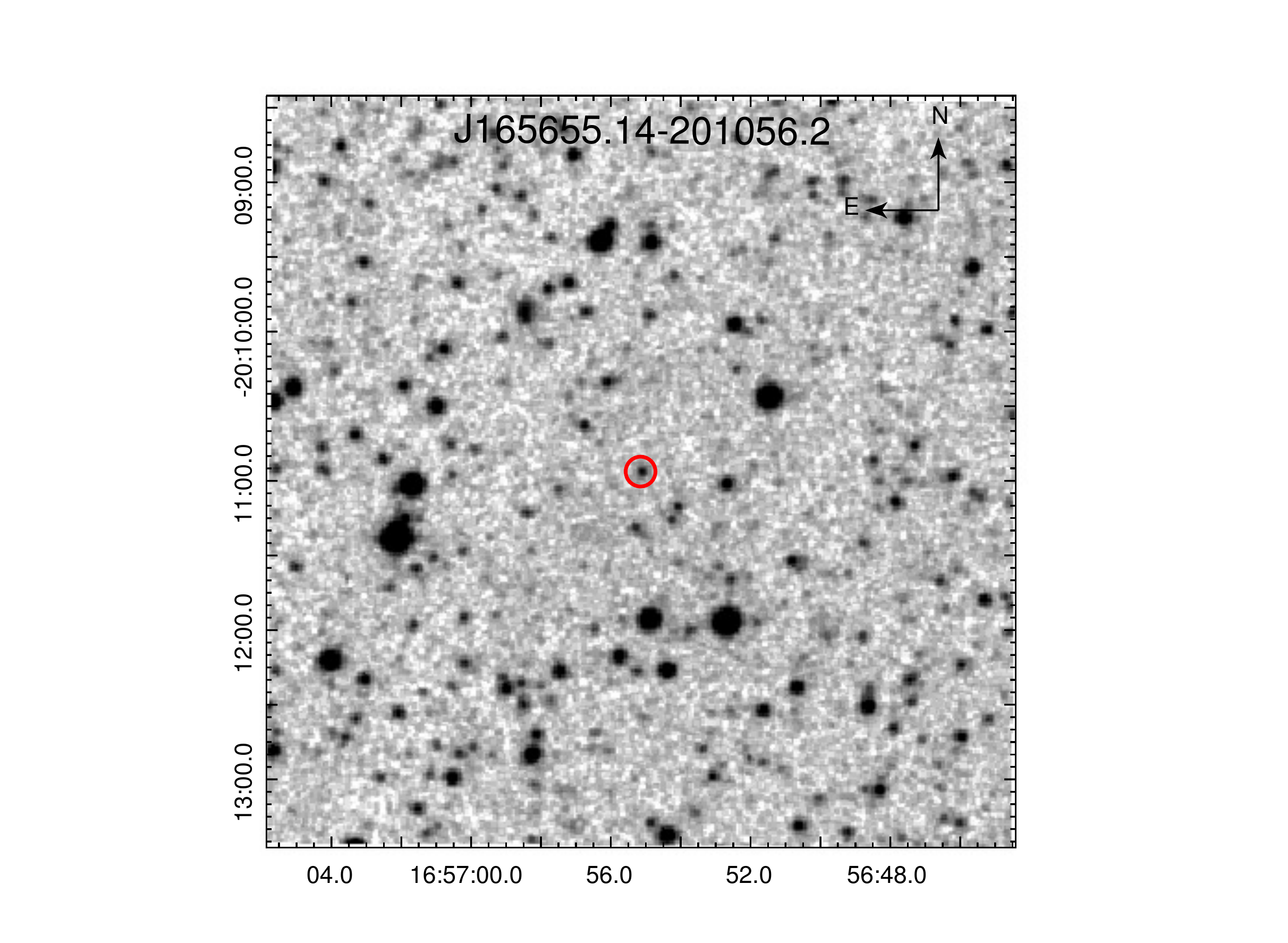} \\
\end{array}$
\end{center}
\caption{(Left panel) Optical spectrum of  WISE J165655.14-201056.2 associated with 3FGL J1656.8-2010, in the upper part it is shown the Signal-to-Noise Ratio of the spectrum. (Right panel) The finding chart ( $5'\times 5'$ ) retrieved from the Digital Sky Survey highlighting the location of the counterpart: WISE J165655.14-201056.2 (red circle).}
\label{fig:J1656}
\end{figure*}

\begin{figure*}{}
\begin{center}$
\begin{array}{cc}
\includegraphics[width=\mywidth,angle=0]{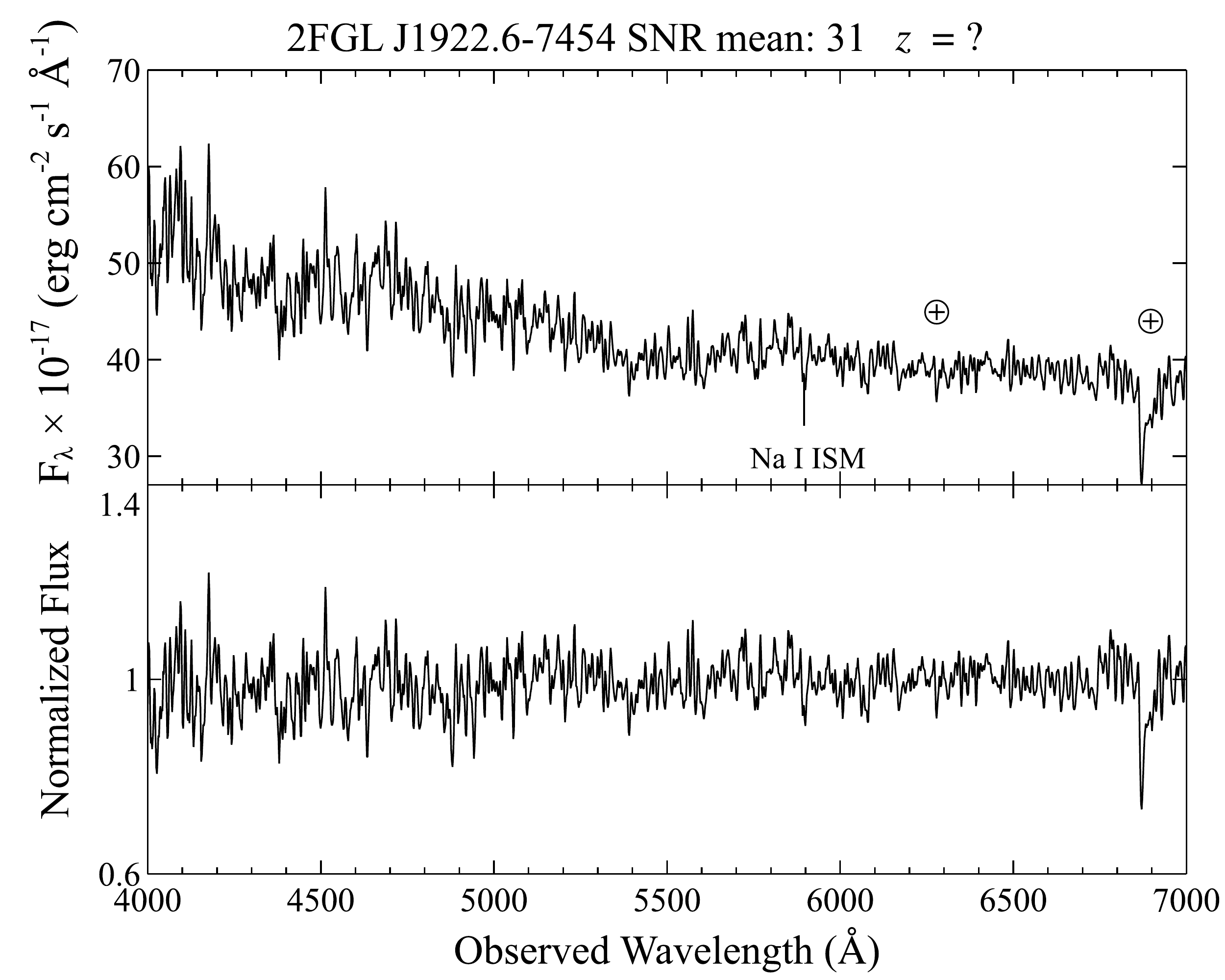} &
\includegraphics[trim=4cm 0cm 4cm 0cm, clip=true, width=7cm,angle=0]{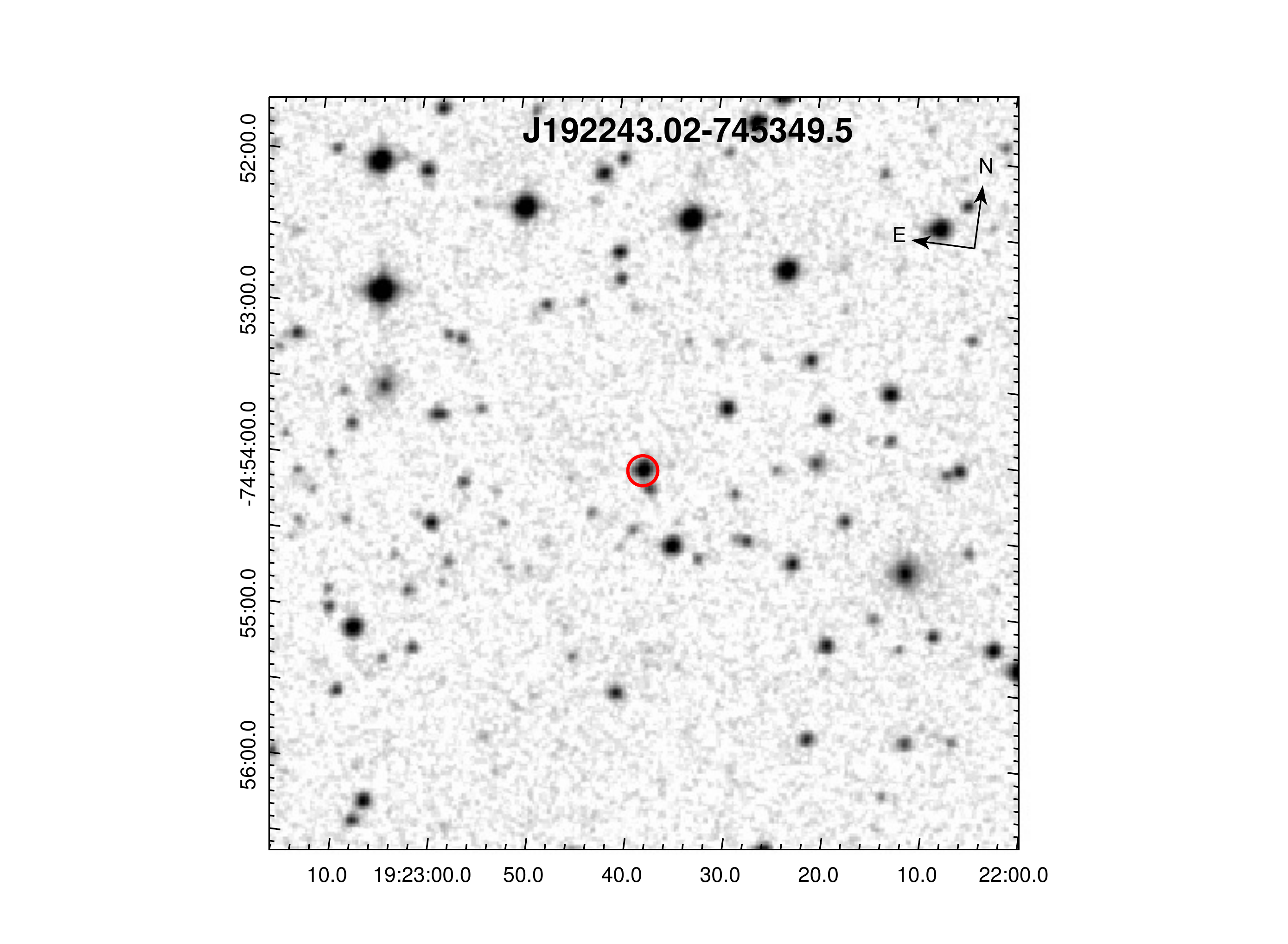} \\
\end{array}$
\end{center}
\caption{(Left panel) Optical spectrum of  WISE J192243.02-745349.5 associated with 2FGL J1922.6-7454, in the upper part it is shown the Signal-to-Noise Ratio of the spectrum. (Right panel) The finding chart ( $5'\times 5'$ ) retrieved from the Digital Sky Survey highlighting the location of the counterpart: WISE J192243.02-745349.5 (red circle).}
\label{fig:J1922}
\end{figure*}

\begin{figure*}{}
\begin{center}$
\begin{array}{cc}
\includegraphics[width=\mywidth,angle=0]{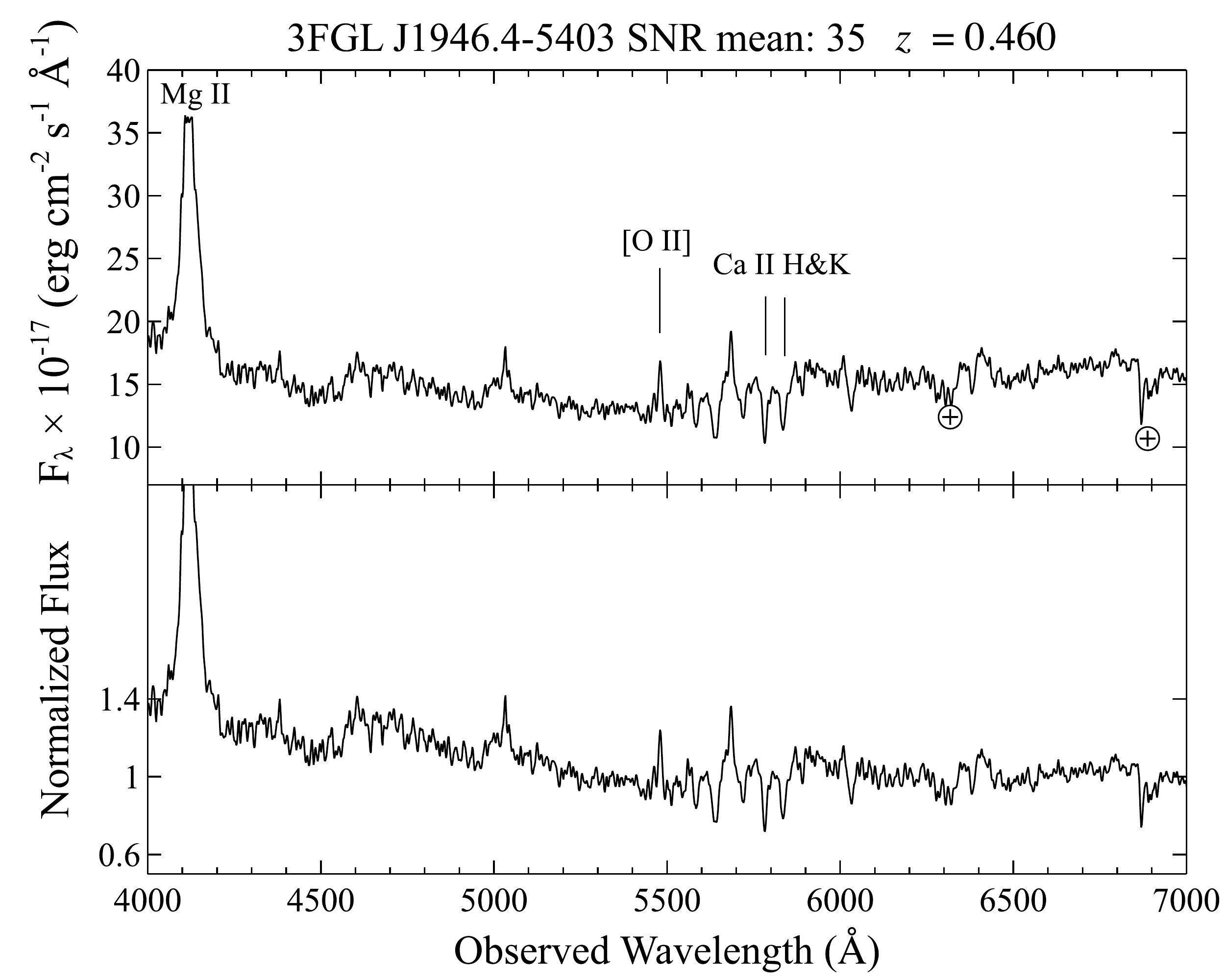} &
\includegraphics[trim=4cm 0cm 4cm 0cm, clip=true, width=7cm,angle=0]{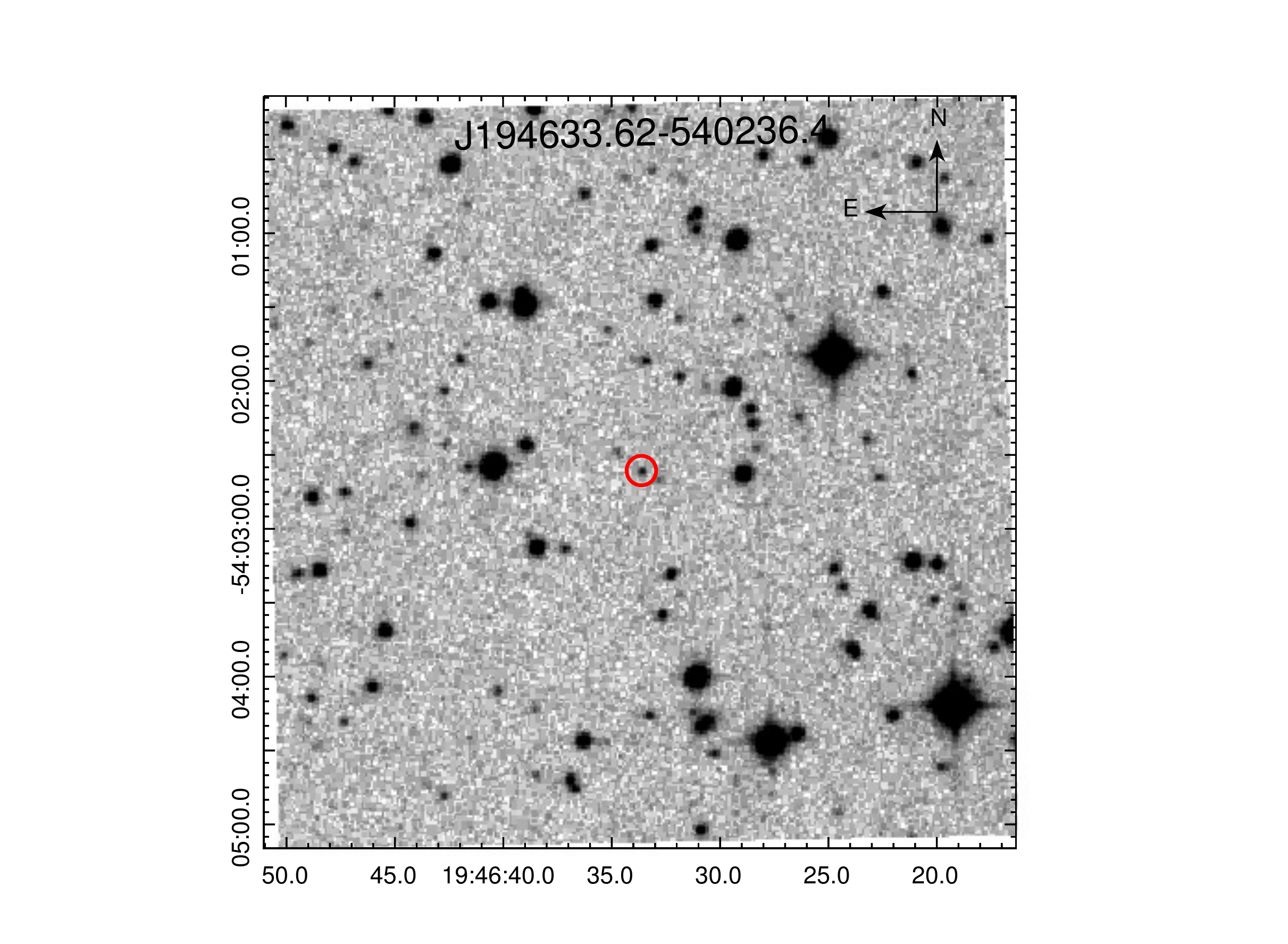} \\
\end{array}$
\end{center}
\caption{(Left panel) Optical spectrum of  WISE J194633.62-540236.4 associated with 3FGL J1946.4-5403. Signal-to-noise ratio is reported in the Figure. (Right panel) The finding chart ( $5'\times 5'$ ) retrieved from the Digital Sky Survey highlighting the location of the potential source: WISE J194633.62-540236.4 (red circle).}
\label{fig:J1946}
\end{figure*}

\begin{figure*}{}
\begin{center}$
\begin{array}{cc}
\includegraphics[width=\mywidth,angle=0]{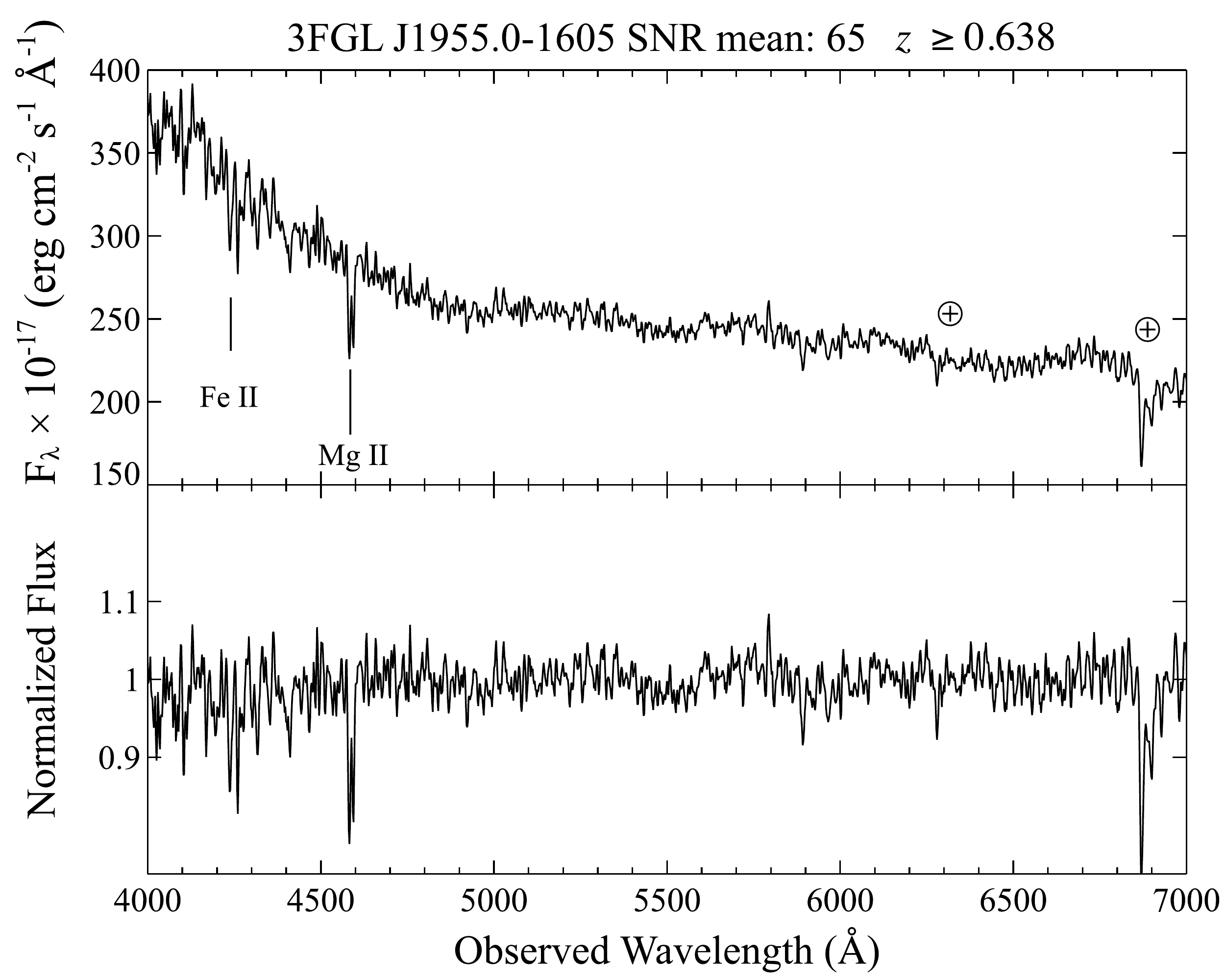} &
\includegraphics[trim=4cm 0cm 4cm 0cm, clip=true, width=7cm,angle=0]{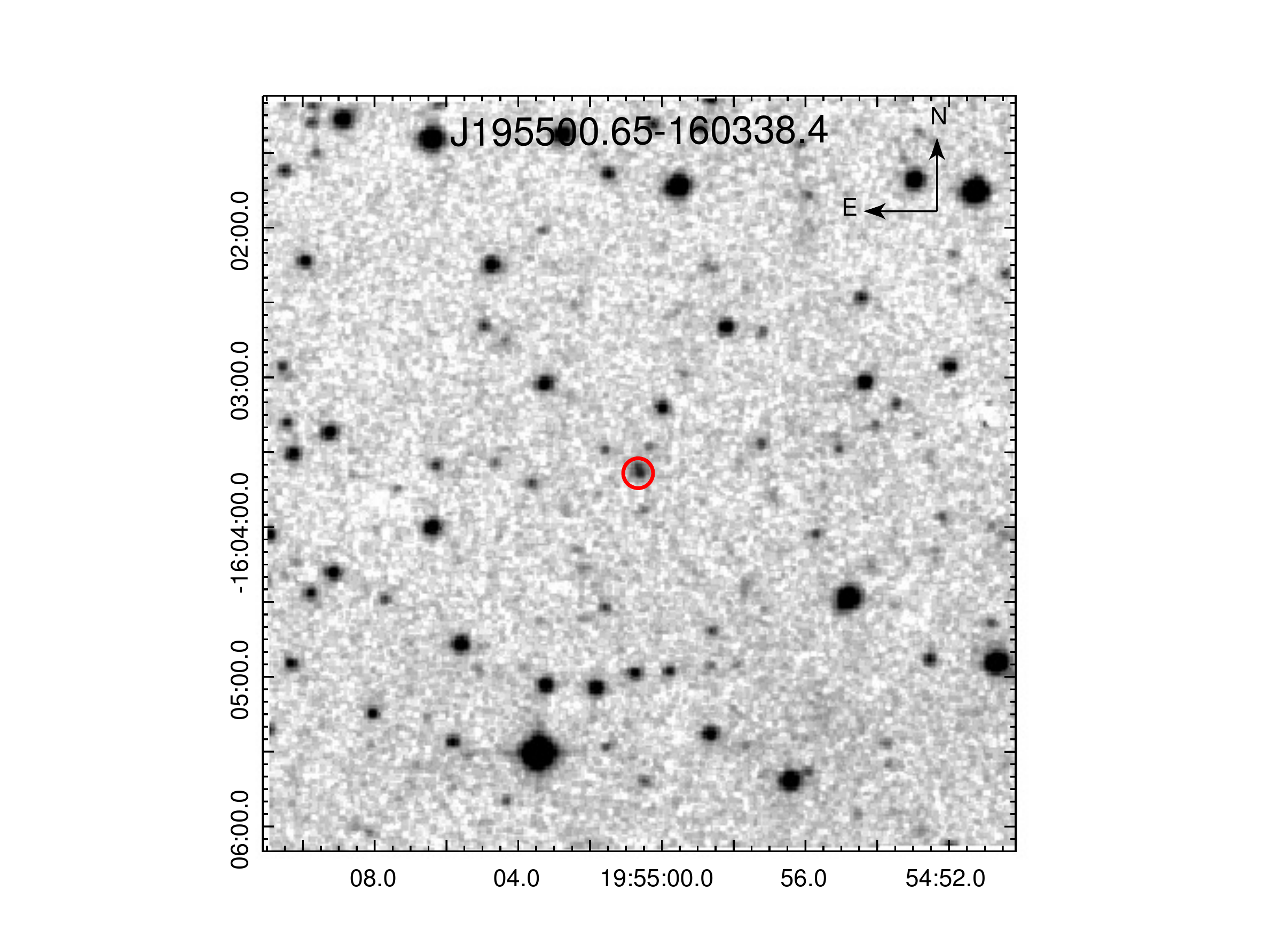} \\
\end{array}$
\end{center}
\caption{(Left panel) Optical spectrum of  WISE J195500.65-160338.4 associated with 3FGL J1955.0-1605, in the upper part it is shown the Signal-to-Noise Ratio of the spectrum. (Right panel) The finding chart ( $5'\times 5'$ ) retrieved from the Digital Sky Survey highlighting the location of the counterpart: WISE J195500.65-160338.4 (red circle).}
\label{fig:J1955}
\end{figure*}

\begin{figure*}{}
\begin{center}$
\begin{array}{cc}
\includegraphics[width=\mywidth,angle=0]{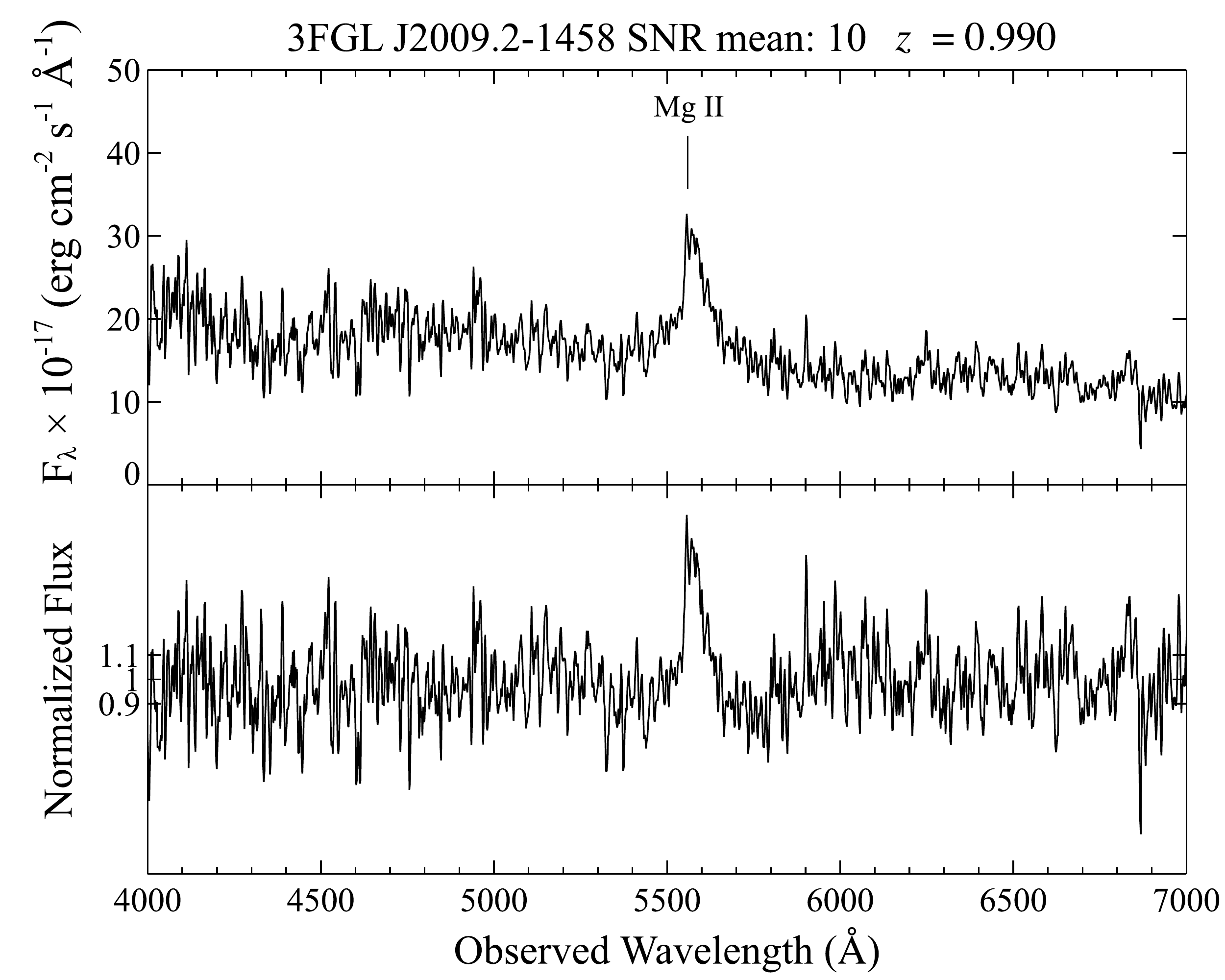} &
\includegraphics[trim=4cm 0cm 4cm 0cm, clip=true, width=7cm,angle=0]{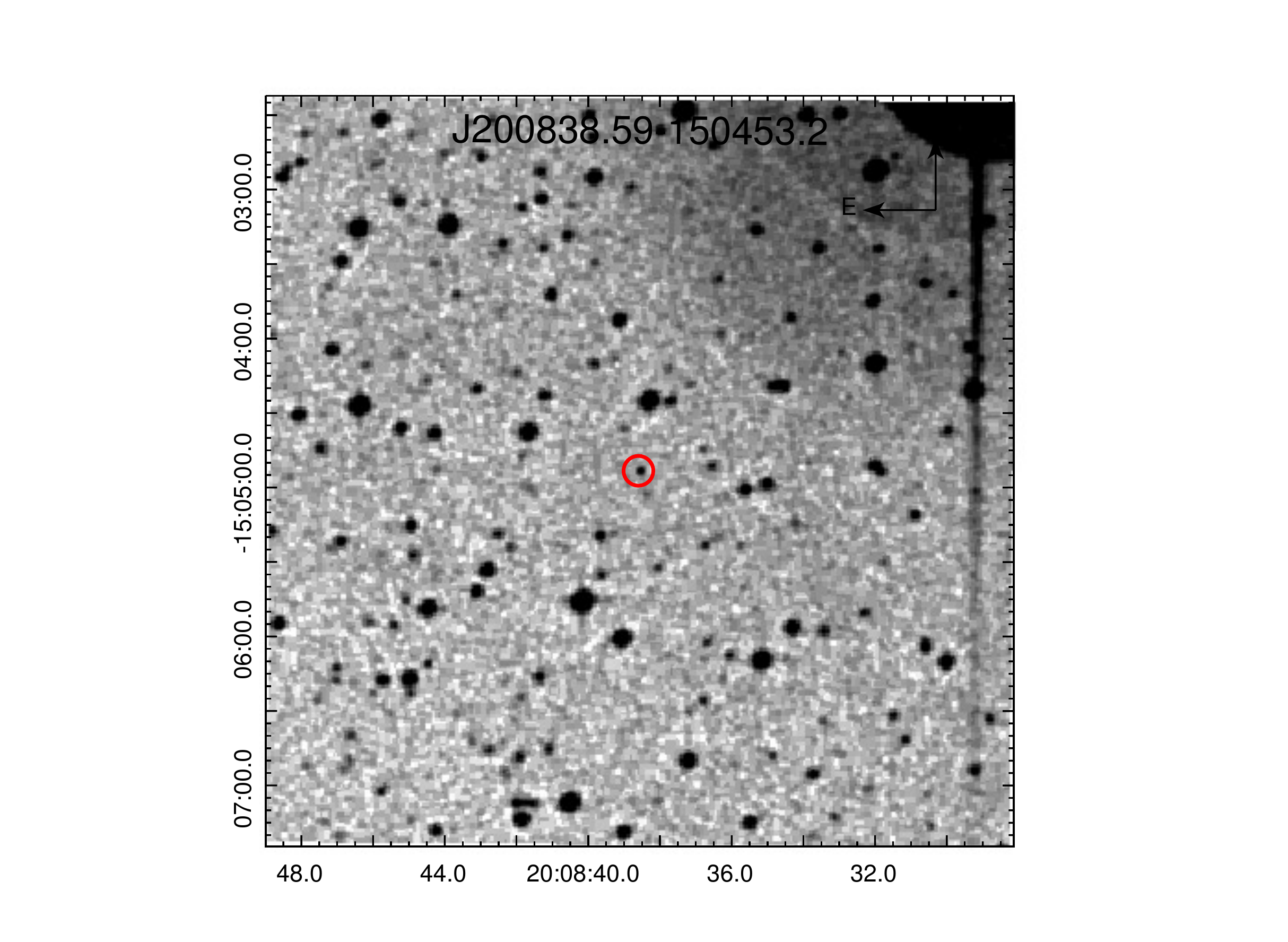} \\
\end{array}$
\end{center}
\caption{(Left panel) Optical spectrum of  WISE J200838.59-150453.2 associated with 3FGL J2009.2-1458. Signal-to-noise ratio is reported in the Figure. (Right panel) The finding chart ( $5'\times 5'$ ) retrieved from the Digital Sky Survey highlighting the location of the potential source: WISE J200838.59-150453.2 (red circle).}
\label{fig:J2009}
\end{figure*}

\begin{figure*}{}
\begin{center}$
\begin{array}{cc}
\includegraphics[width=\mywidth,angle=0]{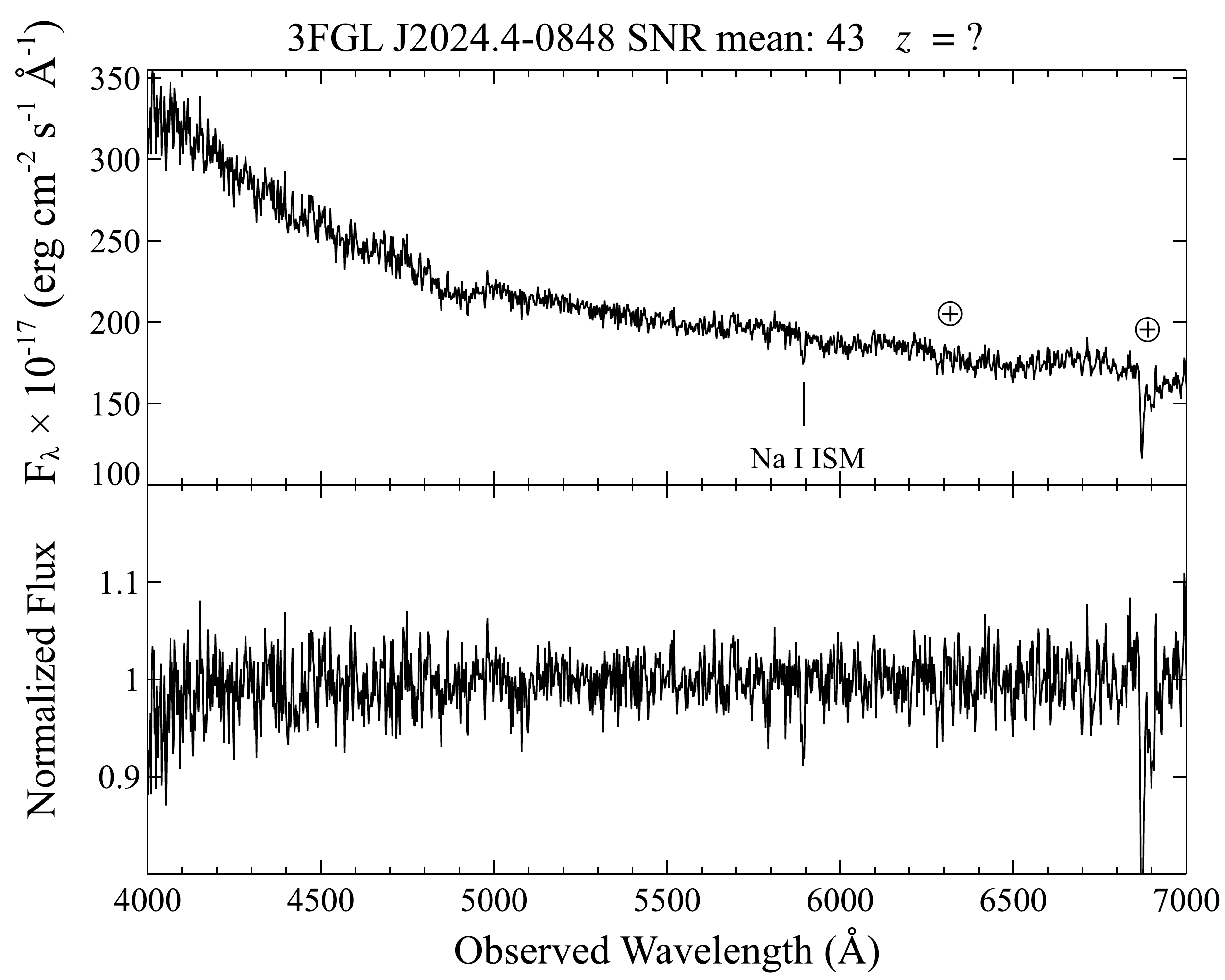} &
\includegraphics[trim=4cm 0cm 4cm 0cm, clip=true, width=7cm,angle=0]{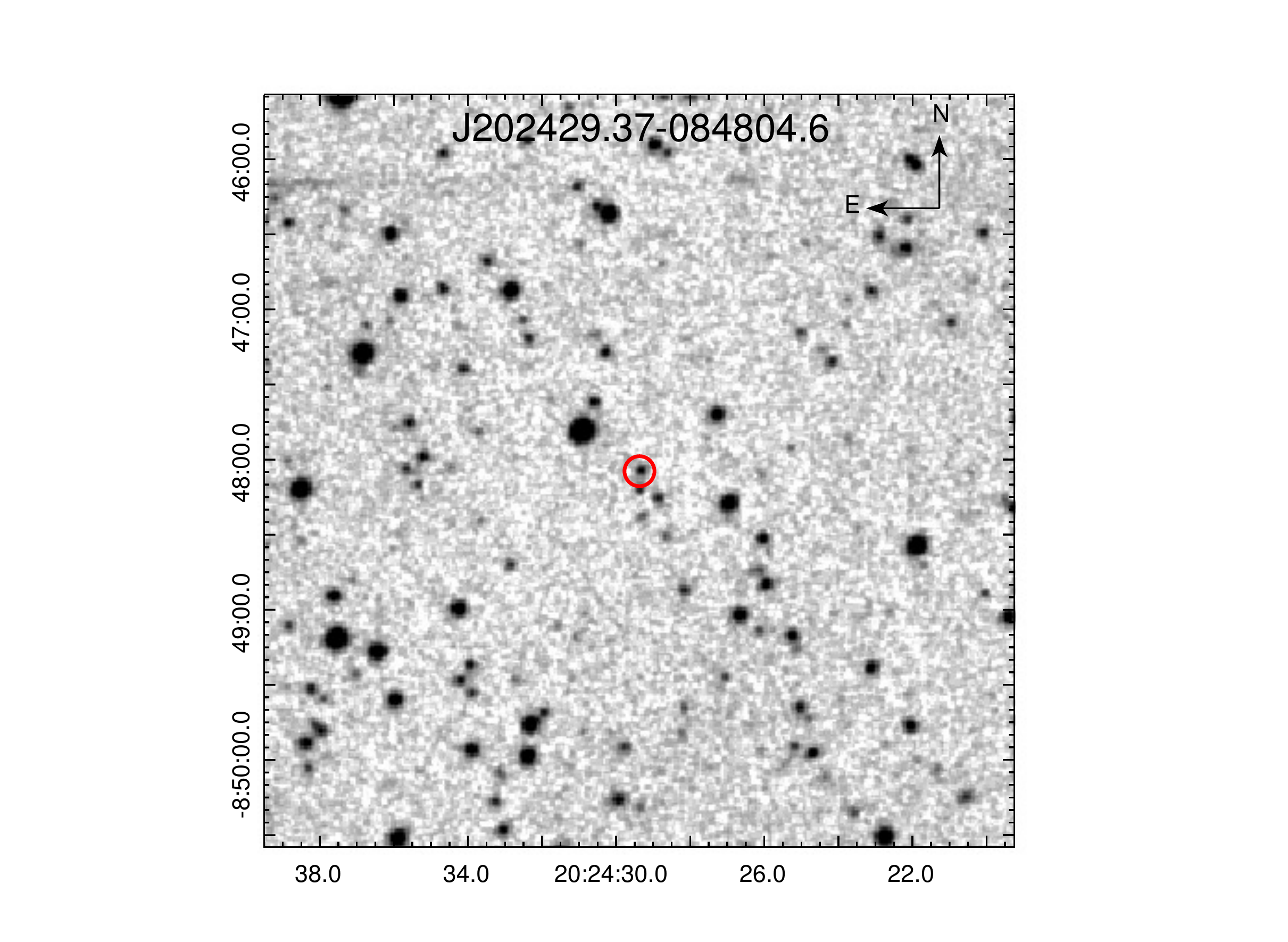} \\
\end{array}$
\end{center}
\caption{(Left panel) Optical spectrum of  WISE J202429.37-084804.6 associated with 3FGL J2024.4-0848. Signal-to-noise ratio is reported in the Figure. (Right panel) The finding chart ( $5'\times 5'$ ) retrieved from the Digital Sky Survey highlighting the location of the counterpart: WISE J202429.37-084804.6 (red circle).}
\label{fig:J2024}
\end{figure*}

\begin{figure*}{}
\begin{center}$
\begin{array}{cc}
\includegraphics[width=\mywidth,angle=0]{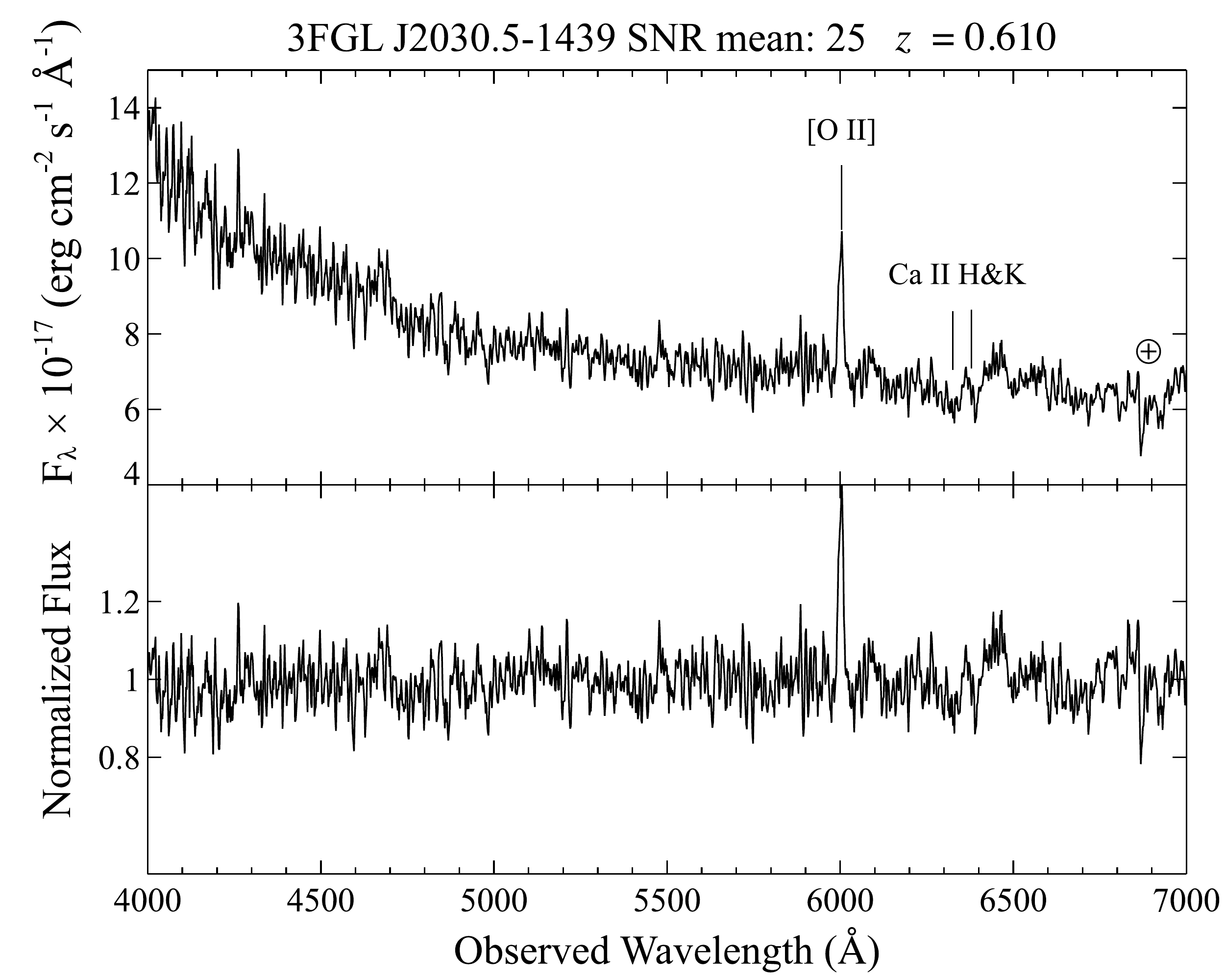} &
\includegraphics[trim=4cm 0cm 4cm 0cm, clip=true, width=7cm,angle=0]{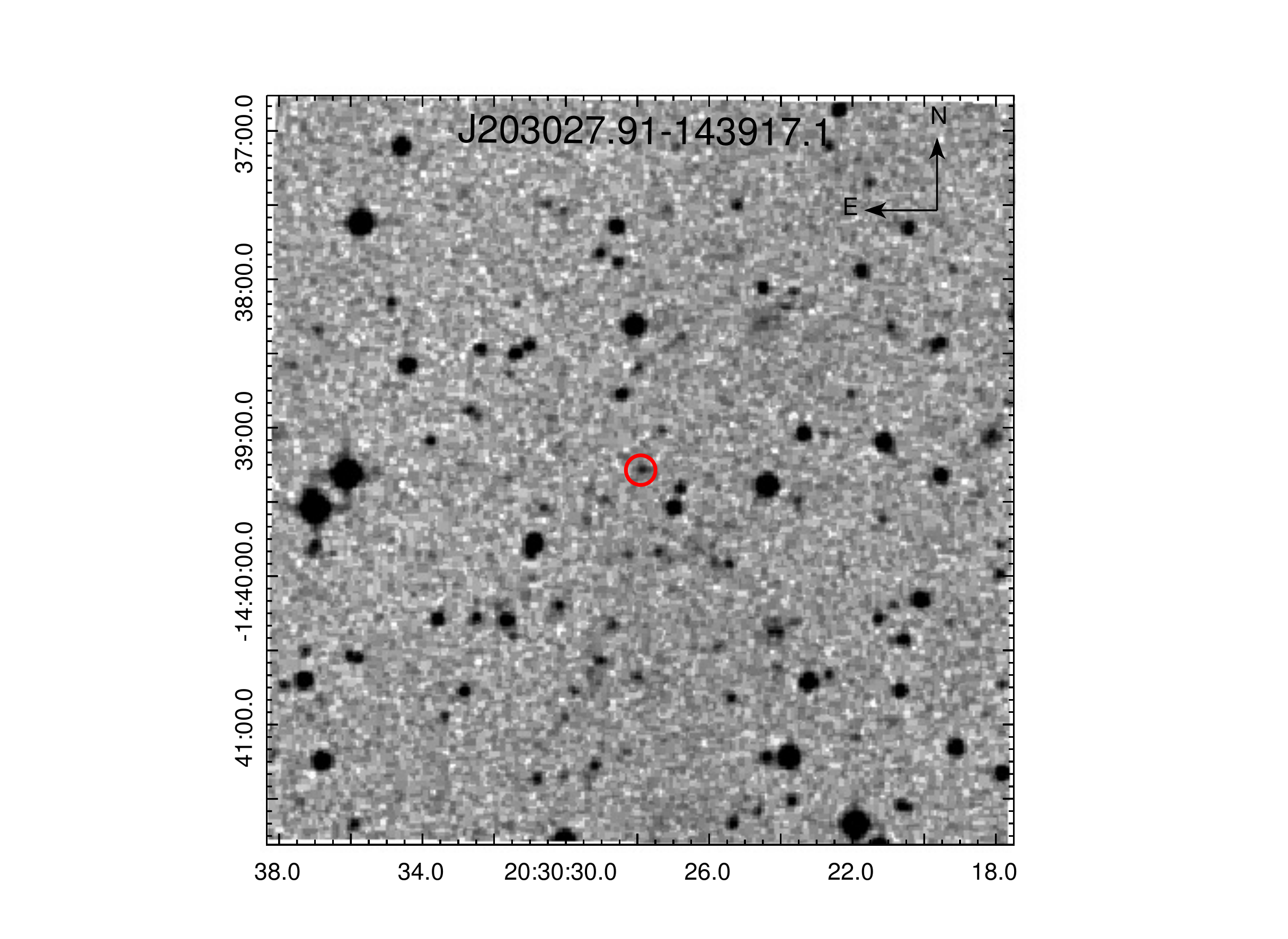} \\
\end{array}$
\end{center}
\caption{(Left panel) Optical spectrum of  WISE J203027.91-143917.1 associated with 3FGL J2030.5-1439. Signal-to-noise ratio is reported in the Figure. (Right panel) The finding chart ( $5'\times 5'$ ) retrieved from the Digital Sky Survey highlighting the location of the potential source: WISE J203027.91-143917.1 (red circle).}
\label{fig:J2030}
\end{figure*}

\begin{figure*}{}
\begin{center}$
\begin{array}{cc}
\includegraphics[width=\mywidth,angle=0]{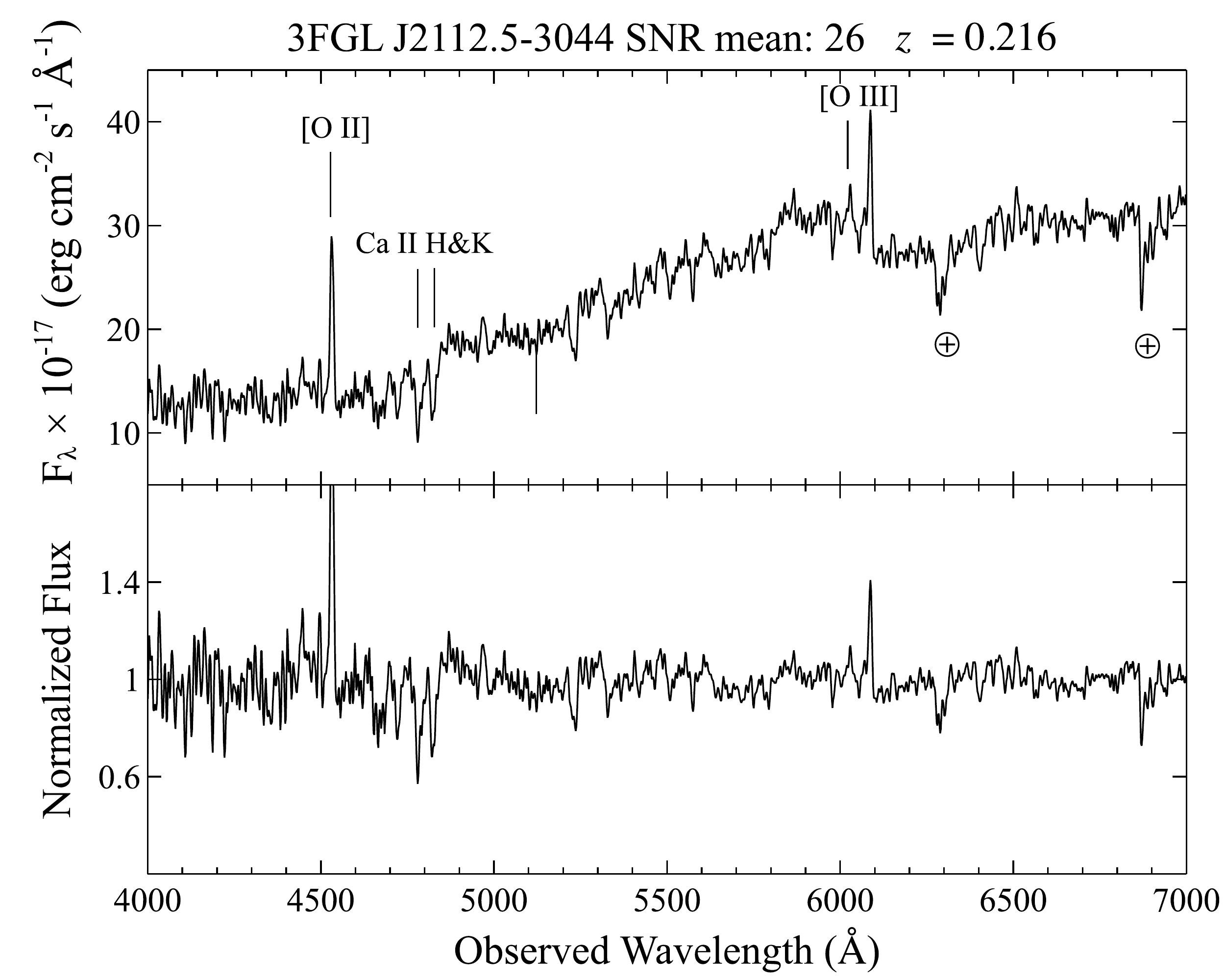} &
\includegraphics[trim=4cm 0cm 4cm 0cm, clip=true, width=7cm,angle=0]{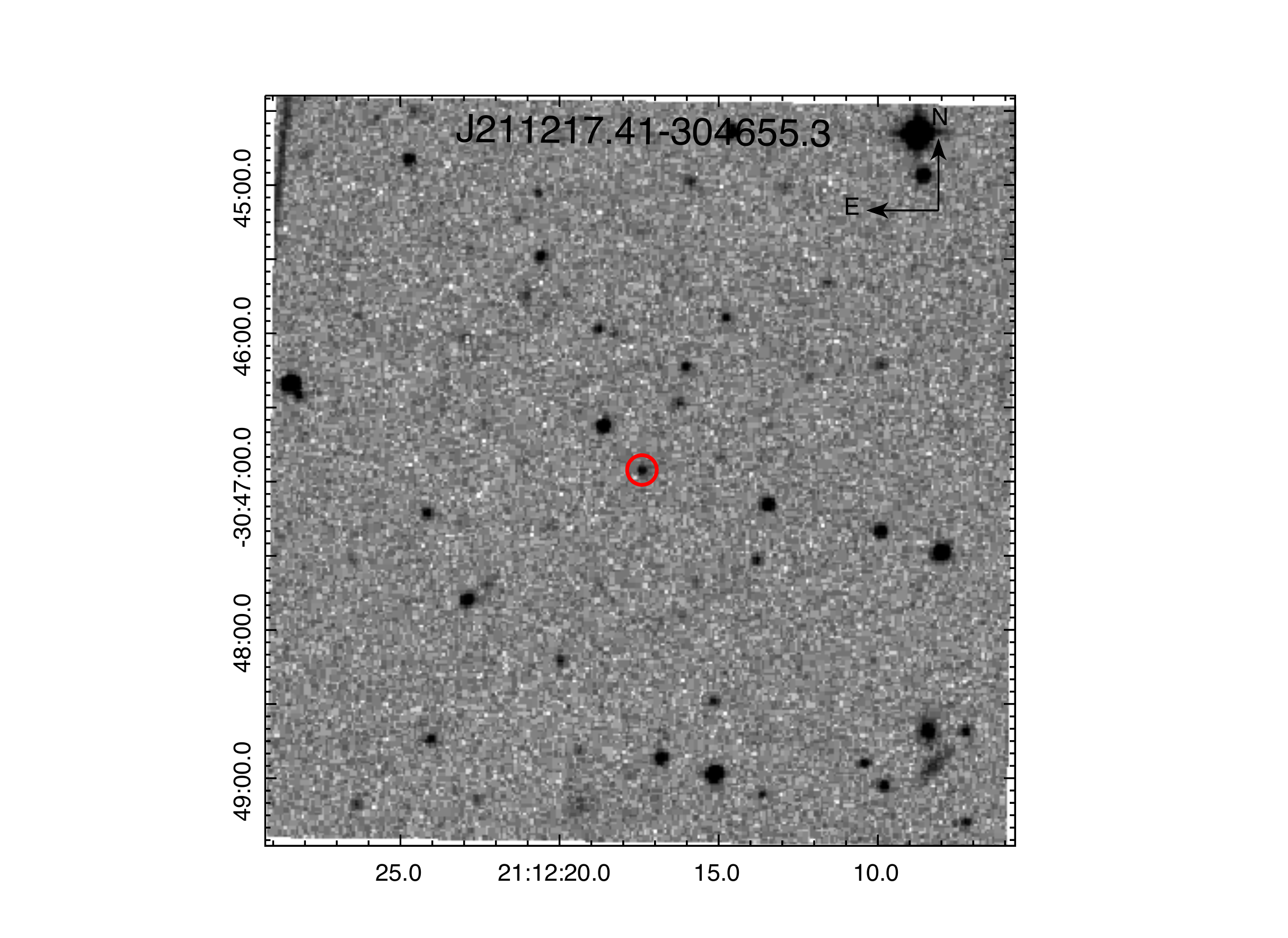} \\
\end{array}$
\end{center}
\caption{(Left panel) Optical spectrum of  WISE J211217.41-304655.3 associated with 3FGL J2112.5-3044. Signal-to-noise ratio is reported in the Figure. (Right panel) The finding chart ( $5'\times 5'$ ) retrieved from the Digital Sky Survey highlighting the location of the potential source: WISE J211217.41-304655.3 (red circle).}
\label{fig:J2112}
\end{figure*}

\begin{figure*}{}
\begin{center}$
\begin{array}{cc}
\includegraphics[width=\mywidth,angle=0]{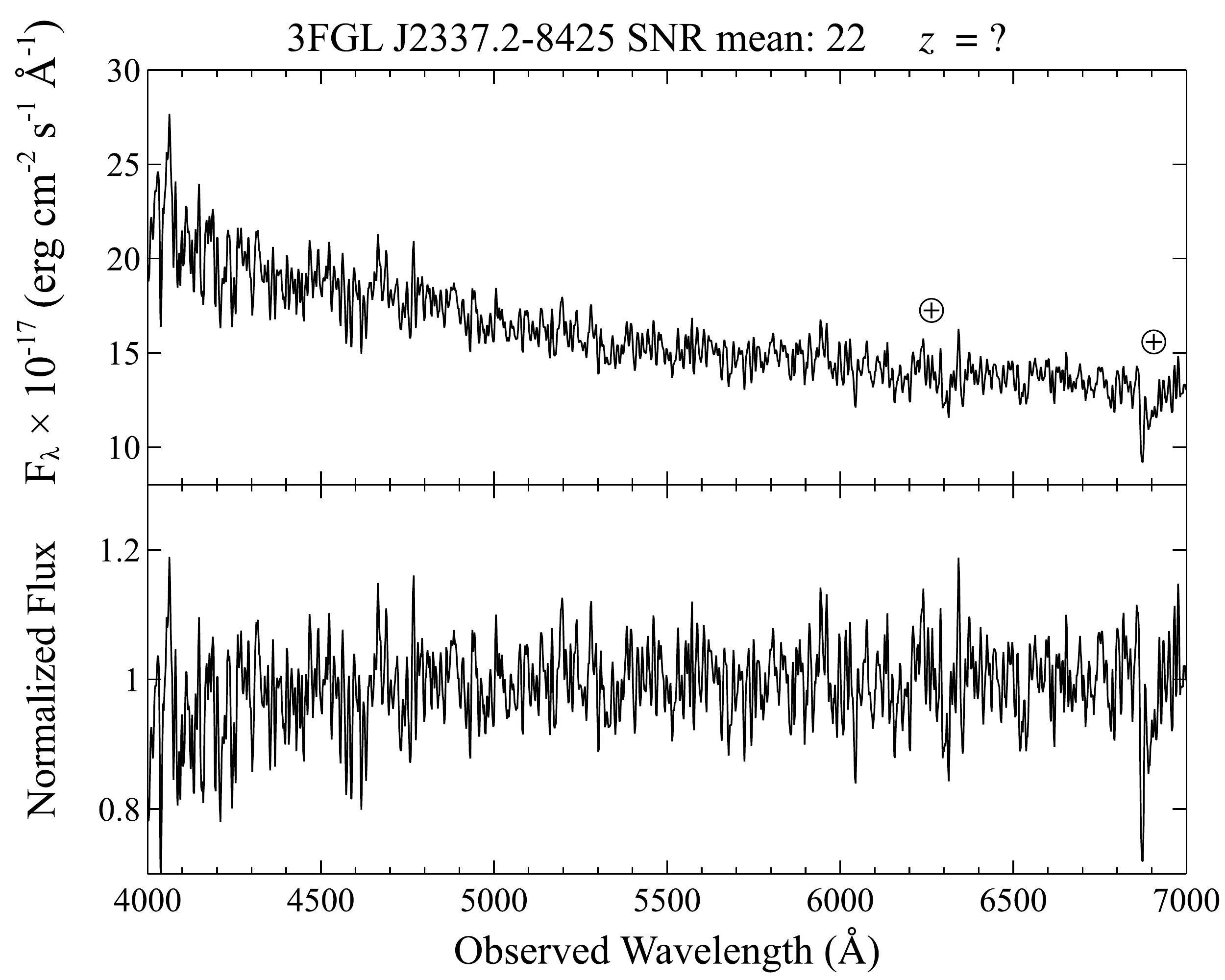} &
\includegraphics[trim=4cm 0cm 4cm 0cm, clip=true, width=7cm,angle=0]{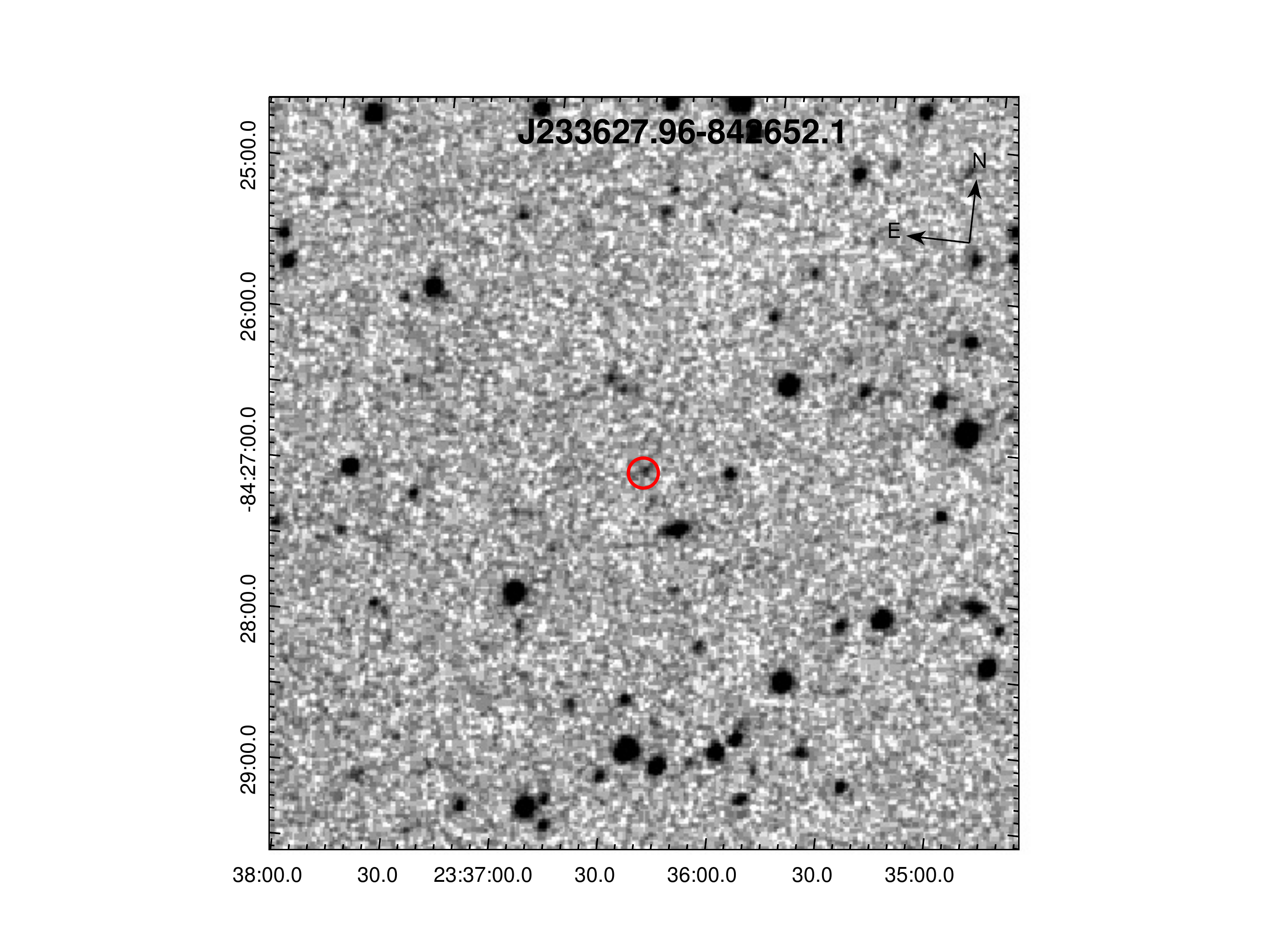} \\
\end{array}$
\end{center}
\caption{(Left panel) Optical spectrum of  WISE J233627.96-842652.1 associated with 3FGL J2337.2-8425. Signal-to-noise ratio is reported in the Figure. (Right panel) The finding chart ( $5'\times 5'$ ) retrieved from the Digital Sky Survey highlighting the location of the potential source: WISE J233627.96-842652.1 (red circle).}
\label{fig:J2337}
\end{figure*}

For each source we acquired at least two exposures, and reduced them using standard IRAF standards reduction techniques \citep{tody86}. We performed the bias and flat fielding corrections, and cosmic rays removal. In addition, we acquired calibration spectra of a Hg-Ar or Fe-Ar lamp for each source to calibrate the dispersion, achieving an accuracy of $0.1-0.5 \, \AA$ rms. During each night we observed at least one spectrophotometric star to perform relative flux calibration. Furthermore, we corrected the spectra by galactic extinction using the reddening law of \citeauthor{cardelli89} \citeyearpar{cardelli89} and values of $E_{\bv}$ computed by \citeauthor{schlegel98} \citeyearpar{schlegel98}. Finally, to highlight the spectroscopic features for visual inspection we normalised the spectra to the local continuum. In the Figures~\ref{fig:J1106}-~\ref{fig:J2337} we report the spectra and the correspondent finding charts, while both the log and the results of our observations are reported in Table~\ref{tab:main}. In addition, we report the results of our spectroscopic observations in Table~\ref{tab:main}.

\section{Results}
\label{sec:results}



All sources are listed in 3FGL with the only exceptions of UGS 1FGL J1129.2-0528 and the BCU 2FGL J1922.6-7454. We present below our results divided in the three categories of our sample.\\

\subsection{Unidentified Gamma-ray Sources}

Out of \textbf{33} UGSs in our sample, \textbf{20} of them show a BZB spectrum. We were not able to estimate the redshift for all of them with a with a single exception, WISE J012152.69-391544.2 which is the potential counterpart of 3FGL J0121.8-3917, with Ca II H \& K lines visible at $\lambda\lambda_{obs} = 5438-5518 \, \AA$ and $EW_{obs}=1.5-1.02 \, \AA$ leading to a redshift of $z=0.390$ (see Table~\ref{tab:main}). The remaining \textbf{19} objects shows featureless blue spectra typical of a BZB. \\

Within the UGSs sample 7 blazar-like candidates have a quasar spectra but none of them has a radio counterpart in NVSS, SUMSS, FIRST (\citeauthor{condon98}\citeyear{condon98}; \citeauthor{sumss}\citeyear{sumss}; \citeauthor{firstb}\citeyear{firstb}; \citeauthor{firstw}\citeyear{firstw}) and in the radio follow up performed by \citeauthor{petrov13} \citeyearpar{petrov13} and \citeauthor{schinzel15}\citeyearpar{schinzel15}. The lack of radio detections does not allow us to classify them as BZQs. 
We also found 6 objects with galaxy-like spectrum. QSOs and the galaxies could be considered contaminants of the selection procedures of UGSs counterparts \citep{ugs1met,ugs3lowfrq,ugs2}. We list all the UGS in Table~\ref{tab:main} along with their classification and redshift estimates.\\

\subsection{Blazar Candidates of Uncertain Type}

In our sample there are 27 BCUs and for all of them we confirm a blazar nature thanks to our follow up spectroscopic observations. Within the sample we found two candidates with QSO spectra. The first one is the candidate WISE J080311.45-033554.5, counterpart of 3FGL J0803.3-0339, showing  [OII]$\lambda3727$ emission line with $EW_{obs}=17 \, \AA$ enabling us to estimate its redshift at z=$0.365$, while the second one is WISE J161717.91-584808.0 counterpart of 3FGL J1617.4-5846 that shows broad MgII$\lambda2798$ and the blending of SiIII]$\lambda1892$ with CIII]$\lambda1909$, respectively, giving a $z=1.423$.\\

The remaining \textbf{25} BCUs show a BZB spectra and for 6 of them it was also possible to estimate their redshifts or a lower limit of their redshift given the detection of interestellar absorption features usually seen in BL Lac spectra (\citeauthor{sbarufatti06}\citeyear{sbarufatti06}). In detail, for WISE J064933.60-313920.3 counterpart of 3FGL J0649.6-3138, and associated with the X-ray source 1RXS J064933.8-313914 for which we estimated a lower limit for its redshift of $z \geq 0.563$ using the Ca II H\&K absorption lines.
Meanwhile for the target J100850.54-313905.5 counterpart of 3FGL J1009.0-3137 we  estimate its redshift at $z=0.534$ based on the [OII]$\lambda3727$ emission line ($EW_{obs}=1.4 \, \AA$) and Ca II H\&K absorption doublet at ($EW_{obs}=0.7-0.4 \, \AA$). \\

The third BCU is WISE J120317.88-392620.9, counterpart of 3FGL J1203.5-3925, for which we estimated a redshift of $0.227$, based on the [OII]$\lambda3727$ emission ($EW_{obs}=3.2 \, \AA$), Ca II H\&K absorption lines ($EW_{obs}=3.3-2.0 \, \AA$) and the [OIII] doublet emission line ($EW_{obs}=2.0-4.2 \, \AA$).
The fourth BCU is the associated to 3FGL J1312.7-2349, WISE J131248.76-235047.3, we estimated the lower limit of its redshift at $z\geq0.462$ showing the Mg II absorption lines ($EW_{obs}=3.2-2.9 \, \AA$) and the doublet of Ca II H\&K absorption lines ($EW_{obs}=0.6-0.7 \, \AA$).
Finally, for the BCU, WISE J195500.65-160338.4, counterpart of 3FGL J1955.0-1605, we estimated a lower limit for its redshift at $z\geq0.630$ using the doublet of Mg II ($EW_{obs}= 2.0-1.6 \, \AA$). \\

We found that WISE J110624.04-364658.9 associated with 3FGL J1106.4-3643 in the 3FGL/3LAC, is a high redshift BZB at $z\geq1.084$, estimated from the multiplet of Fe II absorption lines and the Mg II doublet absorption lines ($EW_{obs}=6.2-5.4  \, \AA$). This is the second most distant BZB to date (\citeauthor{quest16} \citeyear{quest16}).

\subsection{Other targets}

Finally, we report the observation of 3FGL J0814.1-1012 associated with the radio source NVSS J081411-101208 (a.k.a. WISE J081411.69-101210.2). We confirm its nature/classification as BZB but its spectrum does not have any spectroscopic feature that allows to estimate its redshift as occurred in previous observations (\citeauthor{crespo16b} \citeyear{crespo16b}). \\

\onecolumn{
\begin{center}
\begin{landscape}

\begin{longtable}{cccccccccc}
\caption{Observing Log and Summary of Results} \label{tab:main}\\
\tablecaption{Summary of targets} 
\tablewidth{\textwidth} 
\tabletypesize{\footnotesize} \\
\multicolumn{1}{c}{Fermi Name} &
\multicolumn{1}{c}{WISE Counterpart} & 
\multicolumn{1}{c}{R.A. (J2000)} &
\multicolumn{1}{c}{Dec. (J2000)} &
\multicolumn{1}{c}{Obs. Date} &
\multicolumn{1}{c}{Exp. Time} &
\multicolumn{1}{c}{S/N} & 
\multicolumn{1}{c}{z} &
\multicolumn{1}{c}{Class} &
\multicolumn{1}{c}{Notes} \\
\multicolumn{1}{c}{ } &
\multicolumn{1}{c}{ } &
\multicolumn{1}{c}{hh:mm:ss} &
\multicolumn{1}{c}{dd:mm:ss} &
\multicolumn{1}{c}{yyy-mm-dd} &
\multicolumn{1}{c}{$\left( s \right)$} &
\multicolumn{1}{c}{ } &
\multicolumn{1}{c}{ } &
\multicolumn{1}{c}{ } &
\multicolumn{1}{c}{ } \\
\hline
 \endhead
 

\multicolumn{10}{c}{\small \textsc{\textbf{UGSs}}} \\ 
\hline
  3FGL J0121.8-3917 & J012152.69-391544.2 & 01:21:52.70 & -39:15:44.21 & 2016-08-02 & 900 & 41 & 0.390 $\pm$ 0.001 & bzb & N,w,g\\
   \textbf{ 3FGL J0156.5-2423} & J015624.54-242003.7 & 01:56:24.55 & -24:20:03.77 & 2015-08-23 & 600 & 52 & ? & bzb & c,w,x\\
  3FGL J0200.3-4108 & J020020.94-410935.7 & 02:00:20.95 & -41:09:35.70 & 2015-08-26 & 600 & 38 & ? & bzb & w,x\\
  3FGL J0312.7-2222 & J031235.70-222117.2 & 03:12:35.71 & -22:21:17.21 & 2017-01-31 & 1200 & 71 & ? & bzb & N,w\\
  3FGL J0340.4-2423 & J034050.11-242254.6 & 03:40:50.11 & -24:22:54.66 & 2015-08-23 & 600 & 41 & 0.683 $\pm$ 0.002 & qso & w\\
  3FGL J0351.0-2816 & J035051.32-281632.8 & 03:50:51.33 & -28:16:32.81 & 2016-08-29 & 1800 & 45 & ? & bzb & N,w\\
  3FGL J0414.9-0840 & J041433.10-084206.8 & 04:14:33.10 & -08:42:06.86 & 2016-09-11 & 1200 & 30 & ?  & bzb & w, N\\
  3FGL J0420.4-6013 & J042011.02-601505.5 & 04:20:11.03 & -60:15:05.51 & 2015-08-23 & 600 & 38 & ? & bzb & w,x\\
  3FGL J0437.7-7330 & J043837.07-732921.6 & 04:38:37.07 & -73:29:21.60 & 2016-08-30 & 600 & 24 & 0.150 $\pm$ 0.001 & gal & S,w,M,x\\
  3FGL J0704.3-4828 & J070421.81-482647.5 & 07:04:21.82 & -48:26:47.59 & 2016-12-22 & 1800 & 20 & ? & bzb & w,x\\
  3FGL J0721.5-0221 & J072113.90-022055.0 & 07:21:13.90 & -02:20:55.07 & 2014-12-17 & 1800 & 28 & ? & bzb & N,w,x\\
  3FGL J0747.5-4927 & J074724.74-492633.1 & 07:47:24.74 & -49:26:33.18 & 2016-12-30 & 1800 & 36 & ? & bzb & S,w\\
  3FGL J0826.3-6400 & J082627.86-640415.4 & 08:26:27.87 & -64:04:15.45 & 2017-01-30 & 1800 & 51 & ? & bzb & S,w,x\\
  3FGL J1013.4-4008 & J101319.30-400550.4 & 10:13:19.50 & -40:05:49.20 & 2017-01-30 & 2400 & 100 & ?  & bzb & S,N,w\\
  \textbf{ 3FGL J1033.0-5945} & J103332.15-503528.8 & 10:33:32.16 & -50:35:28.82 & 2016-12-22 & 1200 & 27 & ? & bzb & w,x\\
  3FGL J1100.2-2044 & J110028.22-205000.7 & 11:00:28.32 & -20:50:05.60 & 2017-01-30 & 1000 & 33 & 0.239  $\pm$ 0.001 & gal & N,w,s\\
  3FGL J1132.0-4736 & J113209.26-473853.3 & 11:32:09.26 & -47:38:53.31 & 2017-01-30 & 600 & 41 & 0.210 $\pm$ 0.001 & gal & S,w,M,x\\
  3FGL J1325.2-5411 & J132457.35-541503.2 & 13:24:57.36 & -54:15:03.29 & 2016-08-02 & 900 & 20 & 0.218 $\pm$ 0.001 & gal & w\\
  3FGL J1946.4-5403 & J194633.62-540236.4 & 19:46:33.63 & -54:02:36.43 & 2015-06-23 & 900 & 35 & 0.460 $\pm$ 0.002 & qso & w\\
  3FGL J2009.2-1458 & J200838.59-150453.2 & 20:08:38.60 & -15:04:53.27 & 2015-08-23 & 600 & 10 & 0.990 $\pm$ 0.004 & qso & N,w\\
  3FGL J2030.5-1439 & J203027.91-143917.1 & 20:30:27.91 & -14:39:17.17 & 2016-08-02 & 1200 & 25 & 0.234 $\pm$ 0.001 & qso & N,w,s\\
  3FGL J2112.5-3044 & J211217.41-304655.3 & 21:12:17.41 & -30:46:55.33 & 2015-08-18 & 900 & 26 & 0.216 $\pm$ 0.001 & gal & w,x\\
  3FGL J2144.6-5640 & J214429.57-563849.0 & 21:44:29.57 & -56:38:49.08 & 2016-08-02 & 1200 & 25 & ? & bzb & w,x\\
  3FGL J2150.5-1754 & J215046.60-174954.1 & 21:50:46.61 & -17:49:54.18 & 2015-08-26 & 600 & 55 & 0.186 $\pm$ 0.001 & gal & N,w\\
  3FGL J2209.8-0450 & J220941.69-045110.3 & 22:09:41.70 & -04:51:10.33 & 2015-08-23 & 600 & 54 & ? & bzb & F,N,w,s\\
  3FGL J2237.5-8326 & J224201.61-832744.4 & 22:42:01.62 & -83:27:44.45 & 2015-08-18 & 900 & 50 & 0.202 $\pm$ 0.001 & qso & w\\
  3FGL J2244.6+2503 & J224436.66+250343.1 & 22:44:36.67 & +25:03:43.20 & 2015-08-18 & 600 & 36 & ? & bzb & N,w,s\\
  3FGL J2300.1-3547 & J230053.29-355051.0 & 23:00:53.30 & -35:50:51.08 & 2016-12-30 & 1200 & 49 & 0.753 $\pm$ 0.001 & qso & w\\
  3FGL J2321.6-1619 & J232136.98-161928.3 & 23:21:36.98 & -16:19:28.32 & 2015-08-23 & 600 & 47 & ? & bzb & N,w\\
  3FGL J2337.2-8425 & J233627.96-842652.1 & 23:36:27.97 & -84:26:52.17 & 2015-08-18 & 600 & 22 & ? & bzb & w\\
  3FGL J2351.9-7601 & J235116.13-760015.5 & 23:51:16.13 & -76:00:15.53 & 2015-08-26 & 600 & 40 & ? & bzb & S,w,x\\
  3FGL J2358.6-1809 & J235836.83-180717.4 & 23:58:36.84 & -18:07:17.48 & 2015-08-23 & 600 & 51 & ? & bzb & N,w,x\\
  1FGL J1129.2-0528 & J112914.05-052856.3 & 11:29:14.06 & -05:28:56.36 & 2015-05-20 & 1350 & 34 & 0.920 $\pm$ 0.001 & qso & F,N,w,s\\
\hline
\multicolumn{10}{c}{\small \textsc{\textbf{BCUs and AGUs}}} \\ 
\hline

  3FGL J0127.2+0325 & J012713.94+032300.6 & 01:27:13.95 & +03:23:00.64 & 2016-12-22 & 1800 & 44 & ? & bzb & N,w,s,x\\
  3FGL J0310.4-5015 & J031034.72-501631.1 & 03:10:34.72 & -50:16:31.13 & 2014-12-17 & 2100 & 62 & ?  & bzb & S,w,X,x\\
  3FGL J0439.9-1859 & J043949.72-190101.5 & 04:39:49.73 & -19:01:01.57 & 2014-12-17 & 2100 & 57 & ?  & bzb & Pm,w,x\\
  3FGL J0626.6-4259 & J062636.71-425805.9 & 06:26:36.71 & -42:58:05.92 & 2017-01-30 & 900 & 45 & ? & bzb & w,X\\
  3FGL J0649.6-3138 & J064933.60-313920.3 & 06:49:33.60 & -31:39:20.34 & 2014-12-18 & 1050 & 55 & $\geq0.563$ & bzb & w,X,x\\
  3FGL J0703.4-3914 & J070312.65-391418.8 & 07:03:12.66 & -39:14:18.89 & 2014-12-17 & 1800 & 49 & ? & bzb & w, S,N,A,X,x\\
  3FGL J0803.3-0339 & J080311.45-033554.5 & 08:03:11.45 & -03:35:54.57 & 2014-12-18 & 900 & 45 & 0.365 $\pm$ 0.001  & bzq & Pm,V,T N,w,x\\
  3FGL J0827.2-0711 & J082706.16-070845.9 & 08:27:06.17 & -07:08:45.93 & 2015-05-19 & 300 & 52 & ? & bzb & Pm,N,w,M,X\\
  3FGL J0858.1-3130 & J085802.90-313038.3 & 08:58:02.91 & -31:30:38.31 & 2014-12-18 & 1050 & 25 & ? & bzb & w,X\\
  3FGL J0947.1-2542 & J094709.52-254059.9 & 09:47:09.53 & -25:40:59.97 & 2014-12-18 & 600 & 95 & ? & bzb & N,w,6,X,x\\
  3FGL J1009.0-3137 & J100850.54-313905.5 & 10:08:50.55 & -31:39:05.50 & 2015-05-19 & 1350 & 98 & 0.534 $\pm$ 0.001  & bzb & Pm,T,S,N,c,w\\
  3FGL J1106.4-3643 & J110624.04-364658.9 & 11:06:24.04 & -36:46:58.96 & 2015-05-21 & 1500 & 62 & 1.084 $\pm$ 0.001  & bzb & Pm,N,S,A,c,w\\
  3FGL J1125.0-2101 & J112508.62-210105.9 & 11:25:08.63 & -21:01:05.98 & 2015-05-19 & 1200 & 103 & ? & bzb & Pm,N,c,w,X,\\
  3FGL J1203.5-3925 & J120317.88-392620.9 & 12:03:17.89 & -39:26:20.96 & 2015-05-20 & 900 & 63 & 0.227 $\pm$ 0.001  & bzb & Pm,S,N,c,w\\
  3FGL J1218.8-4827 & J121902.26-482627.9 & 12:19:02.27 & -48:26:27.98 & 2015-05-21 & 1800 & 84 & ? & bzb & Pm,S,c,w\\
  3FGL J1307.6-4300 & J130737.98-425938.9 & 13:07:37.98 & -42:59:38.97 & 2015-05-19 & 600 & 115 & ? & bzb & w,X,x\\
  3FGL J1312.7-2349 & J131248.76-235047.3 & 13:12:48.76 & -23:50:47.38 & 2015-05-19 & 1800 & 63 & $\geq0.462$ & bzb & N,w\\
  3FGL J1512.2-2255 & J151212.75-225508.4 & 15:12:12.76 & -22:55:08.47 & 2015-05-21 & 2700 & 67 & ? & bzb & w,X,x\\
  3FGL J1518.0-2732 & J151803.59-273131.1 & 15:18:03.60 & -27:31:31.13 & 2015-05-20 & 150 & 67 & ? & bzb & Pm,T,N,A,c,w,M,g\\
  3FGL J1539.8-1128 & J153941.19-112835.3 & 15:39:41.20 & -11:28:35.36 & 2015-05-20 & 600 & 107 & ? & bzb & Pm,N,w,6,g,X,\\
  3FGL J1547.1-2801 & J154712.13-280221.5 & 15:47:12.13 & -28:02:21.57 & 2015-05-20 & 1350 & 54 & ? & bzb & w,X\\
  3FGL J1617.4-5846 & J161717.91-584808.0 & 16:17:17.91 & -58:48:08.07 & 2015-05-20 & 900 & 115 & 1.414 $\pm$ 0.001  & bzq & Pm,A,w,X\\
  3FGL J1637.6-3449 & NVSS J163750-344915 & 16:37:50.99 & -34:49:15.40 & 2015-05-19 & 300 & 62 & ? & bzb & N,A,X\\
  3FGL J1656.8-2010 & J165655.14-201056.2 & 16:56:55.15 & -20:10:56.30 & 2015-05-21 & 1800 & 43 & ? & bzb & w,X\\
  3FGL J1955.0-1605 & J195500.65-160338.4 & 19:55:00.66 & -16:03:38.41 & 2015-05-19 & 1200 & 65 & $\geq0.630$ & bzb & w,N,6,g,X\\
  3FGL J2024.4-0848 & J202429.37-084804.6 & 20:24:29.37 & -08:48:04.66 & 2015-05-19 & 1800 & 43 & ? & bzb &  N,w,6,g,X\\
  2FGL J1922.6-7454 & J192243.02-745349.5 & 19:22:43.02 & -74:53:49.60 & 2015-08-26 & 600 & 31 & ? & bzb & S,w,X\\
\hline 
 \multicolumn{10}{c}{\small \textsc{\textbf{Known Blazars}}} \\ 
\hline
  3FGL J0814.1-1012 & J081411.69-101210.2 & 08:14:11.69 & -10:12:10.25 & 2014-12-18 & 600 & 93 & ? & bzb & N,A,6,w,x\\
\hline
\end{longtable}
\begin{flushleft}
Column description: 
 (1) 3FGL name; 
 (2) Associated counterpart;  
 (3) R.A. (Equinox J200);
 (4) Dec. (Equinox J200);
 (5) Observation date;
 (6) Exposure time;
 (7) Signal-to-noise ratio;
 (8) redshift, question marks indicate unknown $z$; 
 (9) Source classification;
 (10) Multifrequency notes:
V, VLA Low-frequency Sky Survey Discrete Source Catalog  (VLSS) (\citeauthor{vlss} \citeyear{vlss});
Pm, Parkes-MIT-NRAO Surveys (PMN) \citep{pmn};
T , Texas Survey of Radio Sources (TEXAS) (\citeauthor{texas} \citeyear{texas});
c, Combined Radio All-Sky Targeted Eight-GHz Survey (CRATES) (\citeauthor{crates} \citeyear{crates});
N, NRAO VLA Sky Survey (NVSS) (\citeauthor{condon98} \citeyear{condon98});
F, VLA Faint Images of The Radio Sky at 20 cm (FIRST) (\citeauthor{firstb} \citeyear{firstb});
w, WISE all-sky survey in the Allwise Source catalog Two (WISE)  \citep{wright10};
M, Two Micron All Sky Survey (2MASS) (\citeauthor{2mass} \citeyear{2mass});
s, Sloan Digital Sky Survey Data Release 9 (SDSS DR9) (\citeauthor{ahn12} \citeyear{ahn12});
6, Six-degree-field Galaxy Redshift Survey (6dFGS) (\citeauthor{jones04} \citeyear{jones04}; \citeauthor{jones09}\citeyear{jones09});
\textbf{g, GALaxy Evolution eXplorer All-Sky Survey Source Catalog (GALEX)} (\citeauthor{seibert12}  \citeyear{seibert12});
X, ROSAT Bright Source Catalog ROSAT (RBSC) (\citeauthor{voges99} \citeyear{voges99});
X, ROSAT Faint Source Catalog (RFSC) (\citeauthor{voges00} \citeyear{voges00});
x, XMM-Newton Slew Survey (XMMSL) (\citeauthor{saxton08} \citeyear{saxton08}; \citeauthor{warwick12} \citeyear{warwick12});
x, Deep \textit{Swift} X-Ray Telescope Point Source Catalog (1SXPS) (\citeauthor{evans14} \citeyear{evans14});
x, Chandra Source Catalog (CSC) (\citeauthor{evans10} \citeyear{evans10}).
\end{flushleft}
\end{landscape}
\end{center}}

\twocolumn
\section{Summary and Conclusions}
\label{sec:summary}
We present the spectroscopic observations of \textbf{61} optical targets associated with Fermi-LAT detected sources. The observations were taken between 2014-2017 as part of our follow up optical campaign. Our sample consist of \textbf{33} blazar-like sources lying within the positional uncertainty regions of UGSs, selected on the basis of their IR colors, \textbf{27} BCUs and 1 known 3FGL BL Lac object. Results are summarized as follows.\\

\begin{itemize}

\item For the UGSs, we classified \textbf{20} candidates counterparts as BZBs. An estimate of redshift for WISE J012152.69-391544.2 potential counterpart of the source at 3FGL J0121.8-3917 ($z=0.390$). The remaining targets are thought to be contaminants of the WISE procedure used to search for potential counterparts \citep{ugs1met,ugs3lowfrq,ugs2}, there are 7 QSOs, without radio information, and 6 galaxies.\\

\item We confirm the blazar nature of all the \textbf{27} BCUs. Two of them are BZQs: WISE J080311.45-033554.5 counterpart of 3FGL J0803.3-0339 and WISE J161717.91-584808.0 counterpart of 3FGL J1617.4-5846 at $z=0.365$ and $z=1.414$, respectively. All the others are classified as BZBs with emission or absorption lines in their optical spectra with $EW < 5 \, \AA$. For this subsample of BZBs we obtained 6 redshifts estimates.\\
\end{itemize}

We observed again 3FGL J0814.1-1012, previously observed and classified by \citeauthor{crespo16b} \citeyearpar{crespo16b}. We confirm the blazar nature of its associated counterpart WISE J081411.69-101210.2. The spectrum shows a featureless blue continuum as a classical BZB. \\

To summarize, we identified 50 new blazars, correspondent to an additional $20 \, \%$ of those already classified and confirmed during our campaign to date. Finally, we also highlight the discovery of the BZB with the second most distant BL Lac known to date, 3FGL J1106.4-3643, with $z \geq 1.084$ estimated from the Mg II absorption lines.\\

\clearpage

\clearpage

\clearpage

\clearpage

\begin{figure*}{}
\begin{center}$
\begin{array}{cc}
\includegraphics[width=\mywidth,angle=0]{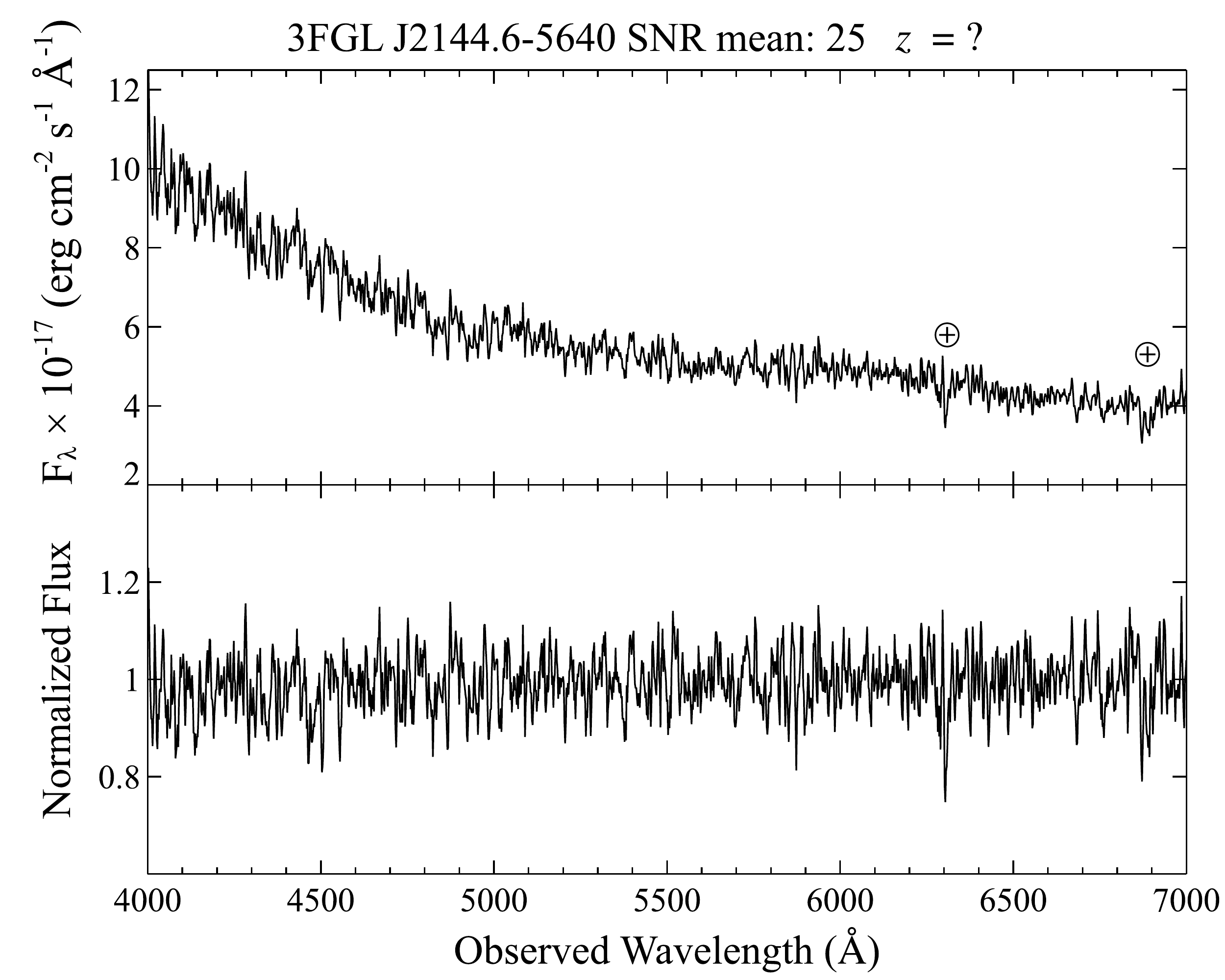} &
\includegraphics[trim=4cm 0cm 4cm 0cm, clip=true, width=7cm,angle=0]{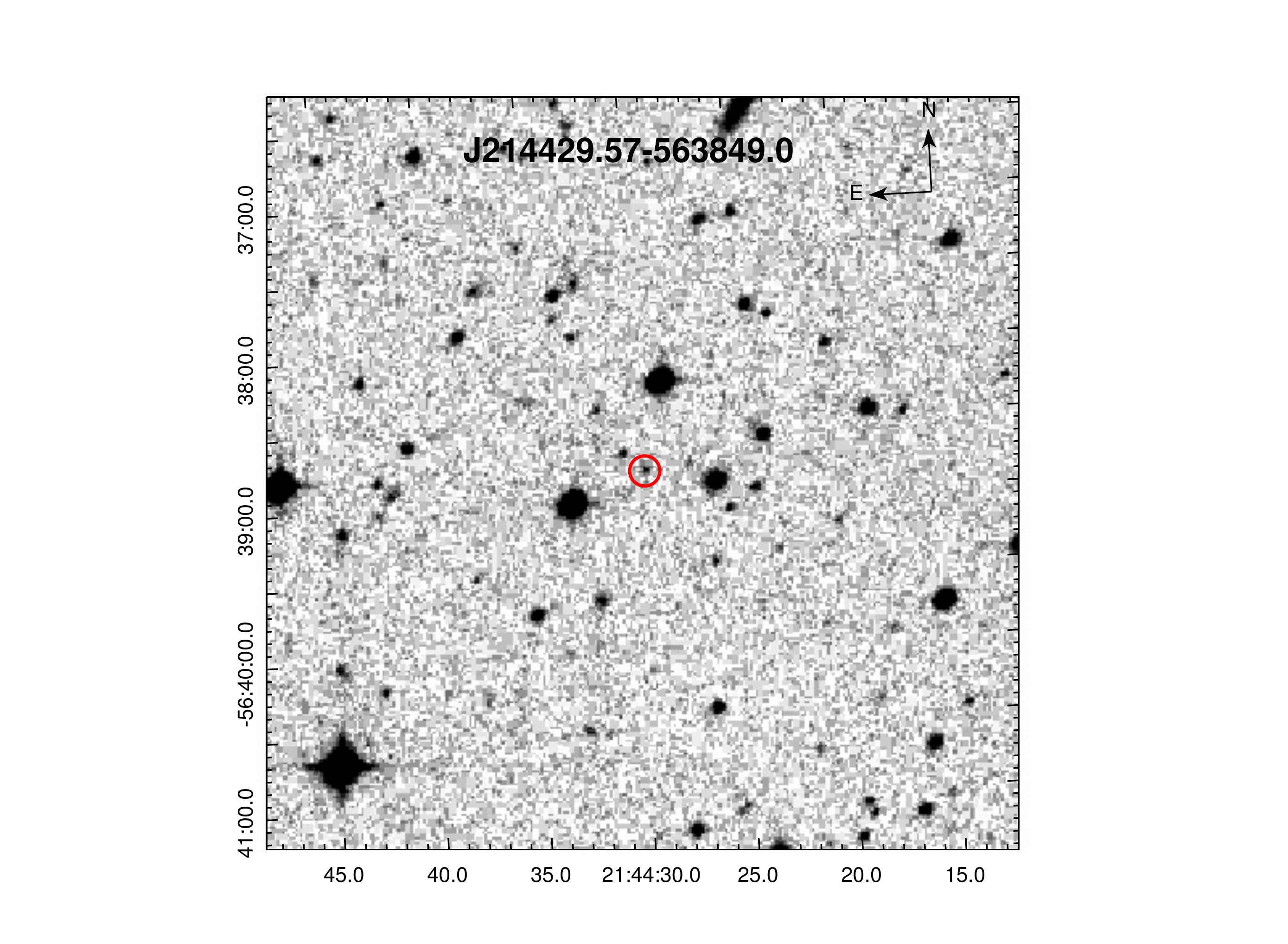} \\
\end{array}$
\end{center}
\caption{(Left panel) Optical spectrum of  WISE J214429.57-563849.0 associated with 3FGL J2144.6-5640. Signal-to-noise ratio is reported in the Figure. (Right panel) The finding chart ( $5'\times 5'$ ) retrieved from the Digital Sky Survey highlighting the location of the potential source: WISE J214429.57-563849.0 (red circle).}
\label{fig:J2144}
\end{figure*}

\begin{figure*}{}
\begin{center}$
\begin{array}{cc}
\includegraphics[width=\mywidth,angle=0]{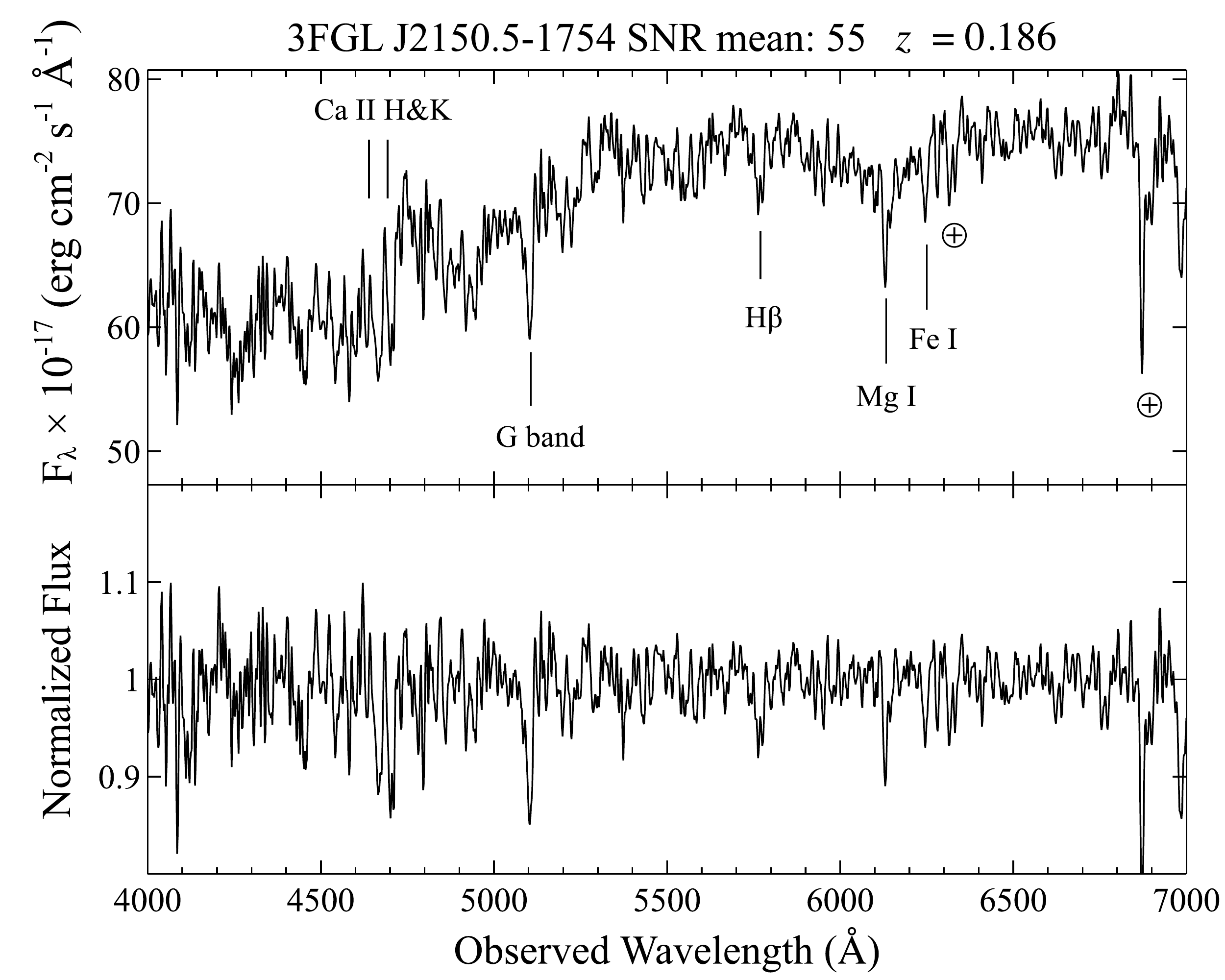} &
\includegraphics[trim=4cm 0cm 4cm 0cm, clip=true, width=7cm,angle=0]{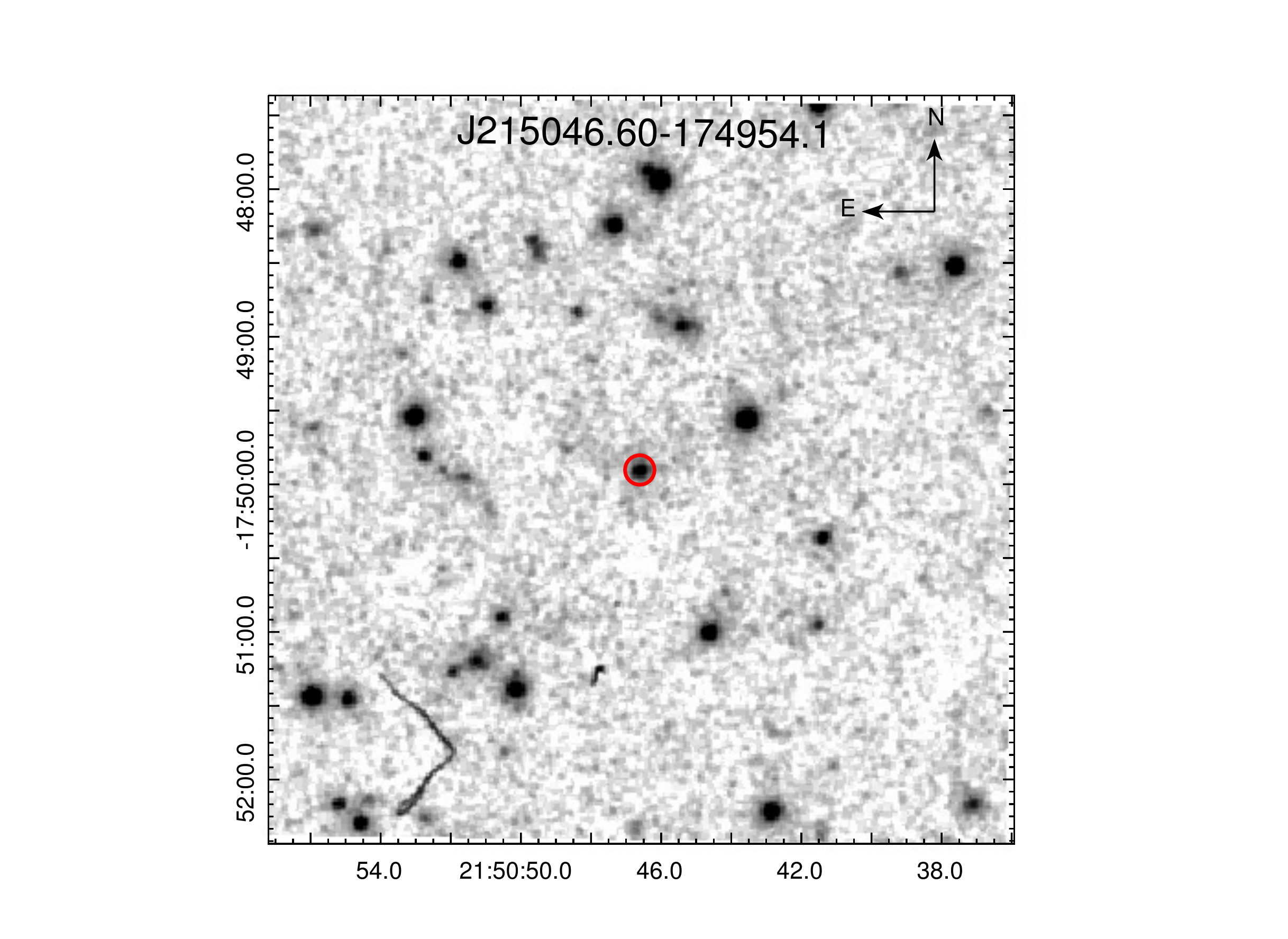} \\
\end{array}$
\end{center}
\caption{(Left panel) Optical spectrum of  WISE J215046.60-174954.1 associated with 3FGL J2150.5-1754. Signal-to-noise ratio is reported in the Figure. (Right panel) The finding chart ( $5'\times 5'$ ) retrieved from the Digital Sky Survey highlighting the location of the potential source: WISE J215046.60-174954.1 (red circle).}
\label{fig:J2150}
\end{figure*}

\begin{figure*}{}
\begin{center}$
\begin{array}{cc}
\includegraphics[width=\mywidth,angle=0]{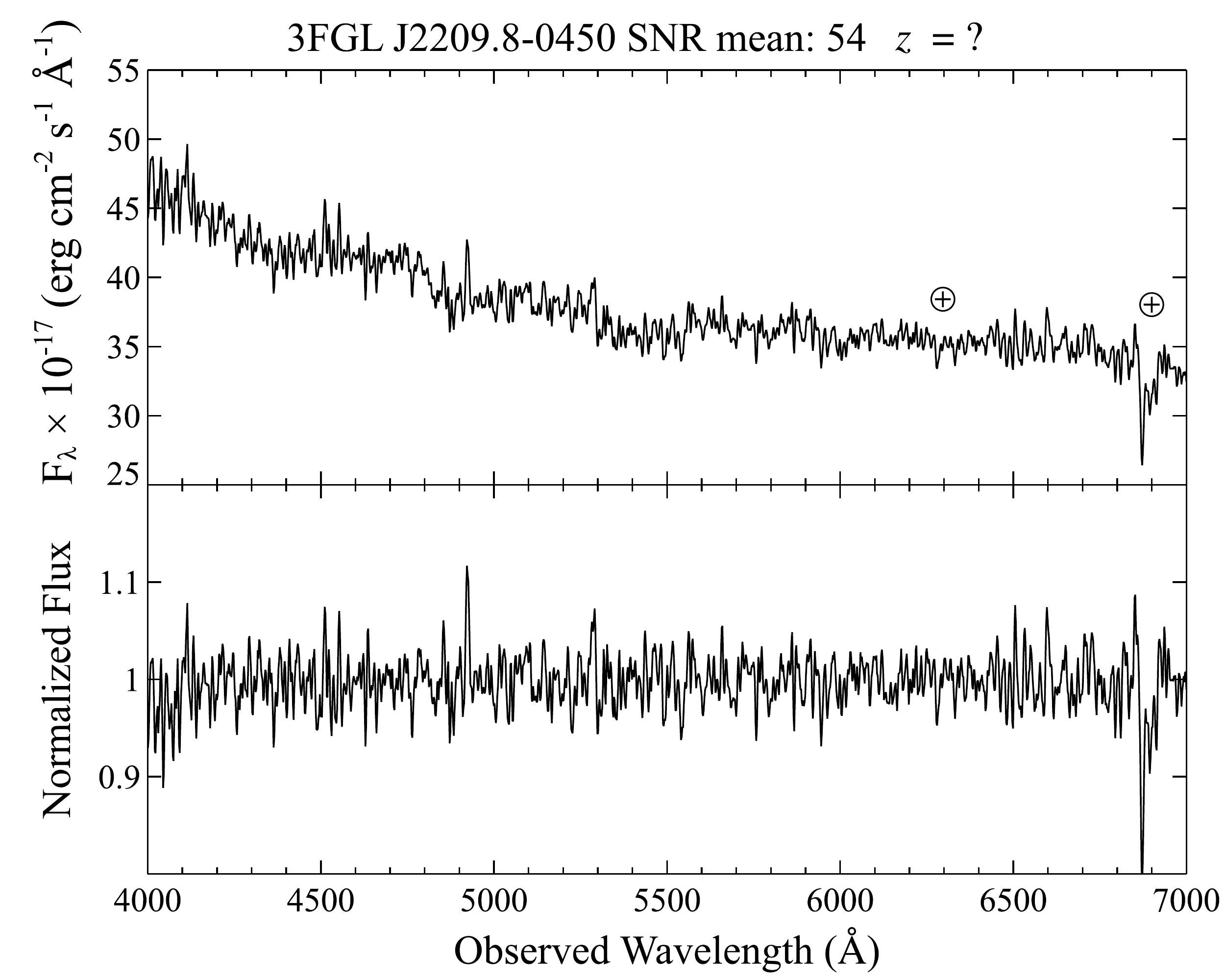} &
\includegraphics[trim=4cm 0cm 4cm 0cm, clip=true, width=7cm,angle=0]{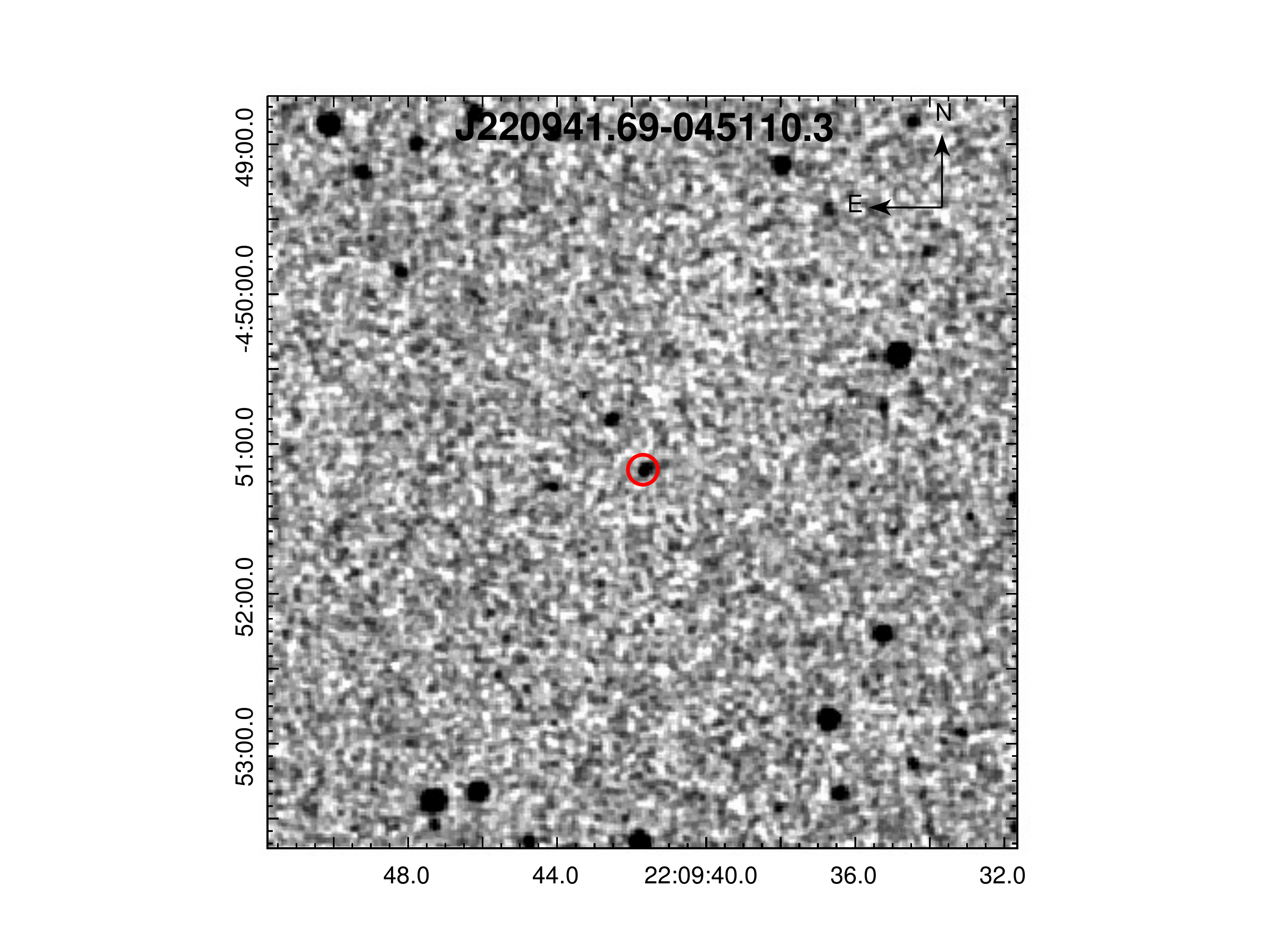} \\
\end{array}$
\end{center}
\caption{(Left panel) Optical spectrum of  WISE J220941.69-045110.3 associated with 3FGL J2209.8-0450. Signal-to-noise ratio is reported in the Figure. (Right panel) The finding chart ( $5'\times 5'$ ) retrieved from the Digital Sky Survey highlighting the location of the potential source: WISE J220941.69-045110.3 (red circle).}
\label{fig:J2209}
\end{figure*}

\begin{figure*}{}
\begin{center}$
\begin{array}{cc}
\includegraphics[width=\mywidth,angle=0]{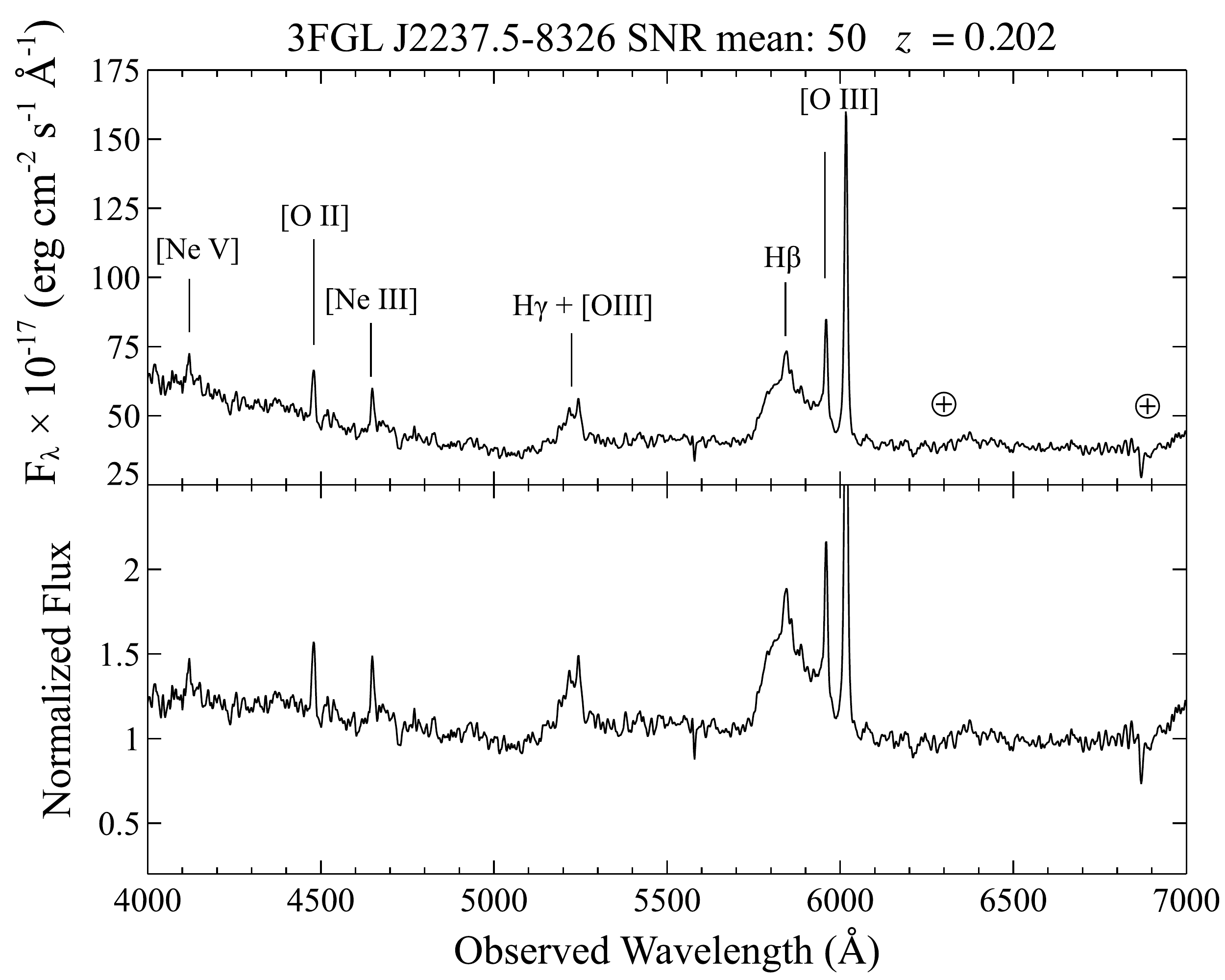} &
\includegraphics[trim=4cm 0cm 4cm 0cm, clip=true, width=7cm,angle=0]{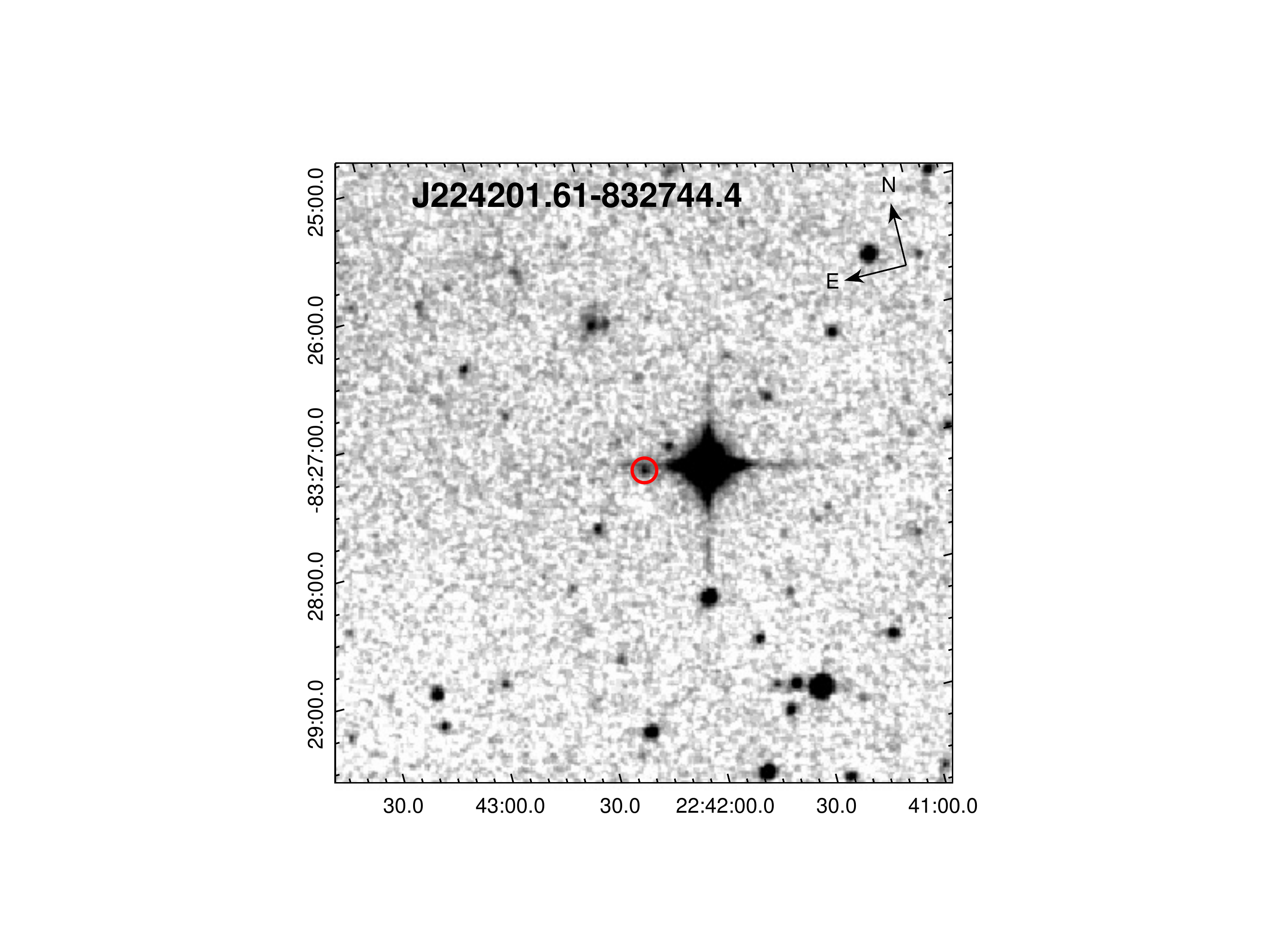} \\
\end{array}$
\end{center}
\caption{(Left panel) Optical spectrum of  WISE J224201.61-832744.4 associated with 3FGL J2237.5-8326. Signal-to-noise ratio is reported in the Figure. (Right panel) The finding chart ( $5'\times 5'$ ) retrieved from the Digital Sky Survey highlighting the location of the potential source: WISE J224201.61-832744.4 (red circle).}
\label{fig:J2237}
\end{figure*}

\begin{figure*}{}
\begin{center}$
\begin{array}{cc}
\includegraphics[width=\mywidth,angle=0]{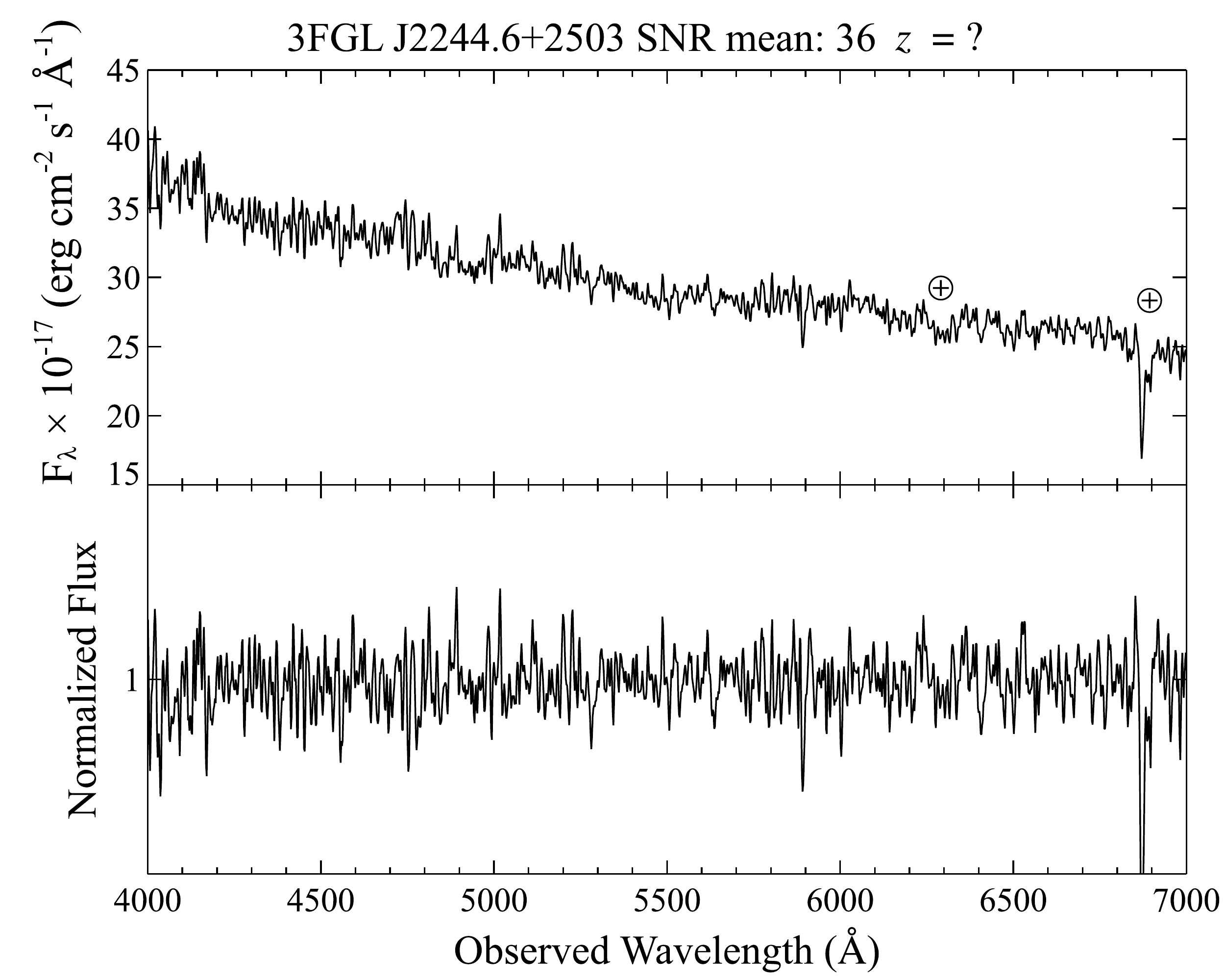} &
\includegraphics[trim=4cm 0cm 4cm 0cm, clip=true, width=7cm,angle=0]{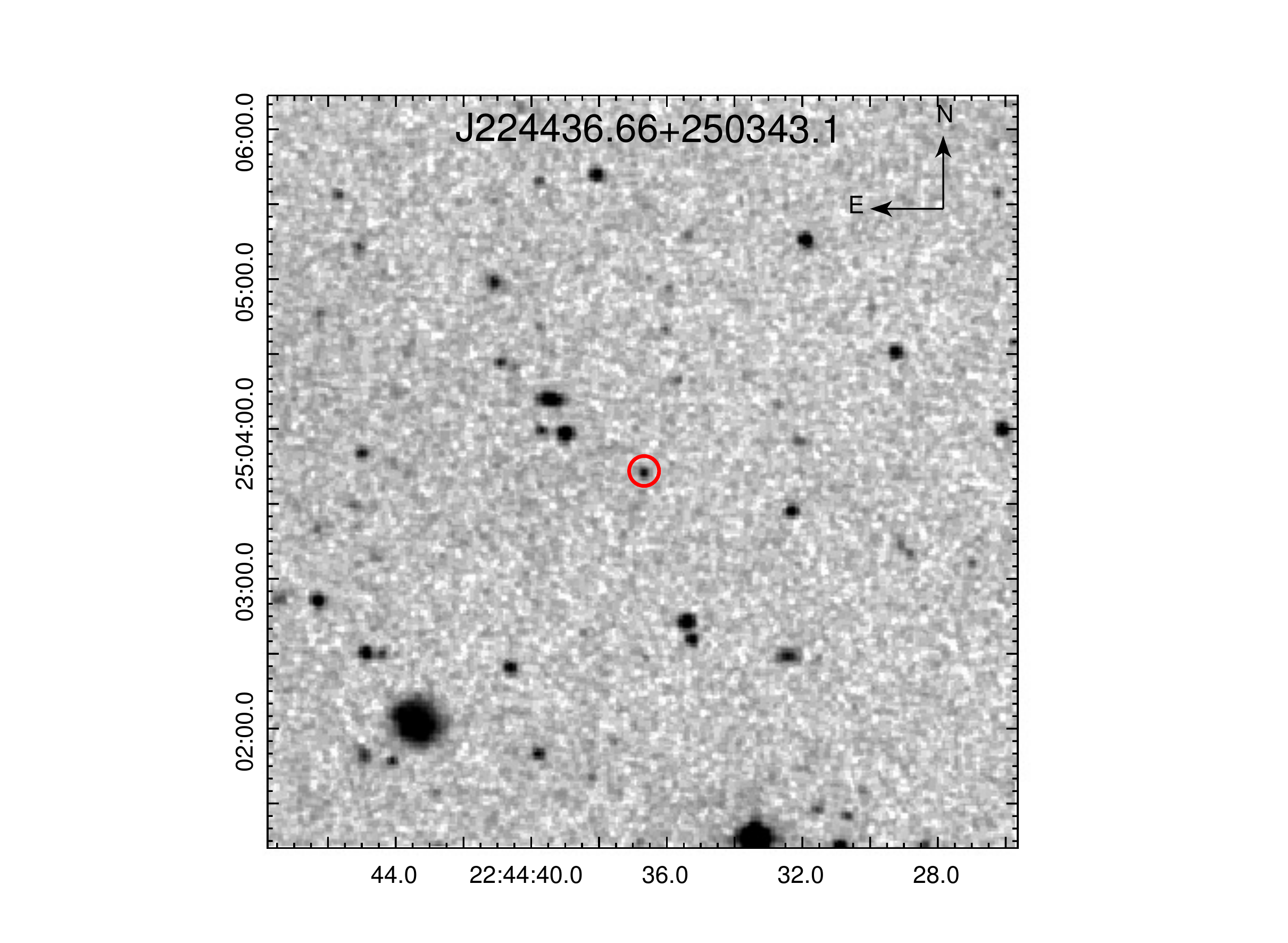} \\
\end{array}$
\end{center}
\caption{(Left panel) Optical spectrum of  WISE J224436.66+250343.1 associated with 3FGL J2244.6+2503. Signal-to-noise ratio is reported in the Figure. (Right panel) The finding chart ( $5'\times 5'$ ) retrieved from the Digital Sky Survey highlighting the location of the potential source: WISE J224436.66+250343.1 (red circle).}
\label{fig:J2244}
\end{figure*}

\begin{figure*}{}
\begin{center}$
\begin{array}{cc}
\includegraphics[width=\mywidth,angle=0]{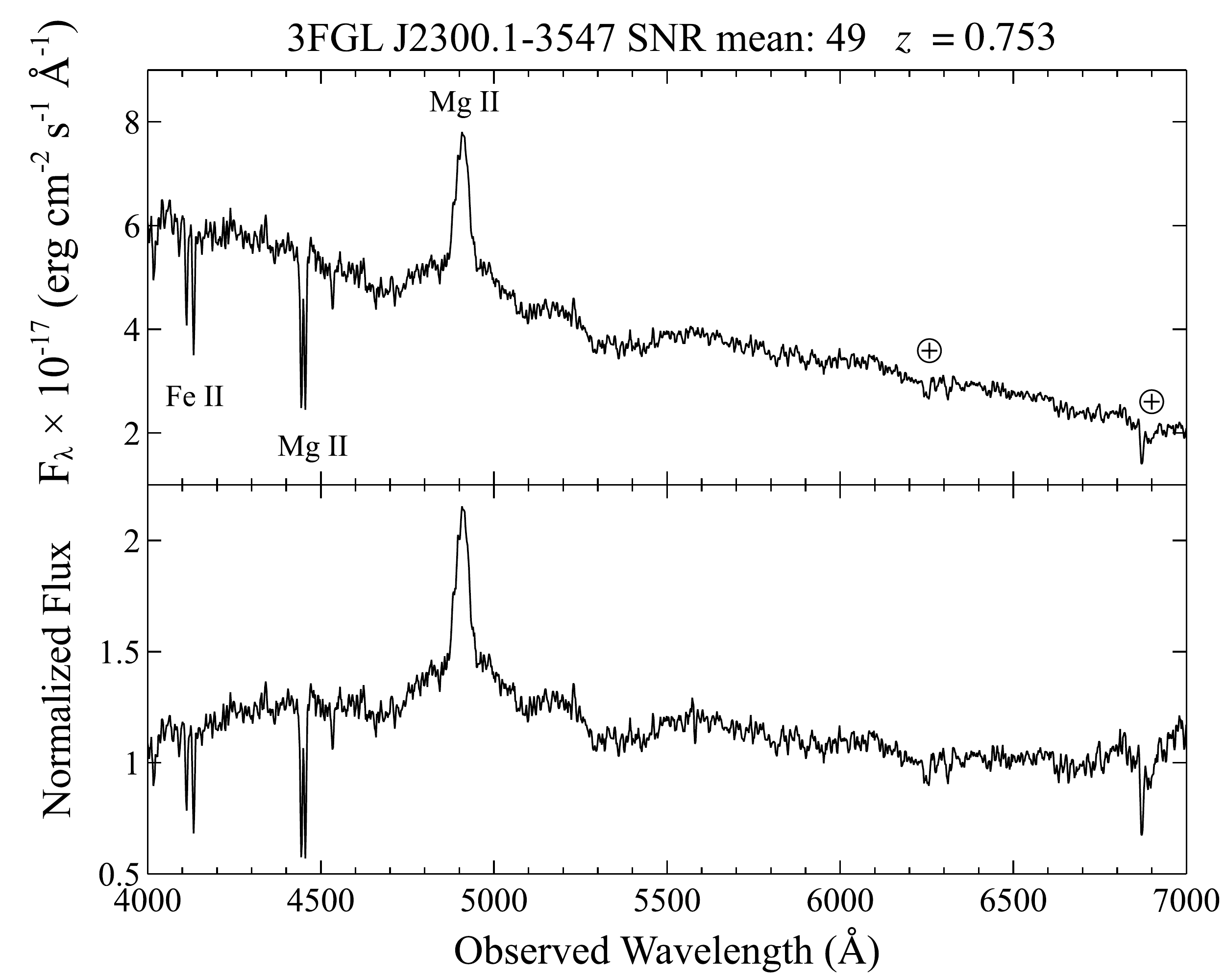} &
\includegraphics[trim=4cm 0cm 4cm 0cm, clip=true, width=7cm,angle=0]{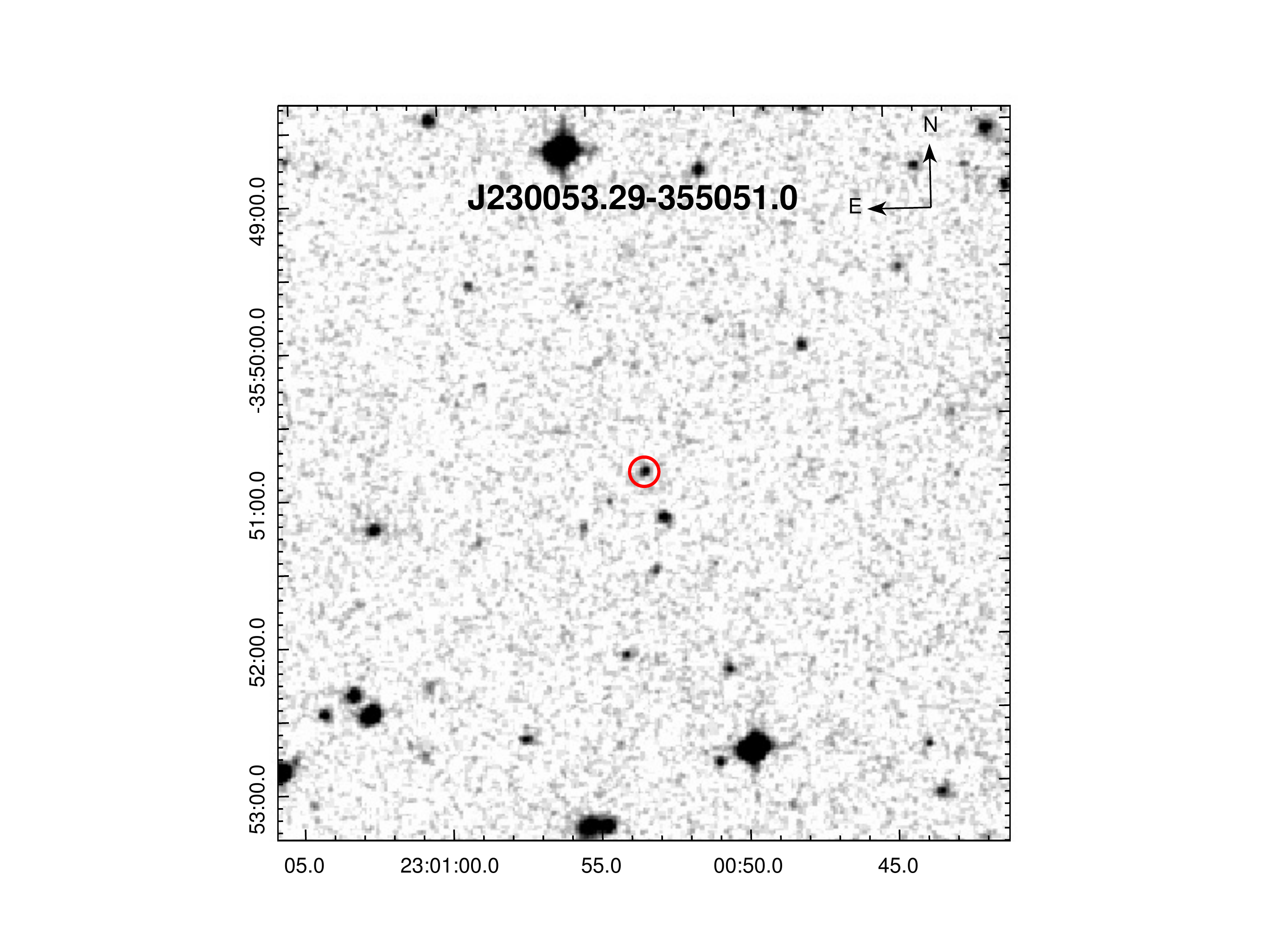} \\
\end{array}$
\end{center}
\caption{(Left panel) Optical spectrum of  WISE J230053.29-355051.0 associated with 3FGL J2300.1-3547. Signal-to-noise ratio is reported in the Figure. (Right panel) The finding chart ( $5'\times 5'$ ) retrieved from the Digital Sky Survey highlighting the location of the potential source: WISE J230053.29-355051.0 (red circle).}
\label{fig:J2300}
\end{figure*}

\clearpage

\begin{figure*}{}
\begin{center}$
\begin{array}{cc}
\includegraphics[width=\mywidth,angle=0]{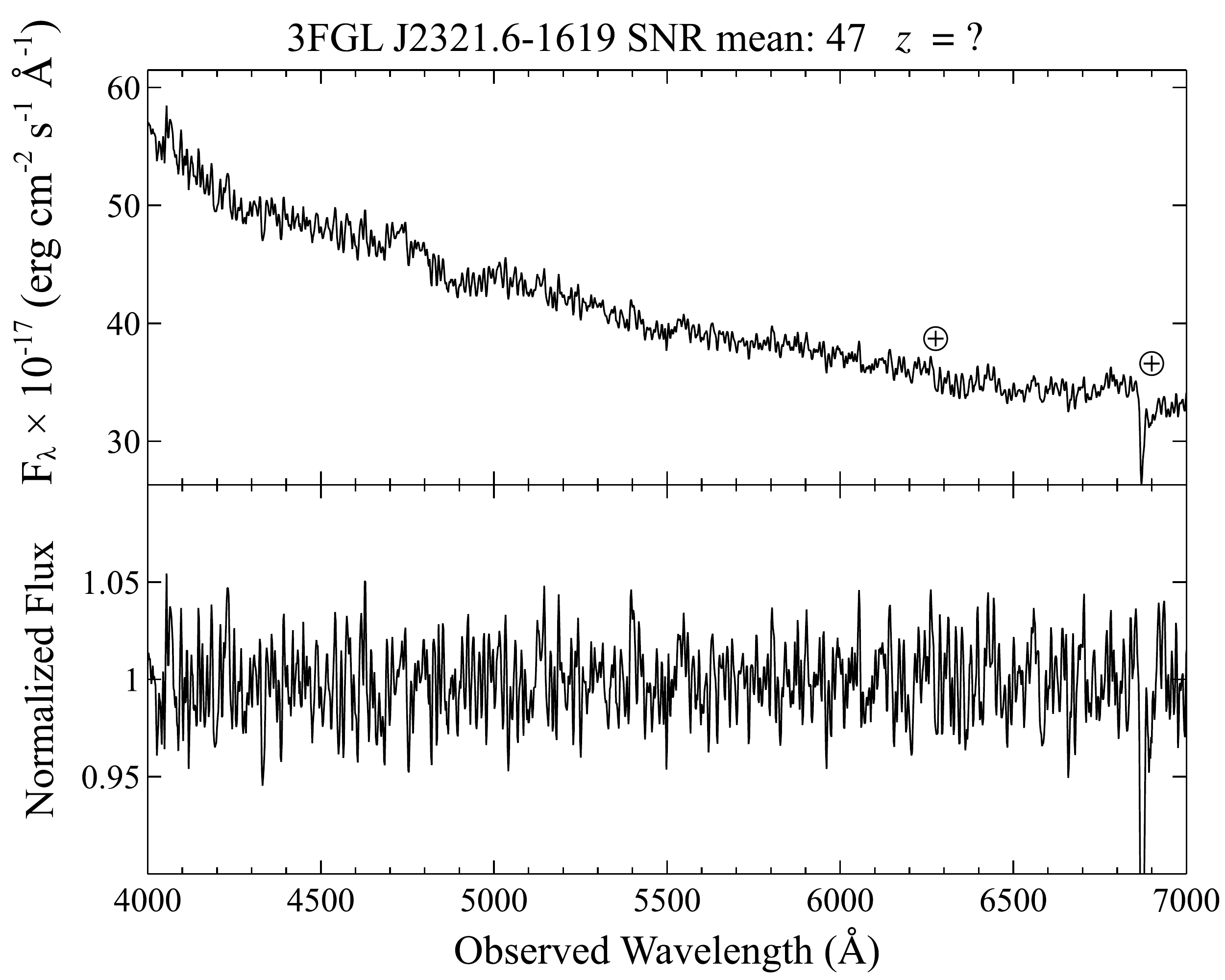} &
\includegraphics[trim=4cm 0cm 4cm 0cm, clip=true, width=7cm,angle=0]{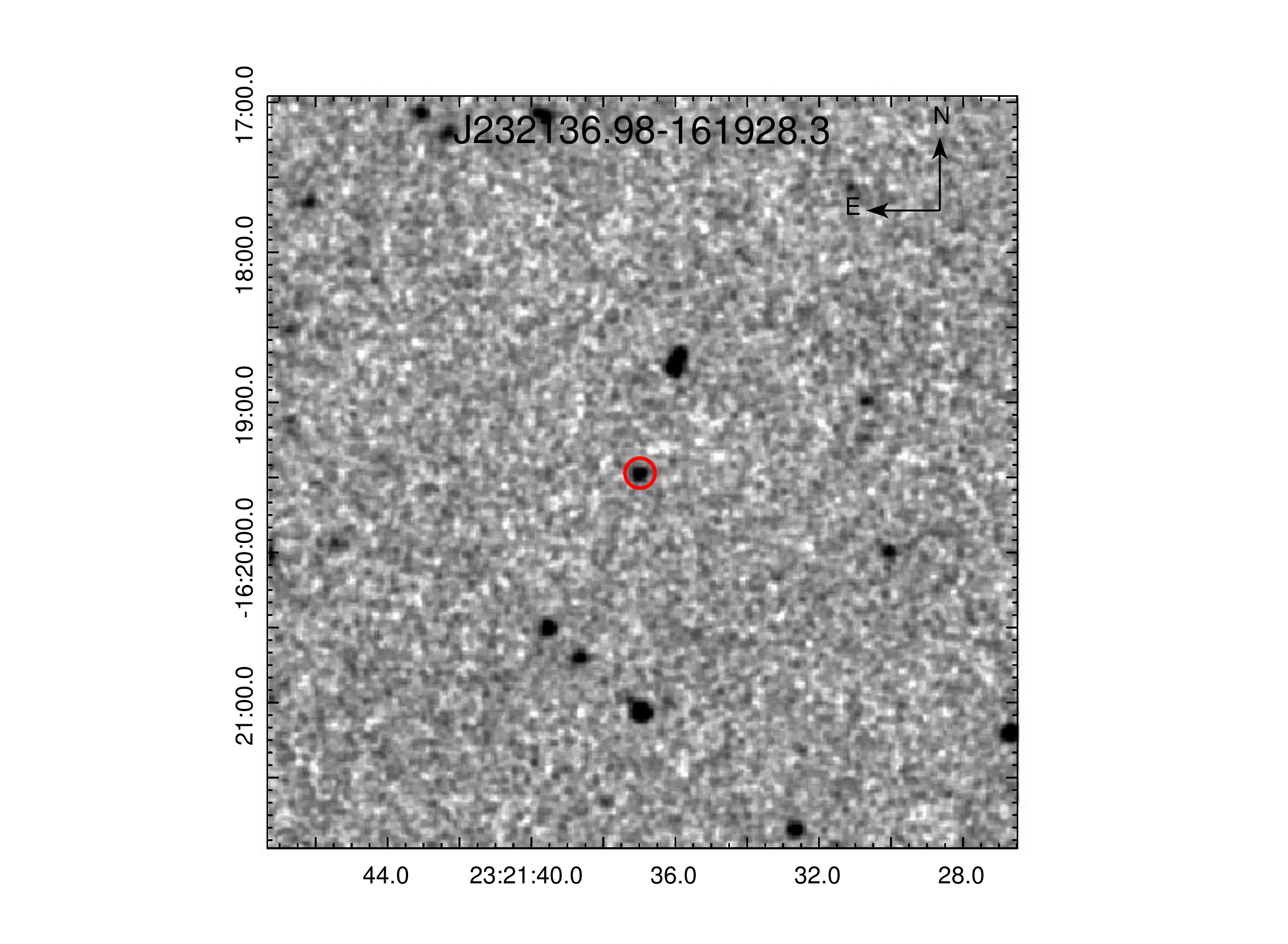} \\
\end{array}$
\end{center}
\caption{(Left panel) Optical spectrum of  WISE J232136.98-161928.3 associated with 3FGL J2321.6-1619. Signal-to-noise ratio is reported in the Figure. (Right panel) The finding chart ( $5'\times 5'$ ) retrieved from the Digital Sky Survey highlighting the location of the potential source: WISE J232136.98-161928.3 (red circle).}
\label{fig:J2321}
\end{figure*}

\begin{figure*}{}
\begin{center}$
\begin{array}{cc}
\includegraphics[width=\mywidth,angle=0]{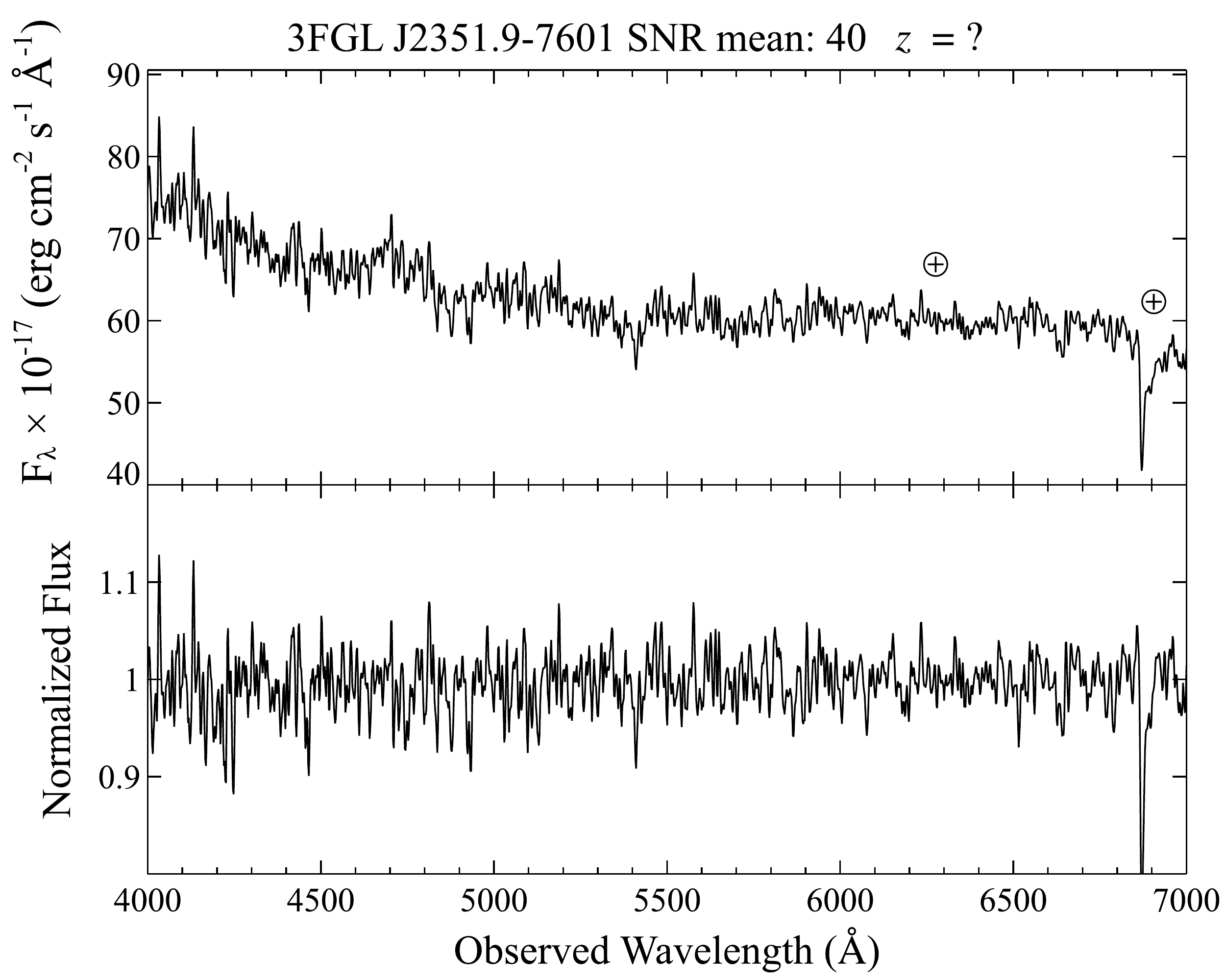} &
\includegraphics[trim=4cm 0cm 4cm 0cm, clip=true, width=7cm,angle=0]{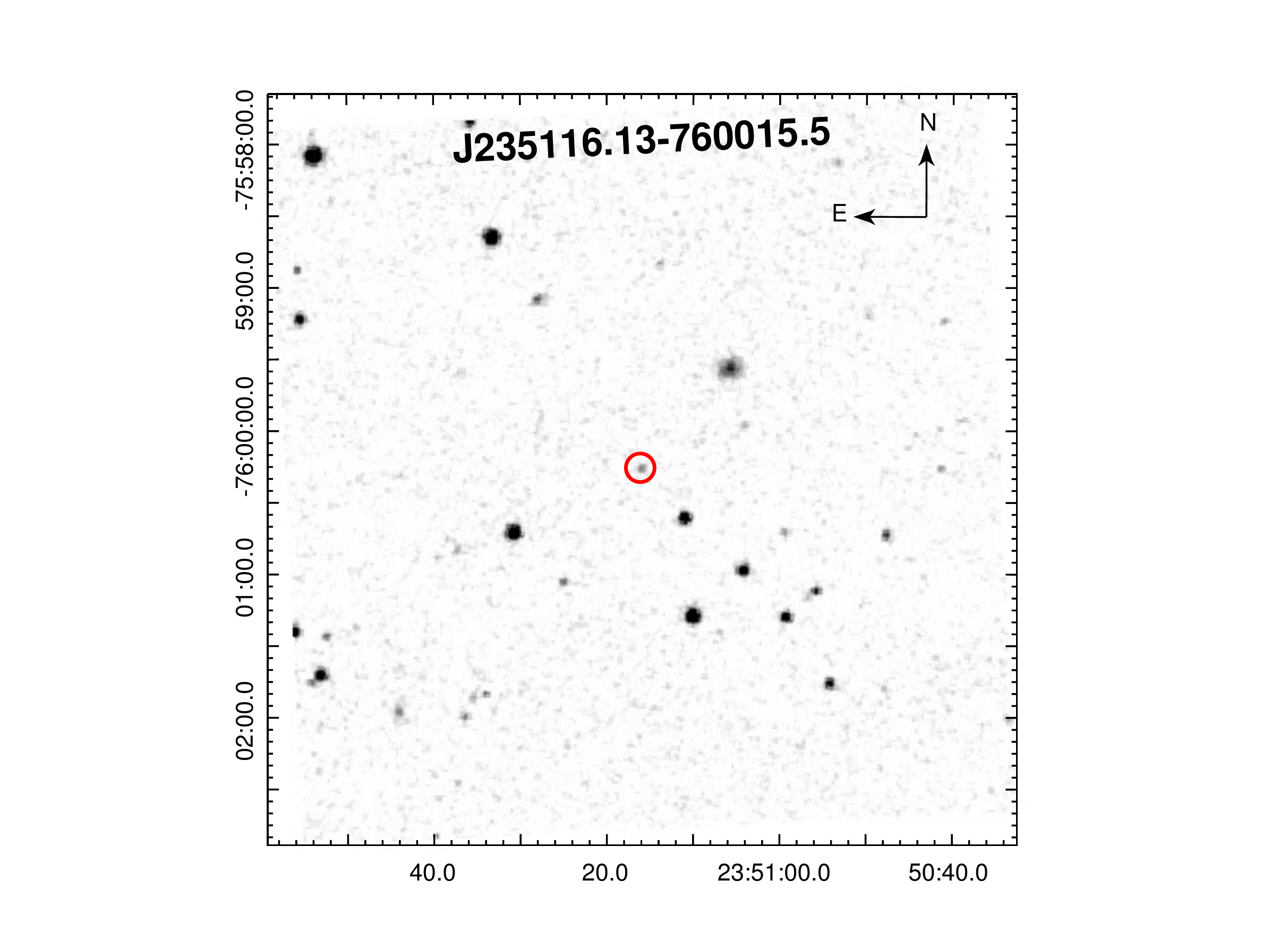} \\
\end{array}$
\end{center}
\caption{(Left panel) Optical spectrum of  WISE J235116.13-760015.5 associated with 3FGL J2351.9-7601. Signal-to-noise ratio is reported in the Figure. (Right panel) The finding chart ( $5'\times 5'$ ) retrieved from the Digital Sky Survey highlighting the location of the potential source: WISE J235116.13-760015.5 (red circle).}
\label{fig:J2351}
\end{figure*}

\begin{figure*}{}
\begin{center}$
\begin{array}{cc}
\includegraphics[width=\mywidth,angle=0]{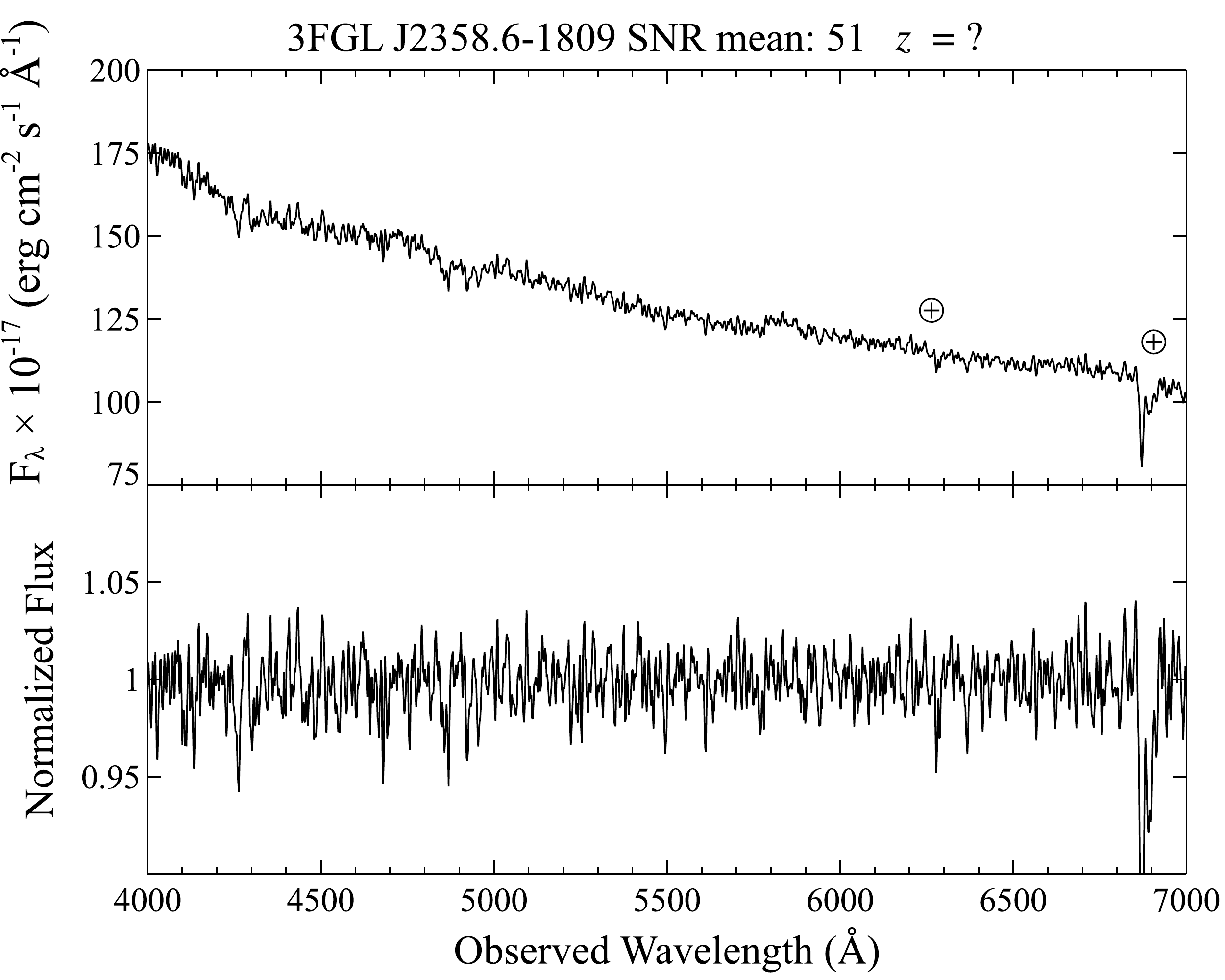} &
\includegraphics[trim=4cm 0cm 4cm 0cm, clip=true, width=7cm,angle=0]{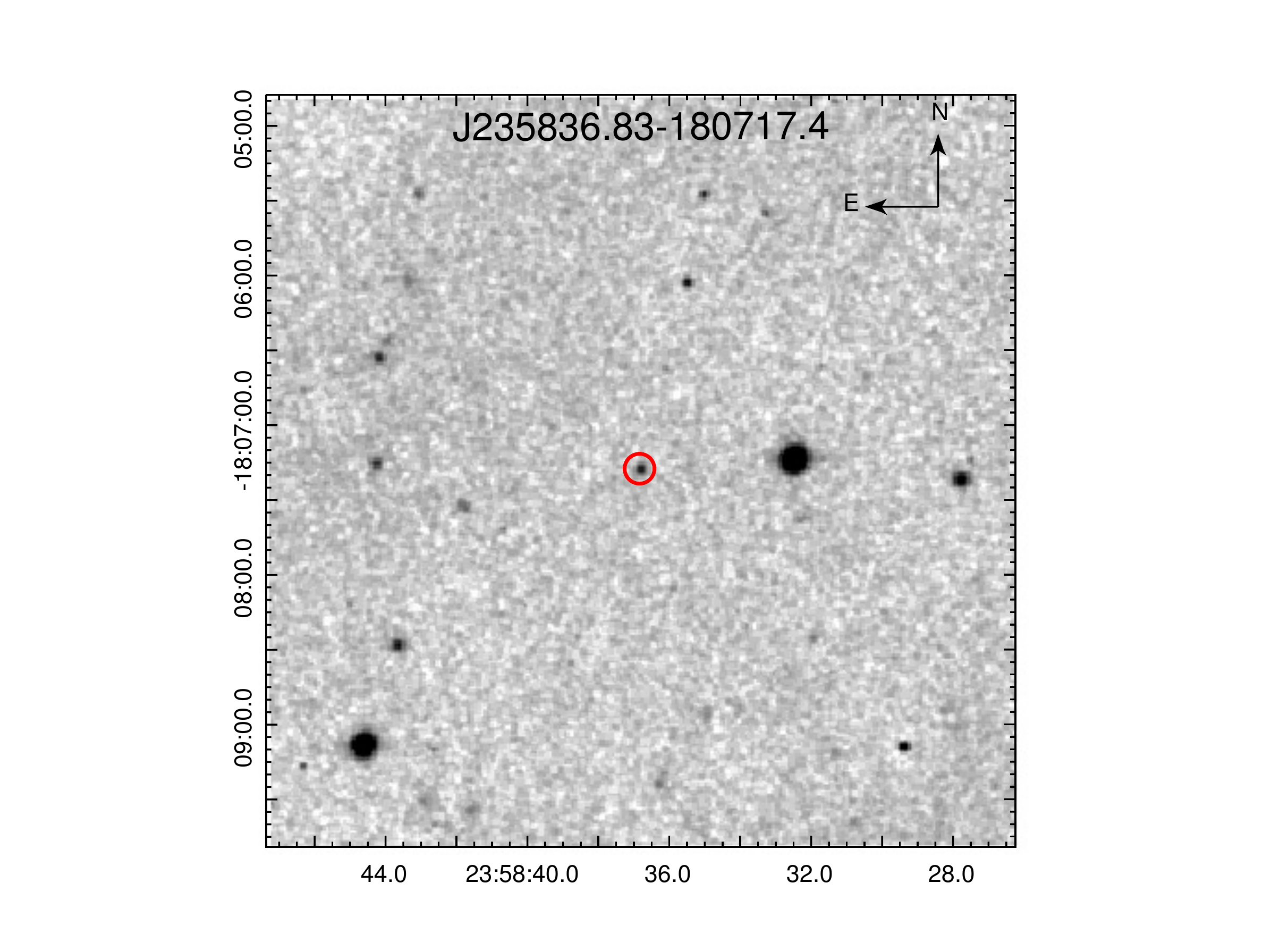} \\
\end{array}$
\end{center}
\caption{(Left panel) Optical spectrum of  WISE J235836.83-180717.4 associated with 3FGL J2358.6-1809. Signal-to-noise ratio is reported in the Figure. (Right panel) The finding chart ( $5'\times 5'$ ) retrieved from the Digital Sky Survey highlighting the location of the potential source: WISE J235836.83-180717.4 (red circle).}
\label{fig:J2358}
\end{figure*}

\clearpage
\acknowledgements

We acknowledge Dr. S. Points for the support while carrying out the SOAR observations. Work by C.C.C. at NRL is supported in part by NASA DPR S-15633-Y.
H. A. Smith acknowledges partial support from NASA grants NNX14AJ61G and NNX15AE56G
JS acknowledges support from NASA grant NNX15AU83G and a Packard Fellowship.
EJB acknowledges support from Programa de Apoyo a Proyectos de Investigaci\'{o}n e Innovaci\'{o}n Tecnol\'{o}gica (IN109217).
This publication makes use of data products from the Wide-field Infrared Survey Explorer, which is a joint project of the University of California, Los Angeles, and 
the Jet Propulsion Laboratory/California Institute of Technology, 
funded by the National Aeronautics and Space Administration.
Based on observations obtained at the Southern Astrophysical Research (SOAR) telescope, which is a joint project of the Minist\'{e}rio da Ci\^{e}ncia, Tecnologia, e Inova\c{c}\~{a}o (MCTI) da Rep\'{u}blica Federativa do Brasil, the U.S. National Optical Astronomy Observatory (NOAO), the University of North Carolina at Chapel Hill (UNC), and Michigan State University (MSU).
TOPCAT\footnote{\underline{http://www.star.bris.ac.uk/$\sim$mbt/topcat/}} 
\citep{taylor05} for the preparation and manipulation of the tabular data.

\bibliographystyle{apj}


\end{document}